\documentclass[a4paper,12pt,onehalfspacing,twoside,openright,tools]{report}
\usepackage{breakcites}
\usepackage{longtable}
\usepackage[utf8]{inputenc}
\usepackage[T1]{fontenc}
\usepackage[frenchb]{babel}
\usepackage{graphicx}
\usepackage{amsfonts}
\usepackage{amsmath}
\usepackage{amssymb}
\usepackage[ruled,vlined,french,titlenumbered]{algorithm2e}
\usepackage{fancyhdr}
\usepackage[usenames]{color}

\newtheorem{theoreme}{Théoréme}
\newtheorem{remarque}{Remarque}
\newtheorem{definition}{Définition}

\newtheorem{proposition}{Proposition}
\newtheorem{exemple}{Exemple}
\newtheorem{propriete}{Propriété}
\newenvironment{Preuve}[1][Preuve]{\noindent\textbf{#1. }\it}{\mbox{}\\\\}

\fancypagestyle{plain}{ % pages de tetes de chapter
\fancyhead{} % supprime l'entete
}
 
\begin{document}
\thispagestyle{empty}
\begin{center}

\textbf{Université de Tunis}\\
\textbf{Institut Sup\'erieur de Gestion}\\
\textbf{\'Ecole Doctorale Sciences de Gestion}
\end{center}
\begin{center}
\begin{figure}[htbp]
    \centering
        \includegraphics[width=3cm,height=3cm]{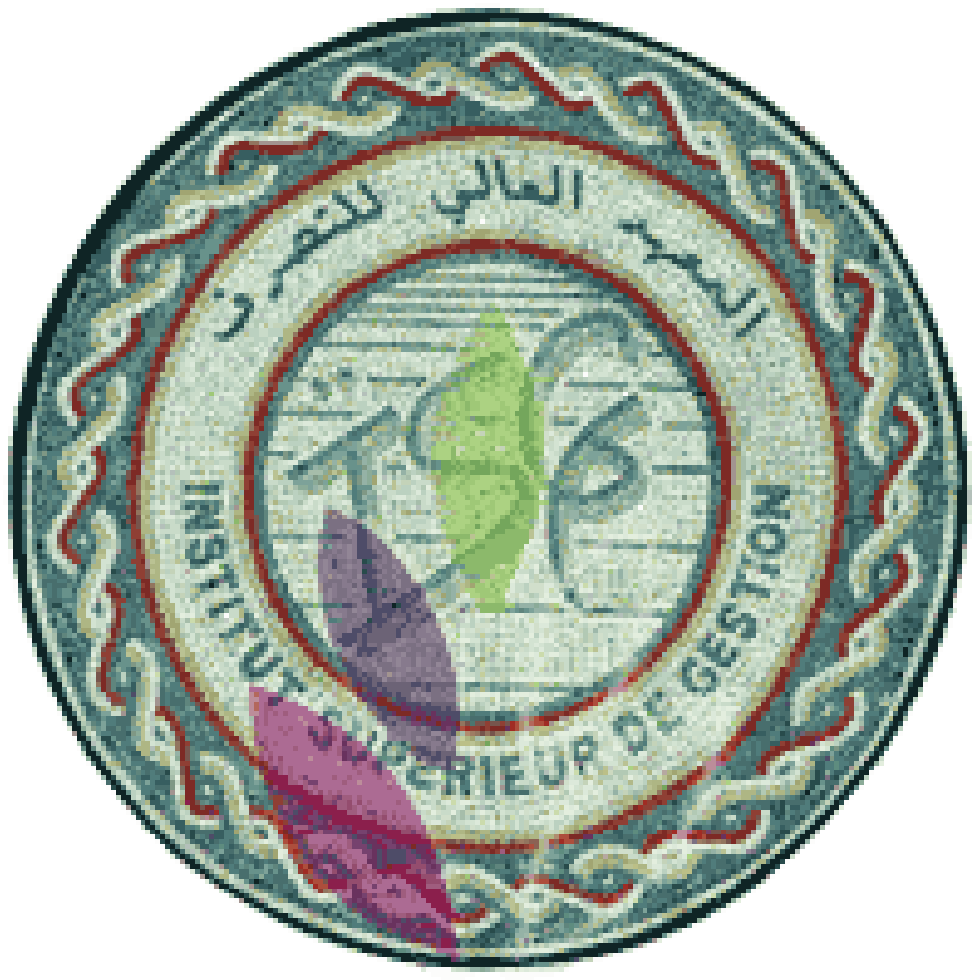}
\end{figure}
\end{center}

\bigskip

\begin{center}
 \LARGE \textbf{\textbf{\textbf{Motifs corrélés rares : Caractérisation et nouvelles représentations concises}}}
\end{center}

 \bigskip

\begin{center}
{\large Mémoire en vue de l'obtention du Mastère en}

{\large \textbf{\emph{Informatique Appliquée à  la Gestion}}}
\end{center}

\vspace{1.cm}

\begin{center}

\bigskip
\normalsize Présenté par :\\

 \bigskip

\textbf{\textbf{\textbf{Souad BOUASKER}}}\\

 \bigskip

 \bigskip

\vspace{0.5cm}

\normalsize Sous la direction de :

\begin{center}
   \textbf{\textbf{Mr. Sadok BEN YAHIA}}\\

   \textbf{\textbf{Mr. Tarek HAMROUNI}}\\
\end{center}

\bigskip

\vspace{1.cm}

\small \textbf{Année Universitaire 2010/2011}
\end{center}

\newpage
\thispagestyle{empty}

\mbox{}

\newpage
\thispagestyle{empty}

\mbox{}

\vspace{2.cm}

\begin{center}
{\Large \textit{\textbf{Dédicaces}}}\end{center}

\bigskip

\bigskip

\bigskip

\begin{center}
\textit{\`A la mémoire de mon père qui est toujours présent dans mon esprit et mon coeur. \`A ton âme papa, et la mémoire du grand amour que tu m'as offert, je dédie ma réussite.}\\
\bigskip

\bigskip

\bigskip

\textit{\`A  ma chére mère à  qui je dois tout et dont l'amour, l'affection, les encouragements et la patience ont été pour moi le meilleur gage de réussite. En témoignage de ma reconnaissance et mon attachement.}\\
\bigskip

\bigskip
\bigskip

\textit{\`A ma famille, à  qui je dois tout mon bonheur et ma prospérité}\\
\bigskip

\bigskip
\bigskip

\textit{\`A mes maîtres, à qui je tiens à leur montrer que je suis et je reste à l'hauteur de
leur espérances}\\
\bigskip

\bigskip

\bigskip
\textit{Et enfin à tous ceux qui m'ont soutenu de prés ou de loin durant mes études}\\
\bigskip
\end{center}

\bigskip

\begin{flushright}
\textit{\textbf{Je dédie ce modeste travail}}\\

\bigskip

\textit{\textbf{Souad}}
\end{flushright}

\newpage
\thispagestyle{empty}

\mbox{}

\newpage
\thispagestyle{empty}

\mbox{}

\vspace{2.cm}

\begin{center}
{\Large \textit{\textbf{Remerciements}}}
\end{center}
\bigskip

\bigskip

\bigskip

\bigskip

Au terme de ce travail, je commence tout d'abord par saluer vivement les membres du jury
pour l'honneur qu'ils me font en acceptant d'évaluer ce modeste travail.

\bigskip

\bigskip
Mes remerciements les plus cordiaux s'adressent à  mon directeur de mémoire Monsieur \textbf{Sadok BEN YAHIA}, Maître de Conférences à  la Faculté des Sciences de Tunis, pour son implication, sa disponibilité incessante et ses conseils rigoureux. Puissent ces lignes être l'expression de ma plus profonde reconnaissance.

\bigskip

\bigskip
Mes remerciements les plus sincères et ma gratitude s'adressent à  mon co-directeur de mémoire Monsieur \textbf{Tarek HAMROUNI}, Assistant à  l'Institut Supérieur des Arts Multimédias de La Mannouba. Sa patience, sa disponibilité, sa rigueur scientifique, son sens critique m'ont été d'une aide précieuse et ses judicieux conseils qui ont contribué à  l'amélioration de ce mémoire.

\bigskip

\bigskip

Je tiens également à  remercier tous ceux qui ont contribué à la réalisation de ce travail.

\newpage

\thispagestyle{empty}

\mbox{}

\newpage

 \pagenumbering{roman}

\thispagestyle{empty} \tableofcontents

\newpage
\thispagestyle{empty}

\mbox{}

\thispagestyle{empty} \listoffigures

\newpage
\thispagestyle{empty}

\mbox{}

\thispagestyle{empty} \listoftables

\newpage
\thispagestyle{empty}

\mbox{}

\vspace{1.cm}

\thispagestyle{empty} \listofalgorithms

\newpage

\thispagestyle{empty}

\mbox{}

\newpage

 \linespread{1.3}
 \normalfont
\pagenumbering{arabic}
\addcontentsline{toc}{chapter}{Introduction générale}
\chapter*{Introduction générale}\label{chapitre_introduction}
\markboth{Introduction générale}{Introduction générale}

L'extraction des règles d'association est une technique très répandue dans la fouille de données et répond aux besoins des experts dans plusieurs domaines d'application
\cite{ayouni2010mining,hamrouni2008succinct,othman2008yet}.Plusieurs travaux se sont ainsi focalisés sur la dérivation des règles d'association à  partir des motifs fréquents. Toutefois, l'utilisation de ces motifs ne constitue pas une solution intéressante pour certaines applications, telles que la détection d'intrusions, la détection des fraudes, l'identification des valeurs extrêmes dans les bases de données, l'analyse des données criminelles, l'analyse du désordre génétique à  partir des données biologiques, l'analyse des maladies rares à  partir des données médicales, l'analyse des données d'apprentissage en ligne, etc. \cite{criminal_data_2009,He02fp,livreIGIGlobal2010,mahmood_less_frequent_patterns_vs_networks,haglin08,romero2010,laszloIJSI2010,gasmi2007extraction}.

En effet, dans de telles situations, un comportement fréquent peut être sans valeur ajoutée pour l'utilisateur final. Par contre, les événements peu fréquents sont les plus intéressants parce qu'ils indiquent qu'un événement inattendu, une exception par exemple \cite{taniar_2008}, est survenue. Une étude doit alors continuer afin de déterminer les causes possibles de ce changement peu commun du comportement normal.

La fouille des motifs rares s'est alors avérée d'une réelle valeur ajoutée \cite{livreIGIGlobal2010,yun_03,weiss04}. En effet, ces motifs, ayant une fréquence d'apparition dans la base inférieure à  un certain seuil donné, permettent de cerner les événements rares, peu communs, inattendus, exceptionnels, cachés, etc. \cite{rare_event_2007,padmanabhan2006,weiss04}.
En effet, la détection des valeurs extrêmes est une tache utile dans plusieurs applications réelles comme la détection des fraudes des cartes de crédit, la découverte des activités criminelles dans le commerce électronique et le marketing.

Comme illustration des applications des motifs rares dans le domaine de la sécurité informatique \cite{brahmi2010mad}, étant donné un fichier log qui représente les tentatives de connexions effectuées sur un serveur Web d'authentification, ces motifs véhiculent les informations liées aux tentatives d'attaques à  savoir par exemple l'origine des attaques, les ports les plus attaqués et les services les plus visés. Considérons par exemple la table \ref{table_fichier_log} qui représente un échantillon réduit d'un tel fichier  \textsc{V} dénote Valide, \textsc{NV} dénote Non Valide, et $d_i$ une date d'accès. Par exemple, si le motif \textsc{(}\textit{197.1.104.19}, 1221, \textsc{NV}\textsc{)} s'avère rare, l'adresse \textit{197.1.104.19} peut étre considérée à  l'origine d'une attaque sur le port 1221. Une analyse détaillée de ses accés est alors à  effectuer.

\begin{tiny}
\begin{table}
\centering
 \begin{tabular}{l c c c}
\hline
Adresse IP& Port  & Authentification & Date \tabularnewline
\hline
197.2.123.87 & 23 &\textsc{NV}& $d_1$\tabularnewline
197.1.104.19 & 1221 &\textsc{NV}& $d_2$\tabularnewline
194.23.22.2  & 80 & \textsc{V}& $d_3$\tabularnewline
197.1.104.19  & 225 & \textsc{V} & $d_4$\tabularnewline
197.2.123.29 & 21 &\textsc{V}& $d_5$\tabularnewline
197.1.104.19 & 1221 &\textsc{NV}& $d_6$\tabularnewline
$\vdots$&$\vdots$&$\vdots$&$\vdots$\tabularnewline
197.1.156.27  & 145 & \textsc{V} & $d_n$\tabularnewline
\hline
\end{tabular}
\caption{Extrait d'un fichier log d'accès à  un serveur Web.}\label{table_fichier_log}
\end{table}
\end{tiny}

Dans la pratique, l'exploitation des motifs rares est confrontée à  diverses contraintes dont les principales sont : \textsc{(}$i$\textsc{)} l'extraction complexe de ces motifs qui ne bénéficient pas des propriétés des motifs fréquents et par conséquent les critères d'élagage appliqués pour ces derniers ne sont pas exploitables; \textsc{(}$ii$\textsc{)} le nombre très important des motifs rares qui ne sont pas aussi rares, dans les applications réelles, que laisserait présager leur qualificatif; et \textsc{(}$iii$\textsc{)} la qualité des motifs rares extraits et qui peuvent comporter des items qui n'ont aucun lien sémantique entre eux. Par exemple, le motif composé par les items ``$Lait$'' et ``$Caviar$'' est un motif rare. Cependant, aucune corrélation n'existe entre le produit ``$Lait$'' très fréquemment acheté et le produit ``$Caviar$'' cher et rarement acheté.

Afin que l'exploitation des motifs extraits soit fructueuse, leur nombre relativement réduit et leur qualité intéressante sont deux critères importants que doit chercher à  faire émerger un processus de fouille. Dans le domaine médical ou encore dans la sécurité des réseaux informatiques par exemple, une information exacte et précise est exigée. Ainsi, l'idée d'extraire les motifs rares tout en intégrant les mesures de corrélations est d'une grande utilité. En effet, l'intégration de telles mesures permet de limiter l'ensemble extrait aux motifs rares ayant une corrélation entre leurs items dépassant un certain seuil de corrélation. Ces motifs \textit{corrélés rares} offrent ainsi un fort lien sémantique entre les items les composant.

Dans \cite{qdc2011}, les auteurs ont proposé des représentations concises \textit{sans perte d'information}, appelées aussi \textit{exactes}, des motifs rares sans aucune considération des mesures de corrélations dans le processus de fouille. Une étude des représentations concises des motifs fréquents \cite{calderssurvey} a été alors menée afin de proposer celles des motifs rares. Elle a prouvé l'intérêt de considérer la notion de classe d'équivalence, associée à  l'opérateur de fermeture conjonctive \cite{ganter99}, permettant de réduire la redondance au sein des motifs en regroupant ensemble ceux caractérisant un même ensemble de transactions. Les éléments minimaux et maximaux de la classe, les générateurs minimaux \textsc{(}appelé aussi itemsets libres\textsc{)} et les itemsets fermés \cite{pasquier99_2005} respectivement, sont ainsi à  la base des représentations des motifs rares proposées. Par ailleurs, un des résultats clés de cette étude est que les représentations basées sur les règles de déduction \cite{calderssurvey} et celles basées sur les identités d'inclusion-exclusion \cite{casalidawak05,tarekdke09} ne sont pas adaptées à  la fouille des motifs rares.

D'autre part, l'approche présentée dans \cite{tarekds2010,tarekds2010} utilise la mesure de corrélation \textit{bond} \cite{Omie03} pour l'extraction de représentations concises exactes des motifs corrélés fréquents. Ceci a permis de ne retenir qu'un sous-ensemble des motifs fréquents, constitué par les motifs présentant une forte corrélation entre les items les constituant. Le choix de cette mesure a été effectué sur la base d'une étude de ses propriétés qui se sont avérées plus intéressantes que celles d'autres mesures de corrélation. Toutefois, l'intégration d'une mesure de corrélation est d'une utilité encore plus grande dans le cas de la fouille des motifs rares. En effet, elle permet d'éviter l'extraction de motifs contenant des items n'ayant aucun lien sémantique entre eux ce qui expliquerait en quelque sorte pourquoi ces motifs sont rares. Ainsi, sans l'utilisation d'une mesure de corrélation, un motif rare peut ne représenter aucune information utile s'il est composé d'items faiblement corrélés entre eux. Un motif rare intéressant serait donc celui qui apparaît un nombre très faible de fois dans la base tout en ayant des items qui sont fortement liés, c-.é.-d. que l'apparition de l'un dépend de celles des autres.

Ainsi, dans ce mémoire, nous allons nous intéresser à  l'extraction des représentations concises exactes des motifs corrélés rares. Dans ce cadre, nous nous intéressons à  la mesure de corrélation \textit{bond} correspondant au rapport entre le support conjonctif d'un motif et son support disjonctif. Notre choix de cette mesure est motivé par le cadre théorique dont elle bénéficie \cite{Omie03} ainsi que l'étude structurelle qui a été effectuée dans \cite{tarekds2010,tarekds2010}. En plus, il a été prouvé dans \cite{surana2010} que la mesure \textit{bond} vérifie les propriétés théoriques que toute mesure de qualité dédiée aux règles d'association rares doit avoir.
La mesure \textit{bond} a été aussi utilisée dans l'approche de fouille de motifs corrélés proposée dans  \cite{borgelt}.
L'extraction des motifs corrélés a été alors montrée plus complexe tout en étant plus informative que celle des motifs fréquents \cite{borgelt}.
Il est toutefois important de noter qu'aucune étude de l'ensemble des motifs corrélés rares n'a été effectuée dans \cite{borgelt,surana2010}.
%%THH : ATTENTION se positionner versus le papier de Surana : qu'est ce qu'on apporte de plus?

Grace à  cette propriété d'anti-monotonie, les motifs corrélés, selon la mesure \textit{bond}, induisent un idéal d'ordre dans le treillis des motifs \textsc{(}tout sous-ensemble d'un motif corrélé est aussi corrélé\textsc{)}. Par opposition à  ces derniers, les motifs rares induisent un filtre d'ordre et vérifient une contrainte monotone \textsc{(}tout sur-ensemble d'un motif rare est aussi rare\textsc{)}. Par conséquent, l'ensemble des motifs corrélés rares que nous visons à  extraire résulte de l'intersection des deux théories \cite{mannila97} associées respectivement aux contraintes de corrélation et de rareté.
%%Il est par conséquent compris entre la bordure associée à  l'idéal d'ordre de la corrélation et celle associée au %%filtre d'ordre de la rareté.
%%%%D'une maniére générale, la localisation de telles bordures est prouvée dans la littérature comme étant un probléme %%%%difficile \cite{boros2002_complexity}.

La nature opposée de ces contraintes permet de différencier l'ensemble des motifs corrélés rares de l'ensemble des motifs induit par une ou plusieurs contraintes de même type \cite{boulicautsurveycontraintes} \cite{pei_contraintes02}. Cette caractéristique rend plus complexe l'extraction de l'ensemble des motifs corrélés rares. \`A cet égard, nous proposons dans ce mémoire une caractérisation de cet ensemble moyennant la notion de classe d'équivalence. Dans notre cas, les classes d'équivalence seront induites par l'opérateur de fermeture associé à  la mesure de corrélation \textit{bond}. Ces classes jouent un rôle clé dans l'élimination de la redondance entre les motifs. Une fois la caractérisation effectuée, nous proposons des représentations exactes des motifs corrélés rares. Ces représentations permettent d'une part de réduire significativement le nombre de motifs corrélés rares extraits. Elles améliorent aussi leur qualité et ce en évitant la redondance entre motifs puisqu'elles ne maintiennent qu'un sous-ensemble sans perte d'information de l'ensemble total des motifs corrélés rares. D'autre part, elles assurent la régénération aisée et efficace de l'ensemble des motifs corrélés rares. Il est aussi important de noter que les représentations proposées permettent non seulement la dérivation du support conjonctif mais aussi des supports disjonctif et négatif des motifs corrélés rares. Ceci permet par exemple d'utiliser ces motifs comme base pour l'extraction des règles généralisées ou les connecteurs de disjonction et de négation sont utilisés en plus de celui classique de conjonction \cite{tarekamai2010}.

Au meilleur de notre connaissance, aucune étude n'a été réalisée dans la littérature dans le but de proposer une représentation concise des motifs corrélés rares. Par ailleurs, une autre originalité de ce travail consiste en la nature des contraintes manipulées et la caractérisation de l'ensemble résultant moyennant une relation d'équivalence. Nous signalons aussi que l'approche proposée dans ce travail n'est pas restreinte aux motifs corrélés rares selon cette mesure. En effet, elle est générique dans le sens qu'elle s'applique à  tout ensemble de motifs corrélés rares selon toute mesure de corrélation vérifiant les mêmes propriétés structurelles que la mesure \textit{bond} telle que par exemple la mesure \textit{all-confidence} \cite{Omie03} $^\textsc{(}$\footnote{Mathématiquement équivalente à  la mesure \textit{h-confidence} \cite{Xiong06hypercliquepattern}.}$^{\textsc{)}}$. Il est toutefois à  noter que la mesure \textit{bond} a pour avantage de permettre la dérivation des supports disjonctif et négatif, ce que ne peut pas permettre les autres mesures.
%Notons que dans ce, nous nous focalisons principalement sur la caractérisation des représentations proposées. L'étude détaillée de l'aspect algorithmique associé à  l'extraction des représentations proposées fera l'objet d'un futur travail.

\section*{Structure du mémoire}

La synthèse des travaux de recherches effectués sera présentée dans le cadre de ce mémoire et répartie dans cinq chapitres comme suit.
\bigskip

Dans le \textbf{premier chapitre}, nous présentons les notions de base
qui seront utilisées tout au long de ce travail.
Ces notions de bases incluent les notions préliminaires relatives à  l'extraction des motifs intéressants et des représentations concises exactes de ces derniers. Les caractéristiques des motifs rares et des motifs corrélés selon la mesure \textit{bond} y seront aussi présentés. De plus, nous intégrons la présentation des opérateurs de fermetures et nous spécifions précisément l'opérateur de fermeture $f_{bond}$ associé à  la mesure de corrélation \textit{bond}.

\bigskip

Le \textbf{deuxième chapitre} décrit les approches d'extraction des motifs rares et les différentes approches d'extraction des motifs corrélés sous contraintes. Nous y étudions et analysons aussi les algorithmes d'extraction de motifs sous la conjonction de contraintes de types opposés.

\bigskip

Dans le \textbf{troisième chapitre}, nous définissons et étudions profondément les propriétés de l'ensemble des
motifs corrélés rares selon la mesure \textit{bond}. Nous présentons aussi les spécificités des classes d'équivalence corrélées rares. Ensuite, nous exposons les nouvelles représentations concises exactes des motifs corrélés rares selon la mesure \textit{bond} à  savoir $\mathcal{RMCR}$, $\mathcal{RMM}$$ax$$\mathcal{F}$ et $\mathcal{RM}$$in$$\mathcal{MF}$. Nous clôturons ce chapitre avec la définition et l'étude de la représentation concise approximative $\mathcal{RM}$$in$$\mathcal{MM}$$ax$$\mathcal{F}$.

\bigskip

Dans le \textbf{quatrième chapitre}, nous introduisons le nouvel algorithme \textsc{CRP\_Miner} d'extraction
de l'ensemble total des motifs corrélés rares dans un premier temps. Ensuite, nous présentons l'algorithme \textsc{CRPR\_Miner} d'extraction des représentations concises proposées. Nous démontrons ses propriétés théoriques de validité et de terminaison.
Par ailleurs, nous décrivons les stratégies d'interrogation et de régénération des motifs corrélés rares à  partir de la représentation concise exacte $\mathcal{RMCR}$ et nous proposons et décrivons les algorithmes dédiés.

\bigskip

Le \textbf{cinquième chapitre} présente les études expérimentaux menées
sur des bases ``benchmark''. Cette étude s'étalera sur deux principaux axes. Le premier axe concerne la quantification et l'analyse des taux de compacité des différentes représentations concises proposées. Le deuxiéme axe concerne la comparaison et l'analyse de la variation des coéts d'extraction des différentes représentations proposées.
De plus, nous décrivons le processus d'extraction des régles de classification corrélées rares à  partir de la représentation concise exacte $\mathcal{RMCR}$. Nous évaluons expérimentalement l'efficacité de la classification 
basée sur ces régles dans le cadre de la détection d'intrusions.

\bigskip

Ce mémoire se termine par une conclusion générale récapitulant l'ensemble de nos contributions et cernant les principales perspectives futures de travaux de recherche.

\bigskip

%%%%%%%%%%%%%%%%%%%%%%%%%%%%%%%%%%%%%%%%%%%%%%%%%%%%%%%%
\chapter{Notions de base}\label{chapitre_notions_base}

\section{Introduction}

Nous repérons, dans ce chapitre, les notions de base que nous estimons primordiales dans la présentation de nos approches. \`A cet égard, la première section sera consacrée à  l'introduction des notions préliminaires sur laquelle est basée l'extraction des motifs. Nous enchaînons, dans la deuxième section, avec la présentation des propriétés des motifs rares.
Ensuite, la troisième section sera dédiée à  la présentation des motifs corrélés
selon la mesure de corrélation \textit{bond}.
La quatrième section sera consacrée à  la présentation de l'opérateur de fermeture $f_{bond}$ \cite{tarekds2010} associé à  la mesure \textit{bond}. Ce chapitre sera clôturé avec une définition des représentations concises associés à  un ensemble de motifs.

\section{Extraction des motifs}

Nous commençons par présenter l'ensemble des notions de base relatives à  l'extraction des motifs, qui seront utilisés tout au long de ce travail. Définissons d'abord une base de transactions à partir de laquelle se fait l'extraction des motifs intéressants.

\begin{definition} \label{definitionbasetransactions} \textsc{(}\textbf{Base de transactions}\textsc{)} Une base de transactions \textsc{(}appelée aussi contexte d'extraction ou simplement contexte\textsc{)} est représentée sous la forme d'un triplet $\mathcal{D}$ = \textsc{(}$\mathcal{T},\mathcal{I},\mathcal{R}$\textsc{)} dans
	lequel $\mathcal{T}$ et $\mathcal{I}$ sont, respectivement, des ensembles finis de transactions \textsc{(}ou objets\textsc{)} et d'items \textsc{(}ou attributs\textsc{)}, et $\mathcal{R}$ $\subseteq$ $\mathcal{T} \times \mathcal{I}$ est une relation binaire entre les transactions et les items. Un couple \textsc{(}$t$, $i$\textsc{)} $\in$ $\mathcal{R}$ dénote le fait que la transaction $t$ $\in$ $\mathcal{T}$ contient l'item $i$ $\in$ $\mathcal{I}$.
\end{definition}
\begin{exemple}
	Un exemple d'une base de transactions $\mathcal{D}$ $=$
	$\textsc{(}$$\mathcal{T},\mathcal{I},\mathcal{R}$$\textsc{)}$
	\textsc{(}\textit{resp.} contexte d'extraction $\mathcal{K}$ $=$
	$\textsc{(}$$\mathcal{O},\mathcal{I},\mathcal{R}$$\textsc{)}$\textsc{)}
	est donné par la table \ref{Base_transactions}. Dans cette base
	\textsc{(}\textit{resp.} ce contexte\textsc{)}, l'ensemble de
	transactions $\mathcal{T} = \{1, 2, 3, 4, 5\}$
	\textsc{(}\textit{resp.} d'objets $\mathcal{O} = \{1, 2, 3, 4,
	5\}$\textsc{)} et l'ensemble d'items  $\mathcal{I}$ $=$
	$\{$\texttt{A}, \texttt{B}, \texttt{C}, \texttt{D}, \texttt{E},$\}$.
	Le couple \textsc{(}2, B\textsc{)} $\in$ $\mathcal{R}$ car  la transaction 2 $\in$ $\mathcal{T}$ contient l'item B $\in$ $\mathcal{I}$.
\end{exemple}
\begin{table}[h]
	\begin{center}
		\footnotesize{
			\begin{tabular}{|c||c|c|c|c|c|c|}
				\hline & \texttt{A}  & \texttt{B}  & \texttt{C}  & \texttt{D} & \texttt{E}  \\
				\hline\hline
				1 & $\times$  &          &$\times$    & $\times$&         \\
				\hline
				2 &           & $\times$ &$\times$    &         & $\times$ \\
				\hline
				3 & $\times$  & $\times$ &$\times$    &         & $\times$ \\
				\hline
				4 &           & $\times$ &            &         & $\times$ \\
				\hline
				5 & $\times$  & $\times$ &$\times$    &         & $\times$  \\
				\hline
		\end{tabular}}
	\end{center}
	\caption{Un exemple d'une base de transactions.}\label{Base_transactions}
\end{table}
\begin{remarque}
	Nous notons, par souci de précision, que les notations de base de
	transactions et de contexte d'extraction seront les mêmes dans la suite. Ils seront
	notés $\mathcal{D}$ $=$ $\textsc{(}\mathcal{T}, \mathcal{I},
	\mathcal{R}\textsc{)}$.
\end{remarque}
\begin{definition} \label{motif}  \textbf{Itemset ou Motif}\\
	Une transaction $t$ $\in$ $\mathcal{T}$, avec un identificateur
	communément noté \textit{TID} $\textsc{(}$Tuple
	IDentifier$\textsc{)}$, contient un ensemble, non vide, d'items de
	$\mathcal{I}$. Un sous-ensemble $I$ de $\mathcal{I}$ ou $k$ $=$
	$\vert I \vert $ est appelé un \textit{$k$-motif} ou simplement un
	\textit{motif}, et $k$ représente la cardinalité de $I$
	Le nombre de transactions $t$ d'une base $\mathcal{D}$ contenant un motif $I$,
	$\vert $ $\{$ $t$ $ \in $ $\mathcal{D}$ $\vert $ $I$ $\subseteq $
	$t$$\}$ $\vert $, est appelé \textit{support absolu} de $I$ et noté
	par la suite $Supp\textsc{(}\wedge I\textsc{)}$. Le \textit{support
		relatif} de $I$ ou la \textit{fréquence} de $I$, notée
	$freq\textsc{(}I\textsc{)}$, est le quotient de son support absolu
	par le nombre total de transactions de $\mathcal{D}$,
	\textit{c.-é.-d.}, $freq\textsc{(}I\textsc{)}$ $=$
	$\displaystyle\frac{\displaystyle{\vert {\{}t \in \mathcal{D} | I \subseteq t
			{\}}\vert}}{\displaystyle{\vert \mathcal{T}\vert}}$.
\end{definition}
\begin{remarque}
	Nous signalons que dans ce travail, nous nous intéressons à  la
	classe des motifs formée par des itemsets, \textit{c.-é.-d.}, des
	ensembles d'items. Par conséquent, nous employons dans la suite une
	forme sans séparateur pour dénoter les itemsets. Par exemple,\texttt{AC} représente l'ensemble d'items
	$\{$\texttt{A}, \texttt{C}$\}$.
\end{remarque}

Introduisons à ce stade la notion de treillis des motifs.
\begin{definition} \label{deftreillis} \textsc{(}\textbf{Treillis des motifs}\textsc{)}
	Un treillis des motifs est un regroupement conceptuel et hiérarchique des motifs. Il est aussi dit
	treillis d'inclusion ensembliste. Toutefois, l'ensemble
	des parties de $\mathcal{I}$ est ordonné par inclusion ensembliste dans le treillis des motifs.
	Le treillis des motifs associé au contexte donné par la table \ref{Base_transactions} est représenté par la figure
	\ref{treillis1}.
\end{definition}

Toutefois, plusieurs mesures sont utilisées pour évaluer l'intérêt d'un motif, dont les plus connues sont présentées à  travers la définition \ref{definitionsupportmotif}.
\begin{definition} \label{definitionsupportmotif} \textsc{(}\textbf{Supports d'un motif}\textsc{)} Soient $\mathcal{D}$ = \textsc{(}$\mathcal{T}, \mathcal{I}, \mathcal{R}$\textsc{)} une base de transactions et un motif non vide $I$ $\subseteq$ $\mathcal{I}$. Nous distinguons trois types de supports correspondants à  $I$ :
	
	- \textbf{\textit{Le support conjonctif :}} \textit{Supp}\textsc{(}$\wedge$$I$\textsc{)} = $\mid$$\{$$t$ $\in$ $\mathcal{T}$ $\mid$ $\forall$ $i$ $\in$ $I$ : \textsc{(}$t$, $i$\textsc{)} $\in$ $\mathcal{R}$$\}$$\mid$
	
	- \textbf{\textit{Le support disjonctif :}} \textit{Supp}\textsc{(}$\vee$$I$\textsc{)} = $\mid$$\{$$t$ $\in$ $\mathcal{T}$ $\mid$ $\exists$ $i$ $\in$ $I$ : \textsc{(}$t$, $i$\textsc{)} $\in$ $\mathcal{R}$$\}$$\mid$
	
	- \textbf{\textit{Le support négatif :}} \textit{Supp}\textsc{(}$\neg$$I$\textsc{)} = $\mid$$\{$$t$ $\in$ $\mathcal{T}$ $\mid$ $\forall$ $i$ $\in$ $I$ : \textsc{(}$t$, $i$\textsc{)} $\notin$ $\mathcal{R}$$\}$$\mid$
\end{definition}
Il est à  noter que la loi de De Morgan assure la transition entre le
support disjonctif et le support négatif de $I$ comme suit :
\textit{Supp}\textsc{(}$\neg$\textit{I}\textsc{)} =
$\mid\mathcal{T}\mid$ - \textit{Supp}\textsc{(}$\vee$\textit{I}\textsc{)}.
\begin{exemple}
	Considérons la base de transactions illustrée par la table \ref{Base_transactions} et qui sera utilisée dans la suite pour les différents exemples. Nous avons \textit{Supp}\textsc{(}$\wedge$\texttt{AD}\textsc{)} = $\mid$$\{$1$\}$$\mid$ = $1$, \textit{Supp}\textsc{(}$\vee$\texttt{AD}\textsc{)} = $\mid$$\{$ 1, 3, 5$\}$$\mid$ = $3$, et, \textit{Supp}\textsc{(}$\neg$\textsc{(}\texttt{AD}\textsc{\textsc{)}\textsc{)}} = $\mid$$\{$2, 4$\}$$\mid$ = $2$ $^{\textsc{(}}$\footnote{Nous employons une forme sans séparateur pour les ensembles d'items : par exemple, \texttt{AD}
		représente l'ensemble $\{$\texttt{A}, \texttt{D}$\}$.}$^{\textsc{)}}$.
\end{exemple}
Dans la suite, s'il n'y a pas de risque de confusion, le \textit{support conjonctif} sera simplement appelé \textit{support}.
%%%%%%%%%%%%%%%%%%%%%%%%%%%%%%%%%%%%
La définition suivante présente le statut de fréquence d'un motif, fréquent ou infréquent, étant donné un seuil minimal de support \cite{Agra94}.
\begin{definition} \label{motiffréq} \textsc{(}\textbf{Motif fréquent/rare}\textsc{)} Soit une base de transactions $\mathcal{D}$ = $\textsc{(}\mathcal{T}, \mathcal{I},\mathcal{R}\textsc{)}$, un seuil minimal de support conjonctif \textit{minsupp}, un motif $I$ $\subseteq$ $\mathcal{I}$ est dit \textit{fréquent} si \textit{Supp}\textsc{(}$\wedge$$I$\textsc{)} $\geq$ \textit{minsupp}. $I$ est dit \textit{infréquent} ou \textit{rare} sinon.
\end{definition}
\begin{exemple} Soit  \textit{minsupp} = 2. \textit{Supp}\textsc{(}$\wedge$\texttt{BCE}\textsc{)}
	= 3, le motif \texttt{BCE} est un motif fréquent. Cependant, le motif \texttt{CD} est non fréquent ou rare puisque \textit{Supp}\textsc{(}$\wedge$\texttt{CD}\textsc{)} = 1 $<$ 2.
\end{exemple}
%%%%%%%%%%%%%%%%%%%%%%
Outre la contrainte de fréquence minimale traduite par le seuil \textit{minsupp}, d'autres contraintes peuvent étre intégrées dans le processus d'extraction des motifs. Ces contraintes admettent différents types, dont les deux principaux sont définis dans ce qui suit \cite{pei_contraintes02}.

\begin{definition}\label{anti-monotone_ET_monotone} \textsc{(}\textbf{Contrainte anti-monotone/monotone}\textsc{)}
	
	$\bullet$ Une contrainte $Q$ est \textit{anti-monotone} si $\forall$ $I$ $\subseteq$ $\mathcal{I}$, $\forall$ $I_1$ $\subseteq$ $I$ : $I$ satisfait $Q$ $\Rightarrow$ $I_1$ satisfait $Q$.
	
	$\bullet$ Une contrainte $Q$ est \textit{monotone} si $\forall$ $I$ $\subseteq$ $\mathcal{I}$, $\forall$ $I_1$ $\supseteq$ $I$ : $I$ satisfait $Q$ $\Rightarrow$ $I_1$ satisfait $Q$.
\end{definition}
\begin{exemple} La \textit{contrainte de fréquence}, c.-é.-d. avoir un support supérieur ou égal à  \textit{minsupp}, est une contrainte anti-monotone. En effet, $\forall$ $I$, $I_1$ $\subseteq$ $\mathcal{I}$, si $I_1$ $\subseteq$ $I$ et \textit{Supp}\textsc{(}$\wedge$$I$\textsc{)} $\geq$ \textit{minsupp}, alors \textit{Supp}\textsc{(}$\wedge$$I_1$\textsc{)} $\geq$ \textit{minsupp} puisque \textit{Supp}\textsc{(}$\wedge$$I_1$\textsc{)} $\geq$ \textit{Supp}\textsc{(}$\wedge$$I$\textsc{)}.
	
	D'une maniére duale, la \textit{contrainte de rareté}, c.-é.-d. avoir un support strictement inférieur à  \textit{minsupp}, est monotone. En effet, $\forall$ $I$, $I_1$ $\subseteq$ $\mathcal{I}$, si $I_1$ $\supseteq$ $I$ et \textit{Supp}\textsc{(}$\wedge$$I$\textsc{)} $<$ \textit{minsupp}, alors \textit{Supp}\textsc{(}$\wedge$$I_1$\textsc{)} $<$ \textit{minsupp} puisque \textit{Supp}\textsc{(}$\wedge$$I_1$\textsc{)} $\leq$ \textit{Supp}\textsc{(}$\wedge$$I$\textsc{)}.
\end{exemple}
%%%%%%%%%%%%%%%%%%%%%%%%%%%%%%%%%%%%%%%%%%
Soit $\mathcal{P}\textsc{(}\mathcal{I}\textsc{)}$ l'ensemble de tous les sous-ensembles de $\mathcal{I}$. Dans ce qui suit, nous introduisons les notions duales d'idéal d'ordre et de filtre d'ordre \cite{ganter99} définis sur $\mathcal{P}\textsc{(}\mathcal{I}\textsc{)}$.
\begin{definition} \textsc{(}\textbf{Idéal d'ordre}\textsc{)} Un sous-ensemble $\mathcal{S}$ de $\mathcal{P}\textsc{(}\mathcal{I}\textsc{)}$ est un idéal d'ordre s'il vérifie les propriétés suivantes :
	\begin{itemize}
		\item Si $I$ $\in$ $\mathcal{S}$, alors $\forall$ $I_1$ $\subseteq$ $I$ : $I_1$ $\in$ $\mathcal{S}$.
		\item Si $I$ $\notin$ $\mathcal{S}$, alors $\forall$ $I$ $\subseteq$ $I_1$ : $I_1$ $\notin$ $\mathcal{S}$.
	\end{itemize}
\end{definition}
\begin{definition} \textsc{(}\textbf{Filtre d'ordre}\textsc{)} Un sous-ensemble $\mathcal{S}$ de $\mathcal{P}\textsc{(}\mathcal{I}\textsc{)}$ est un filtre d'ordre s'il vérifie les propriétés suivantes :
	\begin{itemize}
		\item Si $I$ $\in$ $\mathcal{S}$, alors  $\forall$ $I_1$ $\supseteq$ $I$ : $I_1$ $\in$ $\mathcal{S}$.
		\item Si $I$ $\notin$ $\mathcal{S}$, alors $\forall$ $I$ $\supseteq$ $I_1$ : $I_1$ $\notin$ $\mathcal{S}$.
	\end{itemize}
\end{definition}
Une contrainte anti-monotone telle que la contrainte de fréquence induit un idéal d'ordre. D'une maniére duale, une contrainte monotone telle que la contrainte de rareté forme un filtre d'ordre. L'ensemble des motifs satisfaisant une contrainte donnée est appelé \textit{théorie} dans \cite{mannila97}. Cette théorie est délimitée par deux bordures, une dite \emph{bordure positive} et l'autre appelée \emph{bordure négative}, et qui sont définies comme suit.

\begin{definition}\label{bd} \textsc{(}\textbf{Bordure positive/négative}\textsc{)} \cite{luccheKIS05_MAJ_06}\\
	Pour le cas d'une contrainte anti-monotone $C_{am}$,
	la bordure correspond à  l'ensemble des motifs dont tous les sous ensembles satisfont cette contrainte et dont tous les sur-ensembles ne la satisfont pas.
	Soit un ensemble de motifs $\mathcal{S}$$_{am}$ satisfaisant une contrainte anti-monotone $C_{am}$, la bordure est formellement définie comme suit :
	\begin{center}
		$\mathcal{B}d$\textsc{(}$\mathcal{S}$$_{am}$\textsc{)} =
		$\{$$X$ $|$ $\forall$ $Y$  $\subset$ $X$ : $Y$ $\in$ $\mathcal{S}$$_{am}$ et
		$\forall$ $Z$  $\supset$ $X$ : $Z$ $\notin$ $\mathcal{S}$$_{am}$$\}$
	\end{center}
	Pour le cas d'une contrainte monotone $C_{m}$,
	la bordure correspond à  l'ensemble des motifs dont tous les sur-ensembles satisfont cette contrainte et dont tous les sous ensembles ne la satisfont pas.
	Soit un ensemble de motifs $\mathcal{S}$$_{m}$ satisfaisant une contrainte monotone $C_{m}$, la bordure est formellement définie comme suit :
	\begin{center}
		$\mathcal{B}d$\textsc{(}$\mathcal{S}$$_{m}$\textsc{)} =
		$\{$$X$ $|$ $\forall$ $Y$  $\supset$ $X$ : $Y$ $\in$ $\mathcal{S}$$_{m}$ et
		$\forall$ $Z$ $\subset$ $X$ : $Z$ $\notin$ $\mathcal{S}$$_{m}$$\}$
	\end{center}
	Cependant, il est à  distinguer entre la bordure positive et la bordure négative pour une contrainte $C$ donnée. Soit un ensemble de motifs $\mathcal{S}$ satisfaisant une contrainte $C$. La bordure positive $\mathcal{B}{d}^{+}\textsc{(}\mathcal{S}$\textsc{)} correspond à  l'ensemble des motifs appartenant à  la bordure $\mathcal{B}{d}\textsc{(}\mathcal{S}$\textsc{)}
	et qui satisfont la contrainte $C$.
	La bordure négative $\mathcal{B}{d}^{-}\textsc{(}\mathcal{S}$\textsc{)} correspond à  l'ensemble des motifs
	appartenant à  la bordure $\mathcal{B}{d}\textsc{(}\mathcal{S}$\textsc{)}
	et qui ne satisfont pas la contrainte $C$.
	Ces bordures sont formellement définies comme suit :
	\begin{center}
		$\mathcal{B}{d}^{+}\textsc{(}\mathcal{S}\textsc{)}$ =
		$\mathcal{B}{d}\textsc{(}\mathcal{S}$\textsc{)} $\cap$ $\mathcal{S}$, \\
		$\mathcal{B}{d}^{-}\textsc{(}\mathcal{S}\textsc{)}$ =
		$\mathcal{B}{d}\textsc{(}\mathcal{S}$\textsc{)} $\setminus$ $\mathcal{S}$.
	\end{center}
\end{definition}
\begin{exemple}
	Considérons le contexte donné par la table \ref{Base_transactions}. L'ensemble $\mathcal{MFM}$ des motifs fréquents maximaux est composé des motifs fréquents dont tous les sur-ensembles sont des motifs rares et est égal é,  $\mathcal{MFM}$ = $\{$\texttt{AC}, \texttt{BCE}$\}$.
	Les motifs fréquents maximaux représentent à  la fois la bordure positive des motifs vérifiant la contrainte anti-monotone de fréquence et la bordure négative des motifs vérifiant la contrainte monotone de rareté.
\end{exemple}

Comme dans ce travail, nous nous intéressons principalement aux motifs rares qui sont aussi corrélés, nous présentons certaines propriétés utiles des motifs rares et des motifs corrélés.
%%%%%%%%%%%%%%%%%%%%%%%%%%%%%%%%%%%%%%%%%%
\section{Motifs rares}
L'ensemble des motifs rares, correspondant aux motifs vérifiant la contrainte de rareté, est défini comme suit :
\begin{definition} \label{motifrare} \textsc{(}\textbf{Motifs rares}\textsc{)} L'ensemble des motifs rares est défini par : $\mathcal{MR}$ = $\{$$I$ $\subseteq$ $\mathcal{I}$$|$ \textit{Supp}\textsc{(}$\wedge$$I$\textsc{)} $<$ \textit{minsupp}$\}$.
	Ou d'une maniére équivalente :
	$\mathcal{MR}$ = $\{$$I$ $\subseteq$ $\mathcal{I}$$|$ \textit{Supp}\textsc{(}$\wedge$$I$\textsc{)}
	$\leqslant$ \textit{maxsupp}$\}$
	avec \textit{maxsupp} correspond au seuil maximal du support conjonctif, \textit{maxsupp} = \textit{minsupp} - 1.
\end{definition}
Comme indiqué plus haut, cet ensemble forme un filtre d'ordre dans $\mathcal{P}\textsc{(}\mathcal{I}\textsc{)}$. Il en résulte que tous les sur-ensembles d'un motif rare sont aussi rares.
\begin{exemple} Soit \textit{minsupp} = 4. Le motif \texttt{BC} est rare puisque \textit{Supp}\textsc{(}$\wedge$\texttt{BC}\textsc{)} = 3 $<$ 4, \texttt{BC} $\in$ $\mathcal{MR}$. Aussi, \texttt{ABC} $\in$ $\mathcal{MR}$ puisque \texttt{BC} $\subseteq$ \texttt{ABC}.
\end{exemple}

\begin{remarque}
	Un motif est dit fréquent si son support est supérieur ou égal au seuil minimal de support \textit{minsupp}, et il est rare ou non fréquent si son support est strictement inférieur au seuil \textit{minsupp} ou autrement dit son support est inférieur ou égal au seuil \textit{maxsupp}. Dans ce qui suit nous considérons que le seuil \textit{maxsupp} est calculé à  partir du seuil \textit{minsupp} et est égal à  \textit{minsupp} - $1$.
	Il peut exister un intervalle de valeur entre \textit{minsupp} et \textit{maxsupp} \cite{theselasz}, cependant ,  nous optons pour le cas général \textit{c.-é.-d.}, un motif est rare s'il n'est pas fréquent. Ceci implique l'existence d'une seule bordure entre les motifs fréquents et ceux rares.
\end{remarque}
Nous soulignons que, nous ne considérons pas les motifs rares de support égal à  zéro. En effet, ces motifs n'apparaissent jamais dans la base et ne représentent donc pas un événement rare.

Dans la suite, nous aurons besoin des motifs rares les plus petits, par rapport à  la relation d'inclusion ensembliste. Ces motifs forment la bordure positive de l'ensemble $\mathcal{MR}$ des motifs rares et correspondent aux motifs rares dont tous les sous-ensembles sont fréquents \cite{qdc2011}. Ils sont définis comme suit :

\begin{definition}\label{mrm} \textsc{(}\textbf{Motifs rares minimaux}\textsc{)} L'ensemble $\mathcal{MRM}in$ des motifs rares minimaux correspond aux motifs rares n'ayant aucun sous-ensemble strict rare. Cet ensemble est défini par : $\mathcal{MRM}in$ = $\{$$I$ $\in$ $\mathcal{MR}$$|$ $\forall$ $I_1$ $\subset$ $I$ : $I_1$ $\notin$ $\mathcal{MR}$$\}$, ou d'une maniére équivalente : $\mathcal{MRM}in$ = $\{$$I$ $\in$ $\mathcal{MR}$$|$ $\forall$ $I_1$ $\subset$ $I$ : \textit{Supp}\textsc{(}$\wedge$$I_1$\textsc{)} $\geq$ \textit{minsupp}$\}$.
\end{definition}

\begin {figure}
\parbox{16cm}{
	\hspace{-0.9cm}
	\includegraphics[scale = 0.5]{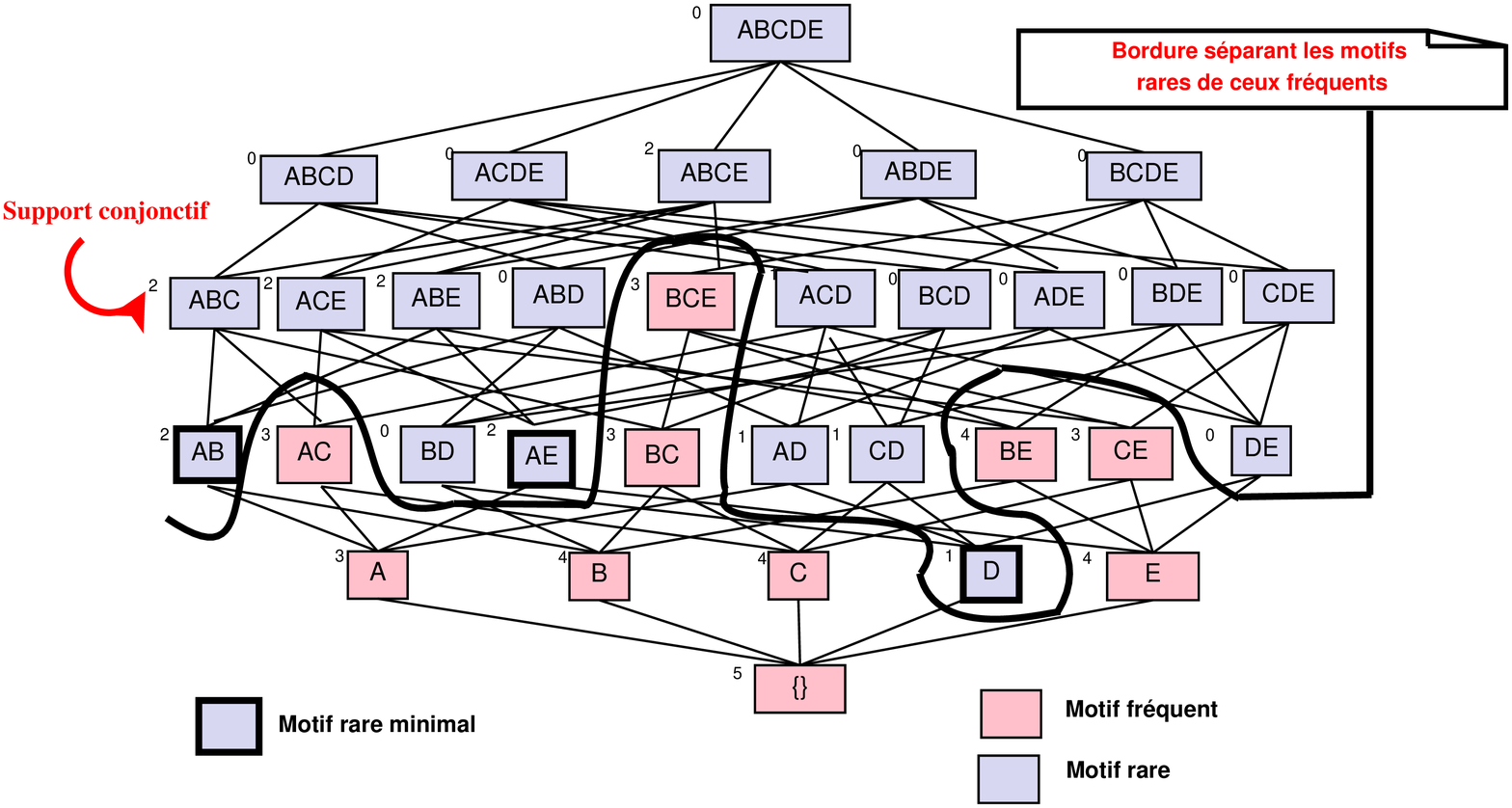}}
\caption{Treillis des motifs associés au contexte $\mathcal{D}$ pour \textit{minsupp} = 3.}
\label{treillis1}
\end{figure}

\begin{exemple}\label{exemple_MRMin}
	Considérons la base illustrée par la table \ref{Base_transactions} pour \textit{minsupp} = 3. Le motif \texttt{AB} $\in$ $\mathcal{MRM}in$ puisque \textit{Supp}\textsc{(}$\wedge$\texttt{AB}\textsc{)} = 2 $<$ 3 et, d'autre part, \textit{Supp}\textsc{(}$\wedge$\texttt{A}\textsc{)} = 3 $\geq$ 3 et \textit{Supp}\textsc{(}$\wedge$\texttt{B}\textsc{)} = 4 $\geq$ 3. Il en est de même pour \texttt{D} et \texttt{AE}. Ainsi, dans ce cas, $\mathcal{MRM}in$ = $\{$\texttt{D}, \texttt{AB}, \texttt{AE}$\}$.
	L'ensemble $\mathcal{MRM}in$ ainsi que l'ensemble $\mathcal{MR}$ se tous les motifs rares sont schématisés par la figure \ref{treillis1}. Le support indiqué en haut à  gauche de chaque cadre représentant un motif est son support conjonctif
\end{exemple}

Aprés avoir présenté les motifs rares, nous étudions dans ce qui suit les propriétés des motifs corrélés selon la mesure \textit{bond}.

\section{Motifs corrélés selon la mesure \textit{bond}}
La mesure \textit{bond} \cite{Omie03} est mathématiquement équivalente aux mesures \textit{coherence} \cite{comine_Lee}, \textit{coefficient de Tanimoto} \cite{Tanimoto1958} et \textit{Jaccard} \cite{jaccard_1901}. Elle a été redéfinie dans \cite{tarekds2010} comme suit :
\begin{definition}\label{La mesure bond} \textsc{(}\textbf{Mesure \textit{bond}}\textsc{)} Soit $I$ $\subseteq$ $\mathcal{I}$. La mesure \textit{bond} de $I$ est définie par :
\begin{center}
	\textit{bond}\textsc{(}$I$ \textsc{)} = $\displaystyle\frac{\displaystyle
		\textit{Supp}\textsc{(}\wedge I\textsc{)}}{\displaystyle
		\textit{Supp}\textsc{(}\vee I\textsc{)}}$
\end{center}
\end{definition}
La mesure \textit{bond} prend ses valeurs sur l'intervalle [0, 1]. En considérant \textit{l'univers} d'un motif $I$ \cite{comine_Lee}, \textit{c.-é.-d.} l'ensemble des transactions contenant un sous-ensemble non-vide de $I$, la mesure \textit{bond} véhicule l'information concernant le taux d'apparition simultanée des items d'un motif dans son univers. Ainsi, plus les items du motif sont dispersés dans son univers \textsc{(}\textit{c.-é.-d.} faiblement corrélés\textsc{)}, plus sera faible la valeur de \textit{bond} puisque \textit{Supp}\textsc{(}$\wedge$$I$\textsc{)} serait nettement plus petit que \textit{Supp}\textsc{(}$\vee$$I$\textsc{)}. Inversement, plus les items de $I$ dépendent les uns des autres \textsc{(}\textit{c.-é.-d.} fortement corrélés\textsc{)}, plus sera élevée la valeur de \textit{bond} puisque \textit{Supp}\textsc{(}$\wedge$$I$\textsc{)} serait proche de \textit{Supp}\textsc{(}$\vee$$I$\textsc{)}.

Aprés avoir présenté la mesure \textit{bond}, nous définissons maintenant les motifs corrélés selon cette mesure.
\begin{definition}\label{motifcorrélé} \textsc{(}\textbf{Motifs corrélés selon la mesure \textit{bond}}\textsc{)}
Soit \textit{minbond} un seuil minimal de corrélation. L'ensemble $\mathcal{MC}$ des motifs corrélés selon la mesure \textit{bond} est défini par : $\mathcal{MC}$ = $\{$$I$ $\subseteq$ $\mathcal{I}$ $\mid$ \textit{bond}\textsc{(}$I$\textsc{)} $\geq$ \textit{minbond}$\}$
\end{definition}
\begin{exemple}
Considérons la base illustrée par la table \ref{Base_transactions}. Pour \textit{minbond} = 0,5, Nous avons \textit{bond}\textsc{(}\texttt{AB}\textsc{)} = $\displaystyle\frac{2}{5}$ = 0,4 $<$ 0,5. Le motif \texttt{AB} est alors non corrélé. Par contre, \textit{bond}\textsc{(}\texttt{BCE}\textsc{)} = $\displaystyle\frac{3}{5}$ = 0,6 $\geq$ 0,5. Ainsi, le motif \texttt{BCE} est corrélé.
\end{exemple}

Il a été démontré dans \cite{tarekds2010} que la mesure \textit{bond} présente diverses propriétés intéressantes. En effet, cette mesure : \textsc{(}$i$\textsc{)} est \textit{symétrique} puisque $\forall$ $I$, $J$ $\subseteq$ $\mathcal{I}$, \textit{bond}\textsc{(}$IJ$\textsc{)} = \textit{bond}\textsc{(}$JI$\textsc{)}; \textsc{(}$ii$\textsc{)} est \textit{descriptive} \textit{c.-é.-d.} insensible au changement du nombre de transactions; \textsc{(}$iii$\textsc{)} vérifie la propriété de \textit{cross support} \cite{Xiong06hypercliquepattern}. Gréce à  cette derniére propriété, pour un motif $I$ $\subseteq$ $\mathcal{I}$ et pour un seuil minimal de corrélation \textit{minbond}, s'il existe un couple d'items $x$, $y$ $\in$ $I$ tel que $\displaystyle\frac{\displaystyle \textit{Supp}\textsc{(}\wedge x\textsc{)}}{\displaystyle \textit{Supp}\textsc{(}\wedge y\textsc{)}}$ $<$ \textit{minbond}, alors $I$ n'est pas corrélé puisque \textit{bond}\textsc{(}$I$ \textsc{)} $<$ \textit{minbond}. $I$ vérifie alors la propriété de cross-support pour le seuil \textit{minbond}; et \textsc{(}$iv$\textsc{)} induit une contrainte anti-monotone du moment que le seuil minimal \textit{minbond} est fixé. En effet, $\forall$ $I$, $I_1$ $\subseteq$ $\mathcal{I}$, si $I_1$ $\subseteq$ $I$, alors \textit{bond}\textsc{(}$I_1$\textsc{)} $\geq$ \textit{bond}\textsc{(}$I$\textsc{)}. Ainsi, l'ensemble $\mathcal{MC}$ des motifs corrélés forme un idéal d'ordre. Autrement dit, si un motif est corrélé, alors tous ses sous-ensembles sont aussi corrélés.

La proposition suivante présente une relation intéressante entre la valeur de la mesure \textit{bond} et les valeurs des supports conjonctifs et disjonctifs pour chaque couple de deux motifs $I$ et $I_1$ tel que $I$ $\subseteq$ $I_1$ \cite{tarekds2010}.
\begin{proposition} \label{PropBond} Soient $I$, $I_1$ $\subseteq$ $\mathcal{I}$ et $I$ $\subseteq$ $I_1$. Si \textit{bond}\textsc{(}$I$\textsc{)} =
\textit{bond}\textsc{(}$I_1$\textsc{)}, alors
\textit{Supp}\textsc{(}$\wedge$$I$\textsc{)} =
\textit{Supp}\textsc{(}$\wedge$$I_1$\textsc{)} et
\textit{Supp}\textsc{(}$\vee$$I$\textsc{)} =
\textit{Supp}\textsc{(}$\vee$$I_1$\textsc{)}.
\end{proposition}

D'aprés la proposition précédente, si \textit{bond}\textsc{(}$I$\textsc{)} =
\textit{bond}\textsc{(}$I_1$\textsc{)}, alors \textit{Supp}\textsc{(}$\neg$$I$\textsc{)} =
\textit{Supp}\textsc{(}$\neg$$I_1$\textsc{)}. En effet, $I$ et $I_1$ ont le même support disjonctif et, par la loi de De Morgan, nous avons le lien suivant entre les supports disjonctif et négatif d'un motif : \textit{Supp}\textsc{(}$\neg$$I$\textsc{)} = $|\mathcal{T}|$ - \textit{Supp}\textsc{(}$\vee$$I$\textsc{)}. D'autre part, si \textit{bond}\textsc{(}$I$\textsc{)} $\neq$
\textit{bond}\textsc{(}$I_1$\textsc{)}, alors
\textit{Supp}\textsc{(}$\wedge$$I$\textsc{)} $\neq$
\textit{Supp}\textsc{(}$\wedge$$I_1$\textsc{)} ou
\textit{Supp}\textsc{(}$\vee$$I$\textsc{)} $\neq$
\textit{Supp}\textsc{(}$\vee$$I_1$\textsc{)} \textsc{(}\textit{c.-é.-d.} un des deux supports est différent ou les deux à  la fois\textsc{)}.\\
\begin{exemple}
Considérons le contexte d'extraction donné par la table \ref{Base_transactions}. Nous avons 
\textit{bond}\textsc{(}\texttt{ACD}\textsc{)} = \textit{bond}\textsc{(}\texttt{CD}\textsc{)} = $\displaystyle\frac{1}{4}$.
Les motifs  \texttt{ACD} et \texttt{CD} ont les mêmes supports conjonctifs et disjonctifs, 
\textit{Supp}\textsc{(}$\wedge$\texttt{ACD}\textsc{)} =
\textit{Supp}\textsc{(}$\wedge$\texttt{CD}\textsc{)} = 1 et \textit{Supp}\textsc{(}$\vee$\texttt{ACD}\textsc{)} =
\textit{Supp}\textsc{(}$\vee$\texttt{CD}\textsc{)} = 4. Nous avons \textit{bond}\textsc{(}\texttt{ABCE}\textsc{)} =
$\displaystyle\frac{2}{5}$ $\neq$ \textit{bond}\textsc{(}\texttt{BCE}\textsc{)} =
$\displaystyle\frac{3}{5}$.  En effet, les motifs \texttt{ABCE} et \texttt{BCE} n'ont pas le même support conjonctif, \textit{Supp}\textsc{(}$\wedge$\texttt{ABCE}\textsc{)} = 2 $\neq$ \textit{Supp}\textsc{(}$\wedge$\texttt{BCE}\textsc{)} = 3.
\end{exemple}
Dans la suite, nous aurons besoin de l'ensemble des motifs corrélés maximaux défini formellement comme suit.
\begin{definition} \label{bdpos} \textsc{(}\textbf{Motifs corrélés maximaux}\textsc{)} L'ensemble des motifs corrélés maximaux constitue la bordure positive des motifs corrélés et correspond aux motifs corrélés n'admettant aucun sur-ensemble strict corrélé. Cet ensemble est défini par : $\mathcal{MCM}ax$ = $\{$$I$ $\in$ $\mathcal{MC}$$\mid$ $\forall$ $I_1$ $\supset$ $I$ : $I_1$ $\notin$ $\mathcal{MC}$$\}$, ou d'une maniére équivalente :
$\mathcal{MCM}ax$ = $\{$$I$ $\in$ $\mathcal{MC}$$\mid$ $\forall$ $I_1$ $\supset$ $I$ :
\textit{bond}\textsc{(}$I_1$\textsc{)} $<$ \textit{minbond}$\}$.
\end{definition}
\begin{exemple}\label{exemple_MCMax}
Soit la base illustrée par la table \ref{Base_transactions}. Pour \textit{minbond} = 0,2, nous avons
$\mathcal{MCM}ax$ = $\{$\texttt{ACD}, \texttt{ABCE}$\}$. En effet, quelque soit le sur-ensemble strict de \texttt{ACD} ou de \texttt{ABCE}, ce sur-ensemble n'est pas corrélé.
\end{exemple}
Nous avons ainsi présenté les notions préliminaires des motifs rares et des motifs corrélés selon la mesure \textit{bond}. Dans la suite, nous enchaénons avec l'opérateur de fermeture $f_{bond}$ associé à  la mesure \textit{bond}. Cet opérateur permettra de caractériser les motifs gréce aux classes d'équivalence induites.

\section{Opérateur de fermeture $f_{bond}$ associé à  la mesure \textit{bond}}

Définissons d'abord le concept de l'opérateur de fermeture.
\begin{definition}\label{Opérateurs de fermeture} \textbf{Opérateur de fermeture} \cite{ganter99}\\
Soit un ensemble partiellement ordonné
\textsc{(}$E$, $\leq$\textsc{)}. Une application $f$ de
\textsc{(}$E$, $\leq$\textsc{)} dans
\textsc{(}$E$, $\leq$\textsc{)} est appelée un \textit{opérateur de
	fermeture}, si et seulement si elle posséde les propriétés
suivantes. Pour tout sous-ensemble $S, S'\subseteq E$ :

1. \textit{Isotonie} : $S\leq S' \Rightarrow f\textsc{(}S\textsc{)}
\leq f\textsc{(}S'\textsc{)}$

2. \textit{Extensivité} : $S\leq f\textsc{(}S\textsc{)}$

3. \textit{Idempotence} :
$f\textsc{(}f\textsc{(}S\textsc{)}\textsc{)}=f\textsc{(}S\textsc{)}$
\end{definition}

L'opérateur de fermeture associé à  la mesure \textit{bond} est défini comme suit \cite{tarekds2010} :
\begin{definition} \label{fermeture_fbond1} \textsc{(}\textbf{Opérateur $f_{bond}$}\textsc{)}  {\setlength\arraycolsep{5pt}
	\begin{eqnarray*}
		\large f_{bond} :
		\mathcal{P}\textsc{(}\mathcal{I}\textsc{)}&
		\rightarrow &
		\mathcal{P}\textsc{(}\mathcal{I}\textsc{)}\\
		I & \mapsto &
		f_{bond}\textsc{(}I\textsc{)} =
		I\cup\{i\in\mathcal{I}\setminus I  | \  \textit{bond}\textsc{(}I\textsc{)} = \textit{bond}\textsc{(} I\cup\{i\}\textsc{)}\}
\end{eqnarray*}}
\end{definition}
L'opérateur $f_{bond}$ a été démontré d'étre un opérateur de fermeture \cite{tarekds2010}. En effet, il vérifie les propriétés d'isotonie, d'extensivité et d'idempotence \cite{ganter99}. Le motif fermé d'un motif $I$ par $f_{bond}$, \textit{c.-é.-d.} $f_{bond}\textsc{(}I$\textsc{)}, est ainsi l'ensemble maximal d'items contenant $I$ et ayant la même valeur de la mesure \textit{bond} que $I$.

\begin{exemple}
Soit la base illustrée par la table \ref{Base_transactions}. Pour minbond = 0,2, nous avons \textit{bond}\textsc{(}\texttt{AB}\textsc{)} = $\displaystyle\frac{2}{5}$, \textit{bond}\textsc{(}\texttt{ABC}\textsc{)} = $\displaystyle\frac{2}{5}$, \textit{bond}\textsc{(}\texttt{ABE}\textsc{)} = $\displaystyle\frac{2}{5}$. Ainsi, \texttt{C} $\in$ $f_{bond}$\textsc{(}\texttt{AB}\textsc{)}, et \texttt{E} $\in$ $f_{bond}$\textsc{(}\texttt{AB}\textsc{)}. Par contre, \textit{bond}\textsc{(}\texttt{ABD}\textsc{)} = $\displaystyle\frac{0}{5}$ = 0. Ainsi, \texttt{D} $\notin$ $f_{bond}$\textsc{(}\texttt{AB}\textsc{)}. Par conséquent, $f_{bond}$\textsc{(}\texttt{AB}\textsc{)} = \texttt{ABCE}.

Illustrons les différentes propriétés d'un opérateur de fermeture. Pour la propriété d'isotonie, nous avons \texttt{AB} $\supset$ \texttt{B}, $f_{bond}$\textsc{(}\texttt{AB}\textsc{)} = \texttt{ABCE} et $f_{bond}$\textsc{(}\texttt{B}\textsc{)} = \texttt{BE}.
Concernant la propriété d'extensivité, nous avons par exemple, $f_{bond}$\textsc{(}\texttt{CD}\textsc{)} = \texttt{ACD},
\texttt{CD} $\subseteq$ $f_{bond}$\textsc{(}\texttt{CD}\textsc{)}.
Pour la propriété d'idempotence de l'opérateur de fermeture $f_{bond}$, nous notons l'exemple du motif fermé \texttt{ABCE}, $f_{bond}$\textsc{(}$f_{bond}$\textsc{(}\texttt{ABCE}\textsc{)}\textsc{)} = \texttt{ABCE}.

\end{exemple}

L'application de l'opérateur $f_{bond}$ partitionne l'ensemble des parties de $\mathcal{I}$ en des classes d'équivalence disjointes définies comme suit.

\begin{definition}\label{clsdéquivbond} \textsc{(}\textbf{Classe d'équivalence associée à  l'opérateur de fermeture $f_{bond}$}\textsc{)} Une classe d'équivalence associée à  l'opérateur de fermeture $f_{bond}$ contient un ensemble de tous les motifs possédant la même fermeture par $f_{bond}$.
\end{definition}

Chaque classe d'équivalence est caractérisée par un élément maximal -- un motif fermé -- et un ou plusieurs éléments minimaux -- des motifs minimaux corrélés. Nous définissons formellement ces motifs.
\begin{definition}\label{fermébondcorrélé} \textsc{(}\textbf{Motifs fermés corrélés}\textsc{)} L'ensemble $\mathcal{MFC}$ des motifs fermés corrélés par $f_{bond}$ est défini par : $\mathcal{MFC}$ = $\{$$I$ $\in$ $\mathcal{MC}$$\mid$ $\nexists$ $I_{1}$ $\supset$ $I$ : \textit{bond}\textsc{(}$I$\textsc{)} = \textit{bond}\textsc{(}$I_{1}\textsc{)}\}$, ou d'une maniére équivalente : $\mathcal{MFC}$ = $\{$$I$ $\in$ $\mathcal{MC}$$\mid$ $\nexists$ $I_{1}$ $\supset$ $I$ : $f_{bond}\textsc{(}I$\textsc{)} = $f_{bond}\textsc{(}I_{1}\textsc{)}\}$.
\end{definition}
\begin{exemple}\label{exempleMFC}
Soit la base illustrée par la table \ref{Base_transactions} pour \textit{minbond} = 0,2. Le motif \texttt{ACD} est corrélé puisque \textit{bond}\textsc{(}\texttt{ACD}\textsc{)} = $\displaystyle\frac{1}{4}$ = 0,25 $\geq$ 0,2. Il est aussi fermé puisque il n'admet pas de sur-ensemble strict de même valeur de \textit{bond}. En effet, \textit{bond}\textsc{(}\texttt{ABCD}\textsc{)} = 0, \textit{bond}\textsc{(}\texttt{ACDE}\textsc{)} = 0, et par conséquent \textit{bond}\textsc{(}\texttt{ABCDE}\textsc{)} = 0.
\end{exemple}

\begin{definition} \label{motifmincorreles} \textsc{(}\textbf{Motifs minimaux corrélés}\textsc{)} L'ensemble $\mathcal{MMC}$ des motifs minimaux corrélés est défini par : $\mathcal{MMC}$ = $\{$$I$ $\in$ $\mathcal{MC}$$\mid$ $\nexists$ $I_{1}$ $\subset$ $I$ : \textit{bond}\textsc{(}$I$\textsc{)} = \textit{bond}\textsc{(}$I_{1}\textsc{)}\}$, ou d'une maniére équivalente : $\mathcal{MMC}$ = $\{$$I$ $\in$ $\mathcal{MC}$$\mid$ $\nexists$ $I_{1}$ $\subset$ $I$ : $f_{bond}\textsc{(}I$\textsc{)} = $f_{bond}\textsc{(}I_{1}\textsc{)}\}$.
\end{definition}
\begin{exemple}\label{exempleMMC}
Soit la base illustrée par la table \ref{Base_transactions} pour \textit{minbond} = 0,2. Le motif \texttt{AB} est corrélé puisque \textit{bond}\textsc{(}\texttt{AB}\textsc{)} = $\displaystyle\frac{2}{5}$ = 0,4 $>$ 0,2. Il est aussi minimal puisque \textit{bond}\textsc{(}\texttt{A}\textsc{)} = \textit{bond}\textsc{(}\texttt{B}\textsc{)} = 1.
%%L'ensemble $\mathcal{MMC}$ et les différents ensembles sus-définis sont schématisés par la figure \ref{treillisMcorrélés}.
\end{exemple}

%\begin {figure}
%\parbox{16cm}{
%\hspace{-0.9cm}
%\includegraphics[scale = 0.5]{treillisMcorrélés.eps}}
%\caption{Treillis des motifs associés au contexte $\mathcal{D}$ pour \textit{minbond} = 0,2.}
%\label{treillisMcorrélés}
%\end{figure}

Nous introduisons dans ce qui suit une propriété intéressante de l'idéal d'ordre des motifs minimaux corrélés.
\begin{propriete}\label{propMMC}
L'ensemble $\mathcal{MMC}$ des motifs minimaux corrélés forme un idéal d'ordre.
\end{propriete}

Toutefois, l'ensemble des motifs minimaux corrélés contient les motifs vérifiant la contrainte anti-monotone ``étre minimal dans sa classe d'équivalence et étre corrélé'' \cite{tarekds2010}. En effet, cette derniére résulte de la conjonction de deux contraintes anti-monotones, à  savoir ``étre un motif minimal'' et ``étre un motif corrélé''.
Par conséquent, la contrainte ``étre un motif minimal corrélé'' est anti-monotone et l'ensemble $\mathcal{MMC}$ des motifs minimaux corrélés forme ainsi un idéal d'ordre dans le treillis des motifs.

Nous avons à  ce stade présenté les motifs fermés et minimaux corrélés associés à  une classe d'équivalence induite par $f_{bond}$. Nous introduisons dans la proposition suivante les propriétés communes à  deux motifs appartenant à  une même classe d'équivalence associée à  l'opérateur de fermeture $f_{bond}$.

\begin{proposition}\label{proprietes_CEq_f_bond}
Soit $\mathcal{C}$ une classe d'équivalence associée à  l'opérateur de fermeture $f_{bond}$ et $I$ et $I_1$ $\in$ $\mathcal{C}$. Nous avons : \textbf{a\textsc{)}} $f_{bond}\textsc{(}I$\textsc{)} = $f_{bond}\textsc{(}I_1$\textsc{)}, \textbf{b\textsc{)}} \textit{bond}\textsc{(}$I$\textsc{)} = \textit{bond}\textsc{(}$I_1$\textsc{)}, \textbf{c\textsc{)}} \textit{Supp}\textsc{(}$\wedge$$I$\textsc{)} = \textit{Supp}\textsc{(}$\wedge$$I_1$\textsc{)}, \textbf{d\textsc{)}} \textit{Supp}\textsc{(}$\vee$$I$\textsc{)} = \textit{Supp}\textsc{(}$\vee$$I_1$\textsc{)}, et, \textbf{e\textsc{)}} \textit{Supp}\textsc{(}$\neg$$I$\textsc{)} = \textit{Supp}\textsc{(}$\neg$$I_1$\textsc{)}.
\end{proposition}
\begin{Preuve}
\begin{description}
\item[a\textsc{)}] Gréce à  la définition \ref{clsdéquivbond}, $I$ et $I_1$ ont la même fermeture par $f_{bond}$. Soit $F$ cette fermeture.
\item[b\textsc{)}] Comme l'opérateur de fermeture préserve la valeur de la mesure \textit{bond} d'un motif \textsc{(}\textit{cf.} Définition \ref{fermeture_fbond1}\textsc{)}, et puisque $I$ et $I_1$ ont la même fermeture $F$, nous avons \textit{bond}\textsc{(}$I$\textsc{)} = \textit{bond}\textsc{(}$F$\textsc{)}, et \textit{bond}\textsc{(}$I_1$\textsc{)} = \textit{bond}\textsc{(}$F$\textsc{)}. Ainsi, \textit{bond}\textsc{(}$I$\textsc{)} = \textit{bond}\textsc{(}$I_1$\textsc{)}.
\item[c\textsc{)}, d\textsc{)}, et e\textsc{)}] Comme $I$ $\subseteq$ $F$ et \textit{bond}\textsc{(}$I$\textsc{)} = \textit{bond}\textsc{(}$F$\textsc{)}, d'aprés la proposition \ref{PropBond}, $I$ et $F$ admettent les mêmes supports conjonctif, disjonctif, et négatif. Il en est de même pour $I_1$ et $F$. Ainsi, $I$ et $I_1$ ont les mêmes supports conjonctif, disjonctif, et négatif.
\end{description}
\end{Preuve}
\begin{exemple}
Soit la base illustrée par la table \ref{Base_transactions} et \textit{minbond} = 0,2. Considérons la classe d'équivalence dont le motif fermé corrélé est \texttt{ABCE}. Les motifs minimaux corrélés associés sont \texttt{AB} et \texttt{AE}. Les motifs corrélés, qui ne sont ni des fermés ni des minimaux, sont \texttt{ABE}, \texttt{ABC}, et \texttt{ACE}. Chacun de ces derniers est compris entre un motif minimal et le fermé corrélé. Les motifs de cette classe d'équivalence, partagent la même valeur de la mesure \textit{bond} égale à  $\displaystyle\frac{2}{5}$, le même support conjonctif égal à  2, le même support disjonctif égal à  5 et le même support négatif égal à  0.
\end{exemple}
%\begin{center}
%\begin {figure}\centering
%\includegraphics[scale = 0.35]{clsdéquiv.eps}
%\caption{Classe d'équivalence induite par $f_{bond}$ correspondant au contexte $\mathcal{D}$ pour \textit{minsupp} = 3 et %\textit{minbond} = 0,2.}
%\label{ClsDéquiv}
%\end{figure}
%\end{center}
Ainsi, tous les motifs d'une classe d'équivalence induite par $f_{bond}$ apparaissent dans les mêmes transactions \textsc{(}gréce à  l'égalité du support conjonctif\textsc{)}. En plus, les items associés aux motifs de la classe caractérisent les mêmes transactions. En effet, chacune de ces derniéres contient nécessairement un sous-ensemble non vide de chaque motif de la classe \textsc{(}gréce à  l'égalité du support disjonctif\textsc{)}. Cet opérateur de fermeture lie ainsi l'espace de recherche conjonctif et celui disjonctif \cite{tarekds2010}. Le motif fermé de la classe offre ainsi l'expression la plus spécifique caractérisant ces transactions, tandis qu'un des motifs minimaux représente une des expression les plus générales.
%%Souad : l'explication de cette phrase est que le motif fermé nous informe sur le contenu spécifique \textsc{(}exact\textsc{)} des %%transactions et chacun des motifs minimaux nous informe sur une expression plus générale du contenu des %%transactions.
Nous avons ainsi cerné les propriétés des motifs rares et des motifs corrélés ainsi que les différentes caractéristiques de l'opérateur de fermeture $f_{bond}$ associé à  la mesure \textit{bond}.
Dans ce qui suit, nous décrivons briévement la notion de représentations concises d'un ensemble donné de motifs.

\section{Représentations concises d'un ensemble de motifs}
En effet, l'extraction des motifs intéressants peut étre coéteuse en espace mémoire et en temps de calcul à  cause du nombre important des candidats générés. \`A cet égard,
l'idée consiste à  extraire des ensembles de taille plus réduite mais capables de régénérer l'ensemble total de motifs, ces ensembles sont dits ``Représentations concises''.
Dans le cas oé la régénération s'effectue sans perte d'information alors la représentation concise est dite \emph{exacte}, sinon elle est dite \emph{approximative}. Ces représentations sont formellement définies dans ce qui suit.

\begin{definition}\label{repConcise}\textbf{Représentations concises} \cite{mannila97}\\
Une représentation concise de l'ensemble des itemsets intéressants est un ensemble représentatif de l'ensemble total permettant de le caractériser d'une maniére exacte ou approximative. D'une maniére générale, une représentation $\mathcal{R}$ constitue une couverture parfaite d'un ensemble $\mathcal{E}$ si et seulement si sa taille ne dépasse jamais la taille de l'ensemble $\mathcal{E}$ à  représenter.
\end{definition}
\begin{exemple}
Soit $\mathcal{R}$ une représentation concise d'un ensemble
$\mathcal{E}$ de motifs fréquents. $\mathcal{R}$ est dite \emph{représentation concise exacte},
si à  partir de $\mathcal{R}$, nous sommes capables de déterminer pour un motif quelconque s'il est fréquent ou non et de retrouver son support exact s'il est fréquent.
Par exemple, les motifs fermés fréquents \cite{pasquier99_2005} constituent une représentation concise exacte de l'ensemble des motifs fréquents.

Cependant, la représentation $\mathcal{R}$ est dite \emph{représentation concise approximative}, si elle
est incapable de déterminer d'une maniére exacte le support de tous les motifs de l'ensemble à  représenter. Elle retourne alors une valeur approximative de ce dernier. Par exemple, les motifs fréquents maximaux \cite{maxminer} constituent une représentation concise approximative de l'ensemble des motifs fréquents. En effet, les motifs fréquents maximaux permettent de déterminer la nature de fréquence \textsc{(}fréquent ou rare\textsc{)} d'un motif quelconque mais ne peuvent pas dériver exactement son support.
\end{exemple}
Toutefois nous nous intéressons, dans ce mémoire, à  l'extraction des représentations concises des motifs corrélés rares. Ces derniers vérifient à  la fois la contrainte monotone de rareté et anti-monotone de corrélation. Par souci de précision, nous signalons que la terminologie ``motifs corrélés rares'' est équivalente à  la terminologie ``motifs rares corrélés''.

\section{Conclusion}

Dans ce chapitre, nous avons cerné l'ensemble des notions de bases relatives à  l'extraction des motifs. Ensuite, nous avons étudié les propriétés des motifs rares et des motifs corrélés selon la mesure \textit{bond}. Nous avons, de plus, détaillé les caractéristiques des classes d'équivalence induites par l'opérateur de fermeture $f_{bond}$ associé à  la mesure de corrélation \textit{bond}. Cet opérateur constitue un élément important qui va nous servir dans l'extraction des représentations concises des motifs corrélés rares par rapport à  la mesure \textit{bond}. Le chapitre suivant est dédié à  la présentation de l'état de l'art des approches traitant de l'extraction des motifs sous contraintes.

%%%%%%%%%%%%%%%%%%%%%%%%%%%%%%%%%%%%
\chapter{\'Etat de l'art de la fouille des motifs sous contraintes}\label{chapitre_EdeA1}

\section{Introduction}

Le processus d'extraction de motifs intéressants souffre souvent de la taille élevée de ces motifs extraits. \`A cet égard, il s'est avéré nécessaire d'intégrer des contraintes dans le processus d'extraction afin de réduire le nombre de motifs extraits. Ces contraintes sont traduites par des traitements qui peuvent étre effectués avant, en cours, ou aprés l'étape de fouille. 

\'Etant donné que dans le cadre de ce mémoire, nous nous intéressons aux motifs corrélés rares, alors nous jugeons intéressant de consacrer ce chapitre à  la présentation et l'étude des approches de l'état de l'art s'inscrivant dans le cadre de notre problématique. Il est important de noter que, dans cette étude, nous nous intéressons principalement aux contraintes monotones et aux contraintes anti-monotones. En effet, les motifs corrélés rares que nous étudions dans ce travail correspondent à  la conjonction entre la contrainte anti-monotone de corrélation et monotone de rareté.

Dans ce chapitre, nous abordons ainsi dans la premiére section la présentation des approches d'extraction des motifs rares. Dans la deuxiéme section, nous effectuons un survol des différentes mesures de corrélation et nous étudions les approches d'extraction des motifs corrélés sous contraintes. La troisiéme section sera consacrée à  l'étude des algorithmes d'extraction de motifs sous la conjonction de contraintes de types opposés.

\section{\'Etat de l'art de la fouille des motifs rares}

\subsection{Extraction des motifs rares}

Toutefois, l'ensemble des motifs rares forme un filtre d'ordre dans le treillis des motifs \cite{laszloIJSI2010} et induit
une contrainte monotone. En effet, tous les sur-ensembles d'un motif rare sont rares. Cette propriété de monotonie rends l'extraction des motifs rares plus difficile que l'extraction d'un ensemble vérifiant une contrainte anti-monotone.
\`A cet égard, différents algorithmes
\cite{laszlo07,Haglin07,adda07,kiran2010,laszloIJSI2010,okubo2010}
ont été dédié à  l'extraction d'un sous ensemble ou de l'ensemble total de tous les motifs rares ont été proposés.

Nous citons par exemple, les approches \textsc{MSApriori} \cite{Liu99} et \textsc{RSAA} \cite{yun_03}. Toutefois, l'idée de l'approche \textsc{MSApriori} 
$^\textsc{(}$\footnote{\textsc{MSApriori} est l'acronyme de
	\textbf{M}ultiple \textbf{S}upport \textbf{A}priori.}$^{\textsc{)}}$
consiste à  définir un seuil minimal de support conjonctif \textit{minsupp} pour chaque item par l'utilisateur et à  extraire l'ensemble des motifs fréquents par rapport aux seuils minimaux posés. \`A cet égard, une partie de l'ensemble des motifs rares sera récupérée pour des seuils bas de \textit{minsupp}.
\begin{exemple}
	Considérons la base de transaction donnée par la table \ref{Base_transactions}.
	Dans le cas oé nous considérons, minsupp\textsc{(}\texttt{A}\textsc{)} = $5$ ,  minsupp\textsc{(}\texttt{C}\textsc{)} = $2$ et  minsupp\textsc{(}\texttt{D}\textsc{)} = $3$.
	Le support de l'itemset \texttt{AD} est égal à  $1$, Supp\textsc{(}\texttt{AD}\textsc{)} = $1$. Ainsi, l'itemset
	\texttt{AD} est rare puisque Supp\textsc{(}\texttt{AD}\textsc{)} $<$ $\min$$\{$minsupp\textsc{(}\texttt{A}\textsc{)}, minsupp\textsc{(}\texttt{D}\textsc{)}$\}$ =
	$\min$$\{$$5$, $3$$\}$ = $3$. Le motif \texttt{AC} est cependant fréquent, Supp\textsc{(}\texttt{AC}\textsc{)} = $3$ $>$ $\min$$\{$minsupp\textsc{(}\texttt{A}\textsc{)}, minsupp\textsc{(}\texttt{C}\textsc{)}$\}$ = $\min$$\{5$, $2\}$ = $2$.
\end{exemple}
Cette approche opére à  la \textsc{Apriori} et les itemsets rares ne seront ainsi identifiés que pour des seuils de \textit{minsupp} trés bas. Dans ce cas, le nombre de régles d'association générées sera énorme. De plus, un probléme lié au choix du seuil \textit{minsupp} adéquat pour chaque item et aux coéts d'évaluation d'un candidat par rapport aux différents seuils, est toujours posé.

Dans le but de pallier cet inconvénient, Yun et \textit{al.} ont proposé dans \cite{yun_03} une nouvelle approche intitulée \textsc{RSAA} 
$^\textsc{(}$\footnote{\textsc{RSAA} est l'acronyme de
	\textbf{R}elative \textbf{S}upport \textbf{A}priori \textbf{A}lgorithm.}$^{\textsc{)}}$.
Cette approche permet de générer des régles d'association englobant des itemsets rares.
La technique proposée est basée sur la notion de support relatif et permet de générer des régles d'association dont le support relatif dépasse un seuil minimal donné.
Le support relatif \textit{RSupp} d'un itemset \texttt{I} de la forme \textsc{(}$i_{1}$, $i_{2}$, $\ldots$, $i_{k}$\textsc{)} est noté RSupp\textsc{(}\texttt{I}\textsc{)} et correspond é,

RSupp\textsc{(}$i_{1}$, $i_{2}$, $\ldots$, $i_{k}$\textsc{)} = $\max$\textsc{(}
$\displaystyle\frac{\displaystyle
	\textit{Supp}\textsc{(}i_{1}, i_{2}, \ldots, i_{k}\textsc{)}}{\displaystyle
	\textit{Supp}\textsc{(}i_{1}\textsc{)}}$, $\displaystyle\frac{\displaystyle
	\textit{Supp}\textsc{(}i_{1}, i_{2}, \ldots, i_{k}\textsc{)}}{\displaystyle
	\textit{Supp}\textsc{(}i_{2}\textsc{)}}$, $\ldots$,
$\displaystyle\frac{\displaystyle
	\textit{Supp}\textsc{(}i_{1}, i_{2}, \ldots, i_{k}\textsc{)}}{\displaystyle
	\textit{Supp}\textsc{(}i_{k}\textsc{)}}$\textsc{)}.
\begin{exemple}
	Considérons la base de transaction donnée par la table \ref{Base_transactions}.
	RSupp\textsc{(}\texttt{ACE}\textsc{)} =
	$\max$\textsc{(}$\displaystyle\frac{\displaystyle
		\textit{Supp}\textsc{(}\texttt{ACE}\textsc{)}}{\displaystyle
		\textit{Supp}\textsc{(}\texttt{A}\textsc{)}}$, $\displaystyle\frac{\displaystyle
		\textit{Supp}\textsc{(}\texttt{ACE}\textsc{)}}{\displaystyle
		\textit{Supp}\textsc{(}\texttt{C}\textsc{)}}$,
	$\displaystyle\frac{\displaystyle
		\textit{Supp}\textsc{(}\texttt{ACE}\textsc{)}}{\displaystyle
		\textit{Supp}\textsc{(}\texttt{E}\textsc{)}}$\textsc{)} =
	$\max$\textsc{(}
	$\displaystyle\frac{2}{3}$,  $\displaystyle\frac{2}{4}$, $\displaystyle\frac{2}{4}$\textsc{)} = $\displaystyle\frac{2}{3}$.
	Nous avons RSupp\textsc{(}\texttt{ACE}\textsc{)} = $\displaystyle\frac{2}{3}$, ainsi pour un seuil minimal de support relatif \textit{minsuppR} = $\displaystyle\frac{1}{2}$, le motif \texttt{ACE} est fréquent. Cependant, pour un seuil \textit{minsuppR} = $\displaystyle\frac{3}{4}$, le motif \texttt{ACE} est considéré comme rare.
\end{exemple}
L'algorithme \textsc{RSAA} permet de repérer une partie des motifs rares tout en résolvant le probléme lié à  la définition de plusieurs seuils minimaux
posé par l'approche \textsc{MSApriori}. Cependant, pour des seuils minimaux de supports relatifs trés bas le nombre d'itemsets générés risque d'exploser et d'engendrer ainsi des coéts énormes d'extraction.

L'approche \textsc{Apriori-Inverse} \cite{Koh05} s'inscrit aussi dans ce même cadre.
En effet, \textsc{Apriori-Inverse} est un algorithme opérant en largeur et permet moyennant un parcours du treillis du bas vers le haut d'extraire pour chaque niveau $n$ les motifs rares de taille $n$ dont tous les sous-ensembles sont rares. Ces motifs sont dits ``\textit{motifs parfaitement rares}''.
Toutefois, l'ensemble extrait n'englobe pas tous les motifs rares puisqu'il élimine les motifs rares minimaux, \textit{c.-é.-d.} les motifs rares dont tous les sous-ensembles sont fréquents.
\begin{exemple}
	Considérons le contexte d'extraction donné par la table \ref{Base_transactions}. Pour un seuil \textit{minsupp} = 4. L'algorithme \textsc{Apriori-Inverse} permet d'extraire uniquement les itemsets rares
	dont le support est à  la fois non nul et inférieur à  \textit{minsupp} et dont tous les sous-ensembles
	sont rares. Par conséquent, seuls les itemsets rares \texttt{D}, \texttt{A} et \texttt{AD} seront extraits.
\end{exemple}

D'autres approches traitant la problématique des motifs rares ont été aussi proposé.
Nous citons par exemple, les approches \textsc{Apriori-Rare} et \textsc{Mrg-Exp} \cite{laszlo06,laszlo07} permettant aussi le repérage d'un sous-ensemble des motifs rares composé uniquement des motifs rares minimaux
\textsc{(}\textit{cf.} Définition \ref{mrm} page \pageref{mrm}\textsc{)}, formant la bordure séparatrice entre les motifs fréquents et les motifs rares. Ces approches effectuent l'exploration du treillis des itemsets du bas vers le haut, en commenéant par l'ensemble vide jusqu'é la localisation de la bordure des motifs rares minimaux.

Les approches \textsc{Apriori-Rare} et \textsc{Mrg-Exp} se reposent sur le même principe cependant l'approche \textsc{Mrg-Exp} se différe de l'approche \textsc{Apriori-Rare} par l'intégration d'un critére d'élagage davantage. Ce critére consiste à  éliminer les motifs rares minimaux aprés leur identification de l'ensemble des motifs candidats afin de ne générer des candidats qu'é partir des motifs fréquents.

Considérons, par exemple, le contexte d'extraction donné par la table \ref{Base_transactions} pour \textit{minsupp} = 3. L'ensemble $\mathcal{MRM}in$ des motifs récupérés par les algorithmes \textsc{Apriori-Rare} et \textsc{Mrg-Exp} correspond é,
$\mathcal{MRM}in$ = $\{$\texttt{D}, \texttt{AB}, \texttt{AE}$\}$.

Une fois cet ensemble est identifié, l'ensemble de tous les motifs rares est alors extrait, moyennant l'approche \textsc{Arima} et ce en générant par tailles croissantes les sur-ensembles des motifs appartenant à  la bordure composée par les motifs rares minimaux.
\begin{exemple}
	Par exemple, considérons l'ensemble $\mathcal{MRM}in$ = $\{$\texttt{D}, \texttt{AB}, \texttt{AE}$\}$ précédemment identifié. L'ensemble $\mathcal{MR}$ de tous les motifs rares correspond é,
	$\mathcal{MR}$ = $\{$\texttt{D}, \texttt{AD}, \texttt{CD}, \texttt{ABC}, \texttt{ABE}, \texttt{ACE}$\}$,
	les motifs rares de support nul, à  savoir
	\texttt{BD}, \texttt{DE}, \texttt{ABD}, \texttt{BCD}, \texttt{BDE}, \texttt{ADE} et \texttt{CDE},
	ne feront pas parti du résultat final.
\end{exemple}

Nous notons que dans \cite{romero2010}, les approches \textsc{Apriori-Inverse} et \textsc{Apriori-Rare}
ont été appliquées sur des données concernant des étudiants dans un site d'apprentissage en ligne. Les régles d'association générées permettent de nous renseigner quant au lien entre les activités en ligne de l'étudiant \textsc{(}navigation sur le site, consultation des forums, durée de navigation\textsc{)} et la mention obtenue \textsc{(}Excellent, Bien, Moyen\textsc{)}. Les régles d'association rares permettent de détecter les comportements infréquents des étudiants et leurs relation avec la mention finale.

L'approche \textsc{Afirm} proposée dans \cite{adda07} s'inscrit aussi dans ce même cadre. Cependant, elle différe des approches proposées par Szathmary et \textit{al.} dans la nature du parcours de l'espace de recherche. En effet, l'approche \textsc{Afirm} suggére un parcours du haut vers le bas du treillis \textit{c.-é.-d} du plus grand motif, par rapport à  la relation d'inclusion ensembliste, jusqu'é la localisation des motifs rares minimaux.

\begin{exemple}
	Considérons le contexte d'extraction de la table \ref{Base_transactions} pour \textit{minsupp} = 3.
	L'algorithme \textsc{Afirm} extrait les motifs rares comme suit.
	Initialement, les plus grands motifs rares seront extraits, $\mathcal{MR}$$_{4}$ = $\{$\texttt{ABCE}, \texttt{ABCD}, \texttt{BCDE}, \texttt{ABDE}$\}$. Ensuite, leurs sous-ensembles rares seront générés et nous aurons, $\mathcal{MR}$$_{3}$ = $\{$ \texttt{ABE},
	\texttt{ABD}, \texttt{BCD}, \texttt{CDE}, \texttt{BDE}, \texttt{ADE},\texttt{ACD}$\}$.
	Les motifs de support nul ne sont pas considérés, donc nous aurons,
	$\mathcal{MR}$$_{4}$ = $\{$\texttt{ABCE}$\}$,
	$\mathcal{MR}$$_{3}$ = $\{$\texttt{ABE}, \texttt{ACD}$\}$.
	$\mathcal{MR}$$_{2}$ = $\{$\texttt{AB}, \texttt{AE}, \texttt{CD}, \texttt{AD}$\}$ et
	$\mathcal{MR}$$_{1}$ = $\{$\texttt{D}$\}$.
\end{exemple}

L'approche \textsc{Minit} 
$^\textsc{(}$\footnote{\textsc{Minit} est l'acronyme de
	\textbf{Min}imal \textbf{I}nfrequent I\textbf{t}emset.}$^{\textsc{)}}$ \cite{Haglin07},
a été aussi proposée et permet de récupérer aussi les motifs rares minimaux dont la taille ne dépasse pas un certain seuil de cardinalité maximale \textit{S}.
L'idée de l'algorithme \textsc{Minit} se déroule de la maniére suivante. D'abord, les items sont triés selon l'ordre croissant du support. Ensuite, nous commenéons par traiter
l'item ayant le plus petit support en lui associant son sous-contexte
et en identifiant la liste des items triés appartenant à  ce sous-contexte et un appel récursif de l'algorithme \textsc{Minit} sera réalisé et le seuil $S$ sera décrémenté. Cette procédure se termine lorsque
le seuil $S$ devient égal à  1.

\begin{figure}
	\begin{center}
		\includegraphics[scale = 0.5]{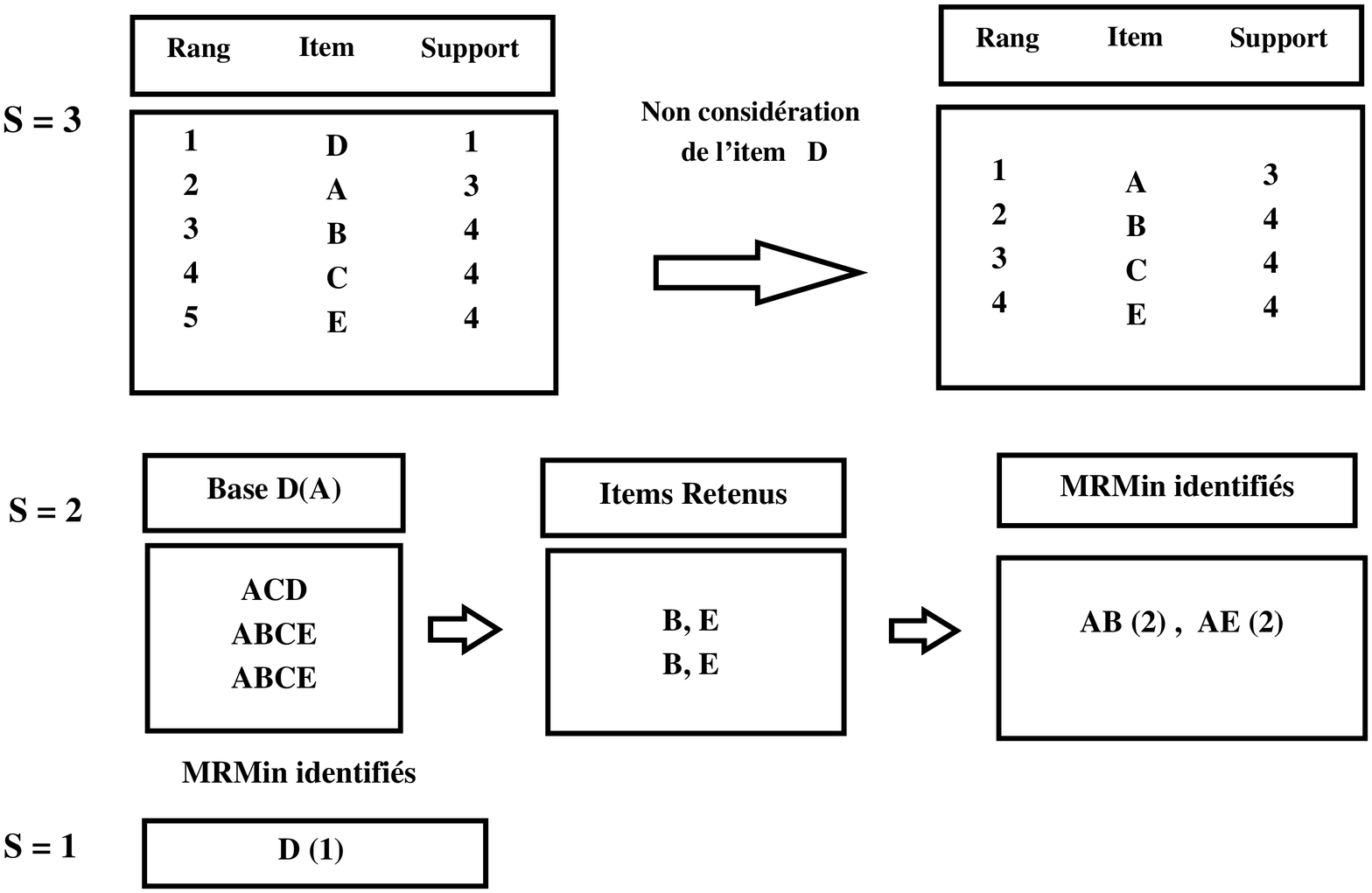}
	\end{center}
	\caption{Trace d'exécution de l'algorithme \textsc{Minit} pour \textit{minsupp} = $3$ et \textit{S} = $3$. \label{minitExp}}
\end{figure}

\begin{exemple}
	Considérons le contexte donné par la table \ref{Base_transactions}. Pour \textit{minsupp} = $3$ et \textit{S} = $3$, le déroulement de l'algorithme \textsc{Minit} est donné par la figure \ref{minitExp}.
	Les motifs rares minimaux de cardinalité inférieur à  $3$ seront ainsi extraits.
	Initialement, l'item \texttt{D} ne sera pas considéré puisqu'il ne donne pas naissance à  un motif rare minimal de taille $2$. Ensuite, nous considérons l'item \texttt{A} et le contexte associé Base D\textsc{(}\texttt{A}\textsc{)} sera crée et la variable \textit{S} sera décrémentée \textit{S} = 2. Les items retenus sont \texttt{B} et \texttt{E}, l'item \texttt{C}  ne sera pas retenu puisqu'il appartient à  toutes les transactions du contexte
	Base D\textsc{(}\texttt{A}\textsc{)}. Les motifs rares minimaux ainsi identifiés sont \texttt{AE} et \texttt{AB} de support $2$ chacun. Par la suite, la variable  \textit{S} sera décrémentée, \textit{S} = $1$,  alors l'algorithme identifie le seul item rare \texttt{D} de support $1$. Les motifs rares minimaux ainsi récupérés correspondent é, $\mathcal{MRM}in$ = $\{$\textsc{(}\texttt{AE}, 2\textsc{)}, \textsc{(}\texttt{AB}, 2\textsc{)}, \textsc{(}\texttt{D}, 1\textsc{)}$\}$.
\end{exemple}

Nous avons ainsi passé en revue les approches les plus représentatives traitant la problématique d'extraction d'un sous-ensemble ou de l'ensemble total des motifs rares.
Toutefois, la fouille des motifs rares souffre d'un deuxiéme probléme lié au nombre des motifs rares qui risque d'étre trés élevé. L'explosion du nombre des motifs rares empéche certainement une exploitation efficace des connaissances cachées dans ces motifs.
Pour pallier cet inconvénient, deux représentations concises exactes des motifs rares ont été défini dans \cite{qdc2011} et sont basées sur la notion de classe d'équivalence.
La premiére représentation concise exacte est composée des éléments minimaux des classes d'équivalences à  savoir les générateurs minimaux rares et est extraite gréce à  l'algorithme \textsc{GMRare}. La deuxiéme représentation est composée des éléments maximaux des classes d'équivalences à  savoir les motifs fermés rares et est extraite moyennant l'algorithme
\textsc{MFRare}.
\subsection{Discussion}

Nous récapitulons dans le tableau \ref{TabCmp0} les caractéristiques des différentes approches étudiées. Cette comparaison couvre les axes suivants :

\begin{enumerate}
	\item \textbf{Motifs extraits} : Cette propriété décrit les motifs générés par l'algorithme.
	\item \textbf{Caractéristiques} :  Cette propriété décrit les caractéristiques de l'approche en question.
\end{enumerate}

\begin{table}[!t]
	\parbox{16cm}{
		\hspace{0.3cm}
		\footnotesize{
			\begin{tabular}{|l||c|c|c|l|}
				\hline
				\multicolumn{1}{|c||}{\textbf{Algorithme}}  & \textbf{Motifs} &\textbf{Caractéristiques}\\
				\multicolumn{1}{|c||}{\textbf{d'extraction}} & \textbf{extraits} &\\
				\hline\hline
				$\bullet$ \textsc{MSApriori}&Uniquement une& $-$ Approches non dédiées\\
				\cite{Liu99}&partie des motifs rares&é la fouille des motifs rares\\
				$\bullet$ \textsc{RSAA}&est extraite pour&$-$ Seule une partie de ces\\
				\cite{yun_03}&un \textit{minsupp} faible &derniers est récupérée\\
				&&$-$ Définition compliquée \\
				&& de la formule du support relatif\\
				\hline\hline	
				$\bullet$ \textsc{Apriori-Inverse}& Motifs parfaitement &$-$ Coéts d'extraction élevés\\
				\cite{Koh05}& rares &cependant seuls les motifs\\
				&&parfaitement rares seront extraits.\\
				\hline
				$\bullet$ \textsc{Apriori-Rare}& Motifs &$-$ Parcours coéteux de\\
				\cite{laszlo06}& rares minimaux&l'ensemble de tous les motifs fréquents\\
				$\bullet$ \textsc{Mrg-Exp}&&$-$ La nécessité d'un deuxiéme algorithme\\
				\cite{laszlo07}&&pour la récupération de tous les motifs rares\\
				\hline	
				$\bullet$ \textsc{Arima}& Motifs rares &$-$ Dans le cas oé le\\
				\cite{laszlo06}&&le nombre de motifs rares\\
				&&est réduit, ces derniers\\
				&&sont localisés dans le plus haut\\
				&&du treillis ainsi le parcours\\
				&&du bas vers le haut proposé\\
				&&n'est pas efficace.\\
				\hline
				$\bullet$ \textsc{Afirm}&Motifs rares&$-$ Nécessité de machines \\
				\cite{adda07}&&de hautes capacités afin de\\
				&&pouvoir exécuter l'algorithme\\
				\hline	
				$\bullet$ \textsc{Minit}\textsc{(}\textit{S}\textsc{)}& Motifs rares&$-$ Extraction d'une partie\\
				\cite{Haglin07} & minimaux de taille&des motifs rares minimaux et\\
				& inférieur a \textit{S}&non la totalité\\
				\hline\hline
				$\bullet$ \textsc{MFRare}& Motifs fermés  &$-$ Les représentations concises proposées\\
				\cite{qdc2011}& et motifs rares&sont sans perte d'information.\\
				&&\\
				$\bullet$ \textsc{GMRare}&Générateurs minimaux.&\\
				\cite{qdc2011}&rares&\\
				\hline						
	\end{tabular}}}
	\caption{Tableau comparatif des algorithmes de fouille des motifs
		rares.}\label{TabCmp0}
\end{table}

D'aprés l'étude des approches de l'état de l'art que nous venons de présenter, nous concluons qu'il n'existe pas dans la littérature un algorithme efficace d'extraction de l'ensemble de tous les motifs rares pour tous les types des contextes ou même pour un type particulier de contextes. Ceci est dé à  la difficulté de localiser la bordure séparant les motifs rares de ceux fréquents dans le treillis. En effet, cette bordure est dépendante du contexte et du seuil minimal du support \textit{minsupp}.
La problématique liée à  l'explosion du nombre de motifs rares a été aussi abordée dans la littérature. En effet, deux représentations concises exactes des motifs rares ont été proposées dans \cite{qdc2011}.
Ces derniéres autorisent une réduction sans perte d'information de l'ensemble total des motifs rares, et constituent ainsi une alternative au probléme de nombre élevé des motifs rares.

Toutefois, un autre probléme lié à  l'absence de corrélation entre les motifs rares retenus, est toujours posé. Par exemple, le motif composé par les items ``$Lait$'' et ``$Caviar$'' est un motif rare cependant aucune corrélation n'existe entre le produit ``$Lait$'' trés fréquemment acheté et le produit ``$Caviar$'' cher et rarement acheté.
En effet, l'information extraite à  partir des motifs rares est souvent appelée à  présenter une valeur ajoutée aux experts dans plusieurs domaines. Nous citons comme exemple, la détection des pannes dans les réseaux informatiques \cite{ma-icdm2001}, la découverte des irrégularités dans les actions boursiéres, la détection de fraudes dans les systémes financiers \cite{Cohen_mcr_2000}, la détection d'intrusions dans les systémes informatiques.

Ainsi, il est intéressant d'intégrer les mesures de corrélation dans l'extraction des motifs rares. Ceci permet, d'une part, d'améliorer la qualité des motifs extraits en ayant un ensemble plus réduit contenant des motifs intéressants qui sont rares mais fortement corrélés. D'autre part, ceci renforce la qualité des régles d'association dérivées à  partir de ces motifs corrélés rares. Par exemple, le motif composé par les items ``Collier en or'' et ``Boucles d'oreilles'' ou aussi l'exemple de
``Télévision'' et ``Lecteur DVD'' correspondent à  des motifs corrélés rares. Ces derniers présentent des articles chers dont la vente est bénéfique pour les commeréants.

Nous constatons, ainsi, que la fouille des motifs corrélés rares est une piste trés intéressante à  exploiter. De  plus, cette problématique n'a pas été abordée auparavant dans la littérature. Par conséquent, nous proposons dans ce mémoire d'étudier profondément les caractéristiques des motifs corrélés rares selon la mesure de corrélation \textit{bond} et de proposer de nouvelles représentations concises exactes de ces motifs.

Comme nous avons passé en revue les différentes approches de fouille des motifs rares dans cette section. Nous nous focalisons alors, dans la section suivante, sur la présentation des approches de fouille des motifs corrélés.

\section{\'Etat de l'art de la fouille des motifs corrélés}

Dans cette section, nous étudions les différentes approches d'extraction des motifs corrélés. Nous commenéons d'abord par un survol non exhaustif des différentes mesures de corrélations et de leurs propriétés.
Nous nous sommes principalement basés sur l'étude des mesures de corrélations présentée dans \cite{CARI2010}.
\subsection{Mesures de corrélation}
L'intégration des mesures de corrélations dans le processus d'extraction de motifs permet
de réduire le nombre de motifs extraits tout en améliorant leur qualité.
En effet, seuls les motifs véhiculant le maximum d'informations concernant la corrélation entre les items les composant seront maintenus. \`A cet égard, différentes mesures de corrélation ont été proposé dans la littérature.\\

$\bullet$ \textbf{La mesure de corrélation \textit{any-confidence}} \cite{Omie03}

Cette mesure est définie, pour tout motif non vide $X$ $\subseteq$ $\mathcal{I}$ comme suit :

\begin{center}
	\textit{any-conf}\textsc{(}$X$\textsc{)} = $\displaystyle\frac{\displaystyle
		\textit{Supp}\textsc{(}\wedge X\textsc{)}}{\displaystyle
		\textit{min} \{ \textit{Supp}\textsc{(}\wedge i\textsc{)} | i\in X
		\}}$
\end{center}
La mesure \textit{any-confidence} ne conserve pas les propriétés intéressantes d'anti-monotonie et de cross-support \cite{CARI2010}.
\begin{exemple}
	Considérons la base des transactions illustrée par la table
	\ref{Base_transactions}. Pour un seuil minimal de
	\textit{any-confidence} de \textit{0,80}.
	La valeur de la mesure \textit{any-confidence} du motif \textsl{AB} est égal é,
	\textit{any-confidence}\textsc{(}\texttt{AB}\textsc{)} =
	$\displaystyle\frac{\displaystyle
		\textit{Supp}\textsc{(}\wedge \texttt{AB}\textsc{)}}{\displaystyle
		\textit{min} \{ \textit{Supp}\textsc{(}\wedge \texttt{A}\textsc{)},
		\textit{Supp}\textsc{(}\wedge \texttt{B}\textsc{)}\}}$
	= $\displaystyle\frac{\displaystyle 2}
	{\displaystyle \textit{min}\{3, 4\}}$ = \textit{0,66}.
	Le motif \textsl{AB} ne satisfait le seuil de
	\textit{any-confidence}, par conséquent il n'est pas un motif corrélé alors que le motif \textsl{AD} est corrélé et sa valeur de la mesure \textit{any-confidence} est égale à  1.
	Nous avons aussi,
	$\displaystyle\frac{\displaystyle
		\textit{Supp}\textsc{(}\wedge\textsl{A}\textsc{)}}{\displaystyle
		\textit{Supp}\textsc{(}\wedge\textsl{C}\textsc{)}}$ $=$
	$\displaystyle\frac{\displaystyle 3}
	{\displaystyle 4}$ = \textit{0,75} $<$ \textit{0,80}, cependant,
	\textit{any-confidence}\textsc{(}\textsl{AD}\textsc{)} $=$
	\textit{1} $>$ \textit{0,80}.
	Cet exemple illustre la non conservation des propriétés d'anti-monotonie et de cross-support.
\end{exemple}

$\bullet$ \textbf{La mesure de corrélation \textit{all-confidence}} \cite{Omie03}

La mesure \textit{all-confidence}  est définie pour tout motif non vide $X$ $\subseteq$ $\mathcal{I}$ comme suit :

\begin{center}
	\textit{all-conf}\textsc{(}$X$\textsc{)} =
	$\displaystyle\frac{\displaystyle\textit{Supp}\textsc{(}\wedge
		X\textsc{)}}{\displaystyle \textit{max} \{
		\textit{Supp}\textsc{(}\wedge i\textsc{)} | i\in X \} }$
\end{center}

La mesure \textit{all-confidence} conserve la propriété d'anti-monotonie \cite{Omie03} ainsi que la propriété de cross-support \cite{Xiong06hypercliquepattern}.

\begin{exemple}
	Considérons la base des transactions illustrée par la table
	\ref{Base_transactions}. Pour un seuil minimal de
	\textit{all-confidence} de \textit{0,40}.
	La valeur de la mesure \textit{all-confidence} du motif \textsl{ABCE} est égal é,
	\textit{all-confidence}\textsc{(}\texttt{ABCE}\textsc{)} =
	$\displaystyle\frac{\displaystyle
		\textit{Supp}\textsc{(}\wedge \texttt{ABCE}\textsc{)}}{\displaystyle
		\textit{max} \{ \textit{Supp}\textsc{(}\wedge \texttt{A}\textsc{)},
		\textit{Supp}\textsc{(}\wedge \texttt{B}\textsc{)},
		\textit{Supp}\textsc{(}\wedge \texttt{C}\textsc{)},
		\textit{Supp}\textsc{(}\wedge \texttt{E}\textsc{)}\}}$
	= $\displaystyle\frac{\displaystyle 2}
	{\displaystyle \textit{max}\{3, 4\}}$ = \textit{0,50}.
	Le motif \textsl{ABCE} est corrélé selon la mesure
	\textit{all-confidence}.
	Tous ses sous-ensembles directs sont aussi des motifs corrélés.
	Nous avons \textit{all-confidence}\textsc{(}\texttt{ABE}\textsc{)} =
	\textit{all-confidence}\textsc{(}\texttt{ACE}\textsc{)} =
	$\displaystyle\frac{\displaystyle 2}
	{\displaystyle 4}$ = \textit{0,50},
	\textit{all-confidence}\textsc{(}\texttt{BCE}\textsc{)} = $\displaystyle\frac{\displaystyle 3}
	{\displaystyle 4}$ = \textit{0,75}.
	
	Pour le motif \texttt{AD}, nous avons
	$\displaystyle\frac{\displaystyle
		\textit{Supp}\textsc{(}\wedge\textsl{D}\textsc{)}}{\displaystyle
		\textit{Supp}\textsc{(}\wedge\textsl{A}\textsc{)}}$ $=$
	$\displaystyle\frac{\displaystyle 1}
	{\displaystyle 3}$ = \textit{0,33} $<$ \textit{0,40} et nous avons
	\textit{all-confidence}\textsc{(}\texttt{AD}\textsc{)} =
	$\displaystyle\frac{\displaystyle 1}
	{\displaystyle 3}$ = \textit{0,33}. Le motif \texttt{AD} ne vérifie pas la propriété de cross-support et il n'est ainsi pas corrélé.
	Cet exemple illustre la conservation des propriétés d'anti-monotonie et de cross-support.
\end{exemple}

$\bullet$ \textbf{La mesure de corrélation \textit{hyper-confidence}} \cite{Xiong06hypercliquepattern}

La mesure \textit{ hyper-confidence} notée \textit{h-conf}
d'un motif $X$ $=$ $\{$
\textit{$i_{1}$}, \textit{$i_{2}$}, \ldots, \textit{$i_{m}$}$\}$
est définie comme suit :

\begin{center}
	\textit{h-conf}\textsc{(}$X$\textsc{)} = min
	$\{$\textit{Conf}\textsc{(}$i_{1}\Rightarrow i_{2}, i_{3}, \ldots,
	i_{m}$\textsc{)}, \ldots, \textit{Conf}\textsc{(}$i_{m}\Rightarrow
	i_{1}, i_{2}, \ldots, i_{m-1}$\textsc{)}$\}$.
\end{center}
Avec \textit{Conf} fait référence à  la mesure \textit{confiance}
associée aux régles d'association $^{\textsc{(}}$\footnote{Soit une régle d'association $R$ : $A$ $\Rightarrow$ $B$, avec $A$, $B$ $\subseteq$ $\mathcal{I}$ : \textit{Conf}\textsc{(}$R$\textsc{)} = $\displaystyle \displaystyle\frac{\textit{Supp}\textsc{(}A \cup B\textsc{)}}{\textit{Supp}\textsc{(}A\textsc{)}}$.}$^{\textsc{)}}$.

La mesure \textit{hyper-confidence} est équivalente à  la mesure \textit{all-confidence} et vérifie ainsi les  propriétés de d'anti-monotonie et de cross-support. \\

$\bullet$ \textbf{Le coefficient de corrélation $\chi^2$} \cite{Brin97}

Le coefficient \textit{$\chi^2$} d'un
motif $Z$ $=$ $ab$, avec $a$ et $b$ $\in$ $\mathcal{I}$, est défini
comme suit $^{\textsc{(}}$\footnote{$\textit{Supp}_r\textsc{(}\wedge\textit{I}\textsc{)}$ = $\displaystyle\frac{\displaystyle \textit{Supp}\textsc{(}\wedge\textit{I}\textsc{)}}{\displaystyle
		|\mathcal{T}|}$ désigne le support relatif de $\textit{I}\subseteq
	\mathcal{I}$.}$^{\textsc{)}}$ :

\begin{center}
	\textit{$\chi^2$}\textsc{(}$Z$\textsc{)} $=$ $|\mathcal{T}|$
	$\times$ $\displaystyle\frac{\displaystyle
		\textsc{(}\textit{Supp}_r\textsc{(}\wedge ab\textsc{)} -
		\textit{Supp}_r\textsc{(}\wedge
		a\textsc{)}\times\textit{Supp}_r\textsc{(}\wedge
		b\textsc{)}\textsc{)}^{2}}{\displaystyle\textit{Supp}_r\textsc{(}\wedge
		a\textsc{)}\times\textit{Supp}_r\textsc{(}\wedge
		b\textsc{)}\times\textsc{(}1 - \textit{Supp}_r\textsc{(}\wedge
		a\textsc{)}\textsc{)}\times\textsc{(}1 -
		\textit{Supp}_r\textsc{(}\wedge b\textsc{)}\textsc{)}}$
\end{center}
\begin{exemple}
	Tout motif corrélé selon la mesure $\chi^2$, tous ses sur-ensembles sont aussi corrélés.
	Le calcul de la valeur de $\chi^2$ pour un motif donné nécessite la construction de la table de contingence associée.
	Considérons par exemple, le motif \texttt{AC}, la table de contingence qui lui est associée est donnée par la table \ref{TC1}. Par $S_{l}$ et $S_{c}$, nous entendons respectivement la somme de la ligne et la somme de la colonne.
	
	Nous avons $\chi^2$ \textsc{(}\texttt{AC}\textsc{)} .
	=
	$\displaystyle\frac{\textsc{(}3 - 4 \times  \displaystyle\frac{3}{5}\textsc{)}^{2}}{4 \times \displaystyle\frac{3}{5}}$
	+
	$\displaystyle\frac{\textsc{(}1 - 4 \times  \displaystyle\frac{2}{5}\textsc{)}^{2}}{4 \times  \displaystyle\frac{2}{5}}$
	+
	$\displaystyle\frac{\textsc{(}0 - 1 \times \displaystyle\frac{3}{5}\textsc{)}^{2}}{1 \times  \displaystyle\frac{3}{5}}$
	+
	$\displaystyle\frac{\textsc{(}1 -  1 \times \displaystyle\frac{2}{5}\textsc{)}^{2}}{1 \times \displaystyle\frac{2}{5}}$
	= \textit{2,325}
\end{exemple}

\begin{table}[h]
	\begin{center}
		\footnotesize{
			\begin{tabular}{|c||c|c|c|c|}
				\hline &  $A$   & $\overline{A}$ & $S_{l}$ \\
				\hline\hline
				$C$           &   3          &   1                     &   4       \\
				\hline
				$\overline{C}$&       0                  &  1    &   1       \\
				\hline
				$S_{c}$&       3       &    2  &     5     \\
				\hline
		\end{tabular}}
	\end{center}
	\caption{Table de contingence de l'itemset `\texttt{AC}' associé au contexte $\mathcal{D}$.}\label{TC1}
\end{table}

D'autres mesures de corrélation existent aussi, comme par exemple la mesure \textit{bond} \cite{tarekds2010} que nous avons décrit dans le premier chapitre, la mesure \textit{cosine} \cite{Cosine05}, la mesure lift \cite{Brin97}, le coefficient de corrélation $\phi$, appelé aussi coefficient de corrélation de Pearson \cite{XiongKDD2004}. 

%%le coefficient $\phi$ et le coefficient $\chi^2$ sont deux mesures statistiques et symétriques et ne %%vérifient pas la propriété de cross-support.\\
Nous avons ainsi présenté quelques mesures de corrélation existantes dans la littérature. Nous enchaénons, dans ce qui suit, avec quelques approches d'extraction des motifs corrélés.

\subsection{Extraction des motifs corrélés}

Trés peu d'approches de la littérature ont abordé la problématique de la fouille des motifs rares corrélés.
Toutefois, une premiére idée naéve de récupération des motifs corrélés rares, consiste à  extraire l'ensemble de tous les motifs corrélés sans aucune intégration de la contrainte de \textit{minsupp}. Cet ensemble englobe bien évidemment tous les motifs corrélés, ceux qui sont rares et ceux qui sont fréquents.

L'approche proposée dans \cite{Cohen_mcr_2000} est fondée sur ce principe. Cette approche permet
d'extraire les paires d'items corrélés selon la mesure de corrélation \textit{Similarity}. En effet, la mesure \textit{Similarity} permet de mesurer la similarité entre deux items et correspond au rapport entre le nombre d'apparitions simultanée des deux items et le nombre de leurs apparitions complémentaires.
Par conséquent, la mesure \textit{Similarity} est équivalente à  la mesure de corrélation \textit{bond}.
Cependant aucune analyse des propriétés de cette mesure n'a été menée.
Toutefois, l'approche proposée permet d'extraire les paires d'items corrélés sans calculer leurs supports.
En effet, elle affecte à  chaque item une signature composée par la liste des identificateurs des transactions auxquelles il appartient. Ensuite, la valeur de la similarité de chaque paire d'items, correspond au nombre d'intersections de leurs signatures divisé par leur union.
Nous notons que, la contrainte de support minimum n'a pas été intégrée et ceci afin d'avoir les itemsets qui sont fortement corrélés et ayant un faible support. \`A partir de ces derniers,
les régles d'association de confiances élevées et de supports faibles seront générées. Ces régles englobent des pépites de connaissances qui trés intéressantes dans plusieurs domaines. Par exemple, l'algorithme proposé a été appliqué sur des articles de presse afin d'extraire les paires de mots apparaissant ensemble et permettant de fournir des informations pertinentes quant au contenu de l'article en question.

Dans ce même cadre, nous citons aussi l'algorithme \textsc{DiscoverMPatterns} proposé dans \cite{ma-icdm2001}. En effet, cet algorithme permet d'extraire tous les motifs corrélés selon la mesure \textit{all-confidence}.
Ces motifs vérifient la propriété d'anti-monotonie.
En effet, une premiére version de l'approche proposée permet d'extraire tous les motifs corrélés sans restriction de la valeur du support conjonctif afin de récupérer les motifs rares fortement corrélés.
Ensuite, dans une deuxiéme version de l'approche, la contrainte anti-monotone de support minimum a été intégrée dans la fouille des motifs corrélés. Par conséquent, l'ensemble de tous les motifs fréquents corrélés résultant de la jointure entre les deux contraintes anti-monotones de fréquence et de corrélation est récupéré.

Une autre idée de récupération d'une partie des motifs corrélés rares consiste
é extraire l'ensemble de tous les motifs fréquents
pour un seuil de \textit{minsupp} trés bas. Cet ensemble englobe une partie des motifs corrélés qui sont trés peu fréquents.

Xiong et \textit{al.} se sont basés sur ce principe et ont introduit l'algorithme \textsc{HypercliqueMiner} dans \cite{Xiong06hypercliquepattern}. Ce dernier permet d'extraire d'une maniére efficace l'ensemble des motifs fréquents corrélés pour des valeurs faibles du seuil minimal \textit{minsupp}.
Par conséquent, une partie de l'ensemble des motifs corrélés rares sera récupérée parmi les motifs extraits.

En effet, les performances considérables de cet algorithme sont justifiés par les propriétés d'anti-monotonie et de cross-support vérifiées par la mesure de corrélation \textit{hyper-confidence}.
Toutefois, ces deux propriétés permettent d'élaguer significativement les candidats et de réduire les coéts inutiles d'évaluation de la contrainte de corrélation.

Dans ce même cadre, se situe aussi l'approche proposée dans \cite{thomo_2010}. Cette approche permet d'extraire des itemsets fréquents et fortement corrélés de taille deux. C'est une approche naéve dont l'idée consiste à  extraire d'abord tous les motifs fréquents de taille deux pour un \textit{minsupp} trés faible gréce à  une solution paralléle. Ensuite, un post traitement a été effectué afin de ne garder que les itemsets présentant une forte corrélation entre les deux items les composants. En effet, le post traitement permet de ne maintenir que les itemsets pour lesquelles
la probabilité conditionnelle d'apparition d'un de ses items en fonction de l'autre dépasse un seuil minimal défini. Cette probabilité correspond, en effet, à  la mesure \textit{all-confidence}. Nous citons aussi l'algorithme \textsc{FT-Miner} \cite{ptminer} d'extraction des motifs fréquents corrélés par rapport à  une mesure de corrélation, nommée \textit{N-confidence}, équivalente à  la mesure \textit{all-confidence}. Cet algorithme permet d'extraire ces motifs pour des seuils de \textit{minsupp} trés bas et effectue l'extraction des régles associatives à  partir de ces motifs.

L'algorithme \textsc{Partition} \cite{Omie03} a été aussi introduit afin d'extraire les motifs qui satisfont soit le seuil minimal de \textit{all-confidence} soit le seuil minimal de \textit{bond} et ce en fonction de l'exigence de l'utilisateur. Cet algorithme, dont le principal atout est d'allier deux mesures, permet d'extraire tous les motifs corrélés indépendamment de leur statut de fréquence.

D'autres travaux de la littérature se sont intéressés à  la fouille des motifs corrélés qui respectent la contrainte de fréquence. \`A cet égard, nous citons l'algorithme \textsc{CoMine} \cite{comine_Lee}. Cet algorithme permet de récupérer l'ensemble des motifs corrélés fréquents. Deux versions de l'algorithme \textsc{CoMine} sont à  distinguer. La premiére version considére la mesure de corrélation \textit{all-confidence} alors que la deuxiéme version traite la mesure \textit{bond}. Dans ce même cadre, nous citons aussi l'algorithme récent \textsc{Jim} \cite{borgelt}. \textsc{Jim} permet d'extraire selon le choix de l'utilisateur, les motifs fréquents corrélés, les motifs fréquents corrélés maximaux et les motifs fermés fréquents corrélés. Les auteurs proposent la mesure de corrélation \textit{bond} et onze autre mesures de corrélations qui sont toutes anti-monotones et l'extraction des motifs se fait selon la mesure de corrélation choisit par l'utilisateur.

La mesure de corrélation \textit{bond} a été aussi utilisée dans les travaux proposés dans \cite{MCROkubo}. En effet, les auteurs dans \cite{MCROkubo} recourent à  la mesure de corrélation \textit{bond} et à  la notion de la fermeture conjonctive afin d'introduire un algorithme en profondeur
permettant de récupérer les motifs fermés conjonctifs rares.
Toutefois, les auteurs se sont basés sur le fait qu'un motif faiblement corrélé par rapport à  la mesure \textit{bond}
est généralement rare dans la base d'extraction.
\`A cet égard, la contrainte de la rareté est exprimée en fonction de la mesure \textit{bond} comme suit,
bond\textsc{(}$M$\textsc{)} $\leqslant$ $\delta$, avec $\delta$ un seuil de corrélation maximal fixé par l'utilisateur.
Cette contrainte de corrélation maximale est monotone puisqu'elle correspond à  l'opposée de la contrainte anti-monotone de corrélation.

Dans le but d'atténuer l'explosion du nombre de motifs rares extraits et afin de se débarrasser des motifs qui sont trés rares dans la base \textit{c.-é.-d.} les motifs qui représentent juste des exceptions et ne sont pas informatifs, les auteurs ont posé une restriction quant à  la rareté des motifs extraits.
Cette restriction est traduite pour tout motif $M$ par une fonction reliant son support conjonctif et le support minimal de ses items, Freq\textsc{(}$M$\textsc{)} = $\alpha$ $\times$ Supp\textsc{(}$M$\textsc{)} + $\beta$ $\times$ $\min$$\{$Supp\textsc{(}$m_{i}$\textsc{)}, $\forall$ $m_{i}$ $\in$ $M$$\}$ avec $\alpha$ et $\beta$ sont deux coefficients fixés par l'utilisateur.
L'idée consiste à  extraire les $N$ premiers motifs respectant la contrainte de corrélation et maximisant la contrainte de fréquence posées. Ces motifs correspondent aux motifs rares les plus informatifs dans la base.

Nous avons ainsi passé en revue différents algorithmes d'extraction de motifs corrélés. Néanmoins, les ensembles des motifs corrélés extraits risque d'étre de taille élevée. Ce qui constitue un obstacle pour une bonne exploitation des connaissances offertes par ces motifs. \`A cet égard, l'idée de concevoir des approches réductrices de l'ensemble des motifs corrélés est d'une grande pertinence. Dans ce sens, nous discutons dans ce qui suit les algorithmes proposés en vue d'extraire des représentations concises des motifs corrélés fréquents.

\subsection{Extraction des représentations concises des motifs corrélés fréquents}

Rares sont les travaux de la littérature traitant la problématique d'extraction des représentations concises des motifs corrélés. Nous citons dans ce cadre uniquement deux approches.
Premiérement, l'algorithme \textsc{CCMine} \cite{ccmine_Kim}
qui permet ainsi d'extraire une représentation
concise exacte des motifs corrélés fréquents associés à  la mesure
\textit{all-confidence}. La représentation concise extraite est basée sur les motifs fermés corrélés fréquents.
Le deuxiéme algorithme \textsc{CCPR\_Miner} a été proposé dans \cite{tarekds2010}. Cet algorithme
permet d'extraire une nouvelle représentation concise exacte des motifs fréquents corrélés associée à  la mesure \textit{bond}. Cette représentation est composée des motifs fermés corrélés fréquents munis de leurs supports conjonctifs et disjonctifs. Elle permet de dériver, pour tout motif corrélé fréquent, exactement le support conjonctif, le support disjonctif et par conséquent la valeur de la mesure \textit{bond} égale au rapport des deux supports conjonctifs et disjonctifs.

Dans ce qui suit nous étudions les algorithmes traitant de la fouille de motifs corrélés sous
la conjonction de contraintes de types opposés.

\subsection{Extraction des motifs corrélés sous la conjonction de contraintes de types opposés}

Dans \cite{Brin97} les auteurs ont étudié les caractéristiques des motifs
contenants des items qui sont fortement dépendants selon la mesure $\chi^2$.
Afin de réduire l'explosion du nombre de candidats, les auteurs proposent d'intégrer la contrainte de fréquence dans la fouille de ces motifs corrélés.
\'Etant donné que la mesure $\chi^2$ est monotone, l'ensemble des motifs fréquents corrélés extraits résulte ainsi de la jointure de deux contraintes de types opposés à  savoir la contrainte monotone de corrélation selon
la mesure $\chi^2$ et la contrainte anti-monotone de fréquence.
Cependant, la problématique d'extraction de l'ensemble total des motifs satisfaisant les deux contraintes de types différents n'a pas été traitée. En effet, les auteurs
se sont limités à  l'extraction des motifs minimaux corrélés fréquents uniquement.
Pour ce faire, un algorithme par niveau effectuant un parcours du bas vers le haut, jusqu'é la localisation de ces motifs minimaux, a été mis en place.

Cet algorithme intitulé
\textsc{BMS} $^\textsc{(}$\footnote{\textsc{BMS} est l'acronyme de
	\textit{\textbf{B}eyond \textbf{M}arket \textbf{B}askets}.}$^{\textsc{)}}$,
a constitué la base de l'approche proposée dans \cite{grahne_correlated_2000}.
En effet, les auteurs dans \cite{grahne_correlated_2000}
proposent d'étendre l'approche proposée dans \cite{Brin97}. Ils étudient la problématique des motifs fréquents corrélés selon la mesure $\chi^2$ et qui satisfont aussi un ensemble de contraintes monotones et de contraintes anti-monotones.
Ils définissent deux ensembles particuliers de motifs. Le premier est composé des motifs minimaux corrélés et qui sont valides quant à  l'ensemble des contraintes. Quant au deuxiéme, il englobe
les motifs minimaux qui satisfont l'ensemble de contraintes \textit{c.-é.-d} les motifs minimaux valides et qui sont corrélés.
Afin d'extraire le premier ensemble, deux algorithmes \textsc{BMS+} et \textsc{BMS++} ont été proposé.
L'algorithme \textsc{BMS+} est une approche naéve qui consiste à  extraire l'ensemble des motifs minimaux corrélés fréquents moyennant le noyau de l'algorithme proposé dans \cite{Brin97} puis à  effectuer un post traitement afin de ne garder que les motifs minimaux corrélés et qui satisfont l'ensemble de contraintes posées. Cependant, l'idée de l'algorithme \textsc{BMS++}
consiste à  intégrer, dans le processus de fouille, les contraintes de corrélation minimale et de fréquence avec l'ensemble des contraintes posées et à  extraire les motifs qui répondent à  toutes ces contraintes à  la fois.\\
De plus, les algorithmes \textsc{BMS*} et \textsc{BMS**} ont été introduit afin de récupérer l'ensemble formée par les motifs minimaux satisfaisant l'ensemble de contraintes et qui sont corrélés. Quant à  l'algorithme \textsc{BMS*}, il est basée sur l'idée naéve d'extraire l'ensemble de tous les motifs minimaux valides par rapport à  l'ensemble de contraintes puis de filtrer les motifs corrélés de cet ensemble. Par opposition à  ce dernier, l'algorithme \textsc{BMS**} est basé sur la même idée stratégique de l'algorithme \textsc{BMS++} qui consiste à  intégrer l'ensemble des contraintes dans le processus de fouille.

%%%Souad : il a été démontré que l'ensemble des motifs minimaux corrélés et qui satisfont l'ensemble de %%contraintes est inclut dans l'ensemble des motifs minimaux qui satisfont l'ensemble de contraintes et %%qui %%sont corrélés. Dans notre cas ces deux ensembles sont égaux et correspondent aux motifs rares %%minimaux %%corrélés.

Nous avons à  ce stade, présenté et analysé les différentes approches d'extraction des ensembles des motifs corrélés sous contraintes.

\subsection{Discussion}

%%%%%%%%%%%%%%%%%
L'étude des approches de l'état de l'art que nous avons présenté dans cette section a été menée selon
trois principaux axes.
Nous avons dans un premier temps analysé les approches d'extraction des motifs corrélés.
Nous constatons que la majorité de ces approches opérent sur la base des
mesures \textit{bond} et \textit{all-confidence}.
Ces derniéres vérifient la propriété d'anti-monotonie.
Cette propriété est pertinente dans le processus de fouille, vu sa capacité à  réduire l'espace de recherche et à  optimiser le temps d'extraction ainsi que la consommation des ressources matérielles.
Ainsi, l'ensemble des motifs corrélés fréquents extraits résulte de la jointure de deux contraintes de même type, à  savoir la contrainte anti-monotone de
corrélation et la contrainte anti-monotone de fréquence.

Toutefois, la récupération de l'ensemble des motifs qui sont à  la fois trés peu fréquents et fortement corrélés est basée sur l'idée naéve d'extraire
l'ensemble de tous les motifs fréquents
pour un seuil de \textit{minsupp} trés bas puis de filtrer ces motifs par la contrainte de corrélation.
Cette opération est trés coéteuse en temps de traitement et en consommation de la mémoire à  cause de l'explosion du nombre de candidats à  évaluer. D'ailleurs, l'approche proposée dans \cite{thomo_2010} est fondée sur ce principe.
Une autre stratégie d'extraction des motifs rares fortement corrélés, consiste é
extraire l'ensemble de tous les motifs corrélés sans aucune intégration de la contrainte de support.
Cette idée permet de récupérer les motifs rares fortement corrélés comme c'est le cas des algorithmes
proposés dans \cite{ma-icdm2001} et dans \cite{Cohen_mcr_2000}.

Nous déduisons pour toutes ces approches, que la contrainte monotone de rareté n'a été jamais incorporée dans la fouille afin de récupérer l'ensemble des motifs rares fortement corrélés.

Dans un deuxiéme volet de cette analyse, nous avons mis l'accent sur les approches d'extraction des motifs corrélés sous la conjonction de contraintes de types opposés. Nous avons analysé les
algorithmes proposés dans \cite{Brin97} et \cite{grahne_correlated_2000}. Ces derniers permettent d'intégrer l'ensemble des contraintes dans le processus de fouille. Ils exploitent ainsi les différentes opportunités d'élagage offertes par l'ensemble des contraintes posées et bénéficient du pouvoir sélectif de chaque type de contrainte.
Cependant, ces approches se limitent à  l'extraction d'un sous ensemble restreint composé uniquement des
motifs minimaux valides \textit{c.-é.-d.} satisfaisant l'ensemble de contraintes posées. De plus, aucune représentation concise des motifs corrélés retenus n'a été proposée.

D'ailleurs, les uniques représentations concises des motifs corrélés ont été proposées dans 
\cite{ccmine_Kim} et dans \cite{tarekds2010}. Ces derniéres, offrent des approches réductrices sans perte d'information de l'ensemble des motifs fréquents corrélés respectivement selon les mesures
\textit{all-confidence} et \textit{bond}.

Les tableaux \ref{TabCmp1} et \ref{TabCmp2} récapitulent les caractéristiques des différentes approches étudiées. Cette récapitulation couvrira les axes suivants :

\begin{enumerate}
	\item \textbf{Mesures de corrélation} : Cette propriété décrit la ou les mesures de corrélation traitée\textsc{(}s\textsc{)} par l'algorithme.
	\item \textbf{Type de motifs extraits} : Cette propriété décrit le type de motifs générés par l'algorithme. Par ``Valide'', nous entendons que le motif respecte l'ensemble des contraintes posées.
	\item \textbf{Natures des contraintes} : Cette propriété décrit les types des contraintes manipulées par l'algorithme.
	\item \textbf{Nature de l'algorithme} : Cette propriété décrit la nature de l'algorithme. Nous distinguons les algorithmes \textit{approximatifs}, algorithmes \textit{paralléles} et les algorithmes \textit{exacts}. Pour les algorithmes exacts, nous spécifions aussi la
	stratégie d'exploration de l'espace de recherche. Autrement dit, la maniére avec laquelle l'espace de recherche sera parcouru par l'algorithme. Ceci peut étre réalisé \textit{en profondeur} \textit{c.-é.-d.} en étendant le candidat en cours de traitement par un item qu'il ne contient pas déjé.
	Ou \textit{en largeur} \textsc{(}appelé aussi \textit{par niveau}\textsc{)}
	et englobe deux type de parcours.
	Le premier se réalise du bas vers le haut du treillis
	\textit{c.-é.-d.} par taille croissante des itemsets en commenéant par l'ensemble vide.
	Le deuxiéme se réalise du haut vers le bas du treillis \textit{c.-é.-d.} par taille décroissante
	en commenéant par le plus grand motif jusqu'a l'ensemble vide.
	\item  \textbf{Principe de déroulement} :  Cette propriété est spécifiée uniquement pour les
	approches d'extraction de motifs corrélés sous la conjonction de contraintes de natures opposés.
	La valeur \textit{Par post traitement} indique
	qu'un seul type de contraintes a été intégré dans la fouille et l'ensemble de motifs intéressant est obtenu par post traitement selon le deuxiéme type de contraintes.
	Cependant, la valeur \textit{Par intégration de contraintes} indique que les contraintes de types opposés ont été intégrées dans la fouille.
\end{enumerate}
%%%%%%%%%%%%%%%%%%%
%%%%%%%%%%%%%%%%%%%
\begin{table}[!t]\small{
		\parbox{16cm}{\hspace{-0.4cm}
			\footnotesize
			\begin{tabular}{|l||c|c|c|c|}
				\hline
				\multicolumn{1}{|c||}{\textbf{Algorithme}}   &  \textbf{Mesure\textsc{(}s\textsc{)} de}  & \textbf{Type de motifs} &\textbf{Nature de\textsc{(}s\textsc{)}}&\textbf{Nature de}\\
				\multicolumn{1}{|c||}{\textbf{d'extraction}} &\textbf{corrélation} & \textbf{extraits} 
				& \textbf{contrainte\textsc{(}s\textsc{)}} &\textbf{l'algorithme}\\
				\hline\hline
				L'approche de & \textit{bond} & Tous les motifs&anti-monotones  &Approximatif\\
				\cite{Cohen_mcr_2000} &       & corrélés de taille deux&    &\\
				\hline	
				\textsc{DiscoverMPattern}& \textit{all-confidence} &Tous les motifs &anti-monotones  &Exact\\
				\cite{ma-icdm2001} &                                &corrélés&    &en largeur\\
				\hline
				\textsc{Partition}  &\textit{all-confidence} &Tous les motifs&anti-monotone &Exact\\
				\cite{Omie03}&\textit{bond}&corrélés&            & en largeur\\
				\hline
				\textsc{CoMine}\textsc{(}$\alpha$\textsc{)}&\textit{all-confidence} &corrélés fréquents &anti-monotones &Exact\\
				\cite{comine_Lee}&         &                      &            &en profondeur\\
				\hline
				\textsc{CoMine}\textsc{(}$\gamma$\textsc{)}&\textit{bond}  &corrélés fréquents &anti-monotones &Exact\\
				\cite{comine_Lee}&         &                      &            &en profondeur\\
				\hline
				\textsc{HypercliqueMiner} & \textit{h-confidence} &corrélés fréquents&anti-monotones&\\
				\cite{Xiong06hypercliquepattern} &          & et une partie des motifs &            &Exact\\
				&          & corrélés rares&                       &en largeur\\	
				\hline
				L'approche de& \textit{all-confidence} &corrélés fréquents&&\\
				\cite{thomo_2010} &                     &et une partie des&anti-monotones&Algorithme\\
				&                      &motifs corrélés rares&&Paralléle\\	
				\hline
				L'approche de & \textit{bond} &faiblement corrélés&monotones  &Exact\\
				\cite{MCROkubo}&               &&              &en profondeur\\
				\hline
				\textsc{CCMine}&\textit{all-confidence} &fermés & anti-monotones &Exact\\
				\cite{ccmine_Kim}&  & corrélés fréquents &      &en profondeur\\
				\hline
				\textsc{CCPR\_Miner} &\textit{bond}&fermés & anti-monotones &Exact\\
				\cite{CAP2010}&   &corrélés fréquents&     &en largeur\\
				\hline
	\end{tabular}}}
	\caption{Tableau comparatif des algorithmes d'extraction des motifs
		corrélés fréquents.}\label{TabCmp1}
\end{table}
%%%%%%%%%%%%%%%%%%

\begin{table}[!t]\small{
		\parbox{16cm}{\hspace{-1.5cm}
			\footnotesize \begin{tabular}{|l||c|c|c|c|c|c|}
				\hline
				\multicolumn{1}{|c||}{\textbf{Algorithme}}   &  \textbf{Mesure\textsc{(}s\textsc{)} de}  & \textbf{Type de motifs} &\textbf{Nature de\textsc{(}s\textsc{)}}  &\textbf{Stratégie}& \textbf{Principe}\\
				\multicolumn{1}{|c||}{\textbf{d'extraction}} &\textbf{corrélation} & \textbf{extraits} 
				& \textbf{contrainte\textsc{(}s\textsc{)}} &\textbf{d'exploration}& \textbf{de déroulement}\\
				\hline\hline
				\textsc{BMS}&  \textit{$\chi^2$} & minimaux corrélés & monotones &largeur&Par intégration\\
				\cite{Brin97}&   &fréquents&   et anti-monotones   &  &de contraintes\\
				\hline
				
				\textsc{BMS+}&    & minimaux corrélés   & && \\
				& \textit{$\chi^2$} &valides    &  monotones&largeur &Par post\\
				\textsc{BMS*}&                    &minimaux valides   & et anti-monotones &&traitement\\
				\cite{grahne_correlated_2000}    &&corrélés           &      &&\\
				\hline\hline
				\textsc{BMS++}&    &minimaux corrélés   &&& \\
				& \textit{$\chi^2$}&valides& monotones   & largeur &Par intégration\\
				\textsc{BMS**}& &minimaux valides           & et anti-monotones &  &de contraintes\\
				\cite{grahne_correlated_2000}& &corrélés    & & &\\
				\hline
	\end{tabular}}}
	\caption{Tableau comparatif des algorithmes d'extraction des motifs
		corrélés sous la conjonction de contraintes de types opposés.}\label{TabCmp2}
\end{table}

Nous avons ainsi présenté dans cette section les différentes approches de fouille de motifs corrélés sous contraintes de types opposés. Dans la section suivante,  nous tenons à  présenter un état de l'art des approches de la littérature traitant de la conjonction entre les contraintes monotones et les contraintes anti-monotones.

\section{\'Etat de l'art des approches traitant de la conjonction de contraintes monotones et de contraintes anti-monotones}

Lors de processus de fouille de motifs, il est plus complexe de localiser les motifs vérifiant des contraintes de natures différentes que de localiser ceux associés à  des contraintes de même nature. En effet, la nature opposée des contraintes fait que les stratégies d'élagage ne sont applicables que pour une partie des contraintes et pas pour les autres. Ceci augmente les coéts des traitements à  effectuer afin de déterminer si un motif donné fait partie de l'ensemble désiré ou non.

Beaucoup de travaux ont été proposés dans la littérature afin de prendre en considération des contraintes de natures différentes lors du processus de fouille de motifs intéressants \cite{boulicautsurveycontraintes,pei_contraintes02}. Toutefois, nous nous limitons dans notre étude aux contraintes monotones et aux contraintes anti-monotones.

Les différentes approches de la littérature traitant de la conjonction de contraintes monotones et de contraintes anti-monotones peuvent étre classés en cinq grandes catégories en fonction de leurs principe de résolution du probléme de fouille. Nous proposons dans ce qui suit une analyse des différentes catégories d'approches.

\subsection{L'approche \textsc{DualMiner}}

\begin{figure}[!t]\centering
	\includegraphics[scale = 0.5]{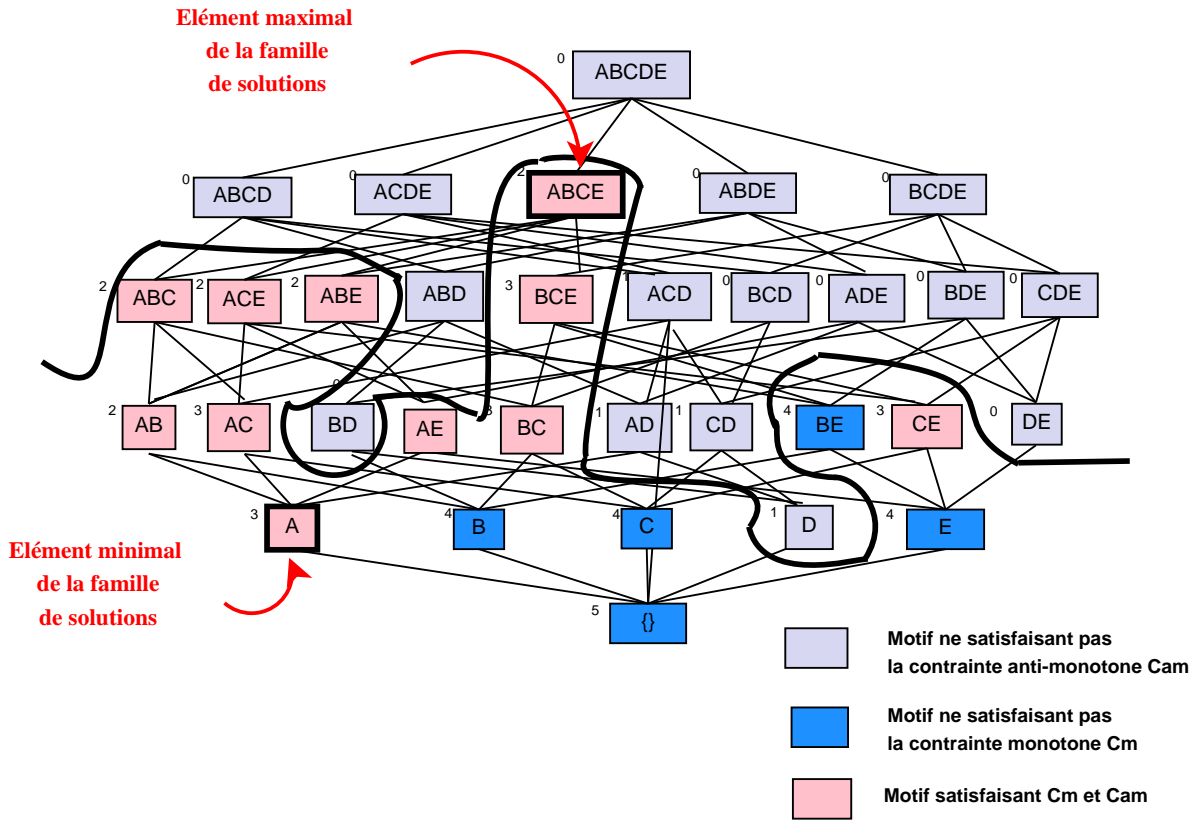}
	\caption{Trace d'exécution de l'algorithme \textsc{DualMiner} pour le contexte $\mathcal{D}$.}
	\label{DM}
\end{figure}

Le premier algorithme se situant dans ce cadre est intitulé \textsc{DualMiner} \cite{Bucila03}.
\textsc{DualMiner} est une approche générique de résolution du probléme d'extraction de motifs sous
la conjonction de contraintes monotones et de contraintes anti-monotones.
Cet algorithme effectue la réduction de l'espace de recherche en incorporant à  la fois les contraintes monotones ainsi que les contraintes anti-monotones.
En effet, \textsc{DualMiner} extrait des familles de solutions valides par rapport à  l'ensemble des contraintes. Chacune de ses familles, est représentée par le couple
$\langle$$T$, $B$$\rangle$ avec $T$ représente l'élément maximal
correspondant à  l'union de tous les éléments de cette famille
et $B$ représente l'élément minimal et correspond é
é l'intersection de tous les éléments de cette famille. Considérons un exemple illustratif dans ce qui suit.
\begin{exemple}
	Considérons le contexte donné par la table \ref{Base_transactions}. La contrainte anti-monotone
	$Cam$ correspond à  Supp\textsc{(}$M$\textsc{)}  $\geq$  \textit{minsupp} et la contrainte monotone $Cm$ correspond é
	Supp\textsc{(}$M$\textsc{)} $\leq$  \textit{maxsupp} avec \textit{minsupp} = 2 et \textit{maxsupp} = 3.
	Par conséquent, tout motif $M$ satisfaisant les contraintes anti-monotone $Cam$ et monotone $Cm$ doit vérifier, 2 $\leq$ Supp\textsc{(}$M$\textsc{)} $\leq$ 3.
	Lors d'un premier parcours de treillis des motifs, l'item \texttt{D} ne vérifie pas la contrainte anti-monotone $Cam$ ainsi il sera supprimé ainsi que tous ses sur-ensembles. Ensuite,
	l'itemset \texttt{BE} ne vérifie pas la contrainte monotone $Cm$, ainsi, il sera supprimé ainsi que ses sous-ensembles \texttt{B} et \texttt{E}. L'item \texttt{C} sera plus considéré puisqu'il viole la contrainte monotone. Par conséquent, la seule partie du treillis correspond à  la famille dont l'élément minimal est \texttt{A} et l'élément maximal est \texttt{ABCE}. Il est à  remarquer que le motif \texttt{A} respecte la contrainte monotone $Cm$, Supp\textsc{(}\texttt{A}\textsc{)} = 3 $\leq$ 3, ainsi, tous ses sur-ensembles le sont aussi. Quant au motif \texttt{ABCE}, il vérifie la contrainte anti-monotone $Cam$, Supp\textsc{(}\texttt{ABCE}\textsc{)} = 2 $\geq$ 2. Par conséquent, la famille de solution délimitée par
	\texttt{A} et \texttt{ABCE} est une famille de solution valide. Cette exécution est représentée par la figure \ref{DM}.
\end{exemple}

Toutefois, \textsc{DualMiner} souffre aussi d'une limite liée au nombre d'évaluation de contraintes. En effet, le temps d'exécution de l'algorithme \textsc{DualMiner} augmente en fonction de la taille de l'espace d'intersection entre les deux contraintes incorporées. Or, cette taille peut augmenter d'une maniére exponentielle en fonction de la taille du probléme \cite{boley2009} ce qui présente une des limites de cette approche.

\subsection{Les approches \textsc{MCP} et \textsc{ACP}}

La problématique de fouille de motifs sous contraintes de types opposés a été abordée aussi dans les travaux de
Boulicaut et Jeudy, qui ont proposé
l'approche \textsc{MCP} 
$^\textsc{(}$\footnote{\textsc{MCP} est l'acronyme de
	\textit{\textbf{M}onotone \textbf{C}onstraint
		\textbf{P}ushing}.}$^{\textsc{)}}$ \cite{mcp2000}.
\textsc{MCP} est une approche permettant l'extraction de motifs satisfaisant à  la fois un ensemble de contraintes monotones et de contraintes anti-monotones.
Cependant, \textsc{MCP} est
basée sur l'hypothése que
les motifs minimaux satisfaisant la contrainte monotone sont facilement récupérables et constituent un paramétre d'entrée de l'approche \textsc{MCP}. Alors que cette hypothése n'est pas toujours valide \cite{Bucila03}. Toutefois, l'ensemble des motifs minimaux valides par rapport à  la contrainte monotone est difficilement repérable vu la propriété de filtre d'ordre induite par toute contrainte monotone.

Dans ce même sens, Bonchi et \textit{al.} ont proposé l'approche \textsc{ACP}
$^\textsc{(}$\footnote{\textsc{ACP} est l'acronyme de
	\textit{\textbf{A}daptive \textbf{C}onstraint
		\textbf{P}ushing}.}$^{\textsc{)}}$
dans \cite{acp2003}.
En effet, l'approche \textsc{ACP} permet d'intégrer
é la fois les contraintes monotones et les contraintes anti-monotones dans un algorithme par niveau.
Par opposition à  l'approche \textsc{MCP}, l'algorithme \textsc{ACP} permet dans un premier temps d'extraire la bordure positive de la contrainte monotone et ensuite de générer les candidats qui correspondent aux sur-ensembles des motifs de cette bordure.
Puisque, les sur-ensembles des motifs de cette bordure vérifient la contrainte monotone, ils seront ainsi évalués uniquement par rapport à  la contrainte anti-monotone.
%La particularité de l'approche \textsc{ACP} réside dans le fait, que l'ensemble des motifs 
%qui ne satisfont pas la contrainte monotone seront quand même maintenus.
%Ces derniers vont servir dans l'élagage des candidats. En effet, tout candidat admettant un sous ensemble qui ne satisfait pas la contrainte monotone sera élagué.

La particularité de l'approche \textsc{ACP} réside dans le fait qu'elle
adopte un comportement adaptatif pour renforcer l'élagage selon une contrainte par rapport à  une autre afin de maximiser l'efficacité. En effet, l'adaptativité se traduit par l'introduction d'un paramétre $\alpha$, initialisé lors du premier balayage de la base puis il est mis à  jour à  chaque niveau en fonction de l'ensemble des nouvelles connaissances collectées lors de l'itération précédente.
Cependant, aucune indication quant à  l'initialisation et la mise à  jour du paramétre $\alpha$ n'a été présentée.
\begin{exemple}
	Soit \textit{minsupp} le seuil minimal de support conjonctif, la contrainte monotone $Cm$ correspondant à  ``Supp\textsc{(}\texttt{X}\textsc{)} $<$ \textit{minsupp}''.
	Considérons le contexte d'extraction donné par la table \ref{Base_transactions} pour \textit{minsupp} = 3. Les motifs minimaux satisfaisant la contrainte monotone $Cm$ extraits moyennant l'approche \textsc{MCP} correspondent aux motifs rares minimaux.
	$\mathcal{MRM}in$ = $\{$\texttt{D}, \texttt{AB}, \texttt{AE}$\}$.
	
	La contrainte anti-monotone $Cam$ est formulée pour chaque itemset $X$ en fonction des prix de ses items $X_{i}$, ``$\min$\textsc{(}$X_{i}$.prix\textsc{)} $<$ 4''. Les prix des items \texttt{A}, \texttt{B}, \texttt{C}, \texttt{D}, \texttt{E} correspondent respectivement
	é 5, 4, 6, 3 et 10.
	
	L'ensemble $\mathcal{E}$ des motifs de support non nul et satisfaisant les contraintes $Cm$ et $Cam$ peut étre récupéré gréce aux approches \textsc{MCP} et \textsc{ACP} et est composé des motifs suivants,
	$\mathcal{E}$ = $\{$\texttt{D}, \texttt{AD}, \texttt{CD}, \texttt{ACD}$\}$.
\end{exemple}

\subsection{Les approches \textsc{VST} et \textsc{FAVST}}

D'autres travaux ont essayé aussi d'étudier cette problématique, comme par exemple l'algorithme \textsc{VST} \cite{vst} d'extraction de motifs de chaénes de caractéres sous la conjonction de contraintes de types différents. Cet algorithme parcourt l'espace de recherche par niveau à  la \textsc{Apriori} \cite{Agra94}. Il se déroule en deux phases principales. La premiére consiste à  effectuer un parcours du bas vers le haut permettant l'élagage des motifs candidats selon la contrainte anti-monotone. Puis, la deuxiéme phase assure un parcours par niveau du haut vers le bas pour l'élagage des motifs candidats selon la contrainte monotone.
Chacune des deux phases inclut ainsi le cycle de génération, évaluation et élagage des candidats. Le cycle se termine quand il n'y a plus de candidats à  générer. \`A la fin de ces deux phases, les motifs retenus sont valides par rapport à  l'ensemble des contraintes considérées.
Toutefois, dans le cas oé la contrainte monotone est trés sélective alors plusieurs évaluations inutiles des candidats selon la contrainte anti-monotone auraient été déjé menés lors de la premiére phase inutilement.
De plus, l'algorithme \textsc{VST} souffre d'une limite liée au nombre de balayages de la base de données. En effet, il réalise au plus $2^{m}$ parcours de la base
avec $m$ correspond à  la taille de la plus grande chaéne valide par rapport à  la contrainte anti-monotone.
\`A cet égard, l'algorithme \textsc{FAVST} \cite{favst} a été introduit afin d'améliorer les performances de l'algorithme \textsc{VST}. L'algorithme \textsc{FAVST} assure une extraction efficace des chaénes de caractéres satisfaisant les contraintes posées et offre de meilleures performances que l'algorithme \textsc{VST}.
\begin{exemple}
	Considérons le contexte d'extraction donné par la table \ref{Base_transactions}.
	Soient la contrainte anti-monotone $Cam$ correspondant à  la conjonction de deux contraintes anti-monotones. La premiére concerne la longueur minimale de la chaéne, Taille\textsc{(}$C$\textsc{)} $\geq$ 3 et la deuxiéme  est traduite par un support non nul, Supp\textsc{(}$C$\textsc{)} $>$  0.
	La contrainte monotone
	$Cm$ permet de ne retenir que les chaénes englobant le caractére \texttt{A}, Appartient\textsc{(}\texttt{A}, $C$\textsc{)}.
	Les approches \textsc{VST} et \textsc{FAVST} permettent d'extraire l'ensemble $\mathcal{E}$
	des motifs satisfaisant à  la fois les contraintes $Cam$ et $Cm$.
	Lors de la premiére phase, tous les motifs satisfaisant la contrainte $Cam$ sont extraits, $\mathcal{E}$ =
	$\{$\texttt{ABC}, \texttt{ABE}, \texttt{ACD}, \texttt{ACE}, \texttt{BCE}$\}$.
	Lors de la deuxiéme phase, les motifs ne respectant pas la contrainte monotone $Cm$ seront élagués. Nous avons ainsi,
	$\mathcal{E}$ = $\{$\texttt{ABC}, \texttt{ABE}, \texttt{ACD}, \texttt{ACE}$\}$. Nous remarquons, que seul le candidat \texttt{BCE} est élagué puisqu'il n'englobe pas l'item \texttt{A}.
\end{exemple}

%%%%%%%%%%%%%%%%%%%%%%%%%%%%%%%%
\subsection{Les approches \textsc{Dpc-Cofi} et \textsc{Bifold-Leap}}

D'autres travaux visant à  améliorer les performances des algorithmes traitant de contraintes de types différents ont aussi vu le jour, tel que \textsc{Dpc-Cofi} 
$^\textsc{(}$\footnote{\textsc{Dpc-Cofi} est l'acronyme de
	\textit{\textbf{D}ual \textbf{P}ushing} of \textbf{C}onstraint in \textbf{Cofi}.}$^{\textsc{)}}$
\cite{dpccofi} et \textsc{BifoldLeap} \cite{bifoldleap}. Ces deux approches
permettent d'extraire l'ensemble des motifs fréquents satisfaisant un ensemble de contraintes monotones et anti-monotones.
Ces approches mettent en place une nouvelle stratégie qui se réalise en deux principales phases.
La premiére phase consiste à  extraire tous les motifs fréquents maximaux moyennant l'algorithme
\textsc{COFI*} \cite{cofi*}. La contrainte monotone est intégrée lors de la premiére phase afin d'élaguer tous les items qui ne vérifient pas cette contrainte puisque tous leurs sur-ensembles ne la vérifient bien évidement pas.
Puis, lors de la deuxiéme phase tous les sous ensembles valides des motifs fréquents maximaux
seront générés selon le principe suivant.
Tout motif fréquent maximal ne satisfaisant pas la contrainte monotone, sera élagué puisque tous ses sous ensembles ne la satisfont pas.
Pour tout motif fréquent maximal satisfaisant la contrainte anti-monotone, alors tous ses sous ensemble satisfont cette contrainte et seront ainsi générés et évalués par rapport à  la contrainte monotone.
\`A la fin de ce traitement, tous les motifs fréquents valides par rapport aux contraintes monotones et anti-monotones posées, seront extraits. Les deux algorithmes proposés présentent des performances meilleures que l'algorithme \textsc{DualMiner} gréce aux techniques de réductions du nombre d'évaluations des contraintes adoptées.
\begin{exemple}
	Considérons le contexte d'extraction donné par la table \ref{Base_transactions}. La contrainte anti-monotone $Cam$ correspond à  la contrainte de corrélation minimale par rapport à  la mesure \textit{bond}, bond\textsc{(}$X$\textsc{)} $\geq$ 0,30. La contrainte monotone $Cm$ correspond, Supp\textsc{(}$X$\textsc{)} $<$ 3.
	
	La premiére phase consiste à  extraire l'ensemble $\mathcal{MFM}$ des motifs maximaux de support non nul, nous avons $\mathcal{MFM}$ = $\{$\textsc{(}\texttt{ABCE}, 2\textsc{)}, \textsc{(}\texttt{ACD}, 1\textsc{)}$\}$. La contrainte de fréquence est traduite par Supp\textsc{(}$X$\textsc{)} $\geq$ 1.
	
	Lors de la deuxiéme étape, le motif \texttt{ACD} ne satisfait pas la contrainte anti-monotone de corrélation, vu que bond\textsc{(}\texttt{ACD}\textsc{)} = $\displaystyle\frac{1}{4}$ = 0,25 $<$ 0,30. Ainsi, tous ses sous-ensembles ne sont pas corrélés, il est ainsi élagué. Par contre, le motif \texttt{ABCE} est corrélé,
	bond\textsc{(}\texttt{ABCE}\textsc{)} = $\displaystyle\frac{2}{5}$ = 0,40 $\geq$ 0,30 et il est rare aussi,
	Supp\textsc{(}\texttt{ABCE}\textsc{)} = 2 $<$ 3. Donc tous ses sous-ensembles sont corrélés, ils seront ainsi générés et évalués uniquement par rapport à  la contrainte monotone de rareté.
	Nous avons ainsi, $\mathcal{S}$ = $\{$\textsc{(}\texttt{ABC}, 2\textsc{)}, \textsc{(}\texttt{ABE}, 2\textsc{)}, \textsc{(}\texttt{BCE}, 3\textsc{)}$\}$. Le motif \texttt{BCE} n'est pas rare, il sera élagué,
	L'ensemble $\mathcal{RES}$ des motifs résultats correspond é, $\mathcal{RES}$ = $\{$\textsc{(}\texttt{ABCE}, 2\textsc{)},
	\textsc{(}\texttt{ABC}, 2\textsc{)}, \textsc{(}\texttt{ABE}, 2\textsc{)}$\}$.
	Les motifs \texttt{ABC} et \texttt{ABE} seront maintenus et tous leurs sous-ensembles sont générés,
	$\mathcal{S}$ = $\{$\textsc{(}\texttt{AB}, 2\textsc{)}, \textsc{(}\texttt{AC}, 3\textsc{)},
	\textsc{(}\texttt{BC}, 3\textsc{)},
	\textsc{(}\texttt{AE}, 2\textsc{)},
	\textsc{(}\texttt{BE}, 4\textsc{)}$\}$.
	Les motifs rares seront ajoutés à  l'ensemble résultat $\mathcal{RES}$ de tous les motifs corrélés rares, $\mathcal{RES}$ = $\{$\textsc{(}\texttt{ABCE}, 2\textsc{)},
	\textsc{(}\texttt{ABC}, 2\textsc{)}, \textsc{(}\texttt{ABE}, 2\textsc{)}
	\textsc{(}\texttt{AE}, 2\textsc{)},
	\textsc{(}\texttt{AB}, 2\textsc{)}$\}$.
\end{exemple}

\subsection{L'approche \textsc{ExAminer}}

\begin{figure}[htbp]
	\parbox{16cm}{
		\hspace{-1.5cm}
		\includegraphics[scale = 0.45]{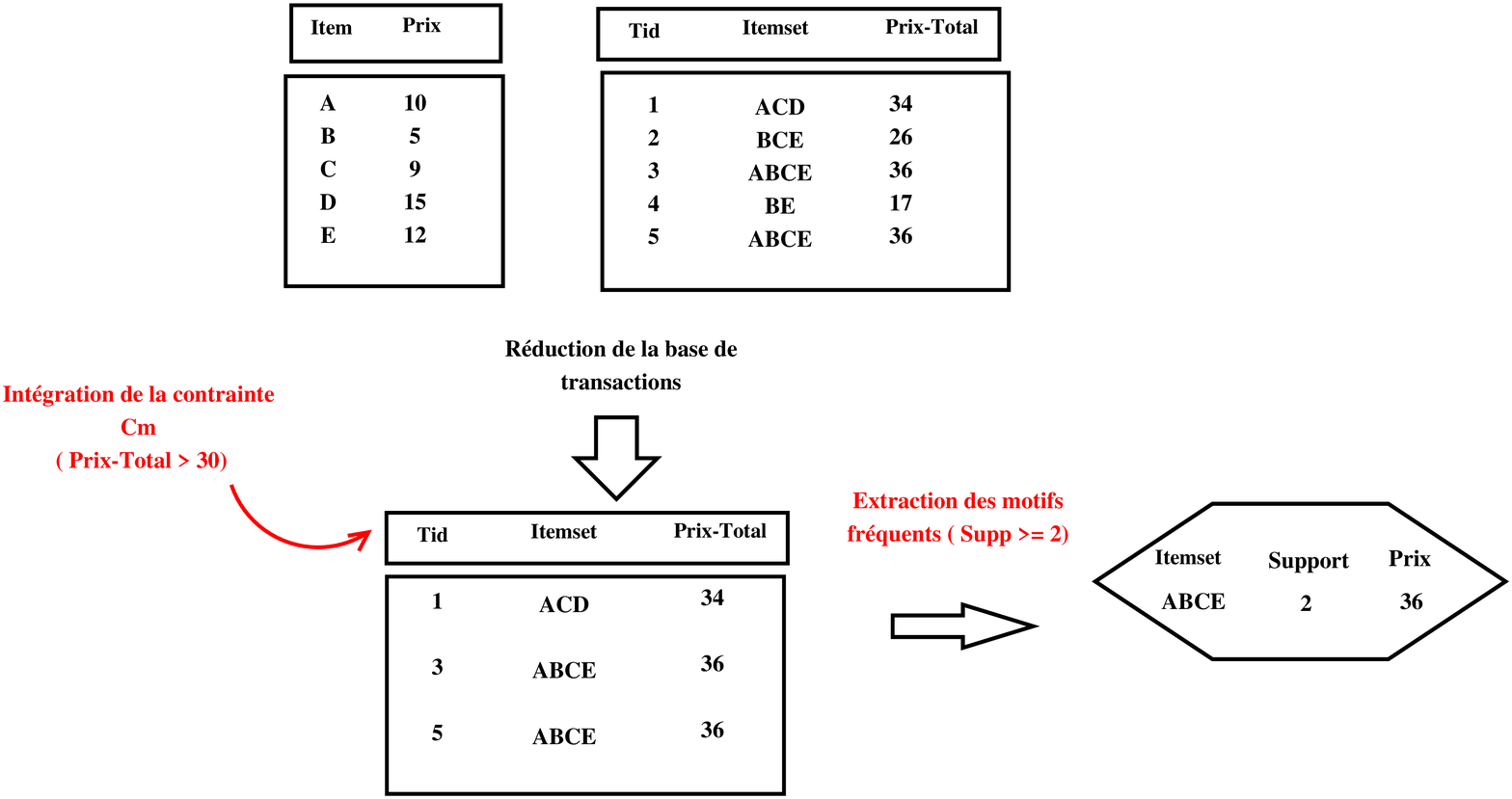}}
	\caption{Le principe de déroulement de l'algorithme \textsc{ExAMiner} pour le contexte $\mathcal{D}$.}
	\label{exanteexp}
\end{figure}

L'algorithme \textsc{ExAMiner} \cite{examiner_kais05} a été aussi proposé pour la fouille des motifs
fréquents sous une contrainte monotone. Les auteurs de cet algorithme prouvent l'existence d'une synergie réelle entre les contraintes de natures opposés. En effet, les contraintes anti-monotones permettent la réduction de l'espace de recherche gréce à  la propriété de fermeture vers le bas, tandis que les contraintes monotones permettent d'éliminer des transactions de la base de transactions. L'élagage des transactions de la base qui ne satisfont pas une contrainte monotone est effectué par l'algorithme
\textsc{ExAnte}.
En effet, la réduction de la base de transaction est justifiée par le fait qu'aucun sous-ensemble d'une transaction élaguée ne satisfait la contrainte monotone considérée.
La synergie est ainsi traduite par le fait que les deux stratégies d'élagages se renforcent en réduisant l'espace de recherche, et simplifient par conséquent la dimension du probléme de fouille.
Il est toutefois important de noter que ces stratégies de réduction du contexte d'extraction,
souffrent d'une limite liée aux coéts des opérations d'entrée et de sortie réalisées lors des processus itératifs de réécriture de la base sur le disque \cite{bifoldleap}.
De plus, ces approches ne sont pas applicables pour le cas d'une contrainte monotone sensible au changement du nombre de transactions. Ceci est le cas de la contrainte de rareté que nous traitons dans ce travail.\\
\begin{exemple} \label{expexaminer}
	Considérons le contexte d'extraction donné par la table \ref{Base_transactions}.
	Pour un motif donné $X$, la contrainte anti-monotone $Cam$ de fréquence est traduite par, Supp\textsc{(}$X$\textsc{)} $\geq$ 3. La contrainte monotone $Cm$ est traduite par, Prix\textsc{(}$X$\textsc{)} $\geq$ 30. Le principe de déroulement de l'algorithme \textsc{ExAMiner} est illustré par la figure \ref{exanteexp}.
\end{exemple}

Nous avons ainsi analysé différentes approches de la littérature abordant la problématique de fouille de motifs sous la conjonction de contraintes de types opposés.
Dans ce qui suit, nous passons en revue les algorithmes d'extraction de représentations concises de motifs fréquents sous la conjonction de contraintes.

\subsection{Les approches d'extraction de représentations concises de motifs fréquents sous la conjonction de contraintes de types opposés}

Trés peu d'approches traitants la problématique d'extraction de représentations concises de motifs fréquents sous la conjonction de contraintes de types opposés ont vu le jour.
En effet, nous distinguons uniquement trois approches.
La premiére approche
a été proposée dans \cite{boulicaut01}
et consiste en une approche d'extraction d'une représentation concise basée sur les générateurs minimaux dits aussi ``itemsets libres'' fréquents et qui satisfont un ensemble donné de contraintes monotones et de contraintes anti-monotones.
En effet, l'algorithme proposé constitue une variation de l'algorithme \textsc{A-Close} \cite{pasquier99_2005} avec  l'incorporation des contraintes dans la phase d'extraction des motifs.
De plus, les motifs libres extraits constituent une représentation concise avec perte d'information \cite{luccheKIS05_MAJ_06}, autrement dit l'ensemble total des motifs fréquents respectant les contraintes posées, ne peut pas étre dérivé à  partir de l'ensemble des motifs libres retenus. \`A cet égard,
des approches d'extraction de représentations concises sous contraintes basée sur les motifs fermés fréquents ont aussi émergé. Nous citons dans ce cadre, l'approche introduite dans \cite{Jia03} et l'algorithme
\textsc{CCI-Miner} \cite{luccheKIS05_MAJ_06}.
Toutefois, l'algorithme
proposé dans \cite{Jia03} opére par post traitement. En effet, tous les motifs fermés fréquents sont extraits dans un premier temps puis, ils sont filtrés par post traitement afin de ne maintenir que les fermés fréquents qui sont valides par rapport à  l'ensemble de contraintes posées.
Par opposition à  ce dernier, \textsc{CCI-Miner} est une approche sophistiquée qui consiste en un algorithme en profondeur basée sur le principe d'incorporation de toutes les contraintes à  la fois.
En effet, \textsc{CCI-Miner} repose sur l'intégration de l'algorithme \textsc{FP-Bonsai} \cite{luccheKIS05_MAJ_06} d'extraction des motifs fréquents sous contraintes monotones avec l'algorithme \textsc{Closet} d'extraction de motifs fermés fréquents.
Ces motifs retournés par \textsc{CCI-Miner} constituent une représentation concise sans perte d'information de l'ensemble de tous les motifs fréquents répondants aux contraintes posées.
En outre, dans \cite{luccheKIS05_MAJ_06} les auteurs ont prouvé expérimentalement que les approches naéves opérants par post-traitement engendrent une perte d'information considérable.
Nous proposons dans ce qui suit un exemple illustratif des différentes représentations concises proposées.

\begin{exemple}
	Considérons le contexte d'extraction donné par la table \ref{Base_transactions} et la table des prix des items utilisée dans l'exemple \ref{expexaminer}.
	Pour un motif donné $X$, la contrainte anti-monotone $Cam$ de fréquence est traduite par, Supp\textsc{(}$X$\textsc{)} $\geq$ 2 et la contrainte monotone $Cm$ est traduite par, Prix\textsc{(}$X$\textsc{)} $\geq$ 12.
	
	L'approche proposée dans \cite{boulicaut01} permet d'extraire l'ensemble $\mathcal{GMF}$ des générateurs minimaux fréquents. Chaque itemset de cet ensemble ne posséde aucun sous-ensemble de même support conjonctif que lui. Nous avons, pour \textit{minsupp} = 2, l'ensemble $\mathcal{GMF}$ des générateurs minimaux fréquents
	muni chacun de son support.
	$\mathcal{GMF}$ =  $\{$\textsc{(}\texttt{E}, 4, 12\textsc{)},
	\textsc{(}\texttt{B}, 4, 5\textsc{)},
	\textsc{(}\texttt{AB}, 2, 15\textsc{)},
	\textsc{(}\texttt{AE}, 2, 22\textsc{)},
	\textsc{(}\texttt{C},  4, 9\textsc{)},
	\textsc{(}\texttt{BC}, 3, 14\textsc{)},
	\textsc{(}\texttt{EC}, 3, 21\textsc{)},
	\textsc{(}\texttt{A},  3, 10\textsc{)}$\}$.
	Tous les éléments de cet ensemble vérifient la contrainte anti-monotone de fréquence, les itemsets ne vérifiant pas la contrainte monotone traduite par le prix minimal seront élagués.
	Ainsi, nous avons  $\mathcal{GMF}$ =  $\{$\textsc{(}\texttt{E}, 4, 12\textsc{)},
	\textsc{(}\texttt{AB}, 2, 15\textsc{)},
	\textsc{(}\texttt{AE}, 2, 22\textsc{)},
	\textsc{(}\texttt{BC}, 3, 14\textsc{)},
	\textsc{(}\texttt{EC}, 3, 21\textsc{)}$\}$.
	L'ensemble $\mathcal{GMF}$ englobe ainsi tous les générateurs minimaux fréquents satisfaisants la contrainte monotone. L'ensemble de tous les motifs fréquents répondant à  la contrainte monotone ne peut pas étre dérivé à  partir de l'ensemble $\mathcal{GMF}$.
	Traitons à  présent l'ensemble $\mathcal{MFF}$ des motifs fermés fréquents extraits gréce à  l'approche \textsc{CCI-Miner} \cite{luccheKIS05_MAJ_06} ou moyennant l'approche proposée dans \cite{Jia03}.
	Tout motif de l'ensemble $\mathcal{MFF}$ ne posséde aucun sur-ensemble de même support que lui. Nous avons
	$\mathcal{MFF}$ l'ensemble de ces motifs munis chacun de son support et de son prix. $\mathcal{MFF}$ = $\{$
	\textsc{(}\texttt{ABCE}, 2, 36\textsc{)},
	\textsc{(}\texttt{BCE}, 3, 26\textsc{)},
	\textsc{(}\texttt{BE}, 4, 17\textsc{)},
	\textsc{(}\texttt{AC}, 3, 19\textsc{)}$\}$.
	L'ensemble $\mathcal{MFF}$  englobe ainsi tous les motifs fermés fréquents dont le prix dépasse 12. L'ensemble $\mathcal{MF}$ de tous les motifs fréquents vérifiant la contrainte monotone de prix minimal peut étre dérivé en générant pour chaque motif fermé, les sous-ensembles appartenant à  la même classe d'équivalence que lui c.-é.-d. possédant le même support que lui.
	$\mathcal{MF}$ = $\{$\textsc{(}\texttt{ABCE}, 2, 36\textsc{)},
	\textsc{(}\texttt{ABC}, 2, 24\textsc{)},
	\textsc{(}\texttt{ABE}, 2, 27\textsc{)},
	\textsc{(}\texttt{AB},  2, 15\textsc{)},
	\textsc{(}\texttt{AE},  2, 22\textsc{)},
	\textsc{(}\texttt{BCE}, 3, 26\textsc{)},
	\textsc{(}\texttt{BC}, 3, 14\textsc{)},
	\textsc{(}\texttt{CE}, 3, 21\textsc{)},
	\textsc{(}\texttt{BE}, 4, 17\textsc{)},
	\textsc{(}\texttt{E}, 4, 12\textsc{)},
	\textsc{(}\texttt{AC}, 3, 19\textsc{)}$\}$.
	
	Nous avons ainsi montré que les représentations concises proposées par \cite{luccheKIS05_MAJ_06}  et \cite{Jia03} autorisent une dérivation compléte de l'ensemble de tous les motifs fréquents satisfaisant la contrainte monotone de prix minimal.
\end{exemple}
\subsection{Discussion}

D'aprés l'étude des différentes approches de la littérature présentée dans cette section, nous constatons que les approches sophistiquées et les approches opérants par post traitement constituent deux solutions pour la problématique d'extraction de motifs sous contraintes.

Les approches opérants par post traitement, consistent à  extraire tous les motifs vérifiant la contrainte anti monotone puis à  effectuer un post traitement pour filtrer ces motifs par rapport à  la contrainte monotone. Autrement dit, aucune intégration de la contrainte monotone n'est réalisée et l'exploration de l'espace de tous les motifs ne vérifiant pas la contrainte monotone s'effectue ainsi inutilement. De plus, les algorithmes actuels risquent de se bloquer à  cause de la consommation excessive de la mémoire centrale.

Quant aux approches sophistiquées, ces derniéres intégrent l'ensemble des contraintes anti-monotones et monotones dans la fouille.
Toutefois, l'intégration des contraintes monotones engendre la réduction des opportunités d'élagage de candidats selon la propriété de l'idéal d'ordre des contraintes anti-monotones.
En effet, les contraintes anti-monotones sont plus faciles à  intégrer dans la réduction de l'espace de recherche en exploitant la propriété trés intéressante de l'idéal d'ordre. Alors que les contraintes monotones sont plus compliquées et difficiles à  incorporer dans le processus de fouille. De plus, elles sont moins efficaces dans la réduction de l'espace de recherche que les contraintes anti-monotones.

Nous constatons, que le choix de la meilleure approche dépend des caractéristiques de la base de données et des spécificités des contraintes posées.

Le tableau \ref{TabCmp3} récapitule les caractéristiques des différentes approches étudiées. Cette récapitulation couvrira les axes suivants :

\begin{enumerate}
	\item \textbf{Type de motifs extraits} : Cette propriété décrit le type de motifs générés par l'algorithme.
	Par \textit{Motifs valides}, nous entendons les motifs qui satisfont l'ensemble des contraintes monotones et anti-monotones posées.
	\item \textbf{Caractéristiques} :  Cette propriété décrit les principales caractéristiques de l'algorithme en question.
\end{enumerate}

\begin{table}[!t]
	\begin{center}
		\footnotesize{ \begin{tabular}{|l||c|c|}
				\hline
				\multicolumn{1}{|c||}{\textbf{Algorithme}}  & \textbf{Type de motifs} &\textbf{Caractéristiques}\\
				\multicolumn{1}{|c||}{\textbf{d'extraction}} & \textbf{extraits}& \\
				\hline\hline
				\textsc{DualMiner}& tous les&$-$Temps d'exécution élevé\\
				\cite{Bucila03}&motifs valides& vu la nature duale\\
				&& du parcours\\ 	
				\hline
				\textsc{MCP}& tous les&$-$ La bordure positive de la contrainte\\
				\cite{mcp2000}&motifs valides      &monotone doit passer\\
				&                                 & en paramétre\\
				\hline
				\textsc{ACP}& tous les&$-$ La notion d'adaptativité proposée\\
				\cite{acp2003}&motifs valides      & n'est pas spécifiée\\
				&                                 & rigoureusement\\	
				\hline
				\textsc{VST}& chaénes de  & $-$ Nombre élevé de balayage \\
				\cite{vst}&caractéres valides   & de la base \\
				&                            &$-$ \'Evaluations \\
				&                            & inutiles des contraintes\\
				\hline
				\textsc{FAVST}& chaénes de&$-$Approche plus efficace \\
				\cite{favst}&caractéres valides&que \textsc{VST}.\\ 	
				\hline
				\textsc{Dpc$\-$Cofi}&&\\
				\cite{dpccofi}&motifs fréquents &$-$Approches sophistiquées\\
				\textsc{Bifold$\-$Leap}&valides&et plus efficaces que \textsc{Dual-Miner}\\
				\cite{bifoldleap}& &\\
				\hline
				\textsc{ExAminer}&motifs fréquents  &$-$ Coéts élevées\\
				\cite{examiner_kais05}&valides      &    de réécriture de la base\\
				&                                   &$-$ Approche applicable que\\
				&                                   &    pour les contraintes\\
				&                                   &    monotones insensibles\\
				&                                   &    au changement de la base\\	
				\hline\hline
				\textsc{CCI$\-$Miner}&motifs fermés        &$-$ Approche sophistiquée \\
				\cite{luccheKIS05_MAJ_06}&fréquents valides&$-$ Représentation concise \\
				&                                         &    et sans perte d'information\\
				\hline
				L'approche de &motifs fermés        & $-$ Approche par\\
				\cite{Jia03}&fréquents valides      &post traitement\\
				&                                  &et engendre une \\
				&                                  &perte d'information\\
				\hline	
				L'approche de &générateurs minimaux &$-$ Représentation concise\\
				\cite{boulicaut01}&fréquents valides&avec perte d'information\\
				\hline
		\end{tabular}}
	\end{center}
	\caption{Tableau comparatif des algorithmes traitant des contraintes de types opposés.}\label{TabCmp3}
\end{table}

Au meilleur de notre connaissance, aucune étude dans la littérature n'a été réalisée afin d'extraire des représentations concises de l'ensemble des motifs corrélés rares selon la mesure \textit{bond} résultant de la conjonction de la contrainte anti-monotone de corrélation et de la contrainte monotone de rareté.
\`A cet égard, nous introduisons dans le chapitre suivant nos approches d'extraction de nouvelles représentations exactes de ces motifs corrélés rares.

\section{Conclusion}

Dans ce chapitre nous avons donné un aperéu de l'extraction des motifs rares et nous avons analysé l'approche récente d'extraction de représentations concises exactes des motifs rares. Ensuite, nous avons détaillé les approches d'extraction des motifs corrélés fréquents et de leurs représentations concises. Nous avons aussi étudié les algorithmes traitants de 
d'extraction des motifs corrélés sous contraintes de types opposés.
Nous avons, par la suite, analysé la panoplie d'approches constituant le cadre générique de la problématique de fouille de motifs sous la conjonction de contraintes de types opposés. Nous avons, à  cet égard, étudié les caractéristiques de ces approches et discuté leurs avantages et limites.
Dans le chapitre suivant, nous allons caractériser l'ensemble des motifs corrélés rares selon la mesure \textit{bond}  et nous présenterons de nouvelles représentations concises de ces motifs.

%%**********************************
\chapter{Nouvelles représentations concises des motifs corrélés rares}
\label{chapitre_representation}

\section{Introduction}

Nous consacrons ce chapitre à  la présentation de nos contributions dans la réduction des motifs corrélés rares. Pour ce faire, nous
étudions profondément les propriétés des motifs corrélés rares dans la premiére section.
Ensuite, nous décrivons dans la deuxiéme section les caractéristiques des classes d'équivalence associées à  ces motifs. Quant à  la troisiéme section, elle sera dédiée à  la justification de nos motivations quant à  l'extraction des représentations concises des motifs corrélés rares.
Ensuite nous proposons, de nouvelles représentations concises des motifs corrélés rares \cite{rnti2011} à  savoir les représentations concises exactes $\mathcal{RMCR}$, $\mathcal{RMM}$$ax$$\mathcal{F}$ et $\mathcal{RM}$$in$$\mathcal{MF}$, ainsi que la représentation concise approximative $\mathcal{RM}$$in$$\mathcal{MM}$$ax$$\mathcal{F}$.

\section{Caractérisation des motifs corrélés rares}\label{section_MCR}

Nous commenéons dans la sous-section suivante par définir l'ensemble des motifs rares corrélés selon la mesure \textit{bond}.

\subsection{Définition des motifs corrélés rares}

\begin{center}
	\begin{figure}[htbp]
		\parbox{16cm}{
			\hspace{-0.6cm}
			\includegraphics[scale = 0.5]{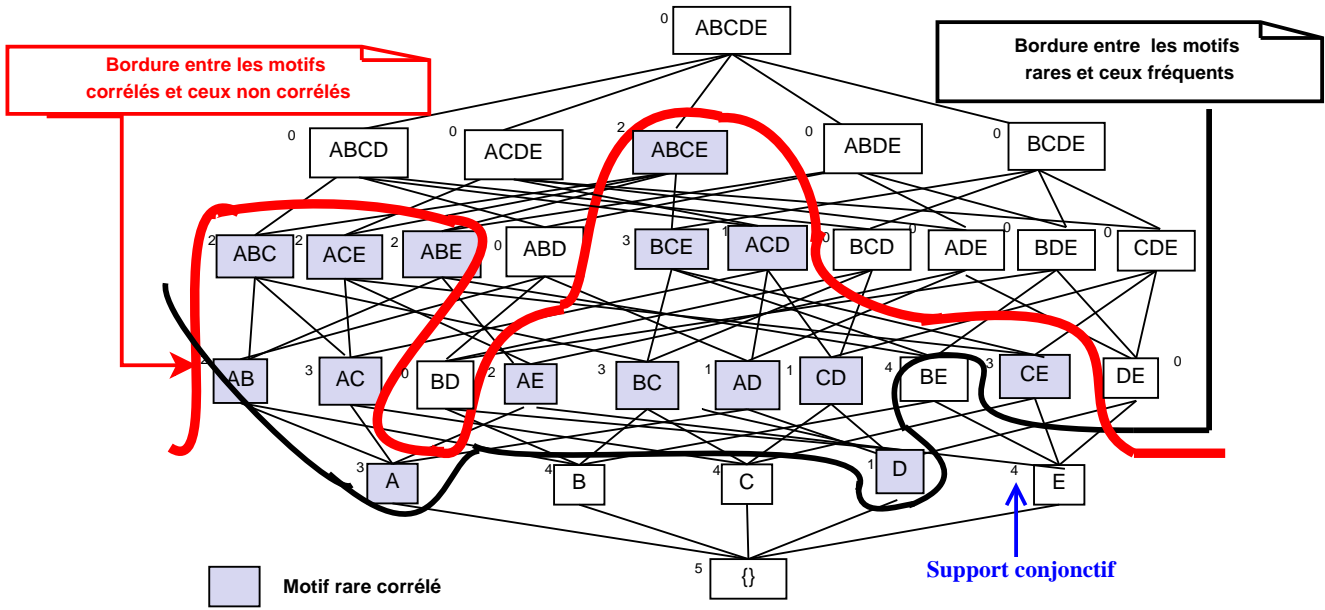}}
		\caption{Espace des motifs corrélés rares pour \textit{minsupp} = 4 et \textit{minbond} = 0,2.}
		\label{figure_MCR}
	\end{figure}
\end{center}

\begin{definition}\label{MCRare} \textsc{(}\textbf{Motifs corrélés rares associés à  la mesure \textit{bond}}\textsc{)} \cite{rnti2011} \\
	étant donnés les seuils minimaux de support conjonctif et de corrélation \textit{minsupp} et \textit{minbond}, respectivement, l'ensemble des motifs corrélés rares, dénoté $\mathcal{MCR}$, est défini comme suit : $\mathcal{MCR}$ = $\{$$I$ $\subseteq$ $\mathcal{I}$$|$ \textit{Supp}\textsc{(}$\wedge$$I$\textsc{)} $<$ \textit{minsupp} et \textit{bond}\textsc{(}$I$\textsc{)} $\geq$ \textit{minbond}$\}$.
\end{definition}
\begin{exemple}\label{exemple_ensemble_MCR}
	Considérons la base illustrée par la table \ref{Base_transactions} pour \textit{minsupp} = 4 et \textit{minbond} = 0,2. L'ensemble $\mathcal{MCR}$ est composé des motifs suivants oé chaque triplet représente le motif, sa valeur de support et sa valeur de \textit{bond} : $\mathcal{MCR}$ = $\{$\textsc{(}\texttt{A}, 3, $\displaystyle\frac{3}{3}$\textsc{)},
	\textsc{(}\texttt{D}, 1, $\displaystyle\frac{1}{1}$\textsc{)},
	\textsc{(}\texttt{AB}, 2, $\displaystyle\frac{2}{5}$\textsc{)},
	\textsc{(}\texttt{AC}, 3, $\displaystyle\frac{3}{4}$\textsc{)},
	\textsc{(}\texttt{AD}, 1, $\displaystyle\frac{1}{3}$\textsc{)},
	\textsc{(}\texttt{AE}, 2, $\displaystyle\frac{2}{5}$\textsc{)},
	\textsc{(}\texttt{BC}, 3, $\displaystyle\frac{3}{5}$\textsc{)},
	\textsc{(}\texttt{CD}, 1, $\displaystyle\frac{1}{4}$\textsc{)},
	\textsc{(}\texttt{CE}, 3, $\displaystyle\frac{3}{5}$\textsc{)},
	\textsc{(}\texttt{ABC}, 2, $\displaystyle\frac{2}{5}$\textsc{)},
	\textsc{(}\texttt{ABE}, 2, $\displaystyle\frac{2}{5}$\textsc{)},
	\textsc{(}\texttt{ACD}, 1, $\displaystyle\frac{1}{4}$\textsc{)},
	\textsc{(}\texttt{ACE}, 2, $\displaystyle\frac{2}{5}$\textsc{)},
	\textsc{(}\texttt{BCE}, 3, $\displaystyle\frac{3}{5}$\textsc{)},
	\textsc{(}\texttt{ABCE}, 2, $\displaystyle\frac{2}{5}$\textsc{)}$\}$.
	Cet ensemble est schématisé par la figure \ref{figure_MCR}. Le support indiqué en haut à  gauche de chaque cadre représentant un motif est son support conjonctif. Comme le montre cette figure, l'ensemble $\mathcal{MCR}$ des motifs corrélés rares correspond aux motifs localisés en dessous de la bordure de la contrainte anti-monotone associée aux motifs corrélés, et au dessus de la bordure de la contrainte monotone associée aux motifs rares.
\end{exemple}

Il résulte de la définition précédente que l'ensemble $\mathcal{MCR}$ correspond à  l'intersection de l'ensemble des motifs corrélés et de l'ensemble des motifs rares : $\mathcal{MCR}$ = $\mathcal{MC}$ $\cap$ $\mathcal{MR}$. La proposition suivante découle de ce résultat.
\begin{proposition}\label{prop_propriete_ens_MCR}
	Soit $I$ $\in$ $\mathcal{MCR}$. Nous avons :
	\begin{itemize}
		\item  D'aprés l'idéal d'ordre de l'ensemble des motifs corrélés selon la mesure \textit{bond}, $\forall$ $I_1$ $\subseteq$ $I$ : $I_1$ $\in$ $\mathcal{MC}$
		\item  D'aprés le filtre d'ordre de l'ensemble des motifs rares, $\forall$ $I_1$ $\supseteq$ $I$ : $I_1$ $\in$ $\mathcal{MR}$.
	\end{itemize}
\end{proposition}
La preuve découle des propriétés induites par les contraintes de corrélation et de rareté. L'ensemble $\mathcal{MCR}$, dont les éléments vérifient la contrainte ``étre un motif corrélé rare'', résulte ainsi de l'intersection de deux ordres résultant de deux contraintes de natures opposés. Cet ensemble n'est ainsi ni un idéal ni un filtre d'ordre. Dans le treillis des motifs, l'espace de recherche des motifs corrélés rares est ainsi délimité, d'une part, par les éléments maximaux vérifiant la contrainte de corrélation et qui sont rares, \textit{c.-é.-d.} les motifs \textit{rares} parmi \textit{l'ensemble $\mathcal{MCM}ax$ des motifs corrélés maximaux} \textsc{(}\textit{cf.} Définition \ref{bdpos}\textsc{)} et, d'autre part, par les éléments minimaux vérifiant la contrainte de rareté et qui sont corrélés, \textit{c.-é.-d.} les motifs corrélés parmi \textit{l'ensemble $\mathcal{MRM}in$ des motifs rares minimaux} \textsc{(}\textit{cf.} Définition \ref{mrm}\textsc{)}. Ainsi, tout motif corrélé rare est nécessairement compris entre un élément de chacun des deux ensembles susmentionnés.

\begin{exemple} Considérons la figure \ref{figure_MCR} pour \textit{minsupp} = 4 et \textit{minbond} = 0,2. L'espace des motifs corrélés rares est délimité par : d'une part les motifs corrélés maximaux pour \textit{minbond} = 0,2, à  savoir \texttt{ACD} et \texttt{ABCE} \textsc{(}\textit{cf.} Exemple \ref{exemple_MCMax}\textsc{)}, et, d'autre part, par les motifs rares minimaux pour \textit{minsupp} = 4, à  savoir \texttt{A}, \texttt{D}, \texttt{BC} et \texttt{CE}. Par exemple, le motif \texttt{AD} est un motif corrélé rare étant donné qu'il est compris entre un motif rare minimal à  savoir \texttt{D} et un motif corrélé maximal à  savoir \texttt{ACD}.
\end{exemple}

Cet espace est ainsi plus difficile à  localiser que les ensembles associés à  des contraintes de même nature. En effet, la conjonction de contraintes anti-monotones \textsc{(}\textit{resp.} monotones\textsc{)} est une contrainte anti-monotone \textsc{(}\textit{resp.} monotone\textsc{)} \cite{luccheKIS05_MAJ_06}. Par exemple, la contrainte ``étre un motif corrélé non rare \textsc{(}\textit{c.-é.-d.} fréquent\textsc{)}'' est une contrainte anti-monotone puisque elle est résultante de la conjonction des contraintes anti-monotones ``étre un motif corrélé'' et ``étre un motif fréquent''. Elle induit donc un idéal d'ordre \cite{tarekds2010}. La contrainte ``étre un motif non corrélé rare'' est une contrainte monotone et l'ensemble associé forme un filtre d'ordre dans le treillis des motifs.

D'un point de vue taille, et étant données les natures des contraintes induites par les seuils minimaux de support et de corrélation, à  savoir respectivement \textit{minsupp} et \textit{minbond}, la taille de l'ensemble $\mathcal{MCR}$ des motifs corrélés rares varie de la maniére indiquée dans la proposition suivante.

\begin{proposition}\mbox{}\label{TailleMCR}
	
	\textbf{a\textsc{)}} Soient \textit{minsupp}$_1$ et \textit{minsupp}$_2$ deux seuils minimaux de support et $\mathcal{MCR}_{s1}$ et $\mathcal{MCR}_{s2}$ les deux ensembles des motifs corrélés rares associés pour une même valeur de \textit{minbond}. Nous avons : si \textit{minsupp}$_1$ $\leq$ \textit{minsupp}$_2$, alors $\mathcal{MCR}_{s1}$ $\subseteq$ $\mathcal{MCR}_{s2}$ et par conséquent $|\mathcal{MCR}_{s1}|$ $\leq$ $|\mathcal{MCR}_{s2}|$.

	\textbf{b\textsc{)}} Soient \textit{minbond}$_1$ et \textit{minbond}$_2$ deux seuils minimaux de corrélation et $\mathcal{MCR}_{b1}$ et $\mathcal{MCR}_{b2}$ les deux ensembles des motifs corrélés rares associés pour une même valeur de \textit{minsupp}. Nous avons : si \textit{minbond}$_1$ $\leq$ \textit{minbond}$_2$, alors $\mathcal{MCR}_{b2}$ $\subseteq$ $\mathcal{MCR}_{b1}$ et par conséquent $|\mathcal{MCR}_{b2}|$ $\leq$ $|\mathcal{MCR}_{b1}|$.
\end{proposition}
\begin{Preuve}
	
	- La preuve de \textbf{a\textsc{)}} dérive du fait que pour $I$ $\subseteq$ $\mathcal{I}$, si \textit{Supp}\textsc{(}$\wedge$$I$\textsc{)} $<$ \textit{minsupp}$_1$, alors \textit{Supp}\textsc{(}$\wedge$$I$\textsc{)} $<$ \textit{minsupp}$_2$. Ainsi, $\forall$ $I$ $\in$ $\mathcal{MCR}_{s1}$, $I$ $\in$ $\mathcal{MCR}_{s2}$. Il en résulte que $\mathcal{MCR}_{s1}$ $\subseteq$ $\mathcal{MCR}_{s2}$.
	
	- La preuve de \textbf{b\textsc{)}} dérive du fait que pour $I$ $\subseteq$ $\mathcal{I}$, si \textit{bond}\textsc{(}$I$\textsc{)} $\geq$ \textit{minbond}$_2$, alors \textit{bond}\textsc{(}$I$\textsc{)} $\geq$ \textit{minbond}$_1$. Ainsi, $\forall$ $I$ $\in$ $\mathcal{MCR}_{b2}$, $I$ $\in$ $\mathcal{MCR}_{b1}$. Par conséquent, $\mathcal{MCR}_{b2}$ $\subseteq$ $\mathcal{MCR}_{b1}$.
\end{Preuve}

Ainsi, la taille de $\mathcal{MCR}$ est proportionnelle à  \textit{minsupp} et inversement proportionnelle à  \textit{minbond}. Il est toutefois à  noter que, dans le cas général, nous ne pouvons rien décider quand les deux seuils varient en même temps et non seulement un à  la fois.

Nous avons ainsi analysé la variation de la taille de l'ensemble $\mathcal{MCR}$. La question qui se pose à  ce stade est, comment pouvons nous caractériser les motifs corrélés rares à  travers leurs sous-ensembles directs?
Autrement dit, pouvons nous affirmer la corrélation et ou la rareté d'un motif quelconque, étant donné la connaissance de ses sous-ensemble? Cette problématique sera étudiée dans ce qui suit.

\subsection{\'Etude des propriétés spécifiques des motifs corrélés rares}

Toutefois, deux cas sont à  distinguer lors de l'affirmation de la nature de corrélation et de rareté d'un motif quelconque $X$ connaissant ses sous-ensembles. Le premier cas se réalise lorsque le motif $X$ posséde un sous-ensemble direct corrélé rare. Quand au deuxiéme cas, il se réalise lorsque le motif $X$ ne posséde aucun sous ensemble corrélé rare. Analysons chaque cas à  part.

Dans le premier cas, le motif $X$  posséde un sous-ensemble direct corrélé rare.
Ce motif est ainsi rare corrélé comme le justifie la proposition suivante,.
\begin{proposition}\label{prop_algo_MCR2}
	Soit $X$ est un motif corrélé. Si $\exists$ $Y$ $\in$ $\mathcal{MCR}$ $|$ $Y$ $\subset$ $X$ et $|Y|$ = $|X|- 1$ alors le motif $X$ est un motif corrélé rare.
\end{proposition}
La preuve de cette proposition découle de la propriété de filtre d'ordre des motifs rares. En effet, tout motif corrélé qui est sur-ensemble d'un motif rare est forcément rare corrélé.

Cependant, dans le deuxiéme cas, \textit{c.-é.-d.} lorsque le motif $X$ ne posséde aucun sous ensemble appartenant à  l'ensemble $\mathcal{MCR}$. Plus précisément, tous les sous-ensembles du motif $X$ sont des motifs corrélés fréquents, alors nous ne pouvons rien décider quant à  la nature de fréquence de ce motif.
La proposition suivante présente une condition nécessaire et suffisante à  l'affirmation
de la nature de fréquence d'un motif corrélé.
\begin{proposition}\label{prop_algo_MCR3}
	Soit $X$ est un motif corrélé. Désignons par \textit{minbond} le seuil minimal de corrélation selon la mesure \textit{bond} et par \textit{minsuppRel} le seuil minimal de support relatif correspondant à  $\displaystyle\frac{\textit{minsupp}}{|\mathcal{T}|}$.
	Si \textit{bond}\textsc{(}$X$\textsc{)} $<$ \textit{minsuppRel} alors le motif corrélé $X$ est un motif corrélé rare.
\end{proposition}
\begin{Preuve}
	Désignons par \textit{SuppRel}\textsc{(}$X$\textsc{)} le support relatif d'un motif $X$.
	Nous avons
	\textit{SuppRel}\textsc{(}$X$\textsc{)} = $\displaystyle\frac{\displaystyle
		\textit{Supp}\textsc{(}\wedge X\textsc{)}}{\displaystyle
		|\mathcal{T}|}$ et
	\textit{bond}\textsc{(}$X$\textsc{)} = $\displaystyle\frac{\displaystyle
		\textit{Supp}\textsc{(}\wedge X\textsc{)}}{\displaystyle
		\textit{Supp}\textsc{(}\vee X\textsc{)}}$.
	
	Or, \textit{Supp}\textsc{(}$\vee$ $X$\textsc{)} $\leq$ $|\mathcal{T}|$
	donc $\displaystyle\frac{\displaystyle 1}{\displaystyle\textit{Supp}\textsc{(}\vee X\textsc{)}}$
	$\geq$ $\displaystyle\frac{\displaystyle 1}{\displaystyle|\mathcal{T}|}$.
	Ainsi, nous avons
	$\displaystyle\frac{\displaystyle
		\textit{Supp}\textsc{(}\wedge X\textsc{)}}{\displaystyle
		\textit{Supp}\textsc{(}\vee X\textsc{)}}$ $\geq$
	$\displaystyle\frac{\displaystyle
		\textit{Supp}\textsc{(}\wedge X\textsc{)}}{\displaystyle
		|\mathcal{T}|}$.
	
	Ceci est équivalent é, \textit{bond}\textsc{(}$X$ \textsc{)} $\geq$ \textit{SuppRel}\textsc{(}$X$\textsc{)}.
	Le motif $X$ étant corrélé, alors \textit{bond}\textsc{(}$X$\textsc{)}
	$\geq$ \textit{minbond}.
	
	Dans le cas oé
	\textit{minbond} $<$ \textit{minsuppRel}
	et \textit{minbond} $\leq$ \textit{bond}\textsc{(}$X$\textsc{)} $<$ \textit{minsuppRel}, nous avons alors
	\textit{SuppRel}\textsc{(}$X$\textsc{)}  $\leq$ \textit{bond}\textsc{(}$X$\textsc{)} $<$ \textit{minsuppRel}.
	Ce qui implique,
	\textit{SuppRel}\textsc{(}$X$\textsc{)}  $<$ \textit{minsuppRel}.
	
	Par conséquent, \textit{SuppRel}\textsc{(}$X$\textsc{)} $\times$ $|\mathcal{T}|$ $<$ \textit{minsuppRel} $\times$ $|\mathcal{T}|$. Ceci est équivalent é,
	\textit{Supp}\textsc{(}$\wedge X$\textsc{)} $<$ \textit{minsupp}, le motif $X$ est par conséquent rare.
\end{Preuve}
D'aprés la proposition précédente, nous concluons qu'un motif corrélé $X$,
pour qu'il soit corrélé rare, il faut que sa valeur de mesure \textit{bond} soit strictement inférieur au seuil \textit{minsuppRel}. Cependant, lorsque la valeur de mesure \textit{bond}
dépasse le seuil \textit{minsuppRel}, alors nous ne pouvons pas affirmer la rareté du motif $X$. Ceci est justifié gréce à  la proposition suivante.
\begin{proposition} \label{prop_algo_MCR4}
	Pour \textit{bond}\textsc{(}$ X$\textsc{)} $\geq$ \textit{minsuppRel},
	nous pouvons rien affirmer quant à  la fréquence du motif corrélé $X$.
\end{proposition}
\begin{Preuve}
	Nous avons
	\textit{Supp}\textsc{(}$\wedge$X\textsc{)} $\geq$
	$\displaystyle\frac{\displaystyle
		\textit{Supp}\textsc{(}\wedge X\textsc{)}}{\displaystyle
		\textit{Supp}\textsc{(}\vee X\textsc{)}}$, ceci est équivalent é
	\textit{Supp}\textsc{(}$\wedge$X\textsc{)} $\geq$ \textit{bond}\textsc{(}$ X$\textsc{)}.
	
	Pour
	\textit{bond}\textsc{(}$ X$\textsc{)} $\geq$ \textit{minsuppRel}, nous avons
	\textit{Supp}\textsc{(}$\wedge$X\textsc{)} $\geq$ \textit{bond}\textsc{(}$X$\textsc{)} $\geq$ \textit{minsuppRel}.
	Ceci donne,
	\textit{minsuppRel} $\leq$ \textit{Supp}\textsc{(}$\wedge$X\textsc{)} et
	\textit{minsuppRel} $\leq$ \textit{minsupp}.
	Par conséquent, nous ne pouvons rien conclure quant à  la fréquence ou la rareté du motif corrélé candidat $X$.
\end{Preuve}

D'aprés les propositions \ref{prop_algo_MCR3} et \ref{prop_algo_MCR4}, nous concluons qu'un motif corrélé  $X$,
pour qu'il soit corrélé rare, il faut que sa valeur de mesure \textit{bond} soit strictement inférieur au seuil \textit{minsuppRel}.
Toutefois, dans le cas oé tous les sous-ensembles du motif $X$ sont fréquents corrélés,
nous ne pouvons pas cerner sa valeur de la mesure \textit{bond} et par conséquent nous ne pouvons
toujours pas, comme l'illustre la proposition suivante, décider de sa fréquence ou de sa rareté.
\begin{proposition} \label{prop_algo_MCR5}
	Soit $X$ un motif corrélé. Si tous les sous-ensembles de $X$ sont corrélés fréquents alors 
	nous pouvons rien affirmer quant à  la fréquence du motif corrélé $X$.
\end{proposition}
\begin{Preuve}
	Tous les sous-ensembles d'un motif corrélé $X$ sont des motifs corrélés fréquents, ainsi nous avons
	pour tout motif $Y$ $\subset$ $X$ et $|Y|$ = $|X|- 1$,
	\textit{bond}\textsc{(}$Y$\textsc{)} $\geq$ \textit{minbond}
	et \textit{SuppRel}\textsc{(}$Y$\textsc{)} $\geq$ \textit{minsuppRel}.
	
	Or, \textit{bond}\textsc{(}$Y$\textsc{)} $\geq$ \textit{SuppRel}\textsc{(}Y\textsc{)}
	donc
	\textit{bond}\textsc{(}$Y$\textsc{)} $\geq$ \textit{minsuppRel}.
	
	Comme $Y$ $\subset$ $X$, alors \textit{bond}\textsc{(}$X$\textsc{)} $\leq$
	\textit{bond}\textsc{(}$Y$\textsc{)} d'aprés la propriété d'anti-monotonie des motifs corrélés.
	
	Puisque \textit{bond}\textsc{(}$Y$\textsc{)} $\geq$ \textit{minsuppRel}, alors deux cas sont possibles :
	
	\textbf{\textsc{(}i\textsc{)}} \textit{bond}\textsc{(}$ X$\textsc{)} $<$
	\textit{minsuppRel} $\leq$ \textit{bond}\textsc{(}$ Y$\textsc{)} : Le motif corrélé candidat $X$ est ainsi rare d'aprés la proposition \ref{prop_algo_MCR3}.
	
	\textbf{\textsc{(}ii\textsc{)}} \textit{minsuppRel} $\leq$ \textit{bond}\textsc{(}$ X$\textsc{)} $\leq$ \textit{bond}\textsc{(}$ Y$\textsc{)} : Nous ne pouvons rien décider quant à  fréquence ou à  la rareté du motif corrélé candidat $X$ d'aprés la proposition \ref{prop_algo_MCR4}.
\end{Preuve}

Il est clair d'aprés la proposition précédente, que même dans le cas oé, tous les sous-ensembles d'un motif corrélé $X$ sont fréquents corrélés, nous devons l'évaluer par rapport à  la contrainte de rareté afin de confirmer sa nature.\\

Nous avons ainsi analysé minutieusement les propriétés des motifs corrélés rares. Dans ce qui suit, nous étudions le mécanisme d'intégration des contraintes dans le processus de fouille.

\section{Mécanisme d'intégration des contraintes de rareté et de corrélation}
L'ordre d'évaluation des contraintes est d'une importance majeure vu la nature opposée des contraintes anti-monotone de corrélation et monotone de rareté que nous considérons dans ce travail. Deux scénarios sont ainsi à  distinguer :

-  \textbf{\textit{Scénario 1 :}} Appliquer la contrainte de rareté, l'opérateur de fermeture associé, et ensuite la contrainte de corrélation, ou,

-  \textbf{\textit{Scénario 2 :}} Appliquer la contrainte de corrélation, l'opérateur de fermeture associé, et ensuite la contrainte de rareté.

Ces deux scénarios sont analysés dans ce qui suit afin de justifier le choix du scénario adéquat dans les approches que nous allons proposer.

\subsection{Premier scénario}
\begin {figure}\parbox{16cm}{
	\hspace{-0.4cm}
	\includegraphics[scale = 0.5]{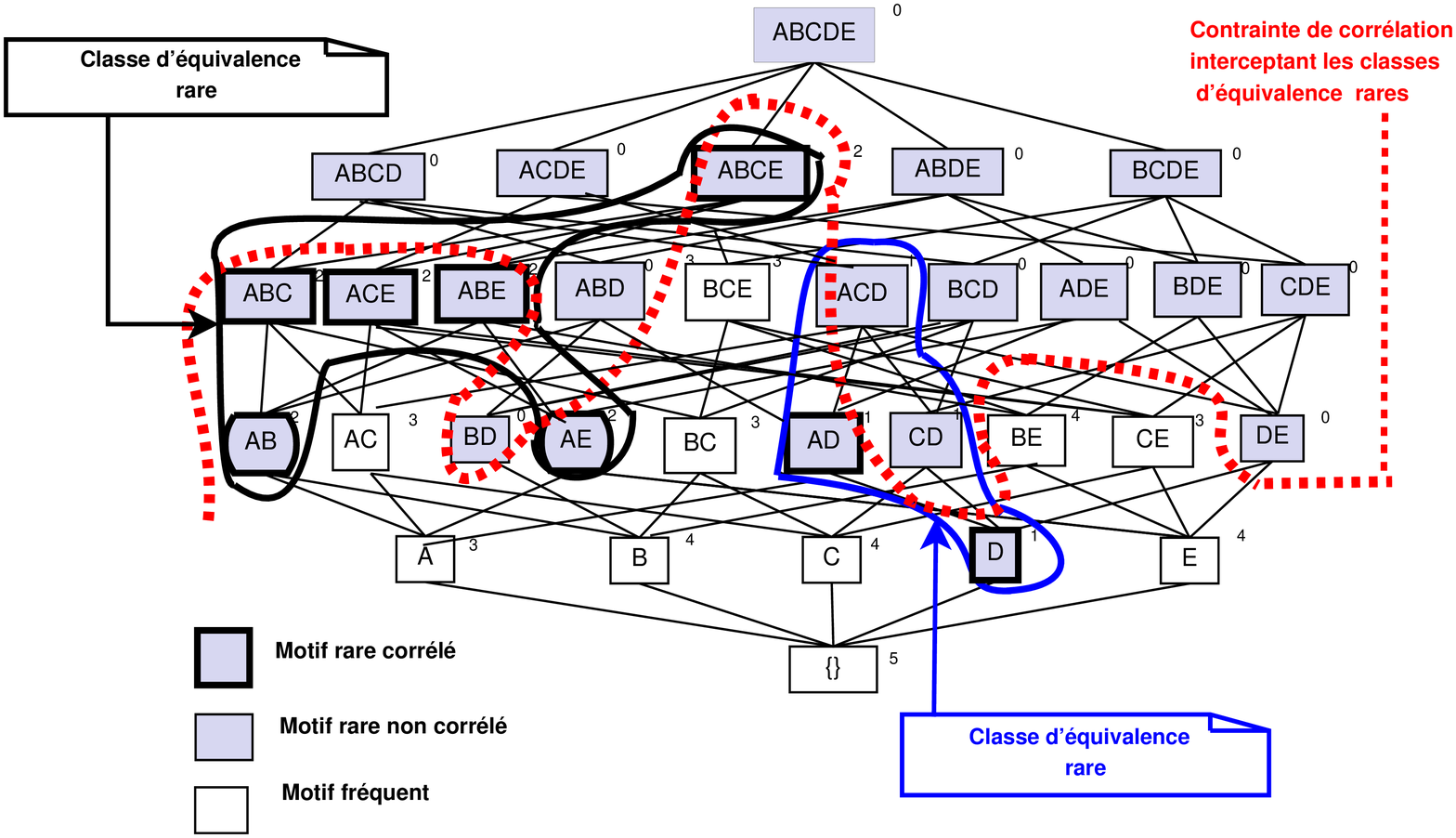}}
\caption{Effet de l'application de la contrainte de corrélation pour \textit{minsupp} = 3 et \textit{minbond} = 0,3.}
\label{Exp1}
\end{figure}

Dans ce cas, l'extraction des motifs rares est effectuée en premier lieu. Ensuite, les motifs retenus seront filtrés en ne gardant que les motifs rares, dont la valeur de la mesure \textit{bond} dépasse le seuil minimal \textit{minbond}. Dans cette situation, afin de réduire la redondance entre les motifs, l'application de l'opérateur de fermeture conjonctive associé au support conjonctif \cite{ganter99} permet de partitionner le treillis en des classes d'équivalence oé, pour un motif donné, cette fermeture ne préserve que le support conjonctif. Il en résulte que tous les motifs qui apparaissent dans les mêmes transactions seront regroupés dans une même classe d'équivalence. Ils ont ainsi le même support conjonctif et la même fermeture conjonctive, mais ont des supports disjonctifs éventuellement différents. Dans ce cas, les classes d'équivalence rares, \textit{c.-é.-d.} celles contenant les motifs rares, seront ensuite évaluées par la contrainte anti-monotone de corrélation. Les motifs d'une même classe seront ainsi divisés en des motifs corrélés rares et des motifs rares non corrélés, comme le montre la figure \ref{Exp1}.

\begin{exemple}
Soit le contexte illustré par la table \ref{Base_transactions}. Pour \textit{minsupp} = 3, nous distinguons les deux classes d'équivalence rares $\mathcal{C}_1$ et $\mathcal{C}_2$ schématisées par la figure \ref{Exp1} et qui sont comme suit :
\begin{itemize}
	\item $\mathcal{C}_1$ contient les motifs \texttt{D}, \texttt{AD}, \texttt{CD} et \texttt{ACD}. Elle admet comme support conjonctif 1 et comme fermeture conjonctive \texttt{ACD}.
	\item $\mathcal{C}_2$ contient les motifs \texttt{AB}, \texttt{AE}, \texttt{ABC}, \texttt{ABE}, \texttt{ACE}, et \texttt{ABCE}. Elle admet comme support conjonctif 2 et comme fermeture conjonctive \texttt{ABCE}.
\end{itemize}
En appliquant la contrainte de corrélation à  travers un seuil minimal \textit{minbond} = 0,3, pour la classe $\mathcal{C}_1$, seuls les motifs $\{$\textsc{(}\texttt{D}, 1, $\displaystyle\frac{1}{1}$\textsc{)}, \textsc{(}\texttt{AD}, 1, $\displaystyle\frac{1}{3}$\textsc{)}$\}$ seront rares corrélés alors que \textsc{(}\texttt{CD}, 1, $\displaystyle\frac{1}{4}$\textsc{)}, \textsc{(}\texttt{ACD}, 1, $\displaystyle\frac{1}{4}$\textsc{)} sont des éléments rares non corrélés. Ceci revient au fait que les motifs de $\mathcal{C}_1$ n'admettent pas le même support disjonctif. Par contre, tous les motifs de la classe $\mathcal{C}_2$ sont corrélés rares.
\end{exemple}

\subsection{Deuxiéme scénario}
Le deuxiéme scénario consiste à  extraire tous les motifs corrélés et les répartir en des classes d'équivalence moyennant l'opérateur de fermeture $f_{bond}$ puis à  les filtrer par rapport à  la contrainte de rareté. En fait,
les éléments d'une même classe d'équivalence partagent bien évidemment la même valeur de la mesure \textit{bond}. Par conséquent, en considérant uniquement la contrainte anti-monotone de corrélation nous distinguons deux types de classes d'équivalence à  savoir les classes corrélées et les classes non corrélées. La question qui se pose à  ce stade, est quel est l'effet de l'ajout de la contrainte monotone de rareté sur ces classes d'équivalence? Autrement dit, Comment s'affectent ces classes d'équivalence lorsqu'elles seront interceptées par la contrainte monotone de rareté?

\begin {figure}\parbox{16cm}{
\hspace{-0.6cm}
\includegraphics[scale = 0.45]{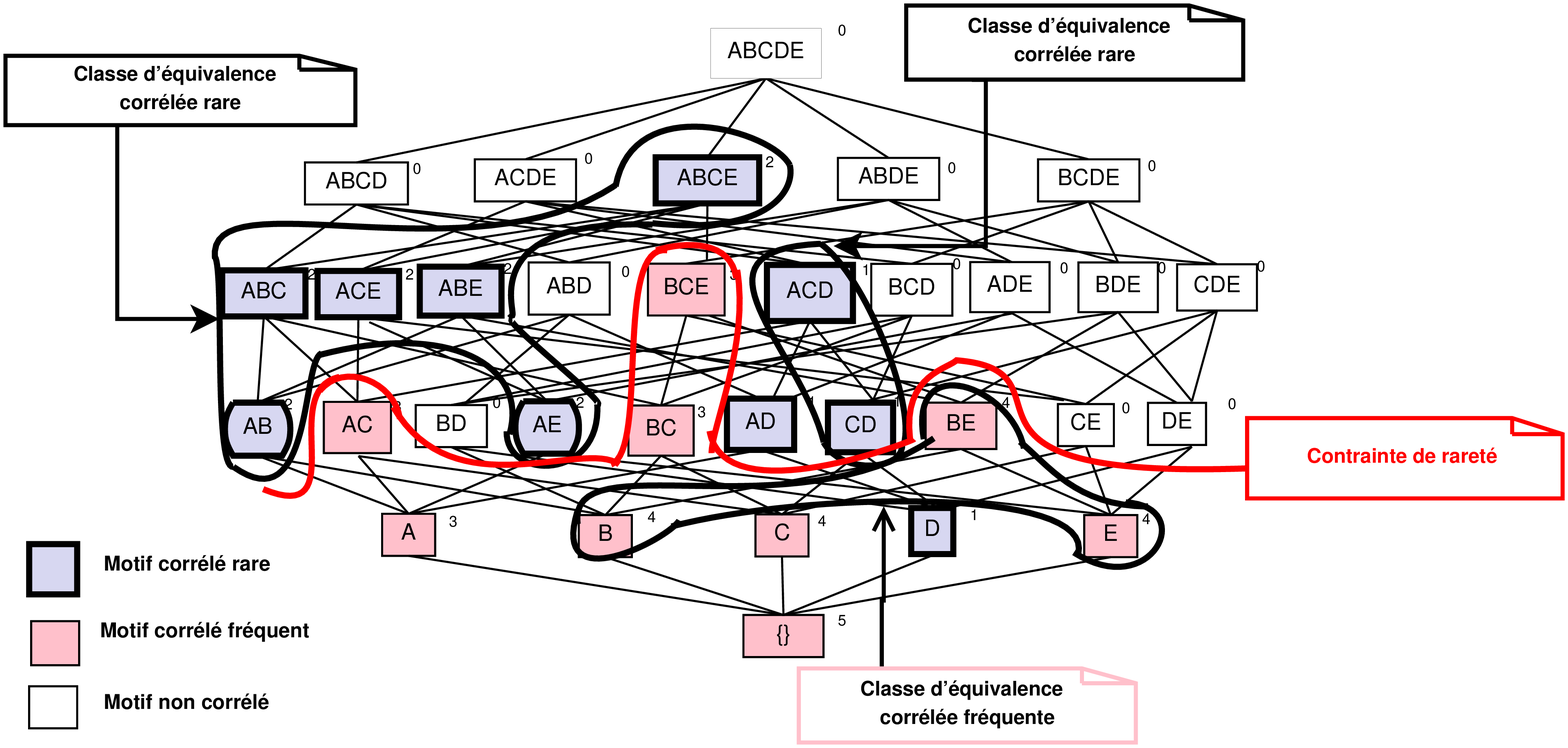}}
\caption{Effet de l'application de la contrainte de rareté pour \textit{minsupp} = 3 et \textit{minbond} = 0,2.}
\label{Exp2}
\end{figure}

En effet, pour chaque classe d'équivalence, la conservation de la valeur de la mesure \textit{bond} conserve bien évidemment le support conjonctif, le support disjonctif et le support négatif. Par conséquent, les éléments d'une même classe d'équivalence ont le même comportement quant aux contraintes de corrélation et de rareté. \`A cet égard, pour une classe d'équivalence corrélée, \textit{c.-é.-d.} celle contenant des motifs corrélés, les éléments sont tous rares ou sont tous fréquents.
Il en est de même pour une classe d'équivalence non corrélée, \textit{c.-é.-d.} dont les motifs associés sont non corrélés. Ainsi, ces classes d'équivalence ne seront pas affectées comme le montre la figure \ref{Exp2}. \`A cet égard, nous distinguons les classes corrélés fréquentes, les classes non corrélés fréquentes, les classes corrélés rares et les classes non corrélés rares.

\subsection{Synthése des deux scénarios}

La caractéristique des classes d'équivalence induites par $f_{bond}$ précédemment analysée, est trés intéressante. En effet, ceci n'est pas le cas de tous les opérateurs de fermeture. Par exemple, l'application de l'opérateur de fermeture associé au support conjonctif, \textit{c.-é.-d.} l'opérateur de fermeture conjonctive \cite{ganter99}, induit des classes d'équivalence oé le comportement d'un motif d'une classe donnée vis-é-vis de la contrainte de corrélation n'est pas représentatif du comportement du reste des motifs de la classe. Pour une classe donnée, chaque motif doit ainsi étre testé indépendamment des autres de la même classe pour savoir s'il est corrélé ou non.
Toutefois, contrairement à  la fermeture conjonctive, l'opérateur de fermeture $f_{bond}$ préserve non seulement le support conjonctif d'un motif mais aussi son support disjonctif, et par conséquent sa valeur de la mesure \textit{bond}.
\`A cet égard, nous optons pour le deuxiéme scénario lors de la conception de l'ensemble des approches que nous proposons. Dans ce qui suit, nous présentons l'étude de ces classes d'équivalence proposée dans \cite{rnti2011}.

\section{Caractérisation des classes d'équivalence corrélées rares} \label{Clsdéquiv}

Nous commenéons par présenter un exemple illustratif des classes d'équivalence corrélées rares.
\begin{exemple}
Considérons la base $\mathcal{D}$ illustrée par la table \ref{Base_transactions} pour \textit{minsupp} = 4 et \textit{minbond} = 0,2. Les classes d'équivalence corrélées rares dont les éléments de chacune sont récapitulés comme suit :
\begin{itemize}
\item $\mathcal{C}_1$ contient le motif \texttt{A}. Elle admet pour support conjonctif 3 et pour valeur de \textit{bond} 1.
\item $\mathcal{C}_2$ contient le motif \texttt{D}. Elle admet pour support conjonctif 1 et pour valeur de \textit{bond} 1.
\item $\mathcal{C}_3$ contient les motifs \texttt{AB}, \texttt{AE}, \texttt{ABC}, \texttt{ABE}, \texttt{ACE}, et \texttt{ABCE}. Elle admet pour support conjonctif 2 et pour valeur de \textit{bond} $\displaystyle\frac{2}{5}$. \texttt{ABCE} est le fermé corrélé de cette classe.
\item $\mathcal{C}_4$ contient le motif \texttt{AC}. Elle admet pour support conjonctif 3 et pour valeur de \textit{bond} $\displaystyle\frac{3}{4}$.
\item $\mathcal{C}_5$ contient le motif \texttt{AD}. Elle admet pour support conjonctif 1 et pour valeur de \textit{bond} $\displaystyle\frac{1}{3}$.
\item $\mathcal{C}_6$  contient les motifs \texttt{CD} et \texttt{ACD}. Elle admet pour support conjonctif 1 et pour valeur de \textit{bond} $\displaystyle\frac{1}{4}$. \texttt{ACD} est le fermé corrélé de cette classe.
\item $\mathcal{C}_7$ contient les motifs \texttt{BC}, \texttt{CE} et \texttt{BCE}. Elle admet pour support conjonctif 3 et pour valeur de \textit{bond} $\displaystyle\frac{3}{5}$. \texttt{BCE} est le fermé corrélé de cette classe.
\end{itemize}
\end{exemple}

%\begin {figure}\centering
%\includegraphics[scale = 0.5]{TreillisClsdéquiv.eps}
%\caption{Classes d'équivalence corrélées rares pour \textit{minsupp} = 4 et \textit{minbond} = 0,2.}
%\label{Exp3}
%\end{figure}

L'ensemble des motifs corrélés rares est ainsi partitionné en classes disjointes, les classes d'équivalence corrélées rares. Dans chaque classe, un motif fermé corrélé rare est alors le motif le plus large au sens de l'inclusion dans la classe. Par contre, les plus petits motifs sont les motifs minimaux corrélés rares incomparables selon la relation d'inclusion. Les motifs minimaux et fermés seront formellement définis dans ce qui suit.

\begin{definition}\label{MFCR} \textsc{(}\textbf{Motifs fermés corrélés rares}\textsc{)} L'ensemble $\mathcal{MFCR}$ des motifs fermés corrélés rares est défini par : $\mathcal{MFCR}$ = $\{$$I$ $\in$ $\mathcal{MCR}$$|$ $\forall$ $I_{1}$ $\supset$ $I$ : \textit{bond}\textsc{(}$I$\textsc{)} $>$ \textit{bond}\textsc{(}$I_{1}\textsc{)}\}$
\end{definition}
En fait, l'ensemble $\mathcal{MFCR}$ résulte de l'intersection entre l'ensemble des motifs corrélés rares et l'ensemble des motifs fermés corrélés. Ainsi, $\mathcal{MFCR}$ = $\mathcal{MCR}$ $\cap$ $\mathcal{MFC}$.

\begin{definition}\label{MMCR} \textsc{(}\textbf{Motifs minimaux corrélés rares}\textsc{)} L'ensemble $\mathcal{MMCR}$ des motifs minimaux corrélés rares est défini par : $\mathcal{MMCR}$ = $\{$$I$ $\in$ $\mathcal{MCR}$$|$ $\forall$ $I_{1}$ $\subset$ $I$ : \textit{bond}\textsc{(}$I$\textsc{)} $<$ \textit{bond}\textsc{(}$I_{1}\textsc{)}\}$.
\end{definition}
Cet ensemble, $\mathcal{MMCR}$, résulte de l'intersection entre l'ensemble des motifs corrélés rares et l'ensemble des motifs minimaux corrélés. Ainsi, $\mathcal{MMCR}$ = $\mathcal{MCR}$ $\cap$ $\mathcal{MMC}$.

\begin{exemple}\label{exemple_MMCR_et_MFCR}
Soit la base illustrée par la table \ref{Base_transactions} pour \textit{minsupp} = 4 et \textit{minbond} = 0,2. Le motif \texttt{ACD} $\in$ $\mathcal{MFCR}$. En effet, il est fermé corrélé \textsc{(}\textit{cf.} Exemple \ref{exempleMFC}\textsc{)}. Il est aussi rare \textsc{(}\textit{cf.} Exemple \ref{exemple_ensemble_MCR}\textsc{)}. Par ailleurs, le motif \texttt{AB} $\in$ $\mathcal{MMCR}$. En effet, il est minimal corrélé \textsc{(}\textit{cf.} Exemple \ref{exempleMMC}\textsc{)}. Il est aussi rare \textsc{(}\textit{cf.} Exemple \ref{exemple_ensemble_MCR}\textsc{)}.

En se référant aux diverses classes d'équivalence corrélées rares, nous avons l'ensemble $\mathcal{MFCR}$ composé des éléments maximaux de ces classes \textit{c.-é.-d.} \texttt{A},  \texttt{D}, \texttt{AC}, \texttt{AD}, \texttt{ACD}, \texttt{BCE} et \texttt{ABCE}. Par ailleurs, l'ensemble $\mathcal{MMCR}$ est composé des éléments minimaux de ces classes \textit{c.-é.-d.} \texttt{A}, \texttt{D}, \texttt{AB}, \texttt{AC}, \texttt{AD}, \texttt{AE}, \texttt{BC}, \texttt{CD} et \texttt{CE}. Les motifs \texttt{A}, \texttt{D}, \texttt{AC} et \texttt{AD} sont à  la fois des motifs fermés et minimaux. Leurs classes associées se réduisent donc chacune à  un unique élément.
\end{exemple}

Nous avons ainsi détaillé les propriétés des classes d'équivalence corrélées rares. Sur la base de ces derniéres, nous décrivons les différentes représentations proposées. Justifions d'abord nos motivations quant à  l'extraction des représentations concises exactes des motifs corrélés rares.

\section{Motivations de l'extraction des représentations concises exactes des motifs corrélés rares} \label{Motiv}

Notre choix d'extraire des représentations concises, et non la totalité de l'ensemble total $\mathcal{MCR}$ des motifs corrélés rares, a été motivé par deux principales raisons. Premiérement, l'extraction des représentations concises est possible et plus efficace dans un certain nombre de cas que l'extraction de l'ensemble total $\mathcal{MCR}$. L'extraction de ce dernier est souvent coéteuse en espace mémoire et en temps d'exécution.

Une représentation exacte des motifs corrélés rares doit permettre de déterminer si un motif arbitraire est corrélé rare ou non, et s'il est corrélé rare, la représentation doit permettre de dériver sans perte d'information son support et sa valeur de la mesure \textit{bond}. Dans ce sens, les représentations proposées dans ce travail seront montrées comme étant toujours de taille plus réduite que l'ensemble total des motifs corrélés rares. Elles permettent ainsi une meilleure exploitation des connaissances extraites. En plus, étant sans perte d'information, elles permettent la dérivation quand ceci est nécessaire de tous les motifs corrélés rares non retenus dans une représentation donnée.

Afin de proposer les représentations concises des motifs corrélés rares, nous nous basons sur les classes d'équivalence. Ces classes permettent de ne retenir que les motifs non-redondants. En effet, parmi les motifs d'une classe donnée, seuls ceux nécessaire à  la régénération de l'ensemble total des motifs corrélés rares seront retenus dans une représentation donnée. Le reste des motifs de la classe ne sera donc pas maintenu, ce qui réduit la redondance dans les connaissances extraites. Ces classes d'équivalence facilitent aussi l'exploration de l'espace de recherche des motifs corrélés rares. En effet, l'application d'un opérateur de fermeture permet de passer de l'élément minimal d'une classe à  son élément maximal sans avoir à  parcourir les niveaux intermédiaires.

Aprés avoir introduit les propriétés des classes d'équivalence corrélées rares, nous introduisons dans la suite les représentations concises proposées.

\section{Représentations concises exactes des motifs corrélés rares}

Une premiére idée intuitive afin de proposer une représentation concise exacte des motifs corrélés rares serait de voir si les éléments minimaux ou les éléments maximaux des classes d'équivalence associées permettraient de représenter sans perte d'information cet ensemble. Dans ce sens, il est important de rappeler que l'ensemble $\mathcal{MCR}$ des motifs corrélés rares résulte de l'intersection de l'idéal d'ordre des motifs corrélés et du filtre d'ordre des motifs rares. Cet ensemble $\mathcal{MCR}$ ne forme donc ni un idéal d'ordre ni un filtre d'ordre. Dans cette situation, pris chacun indépendamment de l'autre, est ce que l'ensemble $\mathcal{MMCR}$ ou l'ensemble $\mathcal{MFCR}$ peut constituer une représentation concise exacte des motifs rares corrélés?

Analysons dans ce qui suit chacun de ces deux ensembles :

- Commenéons par l'ensemble $\mathcal{MMCR}$ des minimaux des classes d'équivalence corrélées rares. De part la nature de ses éléments -- minimaux de leurs classes d'équivalence -- cet ensemble permet pour un motif donné $I$ de vérifier s'il est rare ou non. En effet, il suffit de trouver un élément $J$ $\in$ $\mathcal{MMCR}$, tel que $J$ $\subseteq$ $I$ pour décider que $I$ est un motif rare. Si ce n'est pas le cas, alors $I$ n'est pas un motif rare. Toutefois, l'ensemble $\mathcal{MMCR}$ ne permet pas de déterminer dans le cas général si $I$ est corrélé ou non \textsc{(}ceci n'est possible que si $I$ $\in$ $\mathcal{MMCR}$\textsc{)}. En effet, même s'il existe $J$ $\in$ $\mathcal{MMCR}$, tel que $J$ $\subset$ $I$, et même sachant que $J$ est corrélé, nous ne pouvons rien décider quant au statut de $I$ vis-é-vis de la contrainte de corrélation puisque cette derniére est anti-monotone \textsc{(}le fait que $J$ est corrélé n'implique pas que $I$ l'est aussi\textsc{)}. Ainsi, $\mathcal{MMCR}$ ne peut pas constituer une représentation exacte de $\mathcal{MCR}$.

- Traitons maintenant le cas de l'ensemble $\mathcal{MFCR}$ des maximaux des classes d'équivalence corrélées rares. D'une maniére duale à  $\mathcal{MMCR}$, les éléments de $\mathcal{MFCR}$ permettent de déterminer pour un motif $I$ s'il est corrélé ou non. Il suffit qu'il soit inclus dans un motif $J$ $\in$ $\mathcal{MFCR}$, et sinon $I$ n'est pas corrélé. Toutefois, de part leur nature, les fermés appartenant à  $\mathcal{MFCR}$ ne permettent pas dans le cas général de dériver l'information concernant le statut de rareté d'un motif $I$ quelconque \textsc{(}ceci n'est possible que si $I$ $\in$ $\mathcal{MFCR}$\textsc{)}. En effet, même s'il existe $J$ $\in$ $\mathcal{MFCR}$, tel que $I$ $\subset$ $J$, et même sachant que $J$ est rare, nous ne pouvons pas savoir si $I$ est rare ou non puisque la contrainte de rareté est monotone \textsc{(}le fait que $J$ est rare n'implique pas que $I$ l'est aussi\textsc{)}. Ainsi, $\mathcal{MFCR}$ ne peut pas constituer une représentation exacte de $\mathcal{MCR}$.
Il résulte de l'analyse que nous venons d'effectuer que la complémentarité entre $\mathcal{MMCR}$ et $\mathcal{MFCR}$ peut constituer une représentation exacte des motifs corrélés rares. Cette premiére alternative est étudiée dans la sous-section qui suit, et qui sera suivie par deux optimisations afin de ne retenir que les éléments indispensables à  la régénération sans perte d'information des éléments de $\mathcal{MCR}$.

\subsection{La représentation concise exacte \textbf{$\mathcal{RMCR}$}}

La premiére représentation que nous proposons est définie comme suit.
\begin{definition}\label{rmcr} \textsc{(}\textbf{Représentation $\mathcal{RMCR}$}\textsc{)} \cite{rnti2011}\\
Soit $\mathcal{RMCR}$ la représentation concise exacte des motifs corrélés rares basée sur l'ensemble $\mathcal{MFCR}$ des motifs fermés corrélés rares et sur l'ensemble $\mathcal{MMCR}$ des motifs minimaux corrélés rares. La représentation $\mathcal{RMCR}$ est définie comme suit : $\mathcal{RMCR}$ = $\mathcal{MFCR}$ $\cup$ $\mathcal{MMCR}$. Chaque élément $I$ de $\mathcal{RMCR}$ est muni de son support, \textit{Supp}\textsc{(}$\wedge$$I$\textsc{)}, et sa mesure \textit{bond}, \textit{bond}\textsc{(}$I$\textsc{)}.
\end{definition}

\begin{exemple} \label{exprep1}
Considérons la base de transactions donnée dans la table \ref{Base_transactions}, pour \textit{minsupp} = 4 et \textit{minbond} = 0,2.
En considérant les ensembles $\mathcal{MFCR}$ et $\mathcal{MMCR}$ \textsc{(}\textit{cf.} Exemple \ref{exemple_MMCR_et_MFCR}\textsc{)}, la représentation $\mathcal{RMCR}$ est composée par :
\textsc{(}\texttt{A}, $3$, $\displaystyle\frac{3}{3}$\textsc{)},
\textsc{(}\texttt{D}, $1$, $\displaystyle\frac{1}{1}$\textsc{)},
\textsc{(}\texttt{AB}, $2$, $\displaystyle\frac{2}{5}$\textsc{)},
\textsc{(}\texttt{AC}, $3$, $\displaystyle\frac{3}{4}$\textsc{)},
\textsc{(}\texttt{AD}, $1$, $\displaystyle\frac{1}{3}$\textsc{)},
\textsc{(}\texttt{AE}, $2$, $\displaystyle\frac{2}{5}$\textsc{)},
\textsc{(}\texttt{BC}, $3$, $\displaystyle\frac{3}{5}$\textsc{)},
\textsc{(}\texttt{CD}, $1$, $\displaystyle\frac{1}{4}$\textsc{)},
\textsc{(}\texttt{CE}, $3$, $\displaystyle\frac{3}{5}$\textsc{)},
\textsc{(}\texttt{ACD}, $1$, $\displaystyle\frac{1}{4}$\textsc{)},
\textsc{(}\texttt{BCE}, $3$, $\displaystyle\frac{3}{5}$\textsc{)}
et \textsc{(}\texttt{ABCE}, $2$, $\displaystyle\frac{2}{5}$\textsc{)}.
\end{exemple}

Le théoréme suivant montre que les éléments de $\mathcal{RMCR}$ représentent sans perte d'information les motifs corrélés rares.
\begin{theoreme}\label{representation1_exacte}
La représentation $\mathcal{RMCR}$ est une représentation concise exacte de l'ensemble $\mathcal{MCR}$ des motifs corrélés rares.
\end{theoreme}
\begin{Preuve}
Soit un motif $I$ $\subseteq$ $\mathcal{I}$. Trois cas se présentent :

\textbf{a\textsc{)}} Si $I$ $\in$ $\mathcal{RMCR}$, alors $I$ est un motif corrélé rare et nous avons son support et sa valeur de la mesure \textit{bond}.

\textbf{b\textsc{)}} Si $\nexists$ $J$ $\in$ $\mathcal{RMCR}$ tel que $J$ $\subseteq$ $I$ ou $\nexists$ $Z$ $\in$ $\mathcal{RMCR}$ tel que $I$ $\subseteq$ $Z$, alors $I$ $\notin$ $\mathcal{MCR}$ puisque $I$ n'appartient à  aucune classe d'équivalence corrélée rare.

\textbf{c\textsc{)}} Sinon, $I$ $\in$ $\mathcal{MCR}$. En effet, d'aprés la proposition \ref{prop_propriete_ens_MCR}, $I$ est corrélé puisque inclus dans un motif corrélé, à  savoir $Z$. Il est aussi rare puisque englobant un motif rare, à  savoir $J$. Dans ce cas, il suffit de localiser la fermeture de $I$ par $f_{bond}$, disons $F$. $F$ appartient nécessairement à  $\mathcal{RMCR}$ puisque $I$ est un motif corrélé rare et $\mathcal{RMCR}$ inclut l'ensemble $\mathcal{MFCR}$ des motifs fermés corrélés rares. Par conséquent, $F$ = $min_{\subseteq}$\{$I_1$ $\in$ $\mathcal{RMCR}$$|$ $I$ $\subseteq$ $I_1$\}. Comme l'opérateur $f_{bond}$ préserve la mesure \textit{bond} et par conséquent le support conjonctif \textsc{(}\textit{cf.} Proposition \ref{proprietes_CEq_f_bond}\textsc{)}, nous avons : \textit{bond}\textsc{(}$I$\textsc{)} = \textit{bond}\textsc{(}$F$\textsc{)} et
\textit{Supp}\textsc{(}$\wedge$$I$\textsc{)} = \textit{Supp}\textsc{(}$\wedge$$F$\textsc{)}.
\end{Preuve}
\begin{exemple}\label{exemple_3_cas_RMCR} 
Considérons la représentation $\mathcal{RMCR}$ donnée dans l'exemple précédent. Illustrons chacun des trois cas. Le motif \texttt{AD} $\in$ $\mathcal{RMCR}$. Ainsi, nous avons son support égal à  1 et sa valeur de la mesure bond égale à  $\displaystyle\frac{1}{3}$. 

Considérons le motif \texttt{BE}. Bien qu'il soit inclus dans deux motifs de $\mathcal{RMCR}$, à  savoir \texttt{BCE} et \texttt{ABCE}, \texttt{BE} $\notin$ $\mathcal{MCR}$ puisque aucun élément de $\mathcal{RMCR}$ n'est inclus dans \texttt{BE}. Soit maintenant le motif \texttt{ABC}. Il existe deux motifs de $\mathcal{RMCR}$ qui vérifient la condition faisant de \texttt{ABC} un motif corrélé rare, à  savoir \texttt{AB} et \texttt{ABCE}, puisque \texttt{AB} $\subseteq$ \texttt{ABC} $\subseteq$ \texttt{ABCE}. Le plus petit motif de $\mathcal{RMCR}$ couvrant \texttt{ABC}, \textit{c.-é.-d.} sa fermeture, est \texttt{ABCE}. Ainsi, \textit{bond}\textsc{(}\texttt{ABC}\textsc{)} = \textit{bond}\textsc{(}\texttt{ABCE}\textsc{)} = $\displaystyle\frac{2}{5}$, et \textit{Supp}\textsc{(}$\wedge$\texttt{ABC}\textsc{)} = \textit{Supp}\textsc{(}$\wedge$\texttt{ABCE}\textsc{)} = 2.
\end{exemple}

Il est important de noter que la représentation $\mathcal{RMCR}$ est une \textit{couverture parfaite} de l'ensemble $\mathcal{MCR}$. En effet, la taille de la représentation $\mathcal{RMCR}$ \textit{ne dépasse jamais} celle de l'ensemble $\mathcal{MCR}$ quelle que soit la base et les valeurs de \textit{minsupp} et de \textit{minbond} considérées. En effet, nous avons toujours \textsc{(}$\mathcal{MFCR}$ $\cup$ $\mathcal{MMCR}$\textsc{)} $\subseteq$ $\mathcal{MCR}$.

\begin{remarque} Il est important de noter que nous sommes obligés de maintenir, pour un motif $I$ de la représentation, à  la fois le \textit{Supp}\textsc{(}$\wedge$$I$\textsc{)} et \textit{bond}\textsc{(}$I$\textsc{)}. D'une part, la valeur de \textit{bond}\textsc{(}$I$\textsc{)} étant un rapport entre le support conjonctif et celui disjonctif de $I$ ne permet pas de dériver le support conjonctif de $I$. D'autre part, ayant le support conjonctif d'un motif $I$, ceci n'est pas suffisant pour calculer la valeur de sa mesure \textit{bond}. En effet, ceci nécessite la connaissance de son support disjonctif. Ce dernier ne peut étre dérivé moyennant les identités d'inclusion-exclusion que connaissant les supports conjonctifs de tous les sous-ensembles de $I$ \cite{galambos}. Toutefois, si $I$ est un motif corrélé rare, tous ses-ensembles ne le sont pas forcément et par conséquent nous n'avons pas accés à  leurs supports conjonctifs respectifs. Ainsi, il faut retenir le support conjonctif et la valeur de la mesure \textit{bond} pour chaque élément de la représentation. C'est aussi la raison pour laquelle les représentations concises basées sur les régles de déduction \cite{calderssurvey} et celles basées sur les identités d'inclusion-exclusion \cite{casalidawak05,tarekdke09} ne sont pas applicables pour représenter l'ensemble des motifs corrélés rares. En effet, ces représentations nécessitent pour un motif donné la connaissance du support conjonctif ou disjonctif, suivant la représentation, associé à  tous ses sous-ensembles. Ceci est exigé que ce soit pour retenir les éléments de la représentation ou pour la dérivation des informations associées aux éléments non-retenus dans la représentation.

Dans cette situation, les motifs fermés et minimaux des classes d'équivalence offrent comme montré précédemment
une solution intéressante pour représenter d'une maniére concise l'ensemble des motifs corrélés rares. En effet, la localisation de tels motifs nécessite un voisinage restreint, les sur-ensembles immédiats et les sous-ensembles immédiats respectivement, et non tous leurs sous-ensembles respectifs, comme c'est le cas par exemple des motifs non-dérivables \cite{NDI_2007}. En plus, la dérivation des supports des motifs à  partir des fermés et des minimaux est réalisée d'une maniére directe, contrairement par exemple aux motifs essentiels \cite{casalidawak05} et aux motifs non-dérivables qui nécessitent tous les sous-ensembles du motif dérivé.
\end{remarque}

\begin{remarque}\label{remarque_gestion_MMCR_vs_MFCR} Il est aussi intéressant de signaler que le fait de considérer, dans $\mathcal{RMCR}$, l'union entre les ensembles $\mathcal{MMCR}$ et $\mathcal{MFCR}$ permet d'éviter la redondance -- à  cause de la duplication d'un motif donné -- qui peut apparaétre dans la représentation si nous considérons chacun des ensembles $\mathcal{MMCR}$ et $\mathcal{MFCR}$ séparément. Par exemple, si nous considérons l'exemple \ref{exprep1}, nous remarquons que les éléments
\textsc{(}\texttt{A}, 3, $\displaystyle\frac{3}{3}$\textsc{)}, \textsc{(}\texttt{D}, 1, $\displaystyle\frac{1}{1}$\textsc{)}, \textsc{(}\texttt{AC}, 3, $\displaystyle\frac{3}{4}$\textsc{)}
et \textsc{(}\texttt{AD}, 1, $\displaystyle\frac{1}{3}$\textsc{)} appartiennent à  la fois à  l'ensemble $\mathcal{MMCR}$ et $\mathcal{MFCR}$. Toutefois, un avantage de la gestion de chaque ensemble à  part est la réduction de certains tests d'inclusion lors de l'interrogation de la représentation. Le choix entre tolérer une certaine duplication et réduire éventuellement le coét de la régénération dépend de la nature de l'application oé il sera éventuellement question de privilégier soit l'espace mémoire soit les temps de dérivation. Notons qu'une solution intermédiaire serait de localiser dans un premier groupe les éléments qui sont à  la fois des motifs fermés et des motifs minimaux tels que \texttt{A}, \texttt{D}, \texttt{AC} et \texttt{AD} dans notre cas, dans un second le reste des minimaux à  part, et le reste des fermés formeront un troisiéme groupe. Le premier et le second groupe seront utilisés pour les traitements oé les minimaux seront utiles, tandis que le premier et le troisiéme seront utilisés pour les traitements nécessitant les fermés.
\end{remarque}

Nous proposons dans la suite de cette section deux optimisations de $\mathcal{RMCR}$ permettant de réduire encore plus le nombre de motifs à  retenir dans la représentation tout en garantissant la non-perte d'information concernant l'ensemble $\mathcal{MCR}$ des motifs corrélés rares.

\subsection{La représentation concise exacte $\mathcal{RMM}$$ax$$\mathcal{F}$}

Cette premiére optimisation se base sur le fait que l'ensemble $\mathcal{MMCR}$ des motifs minimaux corrélés rares augmenté seulement des maximaux, par rapport à  l'inclusion ensembliste, parmi les motifs fermés corrélés rares est suffisant pour représenter d'une maniére exacte l'ensemble $\mathcal{MCR}$. L'ensemble $\mathcal{MFCRM}ax$ des motifs fermés corrélés rares maximaux est défini comme suit :

\begin{definition}\label{defMF} \textsc{(}\textbf{Ensemble $\mathcal{MFCRM}ax$ des motifs fermés corrélés rares maximaux}\textsc{)} L'ensemble $\mathcal{MFCRM}ax$ correspond aux motifs qui sont à  la fois des motifs fermés corrélés rares \textsc{(}\textit{cf.} Définition \ref{MFCR}, page \pageref{MFCR}\textsc{)} et des motifs corrélés maximaux \textsc{(}\textit{cf.} Définition \ref{bdpos}, page \pageref{bdpos}\textsc{)}. Ainsi, $\mathcal{MFCRM}ax$ = $\mathcal{MFCR}$ $\cap$ $\mathcal{MCM}ax$.
\end{definition}

L'ensemble $\mathcal{MFCRM}ax$ est donc restreint aux éléments de $\mathcal{MCM}ax$ qui sont aussi rares \textsc{(}en plus d'étre les plus grands motifs corrélés\textsc{)}.

\begin{exemple}\label{exemple_MFCRMax} Soit la base illustrée par la table \ref{Base_transactions}. Pour \textit{minsupp} = 4 et \textit{minbond} = 0,2, $\mathcal{MFCR}$ = $\{$\texttt{A}, \texttt{D}, \texttt{AD}, \texttt{ACD}, \texttt{BCE}, \texttt{ABCE}$\}$ \textsc{(}\textit{cf.} Exemple \ref{exprep1}\textsc{)}.
Par ailleurs, pour \textit{minbond} = 0,2, $\mathcal{MCM}ax$ = $\{$\texttt{ACD}, \texttt{ABCE}$\}$ \textsc{(}\textit{cf.} Exemple \ref{exemple_MCMax}\textsc{)}.
Ainsi, $\mathcal{MFCRM}ax$ = $\mathcal{MFCR}$ $\cap$ $\mathcal{MCM}ax$ = $\{$\texttt{ACD}, \texttt{ABCE}$\}$. En effet, les fermés corrélés \texttt{A}, \texttt{D} et \texttt{AD} ne sont pas retenus puisqu'ils sont inclus dans \texttt{ACD}. Le fermé \texttt{BCE} sera aussi éliminé puisqu'il est inclus dans
\texttt{ABCE}.
\end{exemple}

La définition suivante présente la représentation des motifs corrélés rares basée sur cette optimisation.

\begin{definition} \label{rep2concise exacte def} \textsc{(}\textbf{Représentation $\mathcal{RMM}$$ax$$\mathcal{F}$}\textsc{)} \cite{rnti2011}\\
Soit $\mathcal{RMM}$$ax$$\mathcal{F}$ la représentation basée sur l'ensemble $\mathcal{MMCR}$ et l'ensemble $\mathcal{MFCRM}ax$. Nous avons $\mathcal{RMM}$$ax$$\mathcal{F}$ = $\mathcal{MMCR}$ $\cup$ $\mathcal{MFCRM}ax$. Chaque élément $I$ de $\mathcal{RMM}$$ax$$\mathcal{F}$ est muni de son support, \textit{Supp}\textsc{(}$\wedge$$I$\textsc{)}, et de sa mesure \textit{bond}, \textit{bond}\textsc{(}$I$\textsc{)}.
\end{definition}

\begin{exemple} \label{exprep2}
Considérons la base de transactions donnée dans la table \ref{Base_transactions}. Pour \textit{minsupp} = 4 et
\textit{minbond} = 0,2.
La représentation $\mathcal{RMM}$$ax$$\mathcal{F}$ est composée par :
\textsc{(}\texttt{A}, $3$, $\displaystyle\frac{3}{3}$\textsc{)},
\textsc{(}\texttt{D}, $1$, $\displaystyle\frac{1}{1}$\textsc{)},
\textsc{(}\texttt{AB}, $2$, $\displaystyle\frac{2}{5}$\textsc{)},
\textsc{(}\texttt{AD}, $1$, $\displaystyle\frac{1}{3}$\textsc{)},
\textsc{(}\texttt{AE}, $2$, $\displaystyle\frac{2}{5}$\textsc{)},
\textsc{(}\texttt{CD}, $1$, $\displaystyle\frac{1}{4}$\textsc{)},
\textsc{(}\texttt{BC}, $3$, $\displaystyle\frac{3}{5}$\textsc{)},
\textsc{(}\texttt{CE}, $3$, $\displaystyle\frac{3}{5}$\textsc{)},
\textsc{(}\texttt{AC}, $3$, $\displaystyle\frac{3}{4}$\textsc{)},
\textsc{(}\texttt{ACD}, $1$, $\displaystyle\frac{1}{4}$\textsc{)},
et \textsc{(}\texttt{ABCE}, $2$, $\displaystyle\frac{2}{5}$\textsc{)}.
Nous remarquons que, pour cet exemple, le seul élément appartenant é
la représentation $\mathcal{RMCR}$ et non à  la représentation $\mathcal{RMM}$$ax$$\mathcal{F}$ est le motif
\texttt{BCE}.
En effet, ceci est dé au fait que les motifs fermés éliminés de $\mathcal{MCM}ax$ sont eux mêmes des minimaux corrélés rares, à  savoir \texttt{D}, \texttt{A}, \texttt{AC} et \texttt{AD}. Toutefois, la représentation $\mathcal{RMM}$$ax$$\mathcal{F}$ serait plus réduite que $\mathcal{RMCR}$ si les ensembles $\mathcal{MMCR}$ et $\mathcal{MFCR}$ sont gérés séparément \textsc{(}\textit{cf.} Remarque \ref{remarque_gestion_MMCR_vs_MFCR}\textsc{)}. En effet, il n'y aura plus de duplication de \texttt{A}, \texttt{D}, \texttt{AC} et de \texttt{AD}.
\end{exemple}
Le théoréme \ref{representation2_exacte} montre que $\mathcal{RMM}$ax$\mathcal{F}$ couvre sans perte d'information l'ensemble $\mathcal{MCR}$.
\begin{theoreme}\label{representation2_exacte}
La représentation $\mathcal{RMM}$$ax$$\mathcal{F}$ est une représentation concise exacte de l'ensemble $\mathcal{MCR}$ des motifs corrélés rares.
\end{theoreme}
\begin{Preuve}
Soit un motif $I$ $\subseteq$ $\mathcal{I}$. Trois cas se présentent :
%%THH : ATTENTION_URGENT : trop de redondance avec la démonstration du théoréme 1.

\textbf{a\textsc{)}} Si $I$ $\in$ $\mathcal{RMM}$$ax$$\mathcal{F}$, alors $I$ est un motif corrélé rare et nous avons son support et sa valeur de la mesure \textit{bond}.

\textbf{b\textsc{)}} Si $\nexists$ $J$ $\in$ $\mathcal{RMM}$$ax$$\mathcal{F}$ tel que $J$ $\subseteq$ $I$ ou $\nexists$ $Z$ $\in$ $\mathcal{RMM}$$ax$$\mathcal{F}$ tel que $I$ $\subseteq$ $Z$, alors $I$ $\notin$ $\mathcal{MCR}$ puisque $I$ n'appartient à  aucune classe d'équivalence corrélée rare.

\textbf{c\textsc{)}} Sinon, $I$ $\in$ $\mathcal{MCR}$. En effet, d'aprés la proposition \ref{prop_propriete_ens_MCR}, $I$ est corrélé puisque inclus dans un motif corrélé, à  savoir $Z$. Il est aussi rare puisque englobant un motif rare, à  savoir $J$. Comme $I$ est un motif corrélé rare et la représentation $\mathcal{RMM}$$ax$$\mathcal{F}$ inclut l'ensemble $\mathcal{MMCR}$ contenant les éléments minimaux des différentes classes d'équivalence corrélées rares, cette représentation contient au moins un élément de la classe d'équivalence de $I$, en particulier tous les motifs minimaux de la classe.\\Comme le support conjonctif et la mesure \textit{bond} décroissent avec la taille des motifs, les valeurs du support conjonctif et de la mesure \textit{bond} de $I$ sont égales aux valeurs minimales des mesures associées à  ses sous-ensembles appartenant à  $\mathcal{RMM}$$ax$$\mathcal{F}$. Il en résulte que :

$\bullet$ \textit{Supp}\textsc{(}$\wedge$$I$\textsc{)} = $min$$\{$\textit{Supp}\textsc{(}$\wedge$$I_1$\textsc{)}$|$ $I_1$ $\in$ $\mathcal{RMM}$$ax$$\mathcal{F}$ et $I_1$ $\subseteq$ $I$$\}$, et,

$\bullet$ \textit{bond}\textsc{(}$I$\textsc{)} = $min$$\{$\textit{bond}\textsc{(}$I_1$\textsc{)}$|$ $I_1$ $\in$ $\mathcal{RMM}$$ax$$\mathcal{F}$ et $I_1$ $\subseteq$ $I$$\}$.

%%THH : ATTENTION_URGENT : les propriétés ici nécessitent des clarifications à  l'avance car elles sont difficiles à  %%cerner.
%%
%%%%%THH : la localisation de l'élément appartenant à  la classe d'équivalence de $I$ est un casse téte chinois car
%%%%%il faudra tenir compte de la valeur minimale de bond parmi celle de ses sous-ensembles.
%%Il suffit alors de localiser un motif $K$ de la classe de $I$ appartenant à  $\mathcal{RMM}$$ax$$\mathcal{F}$. Soit %%$\mathcal{MMCR}_{I}$ l'ensemble des motifs de la classe d'équivalence de $I$, inclus dans $I$ et appartenant à  %%$\mathcal{RMM}$$ax$$\mathcal{F}$. Cet ensemble est nécessairement non vide puisque
%%
%%Cet ensemble contient ainsi les éléments de tailles maximales de $\mathcal{RMM}$$ax$$\mathcal{F}$ inclus dans $I$ : %%$\mathcal{MMCR}_{I}$ = $max_{\subseteq}$\{$J$ $\in$ $\mathcal{RMM}$$ax$$\mathcal{F}$$|$ $J$ $\subseteq$ $I$\}. Soit
%%%%%THH : cette définition est fausse car il faudra aussi tenir compte de la valeur de bond.
%%$K$ $\in$ $\mathcal{MMCR}_{I}$. Comme les éléments d'une même classe d'équivalence induite par l'opérateur $f_{bond}$ %%ont la même mesure \textit{bond} et par conséquent le support conjonctif \textsc{(}\textit{cf.} Proposition %%\ref{proprietes_CEq_f_bond}\textsc{)}, nous avons : \textit{bond}\textsc{(}$I$\textsc{)} = %%\textit{bond}\textsc{(}$K$\textsc{)} et \textit{Supp}\textsc{(}$\wedge$$I$\textsc{)} = %%\textit{Supp}\textsc{(}$\wedge$$K$\textsc{)}.
\end{Preuve}

\begin{exemple} Soit la représentation $\mathcal{RMM}$$ax$$\mathcal{F}$ donnée dans l'exemple précédent. Le traitement du premier et du second cas est semblable à  ceux des deux premiers cas de la représentation $\mathcal{RMCR}$ \textsc{(}\textit{cf.} Exemple \ref{exemple_3_cas_RMCR}\textsc{)}.

Considérons donc le motif \texttt{ABE} pour illustrer le troisiéme cas. Il existe deux motifs de $\mathcal{RMM}$$ax$$\mathcal{F}$ qui vérifient la condition faisant de \texttt{ABE} un motif corrélé rare, à  savoir \texttt{AB} et \texttt{ABCE} \textsc{(}\texttt{AB} $\subseteq$ \texttt{ABE} $\subseteq$ \texttt{ABCE}\textsc{)}.
Les motifs de $\mathcal{RMM}$$ax$$\mathcal{F}$ inclus dans \texttt{ABE} sont  \texttt{AB} et \texttt{AE}.
Par conséquent, \textit{Supp}\textsc{(}$\wedge$\texttt{ABE}\textsc{)} = $min$$\{$ \textit{Supp}\textsc{(}$\wedge$AB\textsc{)}, \textit{Supp}\textsc{(}$\wedge$AE\textsc{)}$\}$ = $min$$\{$2, 2$\}$ = 2, et
\textit{bond}\textsc{(}\texttt{ABE}\textsc{)} = $min$$\{$
\textit{bond}\textsc{(}AB\textsc{)}, \textit{bond}\textsc{(}AE\textsc{)}$\}$ = $min$$\{$$\displaystyle\frac{2}{5}$, $\displaystyle\frac{2}{5}$$\}$ = $\displaystyle\frac{2}{5}$.
\end{exemple}

\'Etant incluse dans $\mathcal{RMCR}$, qui a été montrée comme étant une couverture parfaite de $\mathcal{MCR}$, la représentation $\mathcal{RMM}$$ax$$\mathcal{F}$ est aussi une couverture parfaite de $\mathcal{MCR}$.

La sous-section suivante présente une autre optimisation de la représentation $\mathcal{RMCR}$.

\subsection{La représentation concise exacte $\mathcal{RM}$$in$$\mathcal{MF}$}
D'une maniére duale à  la représentation précédente, il suffit de retenir dans $\mathcal{RMCR}$ que les motifs minimaux, par rapport à  l'inclusion ensembliste, parmi ceux de l'ensemble $\mathcal{MMCR}$ des motifs minimaux corrélés rares. L'élagage des autres éléments de $\mathcal{MMCR}$ sera prouvé comme étant sans perte d'information lors de la régénération de l'ensemble $\mathcal{MCR}$ des motifs corrélés rares. L'ensemble $\mathcal{MMCRM}in$ des motifs minimaux parmi ceux de $\mathcal{MMCR}$ est défini comme suit :

\begin{definition}\label{defMM} \textsc{(}\textbf{Ensemble $\mathcal{MMCRM}in$ des éléments minimaux de l'ensemble $\mathcal{MMCR}$}\textsc{)} L'ensemble $\mathcal{MMCRM}in$ contient les motifs qui sont à  la fois des motifs minimaux corrélés rares \textsc{(}\textit{cf.} Définition \ref{MMCR}, page \pageref{mrm}\textsc{)} et des motifs rares minimaux \textsc{(}\textit{cf.} Définition \ref{mrm}, page \pageref{mrm}\textsc{)}. Ainsi, $\mathcal{MMCRM}in$ = $\mathcal{MMCR}$ $\cap$ $\mathcal{MRM}in$.
\end{definition}
L'ensemble $\mathcal{MMCRM}in$ est donc restreint aux éléments de $\mathcal{MRM}in$ qui sont aussi corrélés \textsc{(}en plus d'étre les plus petits motifs rares\textsc{)}.

\begin{exemple} \label{expMM} Soit la base illustrée par la table \ref{Base_transactions}. Pour \textit{minsupp} = 4 et \textit{minbond} = 0,2, $\mathcal{MMCR}$ = $\{$\texttt{A}, \texttt{D}, \texttt{AB}, \texttt{AD}, \texttt{AE}, \texttt{CD}, \texttt{AC}, \texttt{BC}, \texttt{CE}$\}$
\textsc{(}\textit{cf.} Exemple \ref{exprep1}\textsc{)}.
Par ailleurs, pour \textit{minsupp} = 4, $\mathcal{MRM}in$ = $\{$\texttt{A}, \texttt{D}, \texttt{BC}, \texttt{CE}$\}$.
Ainsi, $\mathcal{MMCRM}in$ = $\mathcal{MMCR}$ $\cap$ $\mathcal{MRM}in$ = $\{$\texttt{A}, \texttt{D}, \texttt{BC}, \texttt{CE}$\}$.
Les motifs  minimaux corrélés rares \texttt{AB}, \texttt{AD}, \texttt{AE} et \texttt{AC} ne sont pas retenus puisqu'ils englobent le motif \texttt{A}. De même pour le motif \texttt{CD}, il sera éliminé puisqu'il englobe le motif \texttt{D}.
\end{exemple}
\begin{remarque} Il est important de noter que dans l'exemple précédent, nous avons $\mathcal{MRM}in$ $\subseteq$ $\mathcal{MMCR}$. Toutefois, ceci n'est pas le cas d'une maniére générale. En effet, un motif rare minimal peut bien évidemment ne pas étre corrélé et donc ne pas appartenir à  $\mathcal{MMCR}$. Cette remarque s'applique aussi pour le cas de l'exemple \ref{exemple_MFCRMax} oé $\mathcal{MCM}ax$ $\subseteq$ $\mathcal{MFCR}$. En effet, un motif corrélé maximal peut ne pas étre rare et donc ne pas appartenir à  $\mathcal{MFCR}$.
\end{remarque}

La définition \ref{rep3concise exacte def} présente la représentation résultante de l'utilisation de $\mathcal{MMCRM}in$.
\begin{definition}\label{rep3concise exacte def} \textsc{(}\textbf{Représentation $\mathcal{RM}$$in$$\mathcal{MF}$}\textsc{)} \cite{rnti2011}\\
Soit $\mathcal{RM}$$in$$\mathcal{MF}$ la représentation basée sur l'ensemble $\mathcal{MFCR}$ et l'ensemble $\mathcal{MMCRM}in$. Nous avons $\mathcal{RM}$$in$$\mathcal{MF}$ = $\mathcal{MFCR}$ $\cup$ $\mathcal{MMCRM}in$. Chaque élément $I$ de $\mathcal{RM}$$in$$\mathcal{MF}$ est muni de son support, \textit{Supp}\textsc{(}$\wedge$$I$\textsc{)}, et sa mesure \textit{bond}, \textit{bond}\textsc{(}$I$\textsc{)}.
\end{definition}
\begin{exemple} \label{exprep2}
Considérons la base de transactions donnée dans la table \ref{Base_transactions}, pour \textit{minsupp} = 4 et
\textit{minbond} = 0,2. La représentation $\mathcal{RM}$$in$$\mathcal{MF}$ est composée par : 
\textsc{(}\texttt{A}, 3, $\displaystyle\frac{3}{3}$\textsc{)},
\textsc{(}\texttt{D}, 1, $\displaystyle\frac{1}{1}$\textsc{)},
\textsc{(}\texttt{AC}, 3, $\displaystyle\frac{3}{4}$\textsc{)},
\textsc{(}\texttt{AD}, 1, $\displaystyle\frac{1}{3}$\textsc{)},
\textsc{(}\texttt{BC}, 3, $\displaystyle\frac{3}{5}$\textsc{)},
\textsc{(}\texttt{CE}, 3, $\displaystyle\frac{3}{5}$\textsc{)},
\textsc{(}\texttt{ACD}, 1, $\displaystyle\frac{1}{4}$\textsc{)},
\textsc{(}\texttt{BCE}, 3, $\displaystyle\frac{3}{5}$\textsc{)},
et \textsc{(}\texttt{ABCE}, 2, $\displaystyle\frac{2}{5}$\textsc{)}.

Nous remarquons que, comparée à  $\mathcal{RMCR}$, cette représentation admet trois éléments en moins à  savoir \texttt{AB}, \texttt{AE} et \texttt{CD}.
\end{exemple}

Le théoréme suivant prouve que cette troisiéme représentation est aussi sans perte d'information de l'ensemble $\mathcal{MCR}$ des motifs corrélés rares.
\begin{theoreme}\label{representation3_exacte}
La représentation $\mathcal{RM}$$in$$\mathcal{MF}$ est une représentation concise exacte de l'ensemble $\mathcal{MCR}$ des motifs corrélés rares.
\end{theoreme}

Afin de démontrer ce théoréme, nous adoptons le même principe de démonstration que les
deux autres théorémes précédents.\\
\begin{Preuve}
Soit un motif $I$ $\subseteq$ $\mathcal{I}$. Trois cas se présentent :

\textbf{a\textsc{)}} Si $I$ $\in$ $\mathcal{RM}$$in$$\mathcal{MF}$, alors $I$ est un motif corrélé rare et nous avons son support et sa valeur de la mesure \textit{bond}.

\textbf{b\textsc{)}} Si $\nexists$ $J$ $\in$ $\mathcal{RM}$$in$$\mathcal{MF}$ tel que $J$ $\subseteq$ $I$ ou $\nexists$ $Z$ $\in$ $\mathcal{RM}$$in$$\mathcal{MF}$ tel que $I$ $\subseteq$ $Z$, alors $I$ $\notin$ $\mathcal{MCR}$ puisque $I$ n'appartient à  aucune classe d'équivalence corrélée rare.

\textbf{c\textsc{)}} Sinon, $I$ $\in$ $\mathcal{MCR}$. En effet, d'aprés la proposition \ref{prop_propriete_ens_MCR}, $I$ est corrélé puisque inclus dans un motif corrélé, à  savoir $Z$. Il est aussi rare puisque englobant un motif rare, à  savoir $J$. Comme l'ensemble $\mathcal{MFCR}$ appartient à  $\mathcal{RM}$$in$$\mathcal{MF}$, il suffit de localiser le fermé corrélé de $I$, disons $F$, égal à  : $F$ = $min_{\subseteq}$\{$I_1$ $\in$ $\mathcal{RM}$$in$$\mathcal{MF}$$|$ $I$ $\subseteq$ $I_1$\}. Ainsi, \textit{bond}\textsc{(}$I$\textsc{)} = \textit{bond}\textsc{(}$F$\textsc{)} et
\textit{Supp}\textsc{(}$\wedge$$I$\textsc{)} = \textit{Supp}\textsc{(}$\wedge$$F$\textsc{)}.
\end{Preuve}

%%%%THH : on peut expliquer ici qu'on peut faire la dérivation comme dans le cas de la représentation 2
%%Il est à  noter que pour cette représentation ainsi que pour la premiére, à  savoir $\mathcal{RMCR}$, la %%localisation %%du fermé associé à  $I$ dans le cas c\textsc{)} n'est pas nécessaire. En effet, ...
%%%%\textsc{(}comparaison de supports et de bond avec les éléments englobant $I$\textsc{)}.

\begin{exemple} Soit la représentation $\mathcal{RM}$$in$$\mathcal{MF}$ donnée dans l'exemple précédent. Le traitement du premier et du second cas est semblable à  ceux des deux premiers cas de la représentation $\mathcal{RMCR}$ \textsc{(}\textit{cf.} Exemple \ref{exprep1}\textsc{)}.

Considérons donc le motif \texttt{ABC} pour illustrer le troisiéme cas.
Il existe deux motifs \texttt{ABCE} et \texttt{BC} tel que 
\textsc{(}\texttt{BC} $\subseteq$ \texttt{ABC} et \texttt{ABC} $\subseteq$ \texttt{ABCE}\textsc{)}. Nous concluons donc que le motif \texttt{ABC} est un motif corrélé rare. 
Le fermé associé au motif \texttt{ABC} correspond à  \texttt{ABCE}.
Par conséquent, \textit{Supp}\textsc{(}$\wedge$\texttt{ABC}\textsc{)} = \textit{Supp}\textsc{(}$\wedge$\texttt{ABCE}\textsc{)} = 2
et \textit{bond}\textsc{(}\texttt{ABC}\textsc{)} = \textit{bond}\textsc{(}\texttt{ABCE}\textsc{)}
= $\displaystyle\frac{2}{5}$.
\end{exemple}

\'Etant incluse dans $\mathcal{RMCR}$, $\mathcal{RM}$$in$$\mathcal{MF}$ est comme les deux précédentes représentations une couverture parfaite de $\mathcal{MCR}$.

Nous avons, ainsi, décrit et analysé les caractéristiques des différentes représentations concises exactes des motifs corrélés rares. Dans ce qui suit, nous présentons une représentation concise approximative  résultante de la jointure entre les deux représentations concises exactes  $\mathcal{RM}$$in$$\mathcal{MF}$ et $\mathcal{RMM}$$ax$$\mathcal{F}$.

\section{La représentation concise approximative $\mathcal{RM}$$in$$\mathcal{MM}$$ax$$\mathcal{F}$}

En se basant sur les deux représentations concises exactes $\mathcal{RM}$$in$$\mathcal{MF}$ et $\mathcal{RMM}$$ax$$\mathcal{F}$ précédemment décrites, nous définissons la représentation concise approximative $\mathcal{RM}$$in$$\mathcal{M}$-$\mathcal{M}$$ax$$\mathcal{F}$.
%%%THH : ATTENTION au saut de ligne

Cette derniére est composée de l'ensemble $\mathcal{MFCRM}ax$ des motifs fermés corrélés rares maximaux
\textsc{(}\textit{cf.} Définition \ref{defMF}\textsc{)} et de l'ensemble
$\mathcal{MMCRM}in$ des éléments minimaux de l'ensemble $\mathcal{MMCR}$
\textsc{(}\textit{cf.} Définition \ref{defMM}\textsc{)}. Nous définissons cette représentation formellement comme suit.

\begin{definition} \textbf{\textsc{(}Représentation $\mathcal{RM}$$in$$\mathcal{MM}$$ax$$\mathcal{F}$\textsc{)}} \label{rep4concise approxdef} \cite{rnti2011}\\
Soit $\mathcal{RM}$$in$$\mathcal{MM}$$ax$$\mathcal{F}$ la représentation basée sur
l'ensemble $\mathcal{MFCRM}ax$  et sur l'ensemble $\mathcal{MMCRM}in$.
Nous avons $\mathcal{RM}$$in$$\mathcal{MM}$$ax$$\mathcal{F}$  = $\mathcal{MFCRM}ax$ $\cup$ $\mathcal{MMCRM}in$.
Chaque élément $I$ de $\mathcal{RM}$$in$$\mathcal{MM}$$ax$$\mathcal{F}$ est muni de son support, \textit{Supp}\textsc{(}$\wedge$$I$\textsc{)}, et de sa valeur de la mesure \textit{bond}, \textit{bond}\textsc{(}$I$\textsc{)}.
\end{definition}

\begin{exemple} \label{expRC4}
Nous avons, l'ensemble $\mathcal{MMCRM}in$ = $\{$\texttt{A}, \texttt{D}, \texttt{BC}, \texttt{CE}$\}$
\textsc{(}\textit{cf.} Exemple \ref{expMM}\textsc{)} et
l'ensemble $\mathcal{MFCRM}ax$ = $\{$\texttt{ACD}, \texttt{ABCE}$\}$
\textsc{(}\textit{cf.} Exemple \ref{exemple_MFCRMax}\textsc{)}.
Par conséquent, la représentation $\mathcal{RM}$$in$$\mathcal{MM}$$ax$$\mathcal{F}$ est composée des éléments suivants, $\mathcal{RM}$$in$$\mathcal{MM}$$ax$$\mathcal{F}$ =
$\{$\textsc{(}\texttt{A}, 3, $\displaystyle\frac{3}{3}$\textsc{)},
\textsc{(}\texttt{D}, 1, $\displaystyle\frac{1}{1}$\textsc{)},
\textsc{(}\texttt{BC}, 3, $\displaystyle\frac{3}{5}$\textsc{)},
\textsc{(}\texttt{CE}, 3, $\displaystyle\frac{3}{5}$\textsc{)},
\textsc{(}\texttt{ACD}, 1, $\displaystyle\frac{1}{4}$\textsc{)},
\textsc{(}\texttt{ABCE}, 2, $\displaystyle\frac{2}{5}$\textsc{)}$\}$.
\end{exemple}
La représentation $\mathcal{RM}$$in$$\mathcal{MM}$$ax$$\mathcal{F}$ offre une meilleure réduction que les représentations concises exactes  $\mathcal{RMCR}$, $\mathcal{RM}$$in$$\mathcal{MF}$ et $\mathcal{RMM}$$ax$$\mathcal{F}$.
Dans cet exemple, la représentation $\mathcal{RM}$$in$$\mathcal{MM}$$ax$$\mathcal{F}$
posséde six éléments de moins que la représentation concise exacte $\mathcal{RMCR}$, onze éléments de moins que la représentation concise exacte $\mathcal{RMM}$$ax$$\mathcal{F}$ et un élément de moins que la
représentation concise exacte $\mathcal{RM}$$in$$\mathcal{MF}$.
Cependant, cette représentation n'offre pas le même degré d'exactitude que
les représentations concises exactes précédemment citées. En effet, elle ne permet pas,
comme l'illustre le théoréme suivant, de dériver d'une maniére exacte les données de tout motif corrélé rare.
\begin{theoreme}\label{representation4_approx}
La représentation $\mathcal{RM}$$in$$\mathcal{MM}$$ax$$\mathcal{F}$ n'est qu'une représentation concise approximative de l'ensemble $\mathcal{MCR}$ des motifs corrélés rares.
\end{theoreme}
\begin{Preuve}
En effet, pour un motif arbitraire $I$ $\subseteq$ $\mathcal{I}$, cette représentation permet de déterminer si $I$ est corrélé rare ou non. Il suffit de trouver deux motifs $J$ et $Z$ appartenant à  la représentation tel que $J$ $\subseteq$ $I$ $\subseteq$ $Z$. Si $J$ ou $Z$ n'existe pas alors $I$ $\notin$ $\mathcal{MCR}$. Toutefois, les informations concernant le support et la mesure bond de $I$ ne peuvent étre exactement dérivées que si $I$ $\in$ $\mathcal{RM}$$in$$\mathcal{MM}$$ax$$\mathcal{F}$. Dans le cas contraire, cette représentation ne permet pas de les dériver d'une maniére exacte étant donné qu'elle peut ne contenir aucun élément représentatif de la classe d'équivalence de $I$ \textsc{(}c.-é.-d. ni le fermé associé s'il n'appartient pas à  $\mathcal{MFCRM}ax$ ni les minimaux associés s'il n'appartiennent pas à  $\mathcal{MMCRM}in$\textsc{)}. Seule une approximation des supports de $I$ et de sa valeur de \textit{bond} peut étre effectuée dans ce cas.

\`A cet égard, nous définissons les bornes maximales et minimales du support conjonctif, du support disjonctif et de la valeur de la mesure \textit{bond} d'un motif corrélé rare $I$.

Soient,

$\bullet$ \textit{R1} = $\max$$\{$Supp\textsc{(}$\wedge$$F$\textsc{)}, $F$ $\in$ $\mathcal{MFCRM}ax$ $|$ $I$ $\subseteq$ $F$$\}$,

$\bullet$ \textit{R2} =  $\min$$\{$Supp\textsc{(}$\wedge$$G$\textsc{)}, $G$ $\in$ $\mathcal{MMCRM}in$ $|$ $G$ $\subseteq$ $I$$\}$,

$\bullet$ \textit{R3} = $\min$$\{$Supp\textsc{(}$\vee$$F$\textsc{)}, $F$ $\in$  $\mathcal{MFCRM}ax$ $|$ $I$ $\subseteq$ $F$$\}$ et

$\bullet$ \textit{R4} = $\max$$\{$\textit{Supp}\textsc{(}$\vee$$G$\textsc{)}, $G$ $\in$ $\mathcal{MMCRM}in$ $|$ $G$ $\subseteq$ $I$$\}$.

Alors nous définissons les bornes minimales et maximales du support conjonctif du motif $I$ en fonction de
\textit{R1} et de \textit{R2} comme suit.
Désignons par \textit{MinConj} la borne minimale du support conjonctif du motif $I$,
\textit{MinConj} = $\min$\textsc{(}\textit{R1}, \textit{R2}\textsc{)} et désignons par
\textit{MaxConj} la borne maximale du support conjonctif du motif $I$,
\textit{MaxConj} = $\max$\textsc{(}\textit{R1}, \textit{R2}\textsc{)}.

Concernant le support disjonctif du motif $I$, nous définissons les bornes minimales et maximales du support disjonctif du motif $I$ en fonction de
\textit{R3} et de \textit{R4} comme suit.
Désignons par \textit{MinDisj} la borne minimale du support disjonctif du motif $I$,
\textit{MinDisj} = $\min$\textsc{(}\textit{R3}, \textit{R4}\textsc{)} et désignons par
\textit{MaxDisj} la borne maximale du support disjonctif du motif $I$,
\textit{MaxDisj} = $\max$\textsc{(}\textit{R3}, \textit{R4}\textsc{)}.

Par conséquent le support conjonctif de tout motif corrélé rare $I$ identifié par la représentation
$\mathcal{RM}$$in$$\mathcal{MM}$$ax$$\mathcal{F}$, sera compris entre \textit{MinConj} et \textit{MaxConj}.
Formellement, \textit{Supp}\textsc{(}$\wedge$$I$\textsc{)} $\in$ $[$\textit{MinConj}, \textit{MaxConj}$]$.
Le support disjonctif du motif $I$ sera tout de même cerné par la borne minimale \textit{MinDisj} et la borne maximale \textit{MaxDisj}, \textit{Supp}\textsc{(}$\vee$$I$\textsc{)} $\in$ $[$\textit{MinDisj}, \textit{MaxDisj}$]$.

Ainsi, les bornes minimales et maximales de la valeur de la mesure \textit{bond} du motif corrélé rare $I$ seront définis en fonction de  \textit{MinConj}, \textit{MinDisj}, \textit{MaxConj} et \textit{MaxDisj} comme suit.

Puisque  \textit{MinDisj} $\leq$ \textit{Supp}\textsc{(}$\vee$$I$\textsc{)} $\leq$ \textit{MaxDisj},
ainsi
$\displaystyle\frac{1}{\textit{MaxDisj}}$
$\leq$ $\displaystyle\frac{1}{\textit{Supp}\textsc{(}\vee I\textsc{)}}$
$\leq$  $\displaystyle\frac{1}{\textit{MinDisj}}$.

Comme \textit{Supp}\textsc{(}$\wedge$I\textsc{)} $\>$ 0 alors 
nous déduisons que,
$\displaystyle\frac{\textit{Supp}\textsc{(}\wedge I\textsc{)}}{\textit{MaxDisj}}$
$\leq$
$\displaystyle\frac{\textit{Supp}\textsc{(}\wedge I\textsc{)}}{\textit{Supp}\textsc{(}\vee I\textsc{)}}$  $\leq$
$\displaystyle\frac{\textit{Supp}\textsc{(}\wedge I\textsc{)}}{\textit{MinDisj}}$.
Ceci est équivalent é,
$\displaystyle\frac{\textit{Supp}\textsc{(}\wedge I\textsc{)}}{\textit{MaxDisj}}$  $\leq$
\textit{bond}\textsc{(}$I$\textsc{)}   $\leq$
$\displaystyle\frac{\textit{Supp}\textsc{(}\wedge I\textsc{)}}{\textit{MinDisj}}$.
Or, \textit{MinConj} $\leq$ \textit{Supp}\textsc{(}$\wedge$ $I$\textsc{)}
Ceci implique,
$\displaystyle\frac{\textit{MinConj}}{\textit{MaxDisj}}$  $\leq$
\textit{bond}\textsc{(}$I$\textsc{)}.
De plus, \textit{Supp}\textsc{(}$\wedge$$I$\textsc{)} $\leq$ \textit{MaxConj}
ainsi \textit{bond}\textsc{(}$I$\textsc{)} $\leq$
$\displaystyle\frac{\textit{MaxConj}}{\textit{MinDisj}}$.
Nous avons ainsi,
$\displaystyle\frac{\textit{MinConj}}{\textit{MaxDisj}}$  $\leq$
\textit{bond}\textsc{(}$I$\textsc{)} $\leq$
$\displaystyle\frac{\textit{MaxConj}}{\textit{MinDisj}}$.

Désignons par \textit{Minbond} la borne minimale de la valeur de la mesure \textit{bond},  \textit{Minbond} = $\displaystyle\frac{\textit{MinConj}}{\textit{MaxDisj}}$ et par  \textit{Maxbond} la borne maximale de la valeur de la mesure \textit{bond}, \textit{Maxbond} = $\displaystyle\frac{\textit{MaxConj}}{\textit{MinDisj}}$.
Nous concluons de ce qui précéde, que la valeur de la mesure \textit{bond}  du motif $I$ appartient à  l'intervalle borné par \textit{Minbond} et \textit{Maxbond}, \textit{bond}\textsc{(}$I$\textsc{)} $\in$
$[$\textit{Minbond}, \textit{Maxbond}$]$.
\end{Preuve}

\vspace{-2.cm}

\begin{exemple}
Considérons l'exemple \ref{expRC4}.
Prenons le cas du motif \texttt{ABE}, les deux motifs qui font que l'itemset \texttt{ABE} est un motif corrélé rare  sont \texttt{A} et  \texttt{ABCE} \textsc{(}\texttt{A}  $\subset$  \texttt{ABE} $\subset$   \texttt{ABCE}\textsc{)}. Ainsi le motif \texttt{ABE} est un motif corrélé rare, ses supports conjonctif, disjonctif et sa valeur de la mesure bond seront estimés comme suit.\\
$\bullet$ \textit{R1} = Supp\textsc{(}$\wedge$\texttt{ABCE}\textsc{)} = 2,\\
$\bullet$ \textit{R2} = Supp\textsc{(}$\wedge$\texttt{A}\textsc{)} = 3,\\
$\bullet$ \textit{R3} = Supp\textsc{(}$\vee$\texttt{ABCE}\textsc{)} = 5,\\
$\bullet$ \textit{R4} = Supp\textsc{(}$\vee$\texttt{A}\textsc{)} = 5.\\

Ainsi, nous avons \textit{MinConj} = $\min$\textsc{(}\textit{R1}, \textit{R2}\textsc{)} =  $\min$\textsc{(}2, 3\textsc{)} = 2, \textit{MaxConj} = $\max$\textsc{(}\textit{R1}, \textit{R2}\textsc{)} =  $\max$\textsc{(}2, 3\textsc{)} = 3,
et $\min$\textsc{(}\textit{R3}, \textit{R4}\textsc{)} =  $\max$\textsc{(}\textit{R3}, \textit{R4}\textsc{)}
= $\min$\textsc{(}5, 5\textsc{)} = $\max$\textsc{(}5, 5\textsc{)} = 5.
Donc pour cet exemple, \textit{MinDisj} =  \textit{MaxDisj} = 5.
Ce qui implique que, \textit{Minbond} = $\displaystyle\frac{\textit{MinConj}}{\textit{MaxDisj}}$ = $\displaystyle\frac{2}{5}$ et
\textit{Maxbond} = $\displaystyle\frac{\textit{MaxConj}}{\textit{MinDisj}}$ = $\displaystyle\frac{3}{5}$.

Par conséquent,  nous avons Supp\textsc{(}$\wedge$\texttt{ABE}\textsc{)} $\in$ $[$2, 3$]$,
Supp\textsc{(}$\vee$\texttt{ABE}\textsc{)} $\in$ $[$5, 5$]$ alors  Supp\textsc{(}$\vee$\texttt{ABE}\textsc{)} = 5 et
bond\textsc{(}\texttt{ABE}\textsc{)}  $\in$ $[$$\displaystyle\frac{2}{5}$, $\displaystyle\frac{3}{5}$$]$.

Nous remarquons d'aprés le contexte donné par la table
\ref{Base_transactions}, que les valeurs des supports conjonctif, disjonctif et la valeur de la mesure \textit{bond} du motif \texttt{ABE} correspondent  respectivement à  2, 5 et $\displaystyle\frac{2}{5}$.
Ces valeurs ne contredisent pas les valeurs approximées précédemment calculées.
Nous avons ainsi affirmé que le mécanisme d'approximation offert par la représentation concise approximative $\mathcal{RM}$$in$$\mathcal{MM}$$ax$$\mathcal{F}$ est valide.
\end{exemple}

\section{Conclusion}
Dans ce chapitre, nous avons étudié les caractéristiques des
motifs corrélés rares selon la mesure \textit{bond} et nous avons décrit minutieusement les spécificités des classes d'équivalence corrélées rares. Ensuite, nous avons introduit les différentes représentations concises des motifs corrélés rares selon la mesure \textit{bond} et nous avons prouvé leurs propriétés théoriques d'exactitude et de compacité. Ce chapitre a été cléturé avec la définition et l'analyse de la représentation concise approximative.
Cette derniére sacrifie l'exactitude des résultats au profit d'une meilleure réduction.
Dans le chapitre suivant, nous présenterons l'ensemble des algorithmes dédiés à  l'extraction de l'ensemble des motifs corrélés rares et des différentes représentations concises proposées.

%%%%%%%%%%%%%%%%%%%%%%%%%%%%%%%%%%%%%%%%%%%%%%%%%%%
\chapter{Approches d'extraction des motifs corrélés rares et des représentations proposées}\label{chapitre_algorithme}

\section{Introduction}
Dans ce chapitre, nous introduisons dans la premiére section
l'algorithme
\textsc{CRP\_Miner} d'extraction de l'ensemble $\mathcal{MCR}$ de tous les motifs corrélés rares.
La deuxiéme section sera consacrée à  la présentation de  l'algorithme \textsc{CRPR\_Miner}
d'extraction de la représentation concise exacte $\mathcal{RMCR}$.
Nous enchaénons ensuite avec la démonstration de ses propriétés théoriques.
Par ailleurs, nous suggérons deux algorithmes d'interrogation et de régénération, à  partir
de la représentation $\mathcal{RMCR}$. Le premier, intitulé \textsc{Regenerate}, permettant le requétage de cette représentation. Quant au second algorithme, \textsc{CRPRegeneration}, il rends possible la dérivation de tous les motifs corrélés rares à  partir de la représentation $\mathcal{RMCR}$.
%%%%%%%%%%%%%%%%%%%%%%%%%%%%%%%%%%%%%%%%%%%%%%%%%%%%%%%%%%%%%%%%%%%%%%%%%%
%%%%%%%%%%%%%%%%%%%%%%%%%%%%%%%%%%%%%
\section{\textsc{CRP\_Miner} : Algorithme d'extraction de l'ensemble $\mathcal{MCR}$ de tous les motifs corrélés rares}

Dans cette section, nous introduisons l'algorithme \textsc{CRP\_Miner} $^\textsc{(}$\footnote{\textsc{CRP\_Miner} est l'acronyme de
	\textit{\textbf{C}orrelated\textbf{ R}are \textbf{P}atterns\_Miner}.}$^{\textsc{)}}$ dédié à  l'extraction de l'ensemble $\mathcal{MCR}$ de tous les motifs corrélés rares à  partir d'un contexte d'extraction donné. Les notations utilisées dans l'algorithme \textsc{CRP\_Miner} sont récapitulées dans la table \ref{table_notations_ALGO1}. %%%\textsc{(}\textit{cf.} page \pageref{table_notations_ALGO1}\textsc{)}.

\subsection{Description de l'algorithme \textsc{CRP\_Miner}}
L'ensemble $\mathcal{MCR}$ des motifs corrélés rares est extrait gréce à  l'algorithme \textsc{CRP\_Miner} dont le pseudo-code est donné par l'algorithme \ref{AlgoCRP}.
En effet, l'ensemble $\mathcal{MCR}$ résulte de la conjonction de la contrainte monotone de rareté et de la contrainte anti-monotone de corrélation. Ainsi l'espace de recherche des motifs corrélés rares est délimité d'une part, par les motifs rares parmi l'ensemble $\mathcal{MCM}ax$ des motifs corrélés maximaux \textsc{(}\textit{cf.} Définition \ref{bdpos}\textsc{)} et, d'autre part, par les motifs corrélés parmi l'ensemble $\mathcal{MRM}in$ des motifs rares minimaux \textsc{(}\textit{cf.} Définition \ref{mrm}\textsc{)}.

\begin{table}[htbp]
	\begin{center}
		\begin{tabular}{|lcl|}
			\hline
			$X_{n}$& :&Motif $X$ de taille $n$.\\
			$X$.\textit{Conj}& :&Support conjonctif du motif $X$.\\
			$X$.\textit{Disj}& :&Support disjonctif du motif $X$.\\
			$X$.\textit{bond}& :&Valeur de mesure de corrélation \textit{bond} du motif $X$.\\
			$\mathcal{MC}and_{n}$& :&Ensemble des motifs candidats de taille $n$.\\
			$\mathcal{MCR}$& :&Ensemble des motifs corrélés rares.\\
			$\mathcal{MCM}ax$& :&Ensemble des motifs corrélés maximaux.\\
			$\mathcal{MCM}ax\mathcal{R}$& :&Ensemble des motifs corrélés maximaux rares.\\
			$\mathcal{MCM}ax\mathcal{F}$& :&Ensemble des motifs corrélés maximaux fréquents.\\
			\hline
		\end{tabular}
	\end{center}
	\caption{Notations adoptées dans l'algorithme \textsc{CRP\_Miner}.} \label{table_notations_ALGO1}
\end{table}

Toutefois, il est intéressant de signaler que le repérage de la bordure d'un ensemble donné de motifs est un probléme NP-complet, le temps d'exécution des algorithmes de fouille dédiés est toujours élevé \cite{boley2009}. En effet, il n'existe pas un algorithme permettant de résoudre ce probléme en un temps polynomial. Par conséquent, le probléme d'extraction des motifs corrélés rares, localisés entre deux bordures,  est un probléme complexe.

\`A cet égard, nous nous sommes basés sur l'idée d'extraire une des bordures de cet ensemble dans un premier temps puis de dériver le reste des motifs corrélés rares à  partir de cette bordure.
En effet, les éléments de la bordure formée par les motifs rares minimaux corrélés permet de nous renseigner quant à  la nature de fréquence d'un candidat donné. Tout candidat englobant un motif rare minimal corrélé sera un motif rare gréce à  la propriété de filtre d'ordre des motifs rares. Cependant, on ne peux rien confirmer quant à  sa corrélation.
Toutefois, la bordure composée par les motifs corrélés maximaux rares est plus intéressante quant à  l'identification des motifs corrélés rares potentiels.
En effet, tout candidat inclut dans un motif de l'ensemble des motifs corrélés maximaux rares est corrélé gréce à  la propriété d'idéal d'ordre des motifs corrélés. Ainsi, les candidats seront évalués uniquement par rapport à  la contrainte de rareté.
Par conséquent, nous jugeons intéressant de récupérer la bordure composée par les motifs corrélés maximaux rares moyennant un algorithme dédié puis de dériver le reste des motifs corrélés rares à  partir de cette bordure.
Cette idée constitue le fondement de la conception de
l'algorithme \textsc{CRP\_Miner} d'extraction de l'ensemble $\mathcal{MCR}$ des motifs corrélés rares que nous proposons.

Ce dernier prend en entrée une base de transactions $\mathcal{D}$, un seuil minimal de support \textit{minsupp}, ainsi qu'un seuil de corrélation minimale \textit{minbond} et offre en sortie l'ensemble $\mathcal{MCR}$ de tous les motifs corrélés rares munis de leurs supports conjonctifs et de leur valeur de la mesure \textit{bond}.

L'algorithme \textsc{CRP\_Miner} se déroule en trois principales étapes. La premiére consiste à  extraire l'ensemble des motifs corrélés maximaux puis à  filtrer, par rapport à  \textit{minsupp}, les motifs maximaux corrélés rares et ceux fréquents.
Ensuite, l'ensemble $\mathcal{MCR}$ des motifs corrélés rares sera initialisé à  l'ensemble des motifs corrélés maximaux rares. La deuxiéme principale étape consiste à  extraire les items corrélés rares. Ensuite, la derniére phase correspond à  l'extraction des motifs corrélés rares de taille supérieure ou égale à  deux.

En effet, les stratégies d'élagage mises en place correspondent à  :

\textsc{(}\textbf{\textit{i}}\textsc{)}
\textbf{L'élagage de tout candidat non inclu dans un motif corrélé maximal rare}, puisqu'il sera non corrélé rare.\\
\textsc{(}\textbf{\textit{ii}}\textsc{)}
\textbf{L'élagage de tout candidat inclu dans un motif corrélé maximal fréquent}, puisqu'il sera corrélé fréquent d'aprés la propriété de l'idéal d'ordre des motifs corrélés fréquents.\\
\textsc{(}\textbf{\textit{iii}}\textsc{)}
\textbf{L'élagage par rapport à  la propriété de cross support vérifiée par la mesure \textit{bond}}. En effet, tout candidat possédant deux items vérifiant la propriété de cross-support sera élagué puisqu'il est non corrélé. Expliquons à  présent le déroulement de chaque étape de l'algorithme \textsc{CRP\_Miner}.

\subsection{\'Etape 1 : Extraction des motifs corrélés maximaux}

Lors de la premiére étape, l'ensemble des motifs corrélés maximaux est récupéré gréce à  l'algorithme
\textsc{Algorithme\_Extraction\_MCMax}. Les motifs maximaux corrélés extraits seront filtrés par la contrainte de\textit{ minsupp}. Les motifs rares seront insérés dans l'ensemble $\mathcal{MCM}ax\mathcal{R}$ des motifs corrélés maximaux rares et ceux fréquents seront insérés dans l'ensemble $\mathcal{MCM}ax\mathcal{F}$ des motifs corrélés maximaux fréquents \textsc{(}\textit{cf.} ligne \ref{mcmf} et ligne \ref{mcmr} de l'algorithme \ref{AlgoCRP}\textsc{)}.
Ensuite, les motifs maximaux corrélés rares seront insérés dans l'ensemble des motifs corrélés rares selon leur taille \textsc{(}\textit{cf.}ligne \ref{majmcr1} de l'algorithme \ref{AlgoCRP}\textsc{)}.

\subsection{\'Etape 2 : Extraction des items corrélés rares}

Cette étape consiste à  extraire les items corrélés rares.
L'ensemble des motifs candidats de taille $1$ est composé de l'ensemble de tous les items
\textsc{(}\textit{cf.} ligne \ref{cand1} de l'algorithme \ref{AlgoCRP}\textsc{)}.
Ensuite, les items rares seront insérés dans l'ensemble des motifs corrélés rares
$\mathcal{MCR}$$_{1}$
\textsc{(}\textit{cf.} ligne \ref{majmcr2} de l'algorithme \ref{AlgoCRP}\textsc{)},
puisque la valeur de la mesure \textit{bond} de tout item est égale à  $1$ dépassant ainsi tout seuil minimal de \textit{bond}. L'étape suivante correspond à  l'extraction des  motifs corrélés rares de taille supérieure ou égale à  deux.

\subsection{\'Etape 3 : Extraction des motifs corrélés rares de taille supérieure ou égale à  deux}
Cette étape se réalise moyennant un parcours par niveau du bas vers le haut du treillis et englobe quatre phases, que nous détaillons dans ce qui suit.\\

\textbf{\textsc{(}a\textsc{)} La génération des motifs candidats}\\

La premiére phase correspond à  la phase de génération des motifs candidats de taille $n$ avec $n$ est égale à  $2$. \'Etant donné que les motifs corrélés rares ne forment pas un idéal d'ordre. C'est à  dire pour un motif corrélé rare de taille $n$, ses sous-ensemble ne sont pas forcément tous corrélés rares.
Plutét, tous ses sous-ensembles doivent étre corrélés.
Ainsi, l'ensemble $\mathcal{MC}$$and_{n}$ des motifs candidats de taille $n$ sera généré moyennant la phase combinatoire d'\textsc{Apriori-Gen} à  partir de l'ensemble $\mathcal{MC}$$and_{n-1}$ des motifs corrélés de taille $n-1$ \textsc{(}\textit{cf.} ligne \ref{gener1} de l'algorithme \ref{AlgoCRP}\textsc{)}.\\

\textbf{\textsc{(}b\textsc{)} L'élagage des motifs candidats}\\

L'élagage des motifs candidats se réalise selon deux propriétés. D'abord, l'élagage sera effectué par rapport à  la propriété de cross support vérifiée par la mesure \textit{bond}. 
Cependant, la vérification de la propriété de cross-support pour un candidat donné est conditionné par la valeur du seuil \textit{minbond} comme le montre la proposition suivante.
\begin{proposition}\label{propCrossSupp}
	Désignons d'abord par \textit{MinS} la plus petite valeur du support conjonctif des items d'un contexte donné, \textit{MinS} = $\min$$\{$Supp\textsc{(}$\wedge$$x$\textsc{)} $|$  $x$ $\in$ $\mathcal{I}$$\}$,
	par \textit{MaxS} la plus grande valeur du support conjonctif des items d'un contexte donné,
	\textit{MaxS} = $\max$$\{$Supp\textsc{(}$\wedge$$x$\textsc{)} $|$ $x$ $\in$ $\mathcal{I}$$\}$ et
	par \textit{MinR} la valeur du rapport entre \textit{MinS} et \textit{MaxS},
	\textit{MinR} = $\displaystyle\frac{\textit{MinS}}{\textit{MaxS}}$.
	Pour \textit{minbond} $\leqslant$ \textit{MinR}, aucune paire d'items ne vérifie la propriété de cross-support.
\end{proposition}
\begin{Preuve}
	Soit un contexte d'extraction $\mathcal{D}$ $=$
	$\textsc{(}$$\mathcal{T},\mathcal{I},\mathcal{R}$$\textsc{)}$. Soient $x$
	et $y$ deux items quelconques de $\mathcal{I}$.
	Nous avons \textit{MinS} $\leq$ Supp\textsc{(}$\wedge$$x$\textsc{)} $\leq$  \textit{MaxS} et
	\textit{MinS} $\leq$ Supp\textsc{(}$\wedge$$y$\textsc{)} $\leq$  \textit{MaxS}.
	Ainsi, $\displaystyle\frac{1}{Supp\textsc{(}\wedge y\textsc{)}}$ $\geq$ $\displaystyle\frac{1}{\textit{MaxS}}$.
	Puisque, Supp\textsc{(}$\wedge$ $x$\textsc{)}  $>$ $0$, donc nous avons
	$\displaystyle\frac{Supp\textsc{(}\wedge x\textsc{)}}{Supp\textsc{(}\wedge y\textsc{)}}$ $\geq$ $\displaystyle\frac{Supp\textsc{(}\wedge x\textsc{)}}{\textit{MaxS}}$.
	Ainsi pour toute paire d'items $x$ et $y$, nous avons
	$\displaystyle\frac{Supp\textsc{(}\wedge x\textsc{)}}{Supp\textsc{(}\wedge y\textsc{)}}$ $\geq$
	\textit{MinR}.
	Par conséquent, dans le cas oé le seuil minimal de corrélation \textit{minbond} $\leq$ \textit{MinR}, nous avons \textit{minbond} $\leq$ \textit{MinR} $\leq$ $\displaystyle\frac{Supp\textsc{(}\wedge x\textsc{)}}{Supp\textsc{(}\wedge y\textsc{)}}$, ainsi
	\textit{minbond} $\leq$  $\displaystyle\frac{Supp\textsc{(}\wedge x\textsc{)}}{Supp\textsc{(}\wedge y\textsc{)}}$.
	
	D'une maniére plus sémantique, nous concluons que si \textit{minbond} $\leq$ \textit{MinR}, alors aucune paire d'items ne vérifie la propriété de cross-support traduite par
	$\displaystyle\frac{Supp\textsc{(}\wedge x\textsc{)}}{Supp\textsc{(}\wedge y\textsc{)}}$
	$<$ \textit{minbond}.
\end{Preuve}
Il est à  déduire, d'aprés la proposition précédente, que la vérification de la propriété de cross-support pour un candidat donné ne doit se réaliser que si le seuil minimal de corrélation \textit{minbond}
est strictement supérieur au rapport $\textit{MinR}$.
Dans le cas échéant, tout candidat possédant deux items vérifiant la propriété de cross-support sera élagué \cite{tarekds2010} puisqu'il est non corrélé
\textsc{(}\textit{cf.} ligne \ref{crosssupp} de l'algorithme \ref{AlgoCRP}\textsc{)}.

Le deuxiéme critére d'élagage correspond à  l'élimination de tout candidat non inclut dans un élément de l'ensemble $\mathcal{MCM}ax\mathcal{R}$ des motifs corrélés maximaux rares \textsc{(}\textit{cf.} ligne \ref{prune1} de l'algorithme \ref{AlgoCRP}\textsc{)}.
\`A ce stade, l'ensemble des motifs candidats maintenus sont tous corrélés. Cependant, nous ne pouvons rien déduire quant à  la nature de fréquence de ces candidats. En effet, les candidats qui sont inclus dans un élément de l'ensemble $\mathcal{MCM}ax\mathcal{F}$ des motifs corrélés maximaux fréquents seront fréquents.
\`A cet égard, ses candidats seront élagués, \textsc{(}\textit{cf.} ligne \ref{prunenew} de l'algorithme \ref{AlgoCRP}\textsc{)}.

\textbf{\textsc{(}c\textsc{)} L'évaluation des motifs candidats}\\

L'évaluation des candidats maintenus se réalise par
la fonction dédiée \textsc{(}\textit{cf.} ligne
\ref{calculsuppbond} de l'algorithme \ref{AlgoCRP}\textsc{)},
décrite par l'algorithme \ref{fctcalculsupport}.
Cette fonction est invoquée afin de retourner l'ensemble des motifs corrélés rares de taille $n$.
Ensuite, la variable $n$
sera incrémentée
\textsc{(}\textit{cf.} ligne \ref{majn} de l'algorithme \ref{AlgoCRP}\textsc{)}
et de nouveaux candidats de taille $n$ seront générés à  partir des motifs
corrélés de taille $n-1$
\textsc{(}\textit{cf.} ligne \ref{cand2} de l'algorithme \ref{AlgoCRP}\textsc{)}.
La boucle itérative s'arréte lorsque l'ensemble des motifs candidats est vide
\textsc{(}\textit{cf.} ligne \ref{boucle1} de l'algorithme \ref{AlgoCRP}\textsc{)} alors l'algorithme \textsc{CRP\_Miner} marque sa fin d'exécution et retourne l'ensemble $\mathcal{MCR}$ de tous les motifs corrélés rares \textsc{(}\textit{cf.} ligne \ref{outMcr} de l'algorithme \ref{AlgoCRP}\textsc{)}.

%%\incmargin{0.5em} \linesnumbered
\begin{algorithm}[htbp]\label{AlgoCRP}
	\small{ \small{\caption{\textsc{CRP\_Miner}}}
		%%\SetVline
		%%\setnlskip{-3pt}
		\Donnees{
			\begin{enumerate}
				\item Un contexte d'extraction $\mathcal{D}$.
				\item Le seuil minimal de support conjonctif \textit{minsupp}.
				\item Le seuil minimal de corrélation \textit{minbond}.
			\end{enumerate}
		} \Res{L'ensemble $\mathcal{MCR}$ des motifs corrélés rares
			munis de leurs supports conjonctifs et de leurs valeurs de mesure \textit{bond}.} \Deb{
			
			$\mathcal{MCM}$$ax$ := \textsc{Algorithme\_Extraction\_MCMax}\textsc{(}$\mathcal{D}$,  \textit{minbond}\textsc{)}\;
			$\mathcal{MCM}$$ax\mathcal{F}$ := $\{$ $X$ $\in$  $\mathcal{MCM}$$ax$ $\mid$  $X$.Conj $\geq$
			\textit{minsupp}$\}$ \label{mcmf}\;
			$\mathcal{MCM}$$ax\mathcal{R}$ := $\{$ $X$ $\in$  $\mathcal{MCM}$$ax$ $\mid$  $X$.Conj $<$ \textit{minsupp}$\}$ \label{mcmr}\;
			
			$k$ := $1$\;
			$\mathcal{MCR}$$_{|k|}$ := $\emptyset$\;
			\PourCh{\textsc{(}$M$ $\in$ $\mathcal{MCM}$$ax$$\mathcal{R}$ \textsc{)}}
			{
				$\mathcal{MCR}_{|M|}$ := $\mathcal{MCR}_{|M|}$ $\cup$ \textsc{(}$M$, $M$.\textit{Conj},
				$M$.\textit{bond}\textsc{)} \label{majmcr1}\;
				
				\Si{\textsc{(}$k < |M|$\textsc{)}}{$k$ := $|M|$\;}
			}%%pour
			$\mathcal{MC}$$and_{1}$ := $\{$$i$ $|$ $i$  $\in$ $\mathcal{I}$$\}$ \label{cand1}\;
			$\mathcal{MCR}$$_{1}$ := $\{$$X$ $\in$ $\mathcal{MC}$$and_{1}$  $|$  $X$.\textit{Conj} $<$ \textit{minsupp}$\}$ \label{majmcr2}\;
			$n$ := $2$\;
			$\mathcal{MC}$$and_{n}$ := \textsc{Apriori-Gen}\textsc{(}$\mathcal{MC}$$and_{n-1}$\textsc{)} \label{gener1}\;
			\Tq{\textsc{(}$\mathcal{MC}$$and_{n}$ $\neq$
				$\emptyset$\textsc{)}\label{boucle1}}
			{
				/* \'Elagage des motifs candidats */ \\
				
				$\mathcal{MC}$$and_{n}$ := $\{$ $X_{n}$ $\in$  $\mathcal{MC}$$and_{n}$  $|$
				$\nexists$ x, y $\in$ $X_{n}$ $|$ $\displaystyle\frac{x.\textit{Conj}}
				{y.\textit{Conj}} < \textit{minbond}$\textsc{)}$\}$ \label{crosssupp}\;
				
				$\mathcal{MC}$$and_{n}$ := $\{$$X_{n}$ $\in$  $\mathcal{MC}$$and_{n}$  $|$
				$\exists$ $Z$ $\in$ $\mathcal{MCM}$$ax\mathcal{R}$ $|$ $X_{n}$ $\subset$ $Z$$\}$ \label{prune1}\;
				
				$\mathcal{MC}$$and_{n}$ := $\{$$X_{n}$ $\in$  $\mathcal{MC}$$and_{n}$  $|$
				$\nexists$ $Z$ $\in$ $\mathcal{MCM}$$ax\mathcal{F}$ $|$ $X_{n}$ $\subset$ $Z$$\}$ \label{prunenew}\;
				
				/* \'Evaluation des motifs candidats retenus */\\
				$\mathcal{MCR}$$_{n}$ :=  \textsc{Calcul\_Supports\_Bond}\textsc{(}$\mathcal{D}$, $\mathcal{MC}$$and_{n}$, \textit{minsupp}\textsc{)} \label{calculsuppbond}\;
				/* Génération de nouveaux candidats */\\
				$n$ := $n$ + $1$\label{majn}\;
				$\mathcal{MC}$$and_{n}$ := \textsc{Apriori-Gen}\textsc{(}$\mathcal{MC}$$and_{n-1}$\textsc{)} \label{cand2}\;
			}
			\Retour{ $\displaystyle \bigcup_{i \in  [1, k]}$ $\mathcal{MCR}$$_{i}$\;} \label{outMcr}}}
\end{algorithm}
%%\decmargin{0.5em}
%%{0em} \linesnumbered
%%%%%%%%%%%%%%%%%%%%%%%%%%%%
Expliquons, à  présent, le déroulement de la fonction \textsc{Calcul\_Supports\_Bond}
dont le pseudo code est donné par l'algorithme \ref{fctcalculsupport}.
Cette fonction prend en entrée le contexte d'extraction $\mathcal{D}$, le seuil \textit{minsupp} et l'ensemble $\mathcal{MC}$$and$ des candidats de taille $n$ et fournit en sortie l'ensemble
$\mathcal{MCR}$$_{n}$ des motifs corrélés rares de taille $n$.

Cette fonction opére de la maniére suivante. D'abord, le contexte d'extraction est parcouru d'une maniére séquentielle. Pour chaque transaction $T$ du contexte, nous parcourons l'ensemble de tous les motifs corrélés rares.
Pour chaque candidat $X_{n}$ de l'ensemble $\mathcal{MC}$$and$
\textsc{(}\textit{cf.} ligne \ref{boucleCand} de l'algorithme \ref{fctcalculsupport}\textsc{)}, le support conjonctif, le support disjonctif seront mis à  jour.
En effet, nous désignons par $I$ l'ensemble des items appartenant à  une transaction $T$ et par $\omega$ l'ensemble des items commun entre $I$ et le candidat $X_{n}$.
Deux cas sont alors possibles. \\
\textsc{(}\textit{\textbf{i}}\textsc{)}
Si $\omega$ contient au moins  un seul item du motif candidat $X_{n}$
alors le support disjonctif $X_{n}$.\textit{Disj} sera incrémenté
\textsc{(}\textit{cf.} ligne \ref{calculdisj1} de l'algorithme \ref{fctcalculsupport}\textsc{)}.\\
\textsc{(}\textit{\textbf{ii}}\textsc{)}
Si $\omega$ contient tous les items composant le candidat $X_{n}$
alors le support conjonctif $X_{n}$.\textit{Conj} sera incrémenté
\textsc{(}\textit{cf.} ligne \ref{calculconj1} de l'algorithme \ref{fctcalculsupport}\textsc{)}.

Aprés le balayage de tout le contexte, les supports disjonctifs et conjonctifs de tous les candidats sont calculés.
L'étape suivante consiste é
parcourir l'ensemble des candidats et
é vérifier la rareté du candidat en cours
\textsc{(}\textit{cf.} ligne \ref{testrarete} de l'algorithme \ref{fctcalculsupport}\textsc{)}.
Dans le cas oé, le candidat en cours est rare alors
sa valeur de la mesure bond sera calculée
\textsc{(}\textit{cf.} ligne \ref{calculbond1} de l'algorithme \ref{fctcalculsupport}\textsc{)} et
le candidat sera inséré dans l'ensemble $\mathcal{MCR}$$_{n}$ des motifs corrélés rares de taille $n$
\textsc{(}\textit{cf.} ligne \ref{majMCRn} de l'algorithme \ref{fctcalculsupport}\textsc{)}.
Lorsque tous les candidats sont évalués, la fonction  \textsc{Calcul\_Supports\_Bond} marque sa fin d'exécution et retourne l'ensemble $\mathcal{MCR}$$_{n}$
\textsc{(}\textit{cf.} ligne \ref{res2} de l'algorithme \ref{fctcalculsupport}\textsc{)}
des motifs corrélés rares de taille $n$.

Nous avons ainsi analysé les différentes étapes de l'algorithme \textsc{CRP\_Miner}. Nous enchaénons, dans ce qui suit, par la trace d'exécution de cet algorithme appliquée à  la base de transaction de la table \ref{Base_transactions}.
%%%%%%%%%%%%%%%%%%%%%%%%%%%%
%%\incmargin{0.5em} \linesnumbered
\begin{algorithm}[!t]
	\small{\small{\label{fctcalculsupport} \caption{\textsc{Calcul\_Supports\_Bond}}}
		%%\SetVline
		%%\setnlskip{-3pt}
		\Donnees{
			\begin{enumerate}
				\item  Un contexte d'extraction $\mathcal{D}$.
				\item  L'ensemble $\mathcal{MC}$$and_{n}$ des motifs candidats de taille $n$.
				\item  Le seuil minimal de support conjonctif \textit{minsupp}.
			\end{enumerate}
		} \Res{L'ensemble $\mathcal{MCR}$$_{n}$ des motifs corrélés rares de taille $n$} \Deb{
			\PourCh{\textsc{(}Objet $O$ du contexte  $\mathcal{D}$\textsc{)}\label{boucleBD}}
			{
				\PourCh{\textsc{(}$X_{n}$ de $\mathcal{MC}$$and_{n}$ \textsc{)}\label{boucleCand}}
				{
					$\omega$ := $X_{n}$ $\cap$ $I$  /* $I$ correspond aux items appartenant à  l'objet $O$ */\;
					\Si {\textsc{(}$\omega$ $\neq$ $\emptyset$\textsc{)}}
					{
						$X_{n}$.\textit{Disj} :=  $X_{n}$.\textit{Disj} +1 \label{calculdisj1}\;
						\Si {\textsc{(}$\omega$ $=$ $X_{n}$\textsc{)}}
						{
							$X_{n}$.\textit{Conj} :=  $X_{n}$.\textit{Conj} +1 \label{calculconj1}\;
						}%%si
					}%%si
				}%%Pourcandidat
			}	%%Pourobjet
			%%%%
			\PourCh{\textsc{(}$X_{n}$ de $\mathcal{MC}$$and_{n}$ \textsc{)} \label{bouclecand2}}
			{
				\Si{\textsc{(}$X_{n}$.\textit{Conj} $<$ \textit{minsupp}\textsc{)}\label{testrarete}}
				{
					$X_{n}$.\textit{bond} :=  $\displaystyle\frac{X_{n}.\textit{Conj}}{X_{n}.\textit{Disj}}$\label{calculbond1}\;
					$\mathcal{MCR}$$_{n}$ := $\mathcal{MCR}$$_{n}$  $\cup$ \textsc{(}$X_{n}$, $X_{n}$.\textit{Conj}, $X_{n}$.\textit{bond}\textsc{)}\label{majMCRn}\;
				}%%si
			}%%Pour
			\Retour{$\mathcal{MCR}$$_{n}$\;} \label{res2}
	}}
\end{algorithm}
%%\decmargin{0.5em}
%%\incmargin{0em} \linesnumbered
%%%%%%%%%%%%%%%%%%%%%%%%%%	
\subsection{Trace d'exécution de l'algorithme \textsc{CRP\_Miner}}
Considérons la base de transactions illustrée par la table \ref{Base_transactions}.
Pour \textit{minsupp} = 3 et \textit{minbond} = 0,20, l'application de l'algorithme \textsc{CRP\_Miner} pour l'extraction de l'ensemble $\mathcal{MCR}$ des motifs corrélés rares se réalise de la maniére suivante :
Initialement, les motifs maximaux sont extraits gréce à  l'algorithme \textsc{Algorithme\_Extraction\_MCMax},
$\mathcal{MCM}ax$ = $\{$\textsc{(}\texttt{ACD}, 1, $\displaystyle\frac{1}{4}$\textsc{)},
\textsc{(}\texttt{ABCE}, 2, $\displaystyle\frac{2}{5}$\textsc{)}$\}$.
\'Etant donné que tous les motifs de l'ensemble $\mathcal{MCM}ax$ sont rares, ils sont donc insérés dans l'ensemble  $\mathcal{MCR}$ et répartis en fonction de leur taille comme suit.
$\mathcal{MCR}$$_{3}$ = $\{$\textsc{(}\texttt{ACD}, 1, $\displaystyle\frac{1}{4}$\textsc{)}$\}$, $\mathcal{MCR}$$_{4}$ = $\{$\textsc{(}\texttt{ABCE}, 2, $\displaystyle\frac{2}{5}$\textsc{)}$\}$.

Ensuite, un balayage de la base de transactions est réalisé et l'ensemble des motifs candidats est initialisé,
$\mathcal{MC}$$and_{1}$ =  $\{$\texttt{A}, \texttt{B}, \texttt{C}, \texttt{D}, \texttt{E}$\}$.
Le seul item rare $D$ est identifié,
$\mathcal{MCR}$$_{1}$ = $\{$\textsc{(}\texttt{D}, 1, $\displaystyle\frac{1}{1}$\textsc{)}$\}$.
L'ensemble $\mathcal{MC}$$and_{2}$ des motifs candidats de taille 2 est égal é,
$\mathcal{MC}$$and_{2}$ =  $\{$\texttt{AB}, \texttt{AC}, \texttt{AD}, \texttt{AE}, \texttt{BC}, \texttt{BE}, \texttt{BD}, \texttt{CE}, \texttt{CD}, \texttt{DE}$\}$.\\
\textbf{\underline{Itération 1 :}}\\
La propriété de cross-support n'élaguera aucun des candidats de l'ensemble $\mathcal{MC}$$and_{2}$.
Cependant, les candidats \texttt{BD} et \texttt{DE} sont élagués vu qu'ils ne sont inclus dans aucun
motif maximal corrélé rare. Ainsi, nous avons
$\mathcal{MC}$$and_{2}$ =  $\{$\texttt{AB}, \texttt{AC}, \texttt{AD}, \texttt{AE}, \texttt{BC}, \texttt{BE}, \texttt{CE}, \texttt{CD}$\}$.
Ensuite, les supports conjonctifs, disjonctifs et les valeurs de mesure \textit{bond} des candidats retenus sont calculés par la fonction \textsc{Calcul\_Supports\_Bond}. Ainsi nous distinguons l'ensemble
$\mathcal{MCR}$$_{2}$ = $\{$\textsc{(}\texttt{AD}, 1, $\displaystyle\frac{1}{3}$\textsc{)},
\textsc{(}\texttt{CD}, 1, $\displaystyle\frac{1}{4}$\textsc{)},
\textsc{(}\texttt{AB}, 2, $\displaystyle\frac{2}{5}$\textsc{)},
\textsc{(}\texttt{AE}, 2, $\displaystyle\frac{2}{5}$\textsc{)}$\}$.
La variable $n$ sera ensuite incrémentée, $n$ = 3 et les candidats de taille trois sont générés.
$\mathcal{MC}$$and_{3}$ = $\{$\texttt{ABC}, \texttt{ABD}, \texttt{ABE}, \texttt{ACD},
\texttt{ACE}, \texttt{ADE}, \texttt{BCD}, \texttt{BCE}, \texttt{CDE}$\}$.\\
\textbf{\underline{Itération 2 :}}\\
Les candidats \texttt{ABD}, \texttt{BCD}, \texttt{CDE} et \texttt{ADE} sont élagués
puisqu'ils ne sont pas inclus dans un motif corrélé maximal rare.
Toutefois, le motif \texttt{ABCE} appartient à  l'ensemble des motifs corrélés maximaux rares et il est déjé inséré dans l'ensemble $\mathcal{MCR}$, donc il sera élagué.
$\mathcal{MC}$$and_{3}$ = $\{$\texttt{ABC}, \texttt{ABE}, \texttt{ACE}, \texttt{BCE}$\}$.
Pour les candidats retenus, nous calculons
les supports conjonctifs, disjonctifs et la valeur de la mesure \textit{bond} et nous identifions les motifs corrélés rares suivants,
$\mathcal{MCR}$$_{3}$ = $\{$\textsc{(}\texttt{ACD}, 1, $\displaystyle\frac{1}{4}$\textsc{)},
\textsc{(}\texttt{ABC}, 2, $\displaystyle\frac{2}{5}$\textsc{)},
\textsc{(}\texttt{ABE}, 2, $\displaystyle\frac{2}{5}$\textsc{)},
\textsc{(}\texttt{ACE}, 2, $\displaystyle\frac{2}{5}$\textsc{)}$\}$.
Le déroulement de la boucle itérative se poursuit encore,
la variable $n$ sera incrémentée, $n$ = 4,  et les candidats de taille 4 sont générés.
$\mathcal{MC}$$and_{4}$ =  $\{$\texttt{ABCD}, \texttt{ACDE}, \texttt{ABCE}, \texttt{ABDE}$\}$.\\
\textbf{\underline{Itération 3 :}}\\
Lors de cette itération, tous les candidats sont élagués puisqu'ils ne sont pas des sous ensembles stricts d'aucun motif corrélé maximal.
Toutefois, le motif \texttt{ABCE} appartient à  l'ensemble des motifs corrélés maximaux rares et il est déjé inséré dans l'ensemble $\mathcal{MCR}$, donc il ne sera pas retraité.

Ensuite la variable $n$ sera incrémentée, $n$ = 5. L'ensemble des candidats de taille 5 est vide, $\mathcal{MC}$$and_{5}$ = $\{$$\emptyset$$\}$
alors l'algorithme marque sa fin d'exécution donnant ainsi comme résultat l'ensemble $\mathcal{MCR}$ des motifs corrélés rares.
$\mathcal{MCR}$ =
$\{$\textsc{(}\texttt{D}, 1, $\displaystyle\frac{1}{1}$\textsc{)},
\textsc{(}\texttt{AD}, 1, $\displaystyle\frac{1}{3}$\textsc{)},
\textsc{(}\texttt{AB}, 2, $\displaystyle\frac{2}{5}$\textsc{)},
\textsc{(}\texttt{AE}, 2, $\displaystyle\frac{2}{5}$\textsc{)},
\textsc{(}\texttt{CD}, 1, $\displaystyle\frac{1}{4}$\textsc{)}
\textsc{(}\texttt{ACD}, 1, $\displaystyle\frac{1}{4}$\textsc{)},
\textsc{(}\texttt{ABC}, 2, $\displaystyle\frac{2}{5}$\textsc{)},
\textsc{(}\texttt{ABE}, 2, $\displaystyle\frac{2}{5}$\textsc{)},
\textsc{(}\texttt{ACE}, 2, $\displaystyle\frac{2}{5}$\textsc{)},
\textsc{(}\texttt{ABCE}, 2, $\displaystyle\frac{2}{5}$\textsc{)}$\}$.\\

Il est à  remarquer que dans cet exemple d'exécution de l'algorithme \textsc{CRP\_Miner},
aucun candidat généré n'a été élagué par la propriété de cross-support.
En effet, nous avons pour cet exemple d'exécution
\textit{MinS} = $1$, \textit{MaxS} = $4$  et
\textit{MinR} = $\displaystyle\frac{\textit{MinS}}{\textit{MaxS}}$ = $\displaystyle\frac{1}{4}$ = $0,25$.
Or, \textit{minbond} = $0,20$ $<$ \textit{MinR} = 0,25. Ainsi,
d'aprés la proposition \ref{propCrossSupp}
aucune paire d'items ne vérifie la propriété de cross-support.

%%%%%%%%%%%%%%%%%%%%%%%%%%%%%%%%
Nous avons ainsi présenté et analysé l'algorithme \textsc{CRP\_Miner} d'extraction de l'ensemble $\mathcal{MCR}$
de tous les motifs corrélés rares.
Dans la suite, nous présentons l'algorithme
\textsc{CRPR\_Miner} d'extraction de la représentation concise exacte $\mathcal{RMCR}$.

\section{\textsc{CRPR\_Miner} : Algorithme d'extraction de la représentation concise exacte $\mathcal{RMCR}$}

Dans cette section, nous nous concentrons uniquement sur l'approche d'extraction de la représentation concise exacte $\mathcal{RMCR}$ puisque les autres représentations concises exactes
$\mathcal{RMM}$$ax$$\mathcal{F}$,
$\mathcal{RM}$$in$$\mathcal{MF}$ et la représentation concise approximative
$\mathcal{RM}$$in$$\mathcal{MM}$$ax$$\mathcal{F}$
peuvent y étre facilement dérivées.

\`A cet égard, nous introduisons l'algorithme \textsc{CRPR\_Miner} $^\textsc{(}$\footnote{\textsc{CRPR\_Miner} est l'acronyme de
	\textit{\textbf{C}orrelated \textbf{R}are \textbf{P}atterns \textbf{R}epresentation
		Miner}.}$^{\textsc{)}}$ dédié à  l'extraction de la représentation concise exacte $\mathcal{RMCR}$.
Les notations utilisées dans cet algorithme sont récapitulées dans la table \ref{NotCRPR}. 

\subsection{Description de l'algorithme \textsc{CRPR\_Miner}}

La représentation concise exacte $\mathcal{RMCR}$ \textsc{(}\textit{cf.} Définition \ref{rmcr} page \pageref{rmcr}\textsc{)} composée de l'ensemble
$\mathcal{MMCR}$ des motifs minimaux corrélés rares et de l'ensemble $\mathcal{MFCR}$ des motifs fermés corrélés rares est extraite gréce à  l'algorithme \textsc{CRPR\_Miner}.

Cet algorithme, dont le pseudo-code est donné par l'algorithme \ref{AlgoCRPR}, prend en entrée un contexte d'extraction $\mathcal{D}$, un seuil minimal de support conjonctif \textit{minsupp} ainsi qu'un seuil
minimal de corrélation \textit{minbond}.
L'algorithme \textsc{CRPR\_Miner} permet de déterminer, à  partir du contexte $\mathcal{D}$, l'ensemble des motifs minimaux corrélés rares $\mathcal{MMCR}$ et l'ensemble des motifs fermés corrélés rares $\mathcal{MFCR}$ munis de leurs supports conjonctifs et de leurs valeurs de mesure \textit{bond}.

Les stratégies d'élagage des motifs candidats mises en place sont les suivantes :

\textsc{(}\textbf{\textit{i}}\textsc{)}
\textbf{L'élagage de tout candidat non inclu dans un motif corrélé maximal rare}, puisqu'il sera non corrélé rare.\\
\textsc{(}\textbf{\textit{ii}}\textsc{)}
\textbf{L'élagage de tout candidat inclu dans un motif corrélé maximal fréquent}, puisqu'il sera corrélé fréquent d'aprés la propriété de l'idéal d'ordre des motifs corrélés fréquents.\\
\textsc{(}\textbf{\textit{iii}}\textsc{)}\textbf{\'{E}lagage par rapport à  la propriété de cross-support vérifiée par la mesure de corrélation \textit{bond}}. En effet, tout candidat contenant deux items qui vérifient la propriété de cross-support par rapport au seuil minimal \textit{minbond} est un motif non corrélé et sera ainsi élagué de l'ensemble des motifs candidats.\\
\textsc{(}\textbf{\textit{iv}}\textsc{)}\textbf{ \'{E}lagage par rapport à  la propriété d'idéal d'ordre des motifs minimaux corrélés}. En
effet, les motifs minimaux corrélés vérifient la propriété de l'idéal d'ordre. Ainsi, tout candidat corrélé possédant un sous-ensemble non minimal corrélé, sera non minimal corrélé et il sera ainsi élagué.

\begin{table}
	\begin{tabular}{|lcl|}
		\hline
		$X_{n}$& :&Motif $X$ de taille $n$.\\
		$X$.\textit{Conj}& :&Support conjonctif du motif $X$.\\
		$X$.\textit{Disj}& :&Support disjonctif du motif $X$.\\
		$X$.\textit{bond}& :&Valeur de mesure de corrélation \textit{bond} du motif $X$.\\
		$X$.\textit{$f_{c}$}& :&fermé conjonctif du motif $X$.\\
		$X$.\textit{$f_{d}$}& :&fermé disjonctif du motif $X$.\\
		$X$.\textit{CmpDisj}& :&Ensemble des items n'appartenant à  aucune transaction contenant au \\
		& &   moins un item du motif $X$.\\
		$X$.\textit{$f_{bond}$}& :&fermé associé à  la mesure \textit{bond} du motif $X$.\\
		$\mathcal{MC}and_{n}$& :&Ensemble des motifs candidats de taille $n$.\\
		$\mathcal{MCM}ax$& :&Ensemble des motifs maximaux corrélés.\\
		$\mathcal{MCM}ax$$\mathcal{R}$& :&Ensemble des motifs maximaux corrélés rares.\\
		$\mathcal{MCM}ax$$\mathcal{F}$& :&Ensemble des motifs maximaux corrélés fréquents.\\
		$\mathcal{MMCR}_{n}$& :&Ensemble des motifs minimaux corrélés rares de taille $n$.\\
		$\mathcal{MFCR}_{n}$& :&Ensemble des motifs fermés corrélés rares de taille $n$.\\
		\hline
	\end{tabular}
	\caption{Notations adoptées dans l'algorithme \textsc{CRPR\_Miner}.}\label{NotCRPR}
\end{table}
Nous détaillons dans ce qui suit minutieusement chaque étape de l'algorithme \textsc{CRPR\_-Miner}.

\subsection{\'Etape 1 : Extraction des motifs corrélés maximaux}

Cette premiére étape se réalise d'une maniére similaire à  celle de la premiére étape de l'algorithme
\textsc{CRP\_Miner}. En effet, l'ensemble des motifs corrélés maximaux est récupéré gréce à  l'algorithme
\textsc{Algorithme\_Extraction\_MCMax}. Ces derniers seront filtrés par la contrainte de\textit{ minsupp} et insérés selon leur natures dans les
ensembles $\mathcal{MCM}ax\mathcal{R}$ et $\mathcal{MCM}ax\mathcal{F}$.
Ensuite, les motifs maximaux corrélés rares seront insérés dans l'ensemble des motifs corrélés rares selon leur taille.

\subsection{\'Etape 2 : Extraction de tous les motifs minimaux et fermés corrélés rares}

La deuxiéme étape consiste à  réaliser un parcours par niveau du bas vers le haut du treillis afin d'extraire les motifs minimaux corrélés rares et de calculer leurs fermetures.
Premiérement, l'ensemble des motifs candidats de taille $1$ est composé de l'ensemble de tous les items
\textsc{(}\textit{cf.} ligne \ref{setcand1} de l'algorithme \ref{AlgoCRPR}\textsc{)}.
Ensuite, les items rares seront insérés dans l'ensemble des motifs minimaux corrélés rares
$\mathcal{MMCR}$$_{1}$ et leurs fermetures seront calculés moyennant la fonction
\textsc{Calcul\_Supports\_Fermetures} \textsc{(}\textit{cf.} ligne \ref{appelfct1} de l'algorithme \ref{AlgoCRPR}\textsc{)}.

La phase de génération de candidats s'effectue d'une maniére similaire à  celle de l'algorithme
\textsc{CRP\_Miner}. Toutefois, l'ensemble $\mathcal{MC}$$and_{n}$ des motifs candidats de taille $n$ sera généré moyennant la phase combinatoire d'\textsc{Apriori-Gen} à  partir de l'ensemble $\mathcal{MC}$$and_{n-1}$
\textsc{(}\textit{cf.} ligne \ref{setcand2} de l'algorithme \ref{AlgoCRPR}\textsc{)}.
Par la suite, les candidats vérifiant la propriété de cross-support seront élagués
\textsc{(}\textit{cf.} ligne \ref{crosssupp2} de l'algorithme \ref{AlgoCRPR}\textsc{)}.
De plus, tout candidat non inclu dans un motif maximal corrélé rare sera élagué
\textsc{(}\textit{cf.} ligne \ref{prune2} de l'algorithme \ref{AlgoCRPR}\textsc{)}
puisqu'il n'est pas corrélé d'aprés la proposition \ref{prop_propriete_ens_MCR}.
Ensuite, tout candidat inclu dans un motif corrélé maximal fréquent sera élagué,
\textsc{(}\textit{cf.} ligne \ref{prunenew2} de l'algorithme \ref{AlgoCRPR}\textsc{)}.
Un autre élagage basé sur la propriété de l'idéal d'ordre des motifs minimaux corrélés est réalisé.
En effet, tous les sous-ensembles d'un motif minimal corrélé doivent étre des motifs minimaux corrélés d'aprés la propriété \ref{propMMC}. \`A cet égard, tout candidat de taille $n$ possédant un sous-ensemble de taille $n-1$ non minimal corrélé, sera élagué
\textsc{(}\textit{cf.} ligne \ref{prune3} de l'algorithme \ref{AlgoCRPR}\textsc{)}.

\`A ce stade, l'ensemble $\mathcal{MC}$$and_{n}$  des motifs candidats de taille $n$
englobe tous les motifs minimaux corrélés potentiels. L'étape suivante consiste donc à  identifier les motifs minimaux corrélés qui sont rares et à  calculer leur fermetures par $f_{bond}$.
Ceci est réalisé gréce à  la fonction
\textsc{Calcul\_Supports\_Fermetures}\textsc{(}\textit{cf.} ligne \ref{appelfct2} de l'algorithme \ref{AlgoCRPR}\textsc{)}.
Cette fonction est invoquée afin d'extraire l'ensemble des motifs minimaux corrélés rares de taille $n$, de calculer leurs fermetures et mettre à  jour l'ensemble $\mathcal{MFCR}$
des motifs fermés corrélés rares.
Ensuite, la variable $n$
sera incrémentée
et de nouveaux candidats de taille $n$ seront générés à  partir des motifs
corrélés de taille $n-1$
\textsc{(}\textit{cf.} ligne \ref{setcand3} de l'algorithme \ref{AlgoCRPR}\textsc{)}.
La boucle itérative s'arréte lorsque l'ensemble des motifs candidats est vide
\textsc{(}\textit{cf.} ligne \ref{iter1} de l'algorithme \ref{AlgoCRPR}\textsc{)} alors l'algorithme \textsc{CRPR\_Miner} marque sa fin d'exécution et retourne
la représentation concise exacte $\mathcal{RMCR}$
composée par l'ensemble $\mathcal{MMCR}$ des motifs minimaux corrélés rares
et de l'ensemble  $\mathcal{MFCR}$ des motifs fermés corrélés rares
\textsc{(}\textit{cf.} ligne \ref{output1} de l'algorithme \ref{AlgoCRPR}\textsc{)}.

%%%%%%%%%%%%%%%%%%%%%%%%%%%%
%%%\incmargin{0.5em} 
%%\linesnumbered
\begin{algorithm}[!t]\label{AlgoCRPR}
	\small{ \small{ \caption{\textsc{CRPR\_Miner}}}
		%%\SetVline
		%%\setnlskip{-3pt}
		\Donnees{
			\begin{enumerate}
				\item Un contexte d'extraction $\mathcal{D}$.
				\item Un seuil minimal de corrélation \textit{minbond}.
				\item Un seuil minimal de support conjonctif \textit{minsupp}.
			\end{enumerate}
		} \Res{La représentation concise exacte $\mathcal{RMCR}$ = $\mathcal{MMCR}$ $\cup$ $\mathcal{MFCR}$.} \Deb{
			
			$\mathcal{MCM}$$ax$ := \textsc{Algorithme\_Extraction\_MCMax}\textsc{(}$\mathcal{D}$,  \textit{minbond}\textsc{)}\;
			$\mathcal{MCM}$$ax\mathcal{F}$ := $\{$$X$ $\in$  $\mathcal{MCM}$$ax$ $\mid$  $X$.Conj $\geq$
			\textit{minsupp}$\}$\;
			$\mathcal{MCM}ax\mathcal{R}$ := $\{$$X$ $\in$ $\mathcal{MCM}ax$ $\mid$  $X$.\textit{Conj} $<$ \textit{minsupp}$\}$ \label{getmcmR}\;

			$\mathcal{MC}$$and_{1}$ := $\{$$i$ $\in$ $\mathcal{I}$$\}$ \label{setcand1}\;
			
			$\mathcal{MMCR}$$_{1}$ $\cup$ $\mathcal{MFCR}$
			:= \textsc{Calcul\_Supports\_Fermetures}\textsc{(}$\mathcal{D}$, $\mathcal{MC}$$and_{1}$, \textit{minsupp}\textsc{)} \label{appelfct1}\;
			
			$n$ := $2$\;
			
			$\mathcal{MC}$$and_{n}$ :=
			\textsc{Apriori-Gen}\textsc{(}$\mathcal{MC}$$and_{n-1}$\textsc{)}\label{setcand2}\;
			
			\Tq{\textsc{(}$\mathcal{MC}$$and_{n}$ $\neq$
				$\emptyset$\textsc{)}\label{iter1}}
			{
				/* \'Elagage des motifs candidats */ \\
				
				$\mathcal{MC}$$and_{n}$ := $\{$ $X_{n}$ $\in$  $\mathcal{MC}$$and_{n}$  $|$
				$\nexists$ x, y $\in$ $X_{n}$ $|$ $\displaystyle\frac{x.\textit{Conj}}
				{y.\textit{Conj}} < \textit{minbond}$\textsc{)}$\}$ \label{crosssupp2}\;
				$\mathcal{MC}$$and_{n}$ := $\{$$X_{n}$ $\in$  $\mathcal{MC}$$and_{n}$  $|$
				$\exists$ $Z$ $\in$ $\mathcal{MCR}$ $|$ $X_{n}$ $\subset$ $Z$$\}$ \label{prune2}\;
				
				$\mathcal{MC}$$and_{n}$ := $\{$$X_{n}$ $\in$  $\mathcal{MC}$$and_{n}$  $|$
				$\nexists$ $Z$ $\in$ $\mathcal{MCM}$$ax\mathcal{F}$ $|$ $X_{n}$ $\subset$ $Z$$\}$ \label{prunenew2}\;
				
				$\mathcal{MC}$$and_{n}$ := $\{$$X_{n}$ $\in$  $\mathcal{MC}$$and_{n}$  $|$
				$\nexists$ $Y_{n-1}$ $\subset$ $X_{n}$ et
				$Y_{n-1}$ $\notin$ $\mathcal{MC}and_{n-1}$$\}$\label{prune3}\;
				
				/* Extraction des minimaux et calcul des fermetures */ \\
				
				$\mathcal{MMCR}$$_{n}$ $\cup$ $\mathcal{MFCR}$ := \textsc{Calcul\_Supports\_Fermetures}\textsc{(}$\mathcal{D}$, $\mathcal{MC}$$and_{n}$, \textit{minsupp}\textsc{)}\label{appelfct2}\;
				
				$n$ := $n$ +$1$\;
				
				$\mathcal{MC}$$and_{n}$ :=
				\textsc{Apriori-Gen}\textsc{(}$\mathcal{MC}$$and_{n-1}$\textsc{)} \label{setcand3}\;
			}
			\Retour{$\displaystyle \bigcup_{i \in  [1, n]}$ $\mathcal{MMCR}$$_{i}$ $\cup$
				$\displaystyle \bigcup_{i \in  [1, n]}$ $\mathcal{MFCR}$$_{i}$\;} \label{output1}}}
\end{algorithm}
%%\decmargin{0.5em}
%%\incmargin{0em} 
%%\linesnumbered
%%%%%%%%%%%%%%%%%%%%%%%%%%%%%%%%%%%%%%%
Décrivons, à  présent, le déroulement de la fonction
\textsc{Calcul\_Supports\_Fermetures} dont le pseudo code est donné par l'algorithme \ref{calculSuppF}. Cette fonction prend en entrée le contexte d'extraction $\mathcal{D}$, le seuil \textit{minsupp} et l'ensemble $\mathcal{MC}$$and$ de candidats de taille $n$ et fournit en sortie l'ensemble
$\mathcal{MMCR}$$_{n}$ des motifs minimaux corrélés rares de taille $n$ et l'ensemble
$\mathcal{MFCR}$ des motifs fermés corrélés rares.
Cette fonction opére d'une maniére similaire à  la fonction
\textsc{Calcul\_Supports\_Bond}. De plus du calcul des supports conjonctifs, disjonctifs et la valeur \textit{bond}, la fonction \textsc{Calcul\_Supports\_Fermetures}
assure le calcul des fermetures conjonctives, disjonctives et par conséquent la fermeture par $f_{bond}$ des candidats retenus.
\`A cet égard, le contexte d'extraction est parcouru d'une maniére séquentielle. Pour chaque transaction $T$ du contexte, nous désignons par $I$ l'ensemble des items appartenant à  cette transaction $T$ et par $\omega$ l'ensemble des items commun entre $I$ et le candidat en cours $X_{n}$.
En effet,
Trois cas sont alors possibles. \\
\textbf{\textsc{(}\textit{i}\textsc{)}}
Si $\omega$ est vide, autrement dit aucune intersection n'existe entre le motif candidat $X_{n}$ et la transaction $T$. Alors, le complément de la fermeture disjonctive $X_{n}$.\textit{CmpDisj}, correspond aux items
n'appartenant à  aucune transaction contenant au moins un item du candidat $X_{n}$,
reéoit l'ensemble des items $\omega$
\textsc{(}\textit{cf.} ligne \ref{majcmpdisj} de l'algorithme \ref{calculSuppF}\textsc{)}. \\
\textbf{\textsc{(}\textit{ii}\textsc{)}}
Si $\omega$ contient au moins  un seul item du motif candidat $X_{n}$
alors le support disjonctif $X_{n}$.\textit{Disj} sera incrémenté
\textsc{(}\textit{cf.} ligne \ref{majdisj} de l'algorithme \ref{calculSuppF}\textsc{)}.\\
\textbf{\textsc{(}\textit{iii}\textsc{)}}
Si $\omega$ contient tous les items composant le candidat $X_{n}$
\textsc{(}\textit{cf.} ligne \ref{si-egal} de l'algorithme \ref{calculSuppF}\textsc{)}
alors le support conjonctif $X_{n}$.\textit{Conj} sera incrémenté
\textsc{(}\textit{cf.} ligne \ref{majconj} de l'algorithme \ref{calculSuppF}\textsc{)}
et la fermeture conjonctive du motif candidat $X_{n}$.$f_{c}$ si elle est vide alors elle sera initialisée à  $\omega$
\textsc{(}\textit{cf.} ligne \ref{majfc1} de l'algorithme \ref{calculSuppF}\textsc{)}
sinon elle sera mise à  jour
\textsc{(}\textit{cf.} ligne \ref{majfc2} de l'algorithme \ref{calculSuppF}\textsc{)}.\\
Aprés le balayage de tout le contexte, les supports disjonctifs, conjonctifs, les fermetures conjonctives et les compléments de la fermeture disjonctive de tous les candidats sont mis à  jour.
L'étape suivante consiste à  parcourir l'ensemble des candidats maintenus afin d'identifier les motifs minimaux corrélés rares.
Nous testons d'abord la rareté de chaque candidat
\textsc{(}\textit{cf.} ligne \ref{verifsupp} de l'algorithme \ref{calculSuppF}\textsc{)}.
Si le candidat en cours est rare alors nous
calculons sa valeur de la mesure \textit{bond}
\textsc{(}\textit{cf.} ligne \ref{calculbond} de l'algorithme \ref{calculSuppF}\textsc{)}.
et nous procédons à  la vérification de sa minimalité.
En effet, d'aprés la définition
\ref{motifmincorreles}, un motif minimal corrélé ne doit pas étre inclut dans la fermeture de ses sous-ensembles directs. Autrement dit, un motif minimal corrélé ne doit posséder aucun sous-ensemble direct de même valeur de mesure \textit{bond} que lui
\textsc{(}\textit{cf.} ligne \ref{verifMMC} de l'algorithme \ref{calculSuppF}\textsc{)}.
Dans le cas échéant, le motif candidat est ainsi minimal corrélé rare et sera inséré
dans l'ensemble $\mathcal{MMCR}$$_{n}$ des motifs minimaux corrélés rares de taille $n$
\textsc{(}\textit{cf.} ligne \ref{majMMCR} de l'algorithme \ref{calculSuppF}\textsc{)}.

Nous calculons, ensuite, son fermé disjonctif
$X_{n}$.\textit{$f_{d}$} dans la ligne 20. Le fermé $X_{n}$.\textit{$f_{bond}$}, résultant de l'intersection entre la fermeture conjonctive $X_{n}$.$f_{c}$ et la fermeture disjonctive $X_{n}$.$f_{d}$, sera ensuite dérivé
\textsc{(}\textit{cf.} ligne \ref{calculfbond} de l'algorithme \ref{calculSuppF}\textsc{)}.

Les motifs fermés calculés lors de chaque itération seront insérés dans l'ensemble
$\mathcal{MFCR}$ des motifs fermés corrélés rares et répartis selon leur taille.
\`A cet égard, nous sauvegardons
la taille $l$ de chaque motif fermé, $l$ = $|$$X_{n}$.\textit{$f_{bond}$}$|$
\textsc{(}\textit{cf.} ligne \ref{gettaille} de l'algorithme \ref{calculSuppF}\textsc{)}.
Par conséquent, chaque motif fermé de taille $l$ sera inséré dans l'ensemble
$\mathcal{MFCR}$$_{l}$ des motifs fermés corrélés rares de taille $l$
\textsc{(}\textit{cf.} ligne \ref{majMFCR1} de l'algorithme \ref{calculSuppF}\textsc{)}.
L'ensemble $\mathcal{MFCR}$ de tous les motifs fermés corrélés rares sera, par la suite, mis à  jour \textsc{(}\textit{cf.} ligne \ref{majMFCR2} de l'algorithme \ref{calculSuppF}\textsc{)}.

\`A la fin de l'exécution de la fonction  \textsc{Calcul\_Supports\_Fermetures}, l'ensemble $\mathcal{MMCR}$$_{n}$ des motifs minimaux corrélés rares de taille $n$ sera extrait et l'ensemble $\mathcal{MFCR}$ des motifs fermés corrélés rares sera mis à  jour
\textsc{(}\textit{cf.} ligne \ref{output2} de l'algorithme \ref{calculSuppF}\textsc{)}.

Nous avons ainsi analysé les différentes étapes de l'algorithme \textsc{CRPR\_Miner}. Nous enchaénons dans ce qui suit par sa trace d'exécution appliquée au contexte d'extraction donné par la table \ref{Base_transactions}.
%%%%%%%%%%%%%%%%%%%%%%%%%%%%%%%%%%%%%%%%%%%%%%%%%%%%%%%%%%%%%%%%%%%
%%\incmargin{0.5em} 
%%%\linesnumbered
\begin{algorithm}[htbp]\label{calculSuppF}
	\small{\small{ \caption{\textsc{Calcul\_Supports\_Fermetures}}}
		%%\SetVline
		%%\setnlskip{-3pt}
		\Donnees{
			\begin{enumerate}
				\item Un contexte d'extraction $\mathcal{D}$.
				\item L'ensemble $\mathcal{MC}$$and_{n}$ des motifs candidats de taille $n$.
				\item Le seuil minimal de support conjonctif \textit{minsupp}.
			\end{enumerate}
		} \Res{L'ensemble $\mathcal{MMCR}$$_{n}$ des motifs minimaux corrélés rares sera extrait et l'ensemble  $\mathcal{MFCR}$ des motifs fermés corrélés rares sera mis à  jour} \Deb{
			\PourCh{\textsc{(}Objet $O$ du contexte  $\mathcal{D}$\textsc{)}}
			{
				\PourCh{\textsc{(}$X_{n}$ de $\mathcal{MC}$$and_{n}$ \textsc{)}}
				{
					$\omega$ := $X_{n}$ $\cap$ $I$  /* $I$ correspond aux items appartenant à  l'objet $O$ */\;
					\Si {\textsc{(}$\omega$ $=$ $\emptyset$\textsc{)}}
					{
						
						$X_{n}$.\textit{$CmpDisj$} :=  $X_{n}$.\textit{$CmpDisj$} $\cup$ $I$\label{majcmpdisj}\;
					}
					\Sinon
					{  $X_{n}$.\textit{Disj} :=  $X_{n}$.\textit{Disj} + 1\label{majdisj}\;
						\Si {\textsc{(}$\omega$ $=$ $X_{n}$\textsc{)}\label{si-egal}}
						{
							$X_{n}$.\textit{Conj} :=  $X_{n}$.\textit{Conj} + 1\label{majconj} \;
							\Si {$X_{n}$.$f_{c}$ = $\emptyset$}
							{$X_{n}$.$f_{c}$ := $\omega$\label{majfc1}\;
							}%%si
							\Sinon
							{$X_{n}$.$f_{c}$ := $X_{n}$.$f_{c}$ $\cap$ $\omega$\label{majfc2}\;}
						}%%si
					}%%sinon
				}%%candidat
			}	%%objet
			%%%%
			\PourCh{\textsc{(}$X_{n}$ de $\mathcal{MC}$$and_{n}$\textsc{)}}
			{
				\Si{\textsc{(}$X_{n}$.\textit{Conj} $<$
					\textit{minsupp}\textsc{)}\label{verifsupp}}
				{
					$X_{n}$.\textit{bond} := $\displaystyle\frac{\displaystyle I.\textit{Conj}}
					{\displaystyle I.\textit{Disj}}$ \label{calculbond}\;
					
					\Si{\textsc{(}$\nexists$ $Y_{n}$ $\subset$ $X_{n+1}$ $|$
						\textit{bond}\textsc{(}$Y_{n}$\textsc{)} $=$ \textit{bond}\textsc{(}$X_{n+1}$\textsc{)}\textsc{)}\label{verifMMC}}
					{
						$\mathcal{MMCR}$$_{n}$ := $\mathcal{MMCR}$$_{n}$ $\cup$ \textsc{(}$X_{n}$, $X_{n}$.\textit{Conj}, $X_{n}$.\textit{bond}\textsc{)}\label{majMMCR}\;
						$X_{n}$.$f_{d}$ := $\mathcal{I}$$\setminus$$X_{n}$.$CmpDisj$\label{calculfd}\;      	 
						$X_{n}$.$f_{bond}$ :=  $X_{n}$.$f_{d}$ $\cap$ $X_{n}$.$f_{c}$\label{calculfbond}\;
						$l$ := $|$$X_{n}$.$f_{bond}$$|$ \label{gettaille}\;
						$\mathcal{MFCR}$$_{l}$ := $\mathcal{MFCR}$$_{l}$ $\cup$ \textsc{(}$X_{n}.f_{bond}$, $X_{n}$.\textit{Conj}, $X_{n}$.\textit{bond}\textsc{)}\label{majMFCR1}\;
						$\mathcal{MFCR}$ := $\mathcal{MFCR}$ $\cup$ $\mathcal{MFCR}$$_{l}$\label{majMFCR2}\;
					}%%SiMincorrelated
				}%%siRare
			}%%Pour
			
			\Retour{$\mathcal{MMCR}$$_{n}$ $\cup$ $\mathcal{MFCR}$\;} \label{output2}}}
\end{algorithm}
%%\decmargin{0.5em}
%%\incmargin{0em}
%%\linesnumbered	
%%%%%%%%%%%%%%%%%%%%%%%%%%%%%%%%%%%%%%%%%%%%%%%%%%%%%%%%%%%%%%%%%%
\subsection{Trace d'exécution de l'algorithme \textsc{CRPR\_Miner}}
Considérons la base de transactions \ref{Base_transactions} pour \textit{minsupp} = 3 et \textit{minbond} = 0,20. En utilisant l'algorithme \textsc{CRPR\_Miner}, l'extraction de la représentation $\mathcal{RMCR}$ se fait de la maniére suivante :
Initialement, les motifs maximaux sont récupérés gréce à  l'algorithme \textsc{Algorithme\_Extraction\_MCMax},
$\mathcal{MCM}ax$ = $\{$\textsc{(}\texttt{ACD}, 1, $\displaystyle\frac{1}{4}$\textsc{)},
\textsc{(}\texttt{ABCE}, 2, $\displaystyle\frac{2}{5}$\textsc{)}$\}$.
\'Etant donné que tous les motifs de l'ensemble $\mathcal{MCM}ax$ sont rares,
ils seront ainsi tous insérés dans l'ensemble $\mathcal{MCM}ax$$\mathcal{R}$ des motifs maximaux corrélés rares,
$\mathcal{MCM}ax$$\mathcal{R}$ = $\{$\textsc{(}\texttt{ACD}, 1, $\displaystyle\frac{1}{4}$\textsc{)},
\textsc{(}\texttt{ABCE}, 2, $\displaystyle\frac{2}{5}$\textsc{)}$\}$.

Ensuite, la base de transactions est parcourue et l'ensemble des motifs candidats est initialisé,
$\mathcal{MC}$$and_{1}$ =  $\{$\texttt{A}, \texttt{B}, \texttt{C}, \texttt{D}, \texttt{E}$\}$.
La fonction \textsc{Calcul\_Supports\_Fermetures} est ensuite invoquée.
Le seul item rare $D$ est identifié,
et il est inséré dans l'ensemble $\mathcal{MMCR}$$_{1}$ = $\{$\textsc{(}\texttt{D}, 1, $\displaystyle\frac{1}{1}$\textsc{)}$\}$.
Nous avons \texttt{D}.$f_{c}$ = \texttt{ACD}.$f_{d}$ = \texttt{D} et
\texttt{D}.$f_{bond}$ = \texttt{D}. Le motif fermé \texttt{D} est inséré dans l'ensemble $\mathcal{MFCR}$$_{1}$,
$\mathcal{MFCR}$$_{1}$ = $\{$\textsc{(}\texttt{D}, 1, $\displaystyle\frac{1}{1}$\textsc{)}$\}$.
L'ensemble $\mathcal{MC}$$and_{2}$ des motifs candidats de taille 2 est égal é,
$\mathcal{MC}$$and_{2}$ =  $\{$\texttt{AB}, \texttt{AC}, \texttt{AD}, \texttt{AE}, \texttt{BC}, \texttt{BE}, \texttt{BD}, \texttt{CE}, \texttt{CD}, \texttt{DE}$\}$.
Alors nous commenéons la premiére itération de l'algorithme.\\
\textbf{\underline{Itération 1 :}}\\
Tous les candidats vérifient la propriété de l'idéal d'ordre des motifs minimaux corrélés. De plus, la propriété de cross-support n'élaguera aucun des candidats de l'ensemble $\mathcal{MC}$$and_{2}$.
Cependant, seuls les candidats \texttt{BD} et \texttt{DE} sont élagués vu qu'ils ne sont inclus dans aucun motif maximal corrélé rare.
Ainsi, nous avons
$\mathcal{MC}$$and_{2}$ =  $\{$\texttt{AB}, \texttt{AC}, \texttt{AD}, \texttt{AE}, \texttt{BC}, \texttt{BE}, \texttt{CE}, \texttt{CD}$\}$.
Les supports conjonctifs et disjonctifs des candidats sont calculés.
Pour les candidats rares \texttt{AB},  \texttt{AD}, \texttt{AE} et \texttt{CD}, la valeur de la mesure \textit{bond} est calculée.
Tous ces candidats sont des motifs minimaux corrélés rares, ainsi ils ont insérés dans
l'ensemble $\mathcal{MMCR}$$_{2}$.
$\mathcal{MMCR}$$_{2}$ = $\{$\textsc{(}\texttt{AB}, 2, $\displaystyle\frac{2}{5}$\textsc{)},
\textsc{(}\texttt{AE}, 2, $\displaystyle\frac{2}{5}$\textsc{)}, \textsc{(}\texttt{AD}, 1, $\displaystyle\frac{1}{3}$\textsc{)},  \textsc{(}\texttt{CD}, 1, $\displaystyle\frac{1}{4}$\textsc{)}$\}$.
Nous calculons par la suite leurs fermetures et nous avons
$\mathcal{MFCR}$$_{4}$ = $\{$\textsc{(}\texttt{ABCE}, 2, $\displaystyle\frac{2}{5}$\textsc{)}$\}$,
$\mathcal{MFCR}$$_{3}$ = $\{$\textsc{(}\texttt{ACD}, 1, $\displaystyle\frac{1}{4}$\textsc{)}$\}$,
$\mathcal{MFCR}$$_{2}$ = $\{$\textsc{(}\texttt{AD}, 1, $\displaystyle\frac{1}{3}$\textsc{)}$\}$. La variable $n$ sera ensuite incrémentée, $n$ = 3 et les candidats de taille trois sont générés.
$\mathcal{MC}$$and_{3}$ = $\{$\texttt{ABC}, \texttt{ABD}, \texttt{ABE}, \texttt{ACD},
\texttt{ACE}, \texttt{ADE}, \texttt{BCD}, \texttt{BCE}, \texttt{CDE}$\}$.\\
\textbf{\underline{Itération 2 :}}\\
Lors de la deuxiéme itération, aucun des candidats générés ne sera élagué par la propriété de cross-support. Cependant, les candidats \texttt{ABD}, \texttt{BCD}, \texttt{CDE}, et \texttt{ADE} sont élagués puisqu'ils ne sont pas inclus dans un motif corrélé maximal rare.

$\mathcal{MC}$$and_{3}$ = $\{$\texttt{ABC}, \texttt{ABE}, \texttt{ACD}, \texttt{ACE}, \texttt{BCE}$\}$.
Alors, nous calculons les supports conjonctifs et disjonctifs et la valeur \textit{bond} pour les candidats rares retenus à  savoir
\textsc{(}\texttt{ABC}, 2, $\displaystyle\frac{2}{5}$\textsc{)},
\textsc{(}\texttt{ACD}, 1, $\displaystyle\frac{1}{4}$\textsc{)},
\textsc{(}\texttt{ABE}, 2, $\displaystyle\frac{2}{5}$\textsc{)},
\textsc{(}\texttt{ACE}, 2, $\displaystyle\frac{2}{5}$\textsc{)}.
Nous remarquons qu'aucun de ses candidats n'est rare minimal corrélé. En effet, les candidats \textsc{(}\texttt{ABC}, 2, $\displaystyle\frac{2}{5}$\textsc{)} et
\textsc{(}\texttt{ABE}, 2, $\displaystyle\frac{2}{5}$\textsc{)} ont le sous ensemble \textsc{(}\texttt{AB}, 2, $\displaystyle\frac{2}{5}$\textsc{)} de même mesure bond qu'eux. Pour le candidat \textsc{(}\texttt{ACE}, 2, $\displaystyle\frac{2}{5}$\textsc{)}, il est sur-ensemble du motif \textsc{(}\texttt{AE}, 2, $\displaystyle\frac{2}{5}$\textsc{)} et ont la même valeur de la mesure bond.
Le candidat \textsc{(}\texttt{ACD}, 1, $\displaystyle\frac{1}{4}$\textsc{)} posséde le sous-ensemble
\textsc{(}\texttt{CD}, 1, $\displaystyle\frac{1}{4}$\textsc{)}, avec qui il partage la même valeur de la mesure \textit{bond}.
Ainsi l'ensemble  $\mathcal{MMCR}$$_{3}$ est vide, $\mathcal{MMCR}$$_{3}$= $\{$$\emptyset$$\}$.
La variable $n$ sera ensuite incrémentée $n$ = 4  et l'ensemble des candidats de taille 4 est vide,
$\mathcal{MC}$$and_{4}$ =  $\{$$\emptyset$$\}$ alors
la boucle itérative marque sa fin d'exécution donnant ainsi comme résultat les motifs minimaux corrélés rares $\mathcal{MMCR}$ = $\{$\textsc{(}\texttt{D}, 1, $\displaystyle\frac{1}{1}$\textsc{)},
\textsc{(}\texttt{AB}, 2, $\displaystyle\frac{2}{5}$\textsc{)},
\textsc{(}\texttt{AE}, 2, $\displaystyle\frac{2}{5}$\textsc{)},
\textsc{(}\texttt{AD}, 1, $\displaystyle\frac{1}{3}$\textsc{)},
\textsc{(}\texttt{CD}, 1, $\displaystyle\frac{1}{4}$\textsc{)}$\}$
et leurs fermés  $\mathcal{MFCR}$ = $\{$\textsc{(}\texttt{D}, 1, $\displaystyle\frac{1}{1}$\textsc{)}, 
\textsc{(}\texttt{AD}, 1, $\displaystyle\frac{1}{3}$\textsc{)},
\textsc{(}\texttt{ACD}, 1, $\displaystyle\frac{1}{4}$\textsc{)},
\textsc{(}\texttt{ABCE}, 2, $\displaystyle\frac{2}{5}$\textsc{)}$\}$.
Les éléments de la représentation concise exacte $\mathcal{RMCR}$ extraits sont représentés par la table \ref{execution}.

%%%%%%%%%%%%%%%%%%
\begin{table} 
	\begin{tabular}{|c|c|c|c||c|c|c|c|}
		\hline
		$X$& Type \textsc{(}$X$\textsc{)} & Supp\textsc{(}$\wedge$$X$\textsc{)} & bond\textsc{(}$X$\textsc{)}
		&$X$&  Type \textsc{(}$X$\textsc{)} & Supp\textsc{(}$\wedge$$X$\textsc{)} & bond\textsc{(}$X$\textsc{)} \\
		\hline\hline
		\texttt{D}&MMCR et MFCR& 1 &$\frac{1}{1}$ & \texttt{AB}&MMCR & 2& $\frac{2}{5}$ \\
		\texttt{AD}&MMCR et MFCR & 1& $\frac{1}{3}$ &\texttt{AE}&MMCR & 2& $\frac{2}{5}$ \\
		\texttt{CD}&MMCR & 1& $\frac{1}{4}$ &\texttt{ABCE}&MFCR & 2& $\frac{2}{5}$ \\
		\texttt{ACD}&MFCR & 1& $\frac{1}{4}$ & & & &  \\
		\hline
	\end{tabular}
	\begin{center}
		\caption{Résultat de l'exécution de l'algorithme \textsc{CRPR\_Miner}
			pour \textit{minsupp} $=$ \textit{3} et \textit{minbond} $=$
			\textit{0,20}.} \label{execution}
	\end{center}
\end{table}
Nous avons ainsi présenté et analysé les différentes étapes de l'algorithme \textsc{CRPR\_Miner} d'extraction de la représentation concise exacte $\mathcal{RMCR}$.
Nous signalons que l'extraction des représentations concises exactes
$\mathcal{RMM}$$ax$$\mathcal{F}$ et $\mathcal{RM}$$in$$\mathcal{MF}$
et de la représentation concise approximative
$\mathcal{RM}$$in$$\mathcal{MM}$$ax$$\mathcal{F}$ sont possibles moyennant l'algorithme \textsc{CRPR\_Miner}. \`A cet égard, nous détaillons ce point dans ce qui suit.
%%%%%%%%%%%%%%%%%%%%%%%%%%%%%%%%%%%%%%%%%%%%%%%%%%%%
\subsection{Extraction de la représentation concise exacte $\mathcal{RMM}$$ax$$\mathcal{F}$}
%%%%%%%%%%%%%%%%%%%%
En effet, la représentation $\mathcal{RMM}$$ax$$\mathcal{F}$ est composée de l'ensemble $\mathcal{MMCR}$ des motifs minimaux corrélés rares et de l'ensemble $\mathcal{MFCRM}ax$ des motifs fermés corrélés rares maximaux
\textsc{(}\textit{cf.} Définition \ref{rep2concise exacte def} page \pageref{rep2concise exacte def}\textsc{)}.
Or, lors de chaque itération $n$ l'algorithme \textsc{CRPR\_Miner} permet de mettre à  jour l'ensemble $\mathcal{MFCR}$ des motifs fermés corrélés rares.
Ainsi l'idée consiste à  intégrer dans l'algorithme \textsc{CRPR\_Miner} l'instruction permettant de filtrer les éléments maximaux de l'ensemble des motifs fermés corrélés rares générés.
Or, un motif fermé corrélé rare maximal ne doit posséder aucun sur-ensemble corrélé. \'A cet égard,
l'extraction des motifs fermés corrélés rares maximaux de taille $n-1$ nécessitent la reconnaissance de leurs sur-ensembles de taille $n$
et seront ainsi extraits pendant l'itération $n$ gréce à  l'instruction suivante.\\
$\mathcal{MFCRM}ax$$_{n-1}$ := $\{$$X_{n-1}$ $\in$
$\mathcal{MFCR}$$_{n-1}$ $|$ $\nexists$ $Y_{n}$ $\in$ $\mathcal{MC}$$and_{n}$ avec $Y_{n}$ $\supset$ $X_{n-1}$ et $Y_{n}$.\textit{bond} $\geq$ \textit{minbond}$\}$\;
\subsection{Extraction de la représentation concise exacte $\mathcal{RM}$$in$$\mathcal{MF}$}
Considérons maintenant la représentation concise exacte $\mathcal{RM}$$in$$\mathcal{MF}$ composée de l'ensemble $\mathcal{MFCR}$ des motifs fermés corrélés rares et de l'ensemble $\mathcal{MMCRM}in$
des éléments minimaux de l'ensemble $\mathcal{MMCR}$ des motifs rares
\textsc{(}\textit{cf.} Définition \ref{rep3concise exacte def} page \pageref{rep3concise exacte def}\textsc{)}. En effet, l'ensemble $\mathcal{MMCRM}$$in$ est composé des
motifs qui sont à  la fois minimaux corrélés et rares minimaux. Ces derniers correspondent aux motifs minimaux corrélés dont tous leurs sous ensembles directs sont des motifs fréquents.
Par conséquent, nous intégrons l'instruction suivante permettant de filtrer ces motifs à  partir de l'ensemble
$\mathcal{MMCR}$.
\begin{center}
	$\mathcal{MMCRM}$$in_{n}$ := $\{$$X_{n}$ $\in$
	$\mathcal{MMCR}$$_{n}$ $|$  $\forall$ $Y_{n-1}$ $\subset$ $X_{n}$ , $Y_{n-1}$.\textit{Conj} $\geq$ \textit{minsupp}$\}$\;
\end{center}
\`A présent, nous avons introduit et décrit les instructions permettant d'extraire les
représentations concises exactes $\mathcal{RMM}$$ax$$\mathcal{F}$ et $\mathcal{RM}$$in$$\mathcal{MF}$.
Par conséquent, la représentation concise approximative $\mathcal{RM}$$in$$\mathcal{MM}$$ax$$\mathcal{F}$
\textsc{(}\textit{cf.} Définition \ref{rep4concise approxdef} page
\pageref{rep4concise approxdef}\textsc{)} résultante de l'union des ensembles $\mathcal{MMCRM}$$in$ et $\mathcal{MFCRM}ax$ sera aisément dérivée gréce aux deux instructions précédemment décrites.

Nous avons, à  ce stade, présenté l'algorithme \textsc{CRPR\_Miner} d'extraction de la représentation $\mathcal{MMCR}$ et introduit les instructions à  y intégrer afin d'extraire les autres représentations concises proposées. Dans la suite
nous démontrons les propriétés théoriques de l'algorithme \textsc{CRPR\_Miner}.
\subsection{Preuves théoriques}	
Dans cette section, nous démontrons les propriétés de validité et de terminaison de l'algorithme \textsc{CRPR\_Miner}.

\begin{proposition}
	L'algorithme \textsc{CRPR\_Miner} génére tous les motifs minimaux et fermés corrélés rares munis de leurs supports conjonctifs et de leurs valeurs de la mesure bond.
\end{proposition}

\begin{Preuve}
	L'algorithme \textsc{CRPR\_Miner} est un algorithme par niveau permettant d'extraire avec exactitude tous les éléments de la représentation concise exacte $\mathcal{RMCR}$.
	En effet,
	lors du premier parcours de la base de transactions, les items rares sont identifiés séparément. Les items rares constituent les motifs minimaux corrélés rares de taille $1$ et leurs fermés par $f_{bond}$ seront calculés d'une maniére exacte. Ensuite les éléments minimaux de l'ensemble
	$\mathcal{MMCR}$
	des motifs minimaux corrélés rares seront extraits et leurs fermés respectifs seront calculés d'une maniére itérative.
	En effet, lors de chaque itération, un ensemble de candidats de taille $n$ est généré à  partir des motifs fréquents de taille $n-1$. Ces candidats ne doivent contenir aucun sous ensemble rare non corrélé. De plus, ils ne doivent pas englober des items qui vérifient la propriété de cross-support.
	
	Ensuite, les supports conjonctifs, les supports disjonctifs, les fermetures conjonctives et les fermetures disjonctives de tous les candidats seront calculés moyennant un balayage de 
	la base de transactions. Les candidats fréquents seront sauvegardés dans l'ensemble des motifs fréquents. Cependant, nous calculons la valeur de la mesure \textit{bond} pour les motifs candidats vérifiant la contrainte monotone de rareté.  Dans le cas oé la valeur de \textit{bond} d'un motif rare $X$ dépasse le seuil minimal \textit{minbond}, alors nous vérifions la minimalité du candidat en cours. En effet, s'il ne posséde aucun sous ensemble direct de même mesure bond que lui, alors le candidat en cours est minimal corrélé rare.
	Par conséquent, le motif fermé par $f_{bond}$ correspondant au motif minimal corrélé rare en cours, sera calculé. Il résulte, en effet, de l'intersection entre son fermé conjonctif et son fermé disjonctif.
	\'Etant donné que les supports conjonctifs, disjonctifs et la mesure \textit{bond} d'un fermé sont égaux à  ceux du motif minimal correspondant, alors nous déduisons que les caractéristiques de chaque fermé par l'opérateur de fermeture $f_{bond}$ sont attribués d'une maniére exacte.
	La boucle itérative d'extraction des éléments minimaux de l'ensemble $\mathcal{MMCR}$ et de calcul de leurs fermés respectifs se termine lorsqu'il n'y a plus de motifs candidats à  générer. \`A la fin de cette étape l'ensemble $\mathcal{MMCR}$ est composé de tous les motifs qui sont minimaux corrélés et rares minimaux à  la fois et leurs fermés respectifs sont inclus dans l'ensemble $\mathcal{MFCR}$.\\
	L'étape suivante consiste à  effectuer un traitement itératif afin de dériver le reste des motifs minimaux corrélés rares à  partir des motifs minimaux extraits lors de l'étape précédente et à  calculer leurs fermés par $f_{bond}$.
	En effet, lors de chaque itération, un ensemble de candidats de taille $n$ est généré à  partir des motifs minimaux corrélés rares de taille $n-1$. Ces candidats sont tous rares puisqu'ils sont des sur-ensembles de motifs rares. Ainsi, ils seront testés par rapport à  la propriété de cross support et par rapport à  la contrainte anti-monotone des motifs minimaux corrélés.  Les candidats rares de taille $n$ qui sont minimaux corrélés seront insérés dans l'ensemble $\mathcal{MMCR}$$_{n}$. Par la suite, leurs fermés par $f_{bond}$ seront calculés et insérés dans l'ensemble $\mathcal{MFCR}$ des motifs fermés corrélés rares. L'algorithme \textsc{CRPR\_Miner} marque sa fin d'exécution lorsque l'ensemble de motifs candidats est vide.
	Nous concluons que \textsc{CRPR\_Miner} permet d'extraire avec exactitude tous les éléments des ensembles $\mathcal{MMCR}$ et $\mathcal{MFCR}$ munis de leurs supports conjonctifs et de leurs valeurs de mesure bond exacts. Il est donc correct et complet.
	Ainsi, l'ensemble $\mathcal{MMCR}_{n}$ ne contient que les motifs minimaux corrélés rares de taille $n$ et l'ensemble $\mathcal{MFCR}_{n}$ ne contient que les fermés corrélés rares de taille $n$.
\end{Preuve}

\vspace{-1.cm}

\begin{proposition}
	L'algorithme \textsc{CRPR\_Miner} se termine correctement.
\end{proposition}
\begin{Preuve}
	Le nombre des motifs générés par \textsc{CRPR\_Miner} est fini. En effet, le nombre de motifs candidats pouvant étre générés à  partir d'une base de transactions ayant $n$ items distincts, est au plus égal à  $2^{n}$. De plus, le nombre d'opérations effectuées, afin de traiter chaque motif est fini. Par conséquent, l'algorithme \textsc{CRPR\_Miner} se termine correctement.
\end{Preuve}

Nous avons, ainsi, prouvé les propriétés théoriques de validité et de terminaison de l'algorithme \textsc{CRPR\_Miner} d'extraction de la représentation concise exacte  $\mathcal{RMCR}$.
Dans la section suivante, nous introduisons les
algorithmes d'interrogation et de régénération des motifs corrélés rares à  partir de la représentation concise exacte $\mathcal{RMCR}$.

\section{Algorithmes d'interrogation et de régénération des motifs corrélés rares à  partir de la représentation $\mathcal{RMCR}$}

Nous introduisons, dans cette section, les deux stratégies de régénération à  savoir la régénération d'un seul motif corrélé rare et la régénération de l'ensemble total de tous les motifs corrélés rares à  partir de la représentation concise exacte $\mathcal{RMCR}$.
Les notations utilisées dans les algorithmes de régénération proposés sont introduites dans le tableau \ref{NotAlgoRegenerate}.
\begin{table}
	\begin{center}
		{
			\begin{tabular}{|lcl|}
				\hline
				$I$.\textit{Conj}& :&Support conjonctif du motif $I$.\\
				$I$.\textit{Disj}& :&Support disjonctif du motif $I$.\\
				$I$.\textit{Neg}& :&Support négatif du motif $I$.\\
				$I$.\textit{bond}& :&Valeur de la mesure \textit{bond} du motif I.\\
				$I.f_{bond}$& :& Fermeture du motif $I$ par $f_{bond}$.\\
				\hline
		\end{tabular}}
		\caption{Notations adoptées dans les algorithmes \textsc{Regenerate}
			et \textsc{CRP\_Regeneration}
			.}\label{NotAlgoRegenerate}
	\end{center}
\end{table}

\subsection{Interrogation de la représentation concise exacte $\mathcal{RMCR}$}

Nous introduisons, à  présent, l'algorithme \textsc{Regenerate}
permettant l'interrogation de la représentation $\mathcal{RMCR}$.

%%\incmargin{0.5em} 
%%%\linesnumbered
\begin{algorithm}[!t]
	%%\SetVline 
	\caption{\textsc{Regenerate} \label{Rege1}}
	%%\setnlskip{-3pt}
	\Donnees{\begin{enumerate}
			\item Un motif $I$.
			\item Le nombre de transactions $\mid$$\mathcal{T}$$\mid$.
			\item La représentation $\mathcal{RMCR}$ = $\mathcal{MMCR}$ $\cup$ $\mathcal{MFCR}$.
	\end{enumerate}}
	\Res{Le support conjonctif, disjonctif, négatif et la valeur de la mesure \textit{bond} si le motif $I$ est corrélé rare, sinon l'ensemble vide.}
	\Deb{
		\Si {\textsc{(}$I$ $\in$ $\mathcal{RMCR}$\textsc{)} \label{Apparrmcr}}
		{
			$I$.\textit{Disj} = $\displaystyle\frac{\displaystyle I.\textit{Conj}}
			{\displaystyle I.\textit{bond}}$ \label{calculdisj}\;
			
			$I$.\textit{Neg} = $\mid$$\mathcal{T}$$\mid$ $-$ $I$.\textit{Disj} \label{calculneg}\;
			
			\Retour{ $\{$$I$, $I$.\textit{Conj},
				$I$.\textit{Disj},
				$I$.\textit{Neg},
				$I$.\textit{bond}$\}$ \label{res1}}\;
		}%si
		\Sinon
		{
			\Si{\textsc{(}$\exists$ $J$, $Z$ $\in$ $\mathcal{RMCR}$ $\mid$ $J$ $\subset$ $I$ et $I$ $\subset$ $Z$\textsc{)}\label{si1}}
			{
				
				$F$ := $\min_{\subseteq}$$\{$$I_{1}$ $\in$  $\mathcal{RMCR}$ $\mid$  $I$  $\subset$ $I_{1}$$\}$ \label{fermé}\;
				
				$I$.\textit{Conj} = $F$.\textit{Conj}\;
				$I$.\textit{bond} = $F$.\textit{bond}\;
				$I$.\textit{Disj} = $\displaystyle\frac{\displaystyle I.\textit{Conj}}
				{\displaystyle I.\textit{bond}}$\;
				$I$.\textit{Neg} = $\mid$$\mathcal{T}$$\mid$ $-$ $I$.\textit{Disj}\;
				
				\Retour{ $\{$$I$, $I$.\textit{Conj},
					$I$.\textit{Disj},
					$I$.\textit{Neg},
					$I$.\textit{bond}$\}$ \label{res2}}\;
				\Sinon
				{\Retour{$\emptyset$}\label{res3}\;}
	}}}
\end{algorithm}
%%%\decmargin{1em}

\'Etant donné que nous avons démontré, dans la section précédente, la correction et la complétude de l'algorithme \textsc{CRPR$\_$Miner} d'extraction de la représentation $\mathcal{RMCR}$.
Cette régénération correspond à  l'interrogation de la représentation afin de
déterminer la nature d'un motif quelconque. Dans le cas oé il s'agit d'un motif corrélé rare, alors ses données \textsc{(}support conjonctif, disjonctif, négatif et la valeur de mesure \textit{bond}\textsc{)} seront régénérés gréce à  la représentation $\mathcal{RMCR}$.

Cette opération est achevée gréce à  l'algorithme \textsc{Regenerate} dont le pseudo-code est donné par l'algorithme \ref{Rege1}.
Ce dernier prend en entrée le nombre de transactions $|$$\mathcal{T}$$|$, la représentation $\mathcal{RMCR}$ composée des ensembles $\mathcal{MMCR}$ et $\mathcal{MFCR}$ et le motif en question et fournit en sortie les données
\textsc{(}le support conjonctif, disjonctif, négatif et la valeur de mesure \textit{bond}\textsc{)} du
motif passé en paramétre si'il est corrélé rare sinon il retourne l'ensemble vide.\\
Ainsi, afin de vérifier la nature du motif $I$ passé en paramétre, trois cas se présentent.
Dans le cas oé, le motif $I$ appartient à  la représentation $\mathcal{RMCR}$
\textsc{(}\textit{cf.} ligne \ref{Apparrmcr} de l'algorithme \ref{Rege1}\textsc{)}, alors il est rare corrélé. Nous disposons, alors, de son support conjonctif et de sa valeur de la mesure \textit{bond}. Quant à  son support disjonctif il correspond au rapport de son support conjonctif par la valeur de \textit{bond}
\textsc{(}\textit{cf.} ligne \ref{calculdisj} de l'algorithme \ref{Rege1}\textsc{)}
et son support négatif est extrait à  partir de son support disjonctif
\textsc{(}\textit{cf.} ligne \ref{calculneg} de l'algorithme \ref{Rege1}\textsc{)}. L'algorithme retourne alors les différents supports du motif ainsi que sa valeur de bond
\textsc{(}\textit{cf.} ligne \ref{res1} de l'algorithme \ref{Rege1}\textsc{)}.
Le deuxiéme cas se réalise lorsque le motif $I$ n'appartient pas à  la représentation $\mathcal{RMCR}$ mais il est compris entre deux éléments de la représentation $\mathcal{RMCR}$
\textsc{(}\textit{cf.} ligne \ref{si1}\textsc{)}. Ainsi, le motif fermé associé à  l'itemset $I$
correspond au plus petit, selon l'inclusion ensembliste, sur-ensemble du motif $I$ appartenant à  la représentation $\mathcal{RMCR}$
\textsc{(}\textit{cf.} ligne \ref{fermé} de l'algorithme \ref{Rege1}\textsc{)}. Le motif $I$ partage ainsi les mêmes valeurs des différents supports et de bond que son fermé identifié
\textsc{(}\textit{cf.} ligne \ref{res2}\textsc{)}.
Cependant, lorsque le motif $I$
n'appartient pas à  la représentation $\mathcal{RMCR}$ et il n'est pas compris entre deux éléments de la
représentation $\mathcal{RMCR}$ alors le motif $I$ n'est pas corrélé rare et l'algorithme retourne l'ensemble vide \textsc{(}\textit{cf.} ligne \ref{res3} de l'algorithme \ref{Rege1}\textsc{)}.\\
Ainsi nous avons décrit le déroulement de l'algorithme \textsc{Regenerate}, nous enchaénons par la suite avec un exemple d'exécution.
\begin{exemple}\label{exp_regeneration1}
	Considérons la représentation $\mathcal{RMCR}$ présentée par la table \ref{execution} 
	et considérons le motif \texttt{ACE}.
	En comparant \texttt{ACE} avec les éléments de la représentation $\mathcal{MMCR}$, nous remarquons que
	\texttt{AE} $\subset$ \texttt{ACE} et \texttt{ACE} $\subset$ \texttt{ABCE}. Ainsi, le motif \texttt{ACE}
	est corrélé rare et son fermé est \texttt{ABCE}.  Par conséquent,
	\texttt{ACE}.\textit{Conj} = \texttt{ABCE}.\textit{Conj} = 2,
	\texttt{ACE}.\textit{Disj} = \texttt{ABCE}.\textit{Disj} = 5,
	\texttt{ACE}.\textit{Neg} = $|\mathcal{T}|$ - \texttt{ACE}.\textit{Disj} = 5 - 5 = 0 et
	\texttt{ACE}.\textit{bond} = \texttt{ABCE}.\textit{bond} = $\displaystyle\frac{2}{5}$.\\
	Prenons le cas du motif \texttt{BC}. En effet, \texttt{BC} n'appartient pas à  la représentation
	$\mathcal{RMCR}$ et il n'est pas inclu entre deux éléments de la représentations $\mathcal{RMCR}$. Ainsi, l'algorithme retourne l'ensemble vide pour indiquer que le motif \texttt{BC} n'est pas un motif corrélé rare.
\end{exemple}
Dans ce qui suit nous présentons la procédure de régénération de l'ensemble de tous les motifs corrélés rares
é partir de la représentation concise exacte $\mathcal{RMCR}$.
\subsection{Régénération de l'ensemble total de tous les motifs corrélés rares}

%%%\incmargin{1em} 
%%%\linesnumbered
\begin{algorithm}[!t]\label{CRPRegeneration}\small{
		%%\SetVline
		%%\setnlskip{-3pt}
		\Donnees{\begin{enumerate}
				\item  La représentation concise exacte $\mathcal{RMCR}$ =  $\mathcal{MMCR}$ $\cup$ $\mathcal{MFCR}$.
				\item  Le nombre de transactions : $|$$\mathcal{T}$$|$.
			\end{enumerate}
		}
		\Res{L'ensemble $\mathcal{MCR}$ des motifs corrélés rares munis de leurs valeurs du support conjonctif et de leurs valeurs de la mesure \textit{bond}.}
		\Deb{
			$\mathcal{MCR}$ := $\emptyset$\;
			\PourCh{\textsc{(}$M \in \mathcal{RMCR}$\textsc{)}}
			{
				
				%%$M$.\textit{Disj} = $\displaystyle\frac{\displaystyle M.\textit{Conj}}
				%% {\displaystyle M.\textit{bond}}$ \label{calculdisj2}\;
				
				%%$M$.\textit{Neg} = $\mid$$\mathcal{T}$$\mid$ $-$ $M$.\textit{Disj} \label{calculneg2}\;

				$\mathcal{MCR}$ := $\mathcal{MCR}$ $\cup$ $\{$$M$, $M$.\textit{Conj},
				%%$M$.\textit{Disj},
				%%$M$.\textit{Neg},
				$M$.\textit{bond}$\}$ \label{majMCR1}\;
			}
			
			\PourCh{\textsc{(}$M \in \mathcal{MMCR}$\textsc{)}\label{pour2}}
			{
				$F$ := 	$\min_{\subseteq}$$\{$$M_{1}$ $\in$  $\mathcal{MFCR}$ $\mid$  $M$  $\subset$ $M_{1}$$\}$\label{ferme2}\;	
				
				\PourCh{\textsc{(}$X$ $\mid$ $M$ $\subset$ $X$ et $X$ $\subset$ $F$\textsc{)}\label{pour3}}
				{
					
					$X$.\textit{Conj} = $F$.\textit{Conj}\;
					$X$.\textit{bond} = $F$.\textit{bond}\;
					%%$X$.\textit{Disj} = $\displaystyle\frac{\displaystyle X.\textit{Conj}}
					%%{\displaystyle X.\textit{bond}}$ \label{calculdisj3}\;
					
					%%$X$.\textit{Neg} = $\mid$$\mathcal{T}$$\mid$ $-$ $X$.\textit{Disj} \label{calculneg3}\;
					
					\Si{\textsc{(}$X$ $\notin$ $\mathcal{MCR}$\textsc{)}}
					{
						$\mathcal{MCR}$ := $\mathcal{MCR}$ $\cup$ $\{$$X$, $X$.\textit{Conj},
						%%$X$.\textit{Disj},
						%%$X$.\textit{Neg},
						$X$.\textit{bond}$\}$ \label{majMCR2}\;
					}%%si
				}%%pour1
			}%%pour2
			\Retour{$\mathcal{MCR}$}\;\label{resultat}}
		\caption{\textsc{CRP\_Regeneration}}}
\end{algorithm}
%%%\decmargin{1em}
%%%%
La régénération de l'ensemble $\mathcal{MCR}$ de tous les motifs corrélés rares à  partir de la représentation concise exacte $\mathcal{RMCR}$ s'effectue gréce à  l'algorithme \textsc{CRP\_Regeneration} dont le pseudo-code est donné par l'algorithme \ref{CRPRegeneration}. \\
Cet algorithme prend en entrée la représentation concise exacte $\mathcal{RMCR}$ et fournit en sortie l'ensemble $\mathcal{MCR}$ des motifs corrélés rares munis de leurs supports conjonctifs et de leurs valeurs de mesure \textit{bond}. En effet, la procédure de régénération s'effectue de la maniére suivante.
D'abord, tous les éléments de la représentation  $\mathcal{RMCR}$
seront insérés dans l'ensemble $\mathcal{MCR}$
\textsc{(}\textit{cf.} ligne \ref{majMCR1} de l'algorithme \ref{CRPRegeneration}\textsc{)}. Par la suite,
l'algorithme parcours d'une maniére séquentielle l'ensemble $\mathcal{MMCR}$ des motifs minimaux et
affecte à  chaque minimal $M$ son fermé $F$
\textsc{(}\textit{cf.} ligne \ref{ferme2} de l'algorithme \ref{CRPRegeneration}\textsc{)}.
Puis l'ensemble de motifs compris entre le minimal $M$ et son fermé $F$ est généré
\textsc{(}\textit{cf.} ligne \ref{pour3} de l'algorithme \ref{CRPRegeneration}\textsc{)}.
%%Les supports disjonctifs et négatifs de chaque motif de cet
%%ensemble seront calculés
%%\textsc{(}\textit{cf.} ligne \ref{calculdisj3} et ligne \ref{calculneg3} de l'algorithme \ref{CRPRegeneration}\textsc{)}
Chaque élément de cet ensemble est un motif corrélé rare et partage le même support conjonctif et la même valeur
de bond que son fermé $F$ et sera inséré, si'il n'existe pas déjé, dans
l'ensemble $\mathcal{MCR}$
\textsc{(}\textit{cf.} ligne \ref{majMCR2} de l'algorithme \ref{CRPRegeneration}\textsc{)}.
Lorsque tous les motifs générés sont insérés dans l'ensemble $\mathcal{MCR}$, alors
l'algorithme marque sa fin et retourne, ainsi, l'ensemble total des motifs corrélés rares $\mathcal{MCR}$
\textsc{(}\textit{cf.} ligne \ref{resultat} de l'algorithme \ref{CRPRegeneration}\textsc{)}.\\
Nous avons ainsi analysé l'algorithme \textsc{CRP\_Regeneration}, nous enchaénons dans ce qui suit par la trace d'exécution.
\begin{exemple}
	Reconsidérons la représentation concise exacte
	donnée par la table \ref{execution}. La régénération de tous les motifs corrélés rares par l'algorithme
	\textsc{CRPRegeneration} se réalise de la maniére suivante.
	D'abord, l'ensemble $\mathcal{MCR}$ des motifs corrélés rares est initialisé à  l'ensemble vide. Ensuite, tous les éléments de la représentation concise exacte $\mathcal{RMCR}$ seront insérés dans l'ensemble
	$\mathcal{MCR}$. Ainsi,
	$\mathcal{MCR}$ = $\{$\textsc{(}\texttt{D}, 1, $\displaystyle\frac{1}{1}$\textsc{)},
	\textsc{(}\texttt{AB}, 2, $\displaystyle\frac{2}{5}$\textsc{)},
	\textsc{(}\texttt{AD}, 1, $\displaystyle\frac{1}{3}$\textsc{)},
	\textsc{(}\texttt{AE}, 2, $\displaystyle\frac{2}{5}$\textsc{)},
	\textsc{(}\texttt{CD}, 1, $\displaystyle\frac{1}{4}$\textsc{)},
	\textsc{(}\texttt{ABCE}, 2, $\displaystyle\frac{2}{5}$\textsc{)},
	\textsc{(}\texttt{ACD}, 1, $\displaystyle\frac{1}{4}$\textsc{)}$\}$.
	Par la suite, nous générons les motifs
	\texttt{ABE} et \texttt{ABC}
	compris entre le minimal \textsc{(}\texttt{AB}, 2, $\displaystyle\frac{2}{5}$\textsc{)} et son fermé \textsc{(}\texttt{ABCE}, 2, $\displaystyle\frac{2}{5}$\textsc{)} et le motif
	\texttt{ACE} compris entre le minimal \textsc{(}\texttt{AE}, 2, $\displaystyle\frac{2}{5}$\textsc{)} et son fermé \textsc{(}\texttt{ABCE}, 2, $\displaystyle\frac{2}{5}$\textsc{)}.
	Les motifs \texttt{ABE}, \texttt{ABC} et \texttt{ACE} générés auront ainsi
	le même support conjonctif et la même valeur de bond que leur motif fermé \texttt{ABCE} et seront alors
	insérés dans l'ensemble $\mathcal{MCR}$.
	Ce dernier est par conséquent mis à  jour et englobe tous les motifs corrélés rares.
	$\mathcal{MCR}$ = $\{$\textsc{(}\texttt{D}, 1, $\displaystyle\frac{1}{1}$\textsc{)},
	\textsc{(}\texttt{AD}, 1, $\displaystyle\frac{1}{3}$\textsc{)},
	\textsc{(}\texttt{CD}, 1, $\displaystyle\frac{1}{4}$\textsc{)},
	\textsc{(}\texttt{ACD}, 1, $\displaystyle\frac{1}{4}$\textsc{)},
	\textsc{(}\texttt{AE}, 2, $\displaystyle\frac{2}{5}$\textsc{)},
	\textsc{(}\texttt{AB}, 2, $\displaystyle\frac{2}{5}$\textsc{)},
	\textsc{(}\texttt{ACE}, 2, $\displaystyle\frac{2}{5}$\textsc{)},
	\textsc{(}\texttt{ABE}, 2, $\displaystyle\frac{2}{5}$\textsc{)},
	\textsc{(}\texttt{ABC}, 2, $\displaystyle\frac{2}{5}$\textsc{)},
	\textsc{(}\texttt{ABCE}, 2, $\displaystyle\frac{2}{5}$\textsc{)}$\}$.
\end{exemple}
\section{Conclusion}
Dans ce chapitre, nous avons présenté l'algorithme \textsc{CRP\_Miner} d'extraction de l'ensemble
$\mathcal{MCR}$ de tous les motifs corrélés rares. Ensuite, nous avons introduit l'algorithme
\textsc{CRPR\_Miner} permettant d'extraire la représentation concise
exacte $\mathcal{RMCR}$. Toutefois,
la contrainte monotone de rareté et la contrainte anti-monotone de corrélation sont simultanément intégrées dans le processus de fouille offert par l'algorithme \textsc{CRPR\_Miner}.
Les critéres pertinents de l'élagage
des candidats et de l'optimisation de l'espace de recherche ont été soigneusement incorporés dans cet algorithme.
Ce chapitre a été cloturé avec la présentation des stratégies d'interrogation et de régénération des motifs corrélés rares à  partir de la représentation $\mathcal{RMCR}$ .
Dans le chapitre suivant, nous présenterons les expérimentations réalisées grace aux quelles nous mesurons l'apport bénéfique des représentations concises proposées en terme de taux de réduction et nous étudions également leurs couts d'extractions.

%%%%%%%%%%%%%%%%%%%%%%%%
\chapter{\'{E}tude expérimentale}\label{chapitre_experimentation}

\section{Introduction}
Dans le chapitre précédent, nous avons présenté l'algorithme \textsc{CRP\_Miner} d'extraction de l'ensemble
$\mathcal{MCR}$ des motifs corrélés rares ainsi que
l'algorithme \textsc{CRPR\_Miner} dédié à  l'extraction de la
représentation concise exacte $\mathcal{RMCR}$ basée sur
les motifs minimaux et les motifs fermés corrélés rares.
Dans ce chapitre, nous présentons les expérimentations faites
sur des bases ``benchmark''. \`A cet égard, nous comptabilisons les cardinalités de l'ensemble
$\mathcal{MCR}$ et des différentes représentations concises proposées.
Nous visons à  travers ces expérimentations à  prouver les taux de compacité offerts par les représentations concises proposées. De plus, nous évaluons les performances de l'algorithme  \textsc{CRPR\_Miner} et comparons les coéts d'extraction des différentes représentations. Nous décrivons également le processus de classification basé sur les régles associatives de classification et appliqué dans le cadre de la détection d'intrusions.
\section{Environnement d'expérimentations}
L'ensemble des expérimentations  présentées dans ce chapitre ont été réalisées sur une machine $Acer$ munie d'un
processeur Intel Dual Core $E5400$
avec $2M$ de mémoire cache,
ayant une fréquence d'horloge de $2,7$
$GHz$ et $4$ $Go$ de mémoire vive tournant sur une plateforme Linux Ubuntu
$10.04$.
L'ensemble des algorithmes proposés dans le chapitre précédent ont été implantés en langage C++ et les programmes ont été compilés avec le compilateur gcc 4.3.3.
Nous commenéons dans la suite par la présentation des expérimentations faites sur des bases de test ``benchmark''.
%%%%%%%%%%%%%%%%%
\section{\'{E}valuation expérimentale pour des bases de test ``benchmark''}
%%%%%%%%%%%%%%%%%
\subsection{Description des bases de test}
Les expérimentations, que nous exposons dans cette section, ont été
menées sur les bases ``benchmark'' 
$^\textsc{(}$\footnote{Ces bases sont disponibles à  l'adresse
	suivante : \textsl{http://fimi.cs.helsinki.fi/data}.}$^\textsc{)}$
denses \textsc{Chess},  \textsc{Connect}, \textsc{Mushroom}, \textsc{Pumsb}, \textsc{Pumsb*}
\textsc{BMS-Web-View1} et sur les bases éparses  \textsc{Retail}, \textsc{Accidents},
\textsc{T10I4D100K} et \textsc{T40I10D100K}.
Tous les résultats expérimentaux sont repérés en détail dans les tableaux des annexes.
Les caractéristiques des ces différentes bases sont décrites dans le tableau \ref{Caract_bases_benchmark}.
Ce tableau décrit pour chaque base, le type, le nombre de transactions, le nombre d'items et la taille moyenne des transactions. Par ``Base dense'', nous entendons que les transactions de la base sont fortement corrélées, tandis que, par
``Base éparse'', nous entendons que les transactions de la base sont faiblement corrélées.

\font\xmplbx = cmbx8.5 scaled \magstephalf
\begin{table}[htbp]
	\begin{center}
		\footnotesize{\begin{tabular}{|l||r|r|r|r|} \hline
				\textbf{Base}& \textbf{Type de}&\textbf{Nombre} & \textbf{Nombre de}& \textbf{Taille moyenne}\\
				& \textbf{la base}&\textbf{d'items} &\textbf{transactions}& \textbf{des transactions}\\\hline
				\hline \textsc{Chess} &  \textbf{Dense}  &{\xmplbx75} & {\xmplbx3 196} & {\xmplbx37,00}\\
				\hline \textsc{Connect}& \textbf{Dense}  &{\xmplbx129} & {\xmplbx67 557} & {\xmplbx43,00}\\
				\hline \textsc{Mushroom}&  \textbf{Dense}  &{\xmplbx119} & {\xmplbx8 124} & {\xmplbx23,00}\\
				\hline \textsc{Pumsb}& \textbf{Dense}  &{\xmplbx7 117} & {\xmplbx49 046} & {\xmplbx74,00}\\
				\hline \textsc{Pumsb*}&   \textbf{Dense}  &{\xmplbx7 117} &{\xmplbx49 046} & {\xmplbx50,00}\\
				\hline \textsc{BMS-Web-View1}& \textbf{Dense} &{\xmplbx497} &{\xmplbx59 602} & {\xmplbx2,51}\\
				\hline \textsc{Retail}& \textbf{\'{E}parse} &{\xmplbx16 470} &{\xmplbx88 162} & {\xmplbx10,00}\\
				\hline \textsc{Accidents}& \textbf{\'{E}parse} &{\xmplbx468} &{\xmplbx340 183} & {\xmplbx33,81}\\
				\hline \textsc{T10I4D100K}& \textbf{\'{E}parse} &{\xmplbx870} & {\xmplbx100 000} &{\xmplbx10,10} \\
				\hline \textsc{T40I10D100K}& \textbf{\'{E}parse}&{\xmplbx942} &{\xmplbx100 000} &{\xmplbx39,61}\\\hline
		\end{tabular}}
		\caption{Caractéristiques des bases de test ``benchmark''.}
		\label{Caract_bases_benchmark}
	\end{center}
\end{table}

Nous enchainons dans la suite avec l'étude quantitative des cardinalités. Cette étude s'étalera sur deux principaux axes et ce suivant la nature de la base considérée. Ainsi, nous distinguons l'étude des cardinalités pour des bases denses et l'étude des cardinalités pour des bases éparses.

\subsection{\'Etude des cardinalités pour des bases denses}

\'Etudions d'abord la variation de la taille de l'ensemble $\mathcal{MCR}$ en fonction de la variation des seuils \textit{minbond} et \textit{minsupp} de corrélation et de fréquence respectifs.\\

\textbf{5.3.2.1 Analyse de la variation de la taille de l'ensemble $\mathcal{MCR}$}\\
%%THH : ATTENTION à  la numérotation manuelle

Lors des expérimentations réalisées, nous avons comptabilisé, d'une part,
la cardinalité de l'ensemble $\mathcal{MCR}$ de tous les motifs corrélés
rares et la cardinalité de l'ensemble $\mathcal{MCF}$ de tous les motifs corrélés
fréquents pour différentes valeurs des seuils \textit{minbond} et \textit{minsupp}. \\

\textbf{a. Effet de la variation du seuil \textit{minsupp}} \\

Nous étudions, dans un premier temps, la variation de la taille de l'ensemble $\mathcal{MCR}$ en fonction de la variation de l'ensemble $\mathcal{MCF}$.

Nous concluons d'aprés les résultats donnés par les tableaux \ref{Tab1} et \ref{Tab3}  de l'annexe \ref{appendix}, d'une part, que la taille de l'ensemble $\mathcal{MCR}$ varie dans le sens opposé que la taille de l'ensemble  $\mathcal{MCF}$ qui contient les motifs corrélés qui sont fréquents. Ces derniers correspondent aux motifs corrélés dont le support conjonctif dépasse le seuil minimal \textit{minsupp}.

Prenons le cas de la base \textsc{Mushroom} pour \textit{minbond} = \textit{0,15}. Pour le seuil \textit{minsupp} = \textit{5}$\%$, $|\mathcal{MCR}|$ = \textit{361}, $|\mathcal{MCF}|$ = \textit{100 906}. Cependant, pour \textit{minsupp} = \textit{20}$\%$, le nombre des motifs corrélés rares augmente, $|\mathcal{MCR}|$ = \textit{48 056} alors que le nombre de motifs corrélés fréquents diminue presque de 50$\%$, $|\mathcal{MCF}|$ = \textit{53 211}.

Pour la base \textsc{Chess} pour \textit{minbond} = \textit{0,60}. Pour le seuil \textit{minsupp} = \textit{20}$\%$, $|\mathcal{MCR}|$ = \textit{21}, $|\mathcal{MCF}|$ = \textit{255 023}. Cependant, pour \textit{minsupp} = \textit{70}$\%$, $|\mathcal{MCR}|$ = \textit{206 075} et $|\mathcal{MCF}|$ = \textit{48 969}.

Toutefois, nous remarquons que pour un seuil de corrélation \textit{minbond} fixe, la cardinalité $|\mathcal{MCF}|$ augmente en diminuant le seuil minimal \textit{minsupp} et diminue
en l'augmentant.

Considérons la base \textsc{Connect} pour \textit{minbond} = \textit{0,80}. Pour un seuil \textit{minsupp} faible, \textit{minsupp} \textit{20}$\%$, la taille  $|\mathcal{MCF}|$ est égale à  \textit{534 012}.
Cependant, en variant le seuil \textit{minsupp} de \textit{20}$\%$ à  \textit{86}$\%$, sa taille diminue significativement et passe é
$|\mathcal{MCF}|$ = \textit{105 031}.

%\begin{figure}[!t]
%\begin{center}
%\parbox{7cm}{\includegraphics[scale = 0.90]{MushMCR1.eps}}
%\parbox{7cm}{\includegraphics[scale = 0.90]{ChessMCR1.eps}}
%\parbox{7cm}{\includegraphics[scale = 0.90]{PbMCR1.eps}}
% \parbox{7cm}{\includegraphics[scale = 0.90]{ConnMCR1.eps}}
%  \parbox{7cm}{\includegraphics[scale = 0.90]{PbstarMCR1.eps}}
% \parbox{7cm}{\includegraphics[scale = 0.90]{BMVMCR1.eps}}
%\end{center}
%   \caption{Variation des cardinalités des ensembles $\mathcal{MCR}$ et $\mathcal{MCF}$ en fonction de
%\textit{minsupp} pour des bases denses.}\label{AnalyseMCR1}
%\end{figure}

Nous constatons aussi que, la taille de l'ensemble $\mathcal{MCR}$ varie selon la base. Par exemple considérons
les seuils \textit{minbond} = \textit{0,80}  et \textit{minsupp} = \textit{50}$\%$.
Pour la base \textsc{Connect}, nous avons
$|\mathcal{MCR}|$ = \textit{91}. Cependant, pour ces mêmes seuils de \textit{minbond} et de \textit{minsupp},
et pour la base \textsc{Pumsb}, la taille de l'ensemble $\mathcal{MCR}$ est beaucoup plus importante, nous notons $|\mathcal{MCR}|$ = \textit{3 521}.

Dans ce même sens, nous avons constaté aussi que pour un seuil de corrélation \textit{minbond} fixe, la taille de l'ensemble $\mathcal{MCR}$ est proportionnelle au seuil minimal de fréquence \textit{minsupp}.

Par exemple, pour le cas de la base \textsc{Mushroom} pour \textit{minbond} = \textit{0,15}. Pour le seuil \textit{minsupp} = \textit{5}$\%$, $|\mathcal{MCR}|$ = \textit{361}. Cependant, pour \textit{minsupp} = \textit{20}$\%$, $|\mathcal{MCR}|$ = \textit{48 056}.

De même pour la base \textsc{Chess} et pour \textit{minbond} = \textit{0,60}. Pour le seuil \textit{minsupp} = \textit{20}$\%$, $|\mathcal{MCR}|$ = \textit{21}. Cependant, pour \textit{minsupp} = \textit{70}$\%$ la taille de cet ensemble augmente considérablement et est égale é, $|\mathcal{MCR}|$ = \textit{206 075}.
Ces résultats expérimentaux obtenus renforcent la propriété théorique relative à  la taille de l'ensemble $\mathcal{MCR}$ donnée par la proposition \ref{TailleMCR}\textsc{(}\textit{cf.} page \pageref{TailleMCR}\textsc{)}.

Toutefois, le nombre de motifs corrélés rares augmente en augmentant le seuil \textit{minsupp} et diminue dans le cas opposé.
Il est aussi important de remarquer que l'augmentation considérable de la taille de l'ensemble $\mathcal{MCR}$ des motifs corrélés rares pour un seuil  \textit{minsupp} strictement supérieur au seuil \textit{minbond}.

Par exemple, pour le cas de la base \textsc{Mushroom} pour \textit{minbond} = \textit{0,15}. Pour le seuil \textit{minsupp} = \textit{15}$\%$ nous avons $|\mathcal{MCR}|$ = \textit{3 038}, cependant en augmentant légérement \textit{minsupp} de \textit{5}$\%$, nous aurons  \textit{minsupp} = \textit{20}$\%$, $|\mathcal{MCR}|$ = \textit{48 056}.
Cette constatation est valide aussi pour la base \textsc{Connect} pour \textit{minbond} = \textit{0,80}.
En effet, la taille de l'ensemble $\mathcal{MCR}$ varie de \textit{171} motifs pour \textit{minsupp} = \textit{80}$\%$ é
\textit{212 291} motifs pour \textit{minsupp} = \textit{82}$\%$.

Toutefois pour les seuils \textit{minbond} $<$ \textit{minsupp}, les motifs corrélés rares correspondent aux motifs corrélés dont la valeur de corrélation ne dépasse pas le seuil \textit{minsupp}. D'une maniére formelle, \textit{minbond} $\leq$ \textit{bond} \textsc{(}$X$\textsc{)} $<$ \textit{minsupp}
\textsc{(}\textit{cf.} proposition \ref{prop_algo_MCR3} de la page \pageref{prop_algo_MCR3} \textsc{)}.

Cette variation pertinente de la taille de l'ensemble $\mathcal{MCR}$, s'explique par le fait que la valeur de la corrélation d'une grande tranche de motifs corrélés ne dépasse pas le seuil \textit{minsupp}.

Par exemple, considérons la base \textit{Pumsb} pour \textit{minbond} = \textit{0,80}. Pour \textit{minsupp} = \textit{85}$\%$, $|\mathcal{MCR}|$ =  \textit{125 420}.
Autrement dit, nous avons \textit{125 420} motifs corrélés dont la valeur de la mesure \textit{bond} est supérieure ou égale à  \textit{0,80} mais ne dépasse pas \textit{0,85}.
Pour \textit{minsupp} = \textit{90}$\%$, $|\mathcal{MCR}|$ =  \textit{143 345}.
Ainsi, nous avons \textit{143 345} motifs corrélés rares, dont la corrélation entre \textit{80}$\%$ et \textit{90}$\%$. \`A cet égard, nous déduisons que seuls \textit{17 925} \textsc{(}correspondant à  \textit{143 345} - \textit{125 420}\textsc{)} de motifs présentent un degré de corrélation important compris entre \textit{85}$\%$ et \textit{90}$\%$.

Il est cependant clair que, pour la base \textsc{Pumsb*}, la taille de l'ensemble $\mathcal{MCR}$ ne présente pas une transition aigué lors de la variation du seuil \textit{minsupp} qu'on a constaté précédemment. En effet, pour \textit{minbond} = \textit{0,50}, pour \textit{minsupp} = \textit{50}$\%$, nous avons $|\mathcal{MCR}|$ =  \textit{91 546} et en augmentant
\textit{minsupp} de \textit{5}$\%$, le nombre de motifs corrélés rares est de \textit{91 868}. Ainsi, cette augmentation de \textit{322} itemsets uniquement est relativement négligeable par rapport à  la variation aigué que présentent les autres bases \textsc{Pumsb}, \textsc{Chess}, \textsc{Connect} et \textsc{Mushroom}.

Quant à  la base \textsc{BMS-Web-View1}, elle présente un comportement insensible quant à  la variation du seuil
\textit{minsupp}. En effet, pour un seuil \textit{minbond} fixé à  \textit{0,90}, nous remarquons que plus de \textit{99}$\%$ des motifs corrélés sont rares et leurs supports conjonctifs sont compris entre  \textit{1}$\%$ et \textit{8}$\%$. En effet, pour \textit{minsupp} = \textit{1}$\%$, nous avons \textit{60 002} motifs corrélés rares et uniquement \textit{68}  motifs corrélés fréquents et pour \textit{minsupp} = \textit{8}$\%$, tous les motifs corrélés sont rares et aucun motif corrélé n'est fréquent, $|  \mathcal{MCR}|$ = \textit{60 070} et $|\mathcal{MCF}|$ = \textit{0}.\\

\textbf{b. Effet de la variation du seuil \textit{minbond}} \\	

Nous étudions, à  ce stade pour un seuil \textit{minsupp} fixe, l'effet de la variation du seuil \textit{minbond} sur la variation de la taille des ensembles $\mathcal{MCR}$ et $\mathcal{MCF}$.
Nous nous basons sur les résultats donnés par le tableau \ref{Tab5} de l'annexe \ref{appendix}.

Il nous est clair que la variation de la taille de l'ensemble $\mathcal{MCR}$ est disproportionnelle à  la variation du seuil \textit{minbond}.
Toutefois, le nombre de motifs corrélés rares augmente en diminuant le seuil \textit{minbond} et diminue en augmentant \textit{minbond}.
Considérons la base \textsc{Mushroom} pour \textit{minsupp} = \textit{30}$\%$.  Pour un seuil \textit{minbond} = \textit{0,25}, nous avons $|\mathcal{MCR}|$ = \textit{4 120}
et pour \textit{minbond} = \textit{0,30}, nous avons $|\mathcal{MCR}|$ = \textit{674}.

Nous remarquons aussi que, la taille de l'ensemble  $\mathcal{MCF}$ est aussi disproportionnelle à  la variation du seuil \textit{minbond}.
Nous avons par exemple, pour la base \textsc{Pumsb} pour \textit{minsupp} = \textit{80}$\%$. Le nombre de motifs corrélés fréquents est de \textit{142 156} pour un seuil \textit{minbond} = \textit{0,80}. Ce nombre diminue largement et atteint \textit{20 550} pour une légére augmentation de \textit{5}$\%$ du seuil
\textit{minbond}.

Nous déduisons donc que la taille des ensembles $\mathcal{MCR}$ et $\mathcal{MCF}$ varient dans le même sens pour un seuil \textit{minsupp} fixe. Ceci s'explique par le fait que ces deux ensembles doivent vérifier la contrainte de corrélation minimale et présentent donc des comportements similaires lors de la variation du seuil minimal \textit{minbond}.

%\begin{figure}[!t]
%\begin{center}
%\parbox{7cm}{\includegraphics[scale = 0.90]{MushMCR2.eps}}
% \parbox{7cm}{\includegraphics[scale = 0.90]{PbMCR2.eps}}
%\parbox{7cm}{\includegraphics[scale = 0.90]{ConnMCR2.eps}}
%  \parbox{7cm}{\includegraphics[scale = 0.90]{PbstarMCR2.eps}}
% \parbox{7cm}{\includegraphics[scale = 0.90]{BmvMCR2.eps}}
%\end{center}
% \caption{Variation des cardinalités des ensembles $\mathcal{MCR}$ et $\mathcal{MCF}$ en fonction de
%\textit{minbond} pour des bases denses.}\label{AnalyseMCR2}
%\end{figure}

Cependant, nous remarquons pour la base \textsc{BMS-Web-View1}, que tous les motifs corrélés sont rares.
Pour un seuil \textit{minsupp} fixé à  \textit{10}$\%$, nous avons pour \textit{minbond} = \textit{0,10}, $|$$\mathcal{MCR}$$|$ = \textit{60 161} et
$|$$\mathcal{MCF}$$|$ = \textit{0}. Une augmentation considérable du seuil \textit{minbond} de \textit{0,10} à  \textit{1}, engendre une diminution négligeable de la taille de l'ensemble $\mathcal{MCR}$.
En effet, pour \textit{minbond} = \textit{1}, $|$$\mathcal{MCR}$$|$ = \textit{60 070} et le nombre de motifs corrélés fréquents est toujours nul. Nous déduisons que, cette base présente une forte insensibilité quant à  la variation du seuil \textit{minbond}.

Nous avons ainsi discuté la variation des cardinalités des ensembles $\mathcal{MCR}$ et $\mathcal{MCF}$ en fonction de la variation des seuils \textit{minbond} et \textit{minsupp}. De plus, nous avons étudié, pour différentes bases, le lien de dépendance entre les tailles de ces deux ensembles.
Dans ce qui suit, nous abordons l'analyse de la cardinalité de la représentation concise exacte $\mathcal{RMCR}$.\\

\textbf{5.3.2.2 Analyse de la variation de la taille de la représentation $\mathcal{RMCR}$}\\

Nous nous focalisons à  présent sur l'étude de la cardinalité de la représentation concise exacte  $\mathcal{RMCR}$.  Nous discutons aussi le taux de compacité offert par cette derniére, que nous désignons par  Tx-$\mathcal{RMCR}$. Ce taux est donné en pourcentage et correspond à  la formule suivante : Tx-$\mathcal{RMCR}$ = 1 - $\displaystyle\frac{\displaystyle |\mathcal{RMCR}|}{\displaystyle
	|\mathcal{MCR}|}$.

En effet, nous avons comptabilisé la taille de la représentation concise $|\mathcal{RMCR}|$.
Nous avons tout de même spécifié d'une maniére séparée, afin que l'analyse soit minutieusement établie, la taille des ensembles $\mathcal{MMCR}$ des motifs minimaux corrélés rares et
$\mathcal{MFCR}$ des motifs fermés corrélés rares. Ces cardinalités sont données par les tableaux \ref{Tab2}, \ref{Tab4} et \ref{Tab6} de l'annexe \ref{appendix}.

%\begin{figure}[!t]
%\begin{center}
% \parbox{7cm}{\includegraphics[scale = 0.90]{MushRep1.eps}}
%  \parbox{7cm}{\includegraphics[scale = 0.90]{chessRep1.eps}}
% \parbox{7cm}{\includegraphics[scale = 0.90]{connRep1.eps}}
%\parbox{7cm}{\includegraphics[scale = 0.90]{pbRep1.eps}}
%\parbox{7cm}{\includegraphics[scale = 0.90]{pbstarRep1.eps}}
%\end{center}
%  \caption{Variation des cardinalités en fonction de
%\textit{minsupp} pour des bases denses.}\label{AnalysefctMinsuppBDD}
%\end{figure}

Ces données expérimentaux confirment que la représentation concise exacte $\mathcal{RMCR}$ est une couverture parfaite de l'ensemble
$\mathcal{MCR}$ des motifs corrélés rares. En effet, nous pouvons constater que, pour
toutes les bases et pour tous les seuils minimaux \textit{minsupp} et \textit{minbond}, la taille de la représentation $\mathcal{RMCR}$ ne dépasse jamais celle de l'ensemble $\mathcal{MCR}$.
Par exemple, pour la base \textsc{Mushroom} pour
\textit{minsupp} = \textit{35}$\%$ et \textit{minbond} =
\textit{0,15}, nous avons $|\mathcal{RMCR}|$ = \textit{1 810} $<$
$|\mathcal{MCR}|$ = \textit{100 156}. La taille de la représentation est ainsi réduite par rapport à  la taille de l'ensemble $\mathcal{MCR}$ et le taux de réduction est ainsi trés important,
Tx-$\mathcal{RMCR}$ = \textit{98}$\%$. Ceci s'explique par la nature des classes d'équivalence induites dans ce cas. En effet, nous avons
$|\mathcal{MMCR}|$ = 1 412 et  $|\mathcal{MFCR}|$ = 652. Comme la représentation $\mathcal{RMCR}$ correspond à  l'union sans redondance des ensembles $\mathcal{MMCR}$ et $\mathcal{MFCR}$, nous avons donc toujours $|\mathcal{RMCR}|$ $\leq$ $|\mathcal{MMCR}|$ + $|\mathcal{MFCR}|$.

Cependant, nous constatons d'aprés les résultats obtenus que le taux de réduction de
la représentation $\mathcal{RMCR}$ différe selon la base considérée.
Par exemple, pour la base \textsc{Connect} et pour \textit{minsupp} = \textit{80}$\%$ et \textit{minbond} = \textit{0,80}, le taux de réduction est moyen, Tx-$\mathcal{RMCR}$ = \textit{35}$\%$. Cependant, pour la base \textsc{Chess},
pour un seuil \textit{minsupp} de \textit{70}$\%$ et un seuil \textit{minbond} de
\textit{0,6}, nous avons  $|\mathcal{RMCR}|$ = \textit{862} et
$|\mathcal{MCR}|$ = \textit{206 075}. Ainsi le taux de réduction est important, Tx-$\mathcal{RMCR}$ = \textit{99}$\%$.

%\begin{figure}[!t]
%\begin{center}
%  \parbox{7cm}{\includegraphics[scale = 0.90]{MushTx1.eps}}
%    \parbox{7cm}{\includegraphics[scale = 0.90]{ConnTx1.eps}}
%   \parbox{7cm}{\includegraphics[scale = 0.90]{pbTx1.eps}}
%   \parbox{7cm}{\includegraphics[scale = 0.90]{pbstarTx1.eps}}
%\end{center}
%   \caption{Variation des taux de réduction en fonction de
%\textit{minsupp} pour des bases denses.}\label{tauxminsuppBDD}
%\end{figure}

Nous remarquons que pour la base \textsc{BMS-Web-View1}, la taille de la représentation $\mathcal{RMCR}$ est relativement élevée et égale à  la taille de l'ensemble $\mathcal{MCR}$. En effet, nous avons pour \textit{minsupp} = \textit{10}$\%$ et \textit{minbond} = \textit{0,90} : $|\mathcal{RMCR}|$ = \textit{60 070} $=$ $|\mathcal{MCR}|$ = \textit{60 070}.  Ceci est dé à  la nature des classes d'équivalence induites par l'opérateur $f_{bond}$. En effet, tous les motifs minimaux sont aussi des fermés dans leurs classes d'équivalence. Par exemple, pour ces mêmes seuils de \textit{minsupp} et \textit{minbond}, nous avons
$|\mathcal{MMCR}|$ = $|\mathcal{MFCR}|$ = \textit{60 070} et même en variant le seuil \textit{minbond}, cela est toujours valide. Nous avons pour \textit{minsupp} = \textit{10}$\%$ et \textit{minbond} = \textit{0,30}, $|\mathcal{MMCR}|$ = $|\mathcal{MFCR}|$ = \textit{60 074}.
Ceci implique que le taux Tx-$\mathcal{RMCR}$ offert pour cette base est nul.

Il est a constater aussi que pour toutes les bases et pour tous les seuils de \textit{minsupp} et de
\textit{minbond}, le nombre de motifs fermés corrélés rares  noté $|\mathcal{MFCR}|$ ne dépasse jamais le nombre des motifs minimaux corrélés rares
noté $|\mathcal{MMCR}|$.
Considérons, par exemple la base  \textsc{Pumsb} pour \textit{minsupp} = \textit{30}$\%$ et \textit{minbond} =
\textit{0,80}. Nous avons $|\mathcal{MFCR}|$ = \textit{451} $<$
$|\mathcal{MMCR}|$ = \textit{2 090}. Il en est de même pour
la base  \textsc{Pumsb*},
pour \textit{minsupp} = \textit{20}$\%$ et \textit{minbond} =
\textit{0,50} : $|\mathcal{MFCR}|$ = \textit{882} $<$
$|\mathcal{MMCR}|$ = \textit{2 155}.
Cependant, pour la base  \textsc{Connect},
pour \textit{minsupp} = \textit{20}$\%$ et \textit{minbond} =
\textit{0,80} : $|\mathcal{MFCR}|$  $=$
$|\mathcal{MMCR}|$ = \textit{70}.

Toutefois, ceci s'explique par la nature de l'opérateur de fermeture $f_{bond}$ qui permet de regrouper, dans des classes d'équivalence, les motifs ayant les mêmes caractéristiques. En effet, chaque classe d'équivalence est représentée par un seul motif fermé corrélé et un ou plusieurs minimaux corrélés rares.

Nous avons ainsi présenté et analysé quantitativement la représentation concise exacte $\mathcal{RMCR}$.
Nous enchaénons par la suite avec l'analyse et la comparaison des autres représentations concises exactes proposées.\\

\textbf{5.3.2.3 Analyse de la variation des tailles des représentations concises exactes
	$\mathcal{RMM}$$ax$$\mathcal{F}$ et $\mathcal{RM}$$in$$\mathcal{MF}$}\\

Nous nous focalisons à  présent sur l'étude des cardinalités des représentations concises exactes $\mathcal{RMM}$$ax$$\mathcal{F}$ et $\mathcal{RM}$$in$$\mathcal{MF}$.
Nous discutons aussi les taux de compacité offerts par ces derniéres.
Nous désignons par Tx-$\mathcal{RMM}$$ax$$\mathcal{F}$ le taux de réduction de la représentation concise exacte $\mathcal{RMM}$$ax$$\mathcal{F}$ et par
Tx-$\mathcal{RM}$$in$$\mathcal{MF}$ le taux de réduction de la représentation
concise exacte $\mathcal{RM}$$in$$\mathcal{MF}$. Ces taux sont donnés en pourcentage et correspondent é,
Tx-$\mathcal{RMM}$$ax$$\mathcal{F}$ = 1 - $\displaystyle\frac{\displaystyle |\mathcal{RMM}ax\mathcal{F}|}{\displaystyle
	|\mathcal{MCR}|}$ et
Tx-$\mathcal{RM}$$in$$\mathcal{MF}$ = 1 -
$\displaystyle\frac{\displaystyle |\mathcal{RM}in\mathcal{MF}|}{\displaystyle
	|\mathcal{MCR}|}$.

Toutefois, nous avons comptabilisé les cardinalités de ces représentations. Nous avons aussi spécifié la taille
$|\mathcal{MMCRM}in|$ de l'ensemble des éléments minimaux de l'ensemble $\mathcal{MMCR}$ et la taille
$|\mathcal{MFCRM}ax|$ de l'ensemble des motifs fermés corrélés rares maximaux. Ces cardinalités sont données
par les tableaux
\ref{Tab2}, \ref{Tab4} et \ref{Tab6} de l'annexe \ref{appendix}.
Il est à  remarquer que les tailles des représentations concises exactes
$\mathcal{RMM}$$ax$$\mathcal{F}$ et $\mathcal{RM}$$in$$\mathcal{MF}$
ne dépassent jamais la taille de la représentation concise exacte $\mathcal{RMCR}$.

Par exemple, pour la base  \textsc{Mushroom},
pour \textit{minsupp} = \textit{35}$\%$ et \textit{minbond} =
\textit{0,15}, nous avons  $|\mathcal{RMCR}|$ = \textit{1 810} $>$
$|$$\mathcal{RMM}$$ax$$\mathcal{F}$$|$ = \textit{1 421} $>$
$|$$\mathcal{RM}$$in$$\mathcal{MF}$$|$ = \textit{667}.   
Il en est de même pour la base \textsc{Pumsb}. En effet, pour \textit{minsupp} = \textit{40}$\%$ et \textit{minbond} = \textit{0,80} : $|\mathcal{RMCR}|$ = \textit{2 168} $>$
$|$$\mathcal{RMM}$$ax$$\mathcal{F}$$|$ = \textit{2 136}  $>$  $|$$\mathcal{RM}$$in$$\mathcal{MF}$$|$ = \textit{2 108}.

Ces résultats sont justifiés par les définitions mêmes des
représentations concises exactes $\mathcal{RMM}$$ax$$\mathcal{F}$ et $\mathcal{RM}$$in$$\mathcal{MF}$.
En effet, la représentation $|$$\mathcal{RMM}$$ax$$\mathcal{F}|$ basée sur l'ensemble
$\mathcal{MMCR}$ des minimaux corrélés rares et sur l'ensemble $\mathcal{MFCRM}ax$ des motifs fermés corrélés rares maximaux.

Toutefois, nous remarquons que la taille de l'ensemble
$\mathcal{MFCRM}ax$  ne dépasse jamais la taille de l'ensemble  $\mathcal{MFCR}$.
Par exemple, pour la base  \textsc{Pumsb} pour
\textit{minsupp} = \textit{30}$\%$ et pour \textit{minbond} =
\textit{0,80}, nous avons $|\mathcal{MFCRM}ax|$ = \textit{112} $<$
$|\mathcal{MFCR}|$ = \textit{451}.
Par conséquent, nous avons toujours $|$$\mathcal{RMM}$$ax$$\mathcal{F}|$ $\leq$ $|\mathcal{RMCR}|$.

%\begin{figure}[!t]
%\begin{center}
%  \parbox{7cm}{\includegraphics[scale = 0.90]{MushRep2.eps}}
%   \parbox{7cm}{\includegraphics[scale = 0.90]{pbRep2.eps}}
%   \parbox{7cm}{\includegraphics[scale = 0.90]{pbstarRep2.eps}}
%   \parbox{7cm}{\includegraphics[scale = 0.90]{bmv1Rep2.eps}}
%\end{center}
%  \caption{Variation des cardinalités en fonction de \textit{minbond} pour des bases denses.}\label{AnalysefctMinbondBDD}
%\end{figure}

Concernant la représentation concise $\mathcal{RM}$$in$$\mathcal{MF}$, elle est basée sur l'ensemble  $\mathcal{MMCRM}in$ des éléments minimaux de l'ensemble $\mathcal{MMCR}$ et sur l'ensemble
$\mathcal{MFCR}$ des motifs fermés corrélés rares.
En effet, nous remarquons que la taille de l'ensemble
$\mathcal{MMCRM}in$  ne dépasse jamais la taille de l'ensemble  $|\mathcal{MMCR}|$.
Par exemple, pour la base  \textsc{Mushroom} pour
\textit{minsupp} = \textit{40}$\%$ et pour \textit{minbond} =
\textit{0,15}, nous avons : $|\mathcal{MMCRM}in|$ = \textit{98} $<$
$|\mathcal{MMCR}|$ = \textit{1 491}.
Pour la base \textsc{Chess},
\textit{minsupp} = \textit{20}$\%$ et \textit{minbond} =
\textit{0,60} : $|\mathcal{MMCRM}in|$  $=$
$|\mathcal{MMCR}|$ = \textit{21}.
\`A cet égard, nous déduisons que dans ce cas, la taille de la représentation $\mathcal{RM}$$in$$\mathcal{MF}$ est égale à   la taille de la représentation $\mathcal{RMCR}$, $|\mathcal{RM}$$in$$\mathcal{MF}|$  $=$ $|\mathcal{RMCR}|$.

Toutefois, nous déduisons, d'après les résultats obtenus, que les représentations $\mathcal{RM}$$in$$\mathcal{MF}$ et $\mathcal{RMM}$$ax$$\mathcal{F}$ offrent d'une manière générale des taux de réduction plus intéressants que la représentation $\mathcal{RMCR}$.

Par exemple, la base \textsc{Mushroom} présente des taux de réduction intéressants.
Pour  \textit{minsupp} = \textit{10}$\%$ et \textit{minbond} =
\textit{0,15} :
Tx-$\mathcal{RMCR}$ = \textit{29}$\%$,
Tx-$\mathcal{RMM}$$ax$$\mathcal{F}$ = \textit{48}$\%$,
Tx-$\mathcal{RM}$$in$$\mathcal{MF}$ = \textit{56}$\%$.

Ces taux peuvent étre trés proches et des fois égaux, comme le cas de la base \textsc{Pumsb*} pour \textit{minsupp} = \textit{40}$\%$ et \textit{minbond} = \textit{0,65}, nous avons Tx-$\mathcal{RMCR}$ = \textit{84}$\%$, Tx-$\mathcal{RMM}$$ax$$\mathcal{F}$ = \textit{85}$\%$ et
Tx-$\mathcal{RM}$$in$$\mathcal{MF}$ = \textit{86}$\%$.

Il est important de signaler que nous ne pouvons pas affirmer laquelle de ces deux représentations est plus réduite.
Toutefois, la représentation la plus concise varie en fonction de la base et des seuils minimaux de corrélation et de fréquence posés.
D'ailleurs, pour les exemples considérés précédemment, nous remarquons que la représentation
$\mathcal{RM}$$in$$\mathcal{MF}$ est la plus concise. Cependant,
dans d'autres cas c'est la représentation  $\mathcal{RMM}$$ax$$\mathcal{F}$ qui offre le plus
de réduction.
Considérons la
base \textsc{Pumsb} pour \textit{minbond} =
\textit{0,80}. Nous avons pour \textit{minsupp} = \textit{10}$\%$,
$|\mathcal{RMM}$$ax$$\mathcal{F}|$ est inférieure é
$\mathcal{RM}$$in$$\mathcal{MF}$, $|\mathcal{RMM}$$ax$$\mathcal{F}|$ = \textit{2 041} alors que
$|\mathcal{RM}$$in$$\mathcal{MF}|$ = \textit{2 045}.

Nous avons ainsi analysé les cardinalités des représentations $\mathcal{RM}$$in$$\mathcal{MF}$ et $\mathcal{RMM}$$ax$$\mathcal{F}$ par rapport à  la représentation $\mathcal{RMCR}$. Dans la suite nous nous focalisons sur l'étude de la représentation $\mathcal{RM}$$in$$\mathcal{MM}$$ax$$\mathcal{F}$.\\

\textbf{5.3.2.4 Analyse de la cardinalité de la représentation approximative
	$\mathcal{RM}$$in$$\mathcal{MM}$$ax$$\mathcal{F}$ }\\

Nous constatons d'aprés les résultats expérimentaux présentés par
les tableaux \ref{Tab2}, \ref{Tab4} et \ref{Tab6} de l'annexe \ref{appendix}, que pour les différentes bases denses testées et pour tous les seuils de
\textit{minsupp} et de \textit{minbond}, la taille de la représentation
concise approximative $\mathcal{RM}$$in$$\mathcal{MM}$$ax$$\mathcal{F}$
ne dépasse pas la taille de la représentation concise exacte $\mathcal{RMM}$$ax$$\mathcal{F}$ ou celle de la représentation concise exacte $\mathcal{RM}$$in$$\mathcal{MF}$.

Par exemple, pour la base \textsc{Mushroom}, \textit{minsupp} = \textit{35}$\%$ et \textit{minbond} =
\textit{0,15}, nous avons $|$$\mathcal{RM}$$in$$\mathcal{MM}$$ax$$\mathcal{F}$$|$ = \textit{104} $<$
$|$$\mathcal{RM}$$in$$\mathcal{MF}$$|$ = \textit{667} $<$
$|$$\mathcal{RMM}$$ax$$\mathcal{F}$$|$ = \textit{1 421}.
Pour la base  \textsc{Connect},
\textit{minsupp} = \textit{86}$\%$ et \textit{minbond} =
\textit{0,8} : $|$$\mathcal{RM}$$in$$\mathcal{MM}$$ax$$\mathcal{F}$$|$ = \textit{111} $<$
$|$$\mathcal{RM}$$in$$\mathcal{MF}$$|$ = \textit{263} $<$
$|$$\mathcal{RMM}$$ax$$\mathcal{F}$$|$ = \textit{395}.\\

Ces résultats sont justifiés par le fait que la représentation concise approximative correspond à  l'union sans redondance des ensembles $\mathcal{MMCRM}in$ et $\mathcal{MFCRM}ax$. Ce qui implique que
$|$$\mathcal{RM}$$in$$\mathcal{MM}$$ax$$\mathcal{F}$$|$ $\leq$ $|\mathcal{MMCRM}in|$ + $|\mathcal{MFCRM}ax|$.
Or $|\mathcal{MMCRM}in|$ + $|\mathcal{MFCRM}ax|$ $\leq$  $|\mathcal{MMCR}|$ + $|\mathcal{MFCRM}ax|$
donc $\mathcal{RM}$$in$$\mathcal{MM}$$ax$$\mathcal{F}$ ne dépasse jamais $\mathcal{RMM}$$ax$$\mathcal{F}$.

De plus, $|\mathcal{MMCRM}in|$ + $|\mathcal{MFCRM}ax|$ $\leq$ $|\mathcal{RM}$$in$$\mathcal{MF}|$, donc
$|$$\mathcal{RM}$$in$$\mathcal{MM}$$ax$$\mathcal{F}$$|$ $\leq$ $|\mathcal{RM}$$in$$\mathcal{MF}|$.
Nous déduisons ainsi que la représentation concise approximative est la plus concise.

Désignons par Tx-$\mathcal{RM}$$in$$\mathcal{MM}$$ax$$\mathcal{F}$ le taux de réduction de la représentation $\mathcal{RM}$$in$$\mathcal{MM}$$ax$$\mathcal{F}$. Ce taux est donné en pourcentage et correspond é, Tx-$\mathcal{RM}$$in$$\mathcal{MM}$$ax$$\mathcal{F}$ = 1 -
$\displaystyle\frac{\displaystyle |\mathcal{RM}in\mathcal{MM}ax\mathcal{F}|}{\displaystyle
	|\mathcal{MCR}|}$.

Nous constatons d'après les résultats obtenus que, ce taux diffère d'une base à  une autre. Par exemple, la base \textsc{Mushroom} présente des taux de réduction intéressants.
Pour  \textit{minsupp} = \textit{5}$\%$ et \textit{minbond} =
\textit{0,15} :
Tx-$\mathcal{RM}$$in$$\mathcal{MM}$$ax$$\mathcal{F}$ = \textit{78}$\%$. La base
\textsc{Pumsb} présente des taux de réduction aussi importants.
Pour  \textit{minsupp} = \textit{60}$\%$ et \textit{minbond} =
\textit{0,80} :
Tx-$\mathcal{RM}$$in$$\mathcal{MM}$$ax$$\mathcal{F}$ = \textit{43}$\%$.
Cependant, la base
\textsc{Connect} présente un taux de réduction nul pour
\textit{minsupp} = \textit{50}$\%$ et \textit{minbond} =
\textit{0,80}.

%%\begin{figure}[!t]
%%\begin{center}
%% \parbox{7cm}{\includegraphics[scale = 0.90]{MushTx2.eps}}
%%\parbox{7cm}{\includegraphics[scale = 0.90]{pbTx2.eps}}
%%\parbox{7cm}{\includegraphics[scale = 0.90]{pbstarTx2.eps}}
%%\end{center}
%%  \caption{Variation des taux de réduction en fonction de
%%\textit{minbond} pour des bases denses.}\label{tauxminbondBDD}
%%\end{figure}

Nous avons ainsi étudié les cardinalités des représentations proposées pour différentes bases denses. Dans ce qui suit, nous nous focalisons sur l'étude des représentations proposées pour des bases éparses.

\subsection{\'Etude des cardinalités pour des bases éparses}

Dans cette section, nous analysons et comparons les cardinalités de l'ensemble
$\mathcal{MCR}$ et des différentes représentations concises proposées pour les bases éparses \textsc{T10I4D100K}, \textsc{T40I10D100K}, \textsc{Accidents} et \textsc{Retail}. Nous soulignons que  
nous nous limitons à  une récapitulation de l'analyse de la variation des différentes représentations vu que cette variation est similaire à  celle pour des bases denses.\\

\textbf{5.3.3.1 Effet de la variation du seuil \textit{minsupp}} \\

En effet, d'aprés les résultats expérimentaux présentés par
le tableau \ref{TabXpT101} de l'annexe \ref{appendix}, nous constatons que la taille de l'ensemble $\mathcal{MCR}$ pour les bases éparses est relativement réduite par comparaison à  sa taille pour les bases denses.
Toutefois, la taille  $|\mathcal{MCR}|$ ne dépasse pas \textit{2 703} pour \textsc{T10I4D100K} et
\textit{4 703} pour la base \textsc{T40I10D100K}.
Ces valeurs réduites sont dues à  la nature éparse des bases \textsc{T40I10D100K} et \textsc{T10I4D100K}.
Cependant, la taille de l'ensemble $\mathcal{MCR}$ est plus importante pour les bases  \textsc{Accidents}
et \textsc{Retail}. En effet, cet ensemble englobe \textit{29 834} motifs pour la base \textsc{Retail}
pour un seuil \textit{minsupp} de \textit{20}$\%$ et pour un seuil \textit{minbond} faible de \textit{0,10}.
Pour la base \textsc{Accidents}, le nombre de motifs corrélés rares est plus élevé. Nous avons pour \textit{minsupp} = \textit{50}$\%$ et \textit{minbond} = \textit{0,30}, $|\mathcal{MCR}|$ = \textit{142 275}.

%%\begin{figure}[!t]
%%\begin{center}
%%  \parbox{7cm}{\includegraphics[scale = 0.90]{T10Rep1.eps}}
%% \parbox{7cm}{\includegraphics[scale = 0.90]{T40Rep1.eps}}
%%  \parbox{7cm}{\includegraphics[scale = 0.90]{AccRep1.eps}}
%%  \parbox{7cm}{\includegraphics[scale = 0.90]{RetRep1.eps}}
%%\end{center}
%% \caption{Variation des cardinalités en fonction de
%%\textit{minsupp} pour des bases éparses.}\label{AnalysefctMinsuppBDE}
%%\end{figure}

Nous constatons aussi que pour un seuil \textit{minbond} fixe, la taille de l'ensemble $\mathcal{MCR}$ croit en augmentant \textit{minsupp}. Par exemple, pour la base \textsc{T40I10D100K}, pour \textit{minsupp} = \textit{5}$\%$ et \textit{minbond} \textit{0,10}, nous avons $|\mathcal{MCR}|$ = \textit{4 387} et pour
\textit{minsupp} = \textit{30}$\%$  et \textit{minbond} =
\textit{0,10}, nous avons $|\mathcal{MCR}|$ = \textit{4 703}.

Il est à  remarquer aussi, que la taille de l'ensemble $\mathcal{MCR}$ varie légèrement pour une variation importante du seuil \textit{minsupp} pour les bases \textsc{T40I10D100K}, \textsc{T10I4D100K} et \textsc{Retail}. Cependant la base \textsc{Accidents}, présente un comportement différent.
En effet, elle assure une meilleure sensibilité quant au seuil \textit{minsupp}.
Par exemple, pour \textit{minbond} = \textit{0,30} nous avons pour
\textit{minsupp} = \textit{40}$\%$, $|\mathcal{MCR}|$ = \textit{117 805}
et pour \textit{minsupp} = \textit{50}$\%$ nous avons  $|\mathcal{MCR}|$ = \textit{142 275}.

Analysons à  présent les cardinalités des différentes représentations proposées en se basant sur les résultats donnés par le tableau \ref{TabXpT102} de l'annexe \ref{appendix}.

%%\begin{figure}[!t]
%%\begin{center}
%% \parbox{7cm}{\includegraphics[scale = 0.90]{T10Tx1.eps}}
%%  \parbox{7cm}{\includegraphics[scale = 0.90]{T40Tx1.eps}}
%%  \parbox{7cm}{\includegraphics[scale = 0.90]{AccTx1.eps}}
%%  \parbox{7cm}{\includegraphics[scale = 0.90]{RetTx1.eps}}
%%\caption{Variation des taux de réduction en fonction de
%%\textit{minsupp} pour des bases éparses.}\label{tauxminsuppBDE}
%%\end{center}
%%\end{figure}

Nous constatons que les cardinalités des différentes représentations sont presque stables pour chacune des trois bases \textsc{T10I4D100K}, \textsc{T40I10D100K} et \textsc{Retail}.
Par conséquent, les taux de réduction  de ces représentations ne présentent pas une variation marquante.
Prenons l'exemple de la base \textsc{Retail} pour un seuil \textit{minbond} fixe à  \textit{0,15}.
Nous avons pour \textit{minsupp} = \textit{5}$\%$,
$|\mathcal{RMCR}|$ = \textit{19 791} et $|\mathcal{RMM}$$ax$$\mathcal{F}|$ = \textit{18 898}.
Ainsi nous avons les taux suivants,  Tx-$\mathcal{RMCR}$ = \textit{13}$\%$ et
Tx-$\mathcal{RMM}$$ax$$\mathcal{F}$ = 9$\%$.
En augmentant le seuil \textit{minsupp} 0 \textit{50}$\%$, nous aurons
$|\mathcal{RMCR}|$ = \textit{19 802} et $|\mathcal{RMM}$$ax$$\mathcal{F}|$ = \textit{18 909}.
Ainsi, les taux demeurent toujours constants, Tx-$\mathcal{RMCR}$ = \textit{13}$\%$ et
Tx-$\mathcal{RMM}$$ax$$\mathcal{F}$ = 9$\%$.

Il est aussi remarquable que les cardinalités des différentes représentations sont relativement plus variable pour la base \textsc{Accidents}. En effet, une petite augmentation du seuil \textit{minsupp} engendre une augmentation considérable des cardinalités des différentes bases.
Par exemple, pour un seuil \textit{minbond} fixe à  \textit{0,30}.
Nous avons pour \textit{minsupp} = \textit{30}$\%$,
$|\mathcal{RMCR}|$ = \textit{686} et $|\mathcal{RM}$$in$$\mathcal{MF}|$ = \textit{614}.
Ainsi nous avons les taux suivants,  Tx-$\mathcal{RMCR}$ = \textit{12}$\%$ et
Tx-$\mathcal{RM}$$in$$\mathcal{MF}$ = 22$\%$.
En augmentant le seuil \textit{minsupp} de \textit{10}$\%$, nous aurons
$|\mathcal{RMCR}|$ = \textit{1 722} et $|\mathcal{RM}$$in$$\mathcal{MF}|$  = \textit{754}.
Ainsi les taux augmentent, Tx-$\mathcal{RMCR}$ = \textit{98}$\%$ et
Tx-$|\mathcal{RM}$$in$$\mathcal{MF}|$  = 99$\%$.\\

\textbf{5.3.3.2 Effet de la variation du seuil \textit{minbond}}\\

Nous nous basons dans cette partie sur les résultats expérimentaux donnés par les tableaux \ref{TabXpT104} et \ref{TabXpT105} de l'annexe \ref{appendix}.

Nous constatons, pour toutes les bases éparses, qu'en augmentant le seuil \textit{minbond} pour un seuil fixe de \textit{minsupp}, la taille de l'ensemble $\mathcal{MCR}$ diminue.
Par exemple, considérons la base \textsc{T40I10D100K} pour
\textit{minsupp} = \textit{25}$\%$. Pour  \textit{minbond} =
\textit{0,05}, nous avons  $|\mathcal{MCR}|$ = \textit{40 533}. En augmentant \textit{minbond} de \textit{0,50} uniquement, la taille  $|\mathcal{MCR}|$  décroét presque de \textit{90}$\%$, nous avons pour \textit{minbond} = \textit{0,10}, $|\mathcal{MCR}|$ = \textit{4 702}.

L'augmentation du seuil \textit{minbond} engendre aussi une diminution des cardinalités des différentes représentations proposées. Par conséquent, les taux de réduction se dégradent.
Considérons, par exemple la base \textsc{Accidents} pour \textit{minsupp} = \textit{50}$\%$. Pour \textit{minbond} =
\textit{0,20}, nous avons
Tx-$\mathcal{RMCR}$ = \textit{100}$\%$. Ce taux diminue progressivement, pour \textit{minbond} =
\textit{0,30}, nous avons
Tx-$\mathcal{RMCR}$ = \textit{98}$\%$ et pour \textit{minbond} =
\textit{0,40}, nous avons
Tx-$\mathcal{RMCR}$ = \textit{95}$\%$.

Par opposition à  la base \textsc{Accidents}, les bases éparses  \textsc{T10I4D100K}, \textsc{T40I10D100K} et \textsc{Retail} présentent toujours des taux de réduction stables même pour une variation importante du seuil \textit{minbond}.

%\begin{figure}[!t]
%%\begin{center}
%% \parbox{7cm}{\includegraphics[scale = 0.90]{T10Rep2.eps}}
%%  \parbox{7cm}{\includegraphics[scale = 0.90]{T40Rep2.eps}}
%%   \parbox{7cm}{\includegraphics[scale = 0.90]{AccRep2.eps}}
%%   \parbox{7cm}{\includegraphics[scale = 0.90]{RetRep2.eps}}
%%\end{center}
%%  \caption{Variation des cardinalités en fonction de \textit{minbond} pour des bases éparses.}\label{AnalysefctMinbondBDE}
%%\end{figure}

Par exemple, la base \textsc{T40I10D100K},
pour  \textit{minsupp} = \textit{25}$\%$ et \textit{minbond} =
\textit{0,15} :
Tx-$\mathcal{RMCR}$ = 1$\%$,
Tx-$\mathcal{RMM}$$ax$$\mathcal{F}$ = 8$\%$,
Tx-$\mathcal{RM}$$in$$\mathcal{MF}$ = 9$\%$ et
Tx-$\mathcal{RM}$$in$$\mathcal{MM}$$ax$$\mathcal{F}$ = 38$\%$.
Ces valeurs sont trés proches pour la base
\textsc{T10I4D100K}.
Pour  \textit{minsupp} = \textit{25}$\%$ et \textit{minbond} =
\textit{0,15} :
Tx-$\mathcal{RMCR}$ = 1$\%$,
Tx-$\mathcal{RMM}$$ax$$\mathcal{F}$ = 8$\%$,
Tx-$\mathcal{RM}$$in$$\mathcal{MF}$ = 9$\%$ et
Tx-$\mathcal{RM}$$in$$\mathcal{MM}$$ax$$\mathcal{F}$ = 40$\%$.

Nous remarquons aussi que pour une même base de données, les taux Tx-$\mathcal{RMM}$$ax$$\mathcal{F}$ et
Tx-$\mathcal{RM}$$in$$\mathcal{MF}$ sont plus intéressants que le taux  Tx-$\mathcal{RMCR}$.
Le taux  Tx-$\mathcal{RM}$$in$$\mathcal{MM}$$ax$$\mathcal{F}$ est dans la majorité des cas plus élevé que les deux  taux précédents.

Considérons par exemple la base \textsc{Retail} pour \textit{minsupp} = \textit{20}$\%$. Pour  \textit{minbond} = \textit{0,10}, nous avons
Tx-$\mathcal{RMCR}$ = \textit{26}$\%$,
Tx-$\mathcal{RMM}$$ax$$\mathcal{F}$ = \textit{31}$\%$,
Tx-$\mathcal{RM}$$in$$\mathcal{MF}$ = \textit{31}$\%$ et
Tx-$\mathcal{RM}$$in$$\mathcal{MM}$$ax$$\mathcal{F}$ = \textit{45}$\%$.

%%\begin{figure}[!t]
%%\begin{center}
%%  \parbox{7cm}{\includegraphics[scale = 0.90]{T10Tx2.eps}}
%%  \parbox{7cm}{\includegraphics[scale = 0.90]{T40Tx2.eps}}
%%  \parbox{7cm}{\includegraphics[scale = 0.90]{AccTx2.eps}}
%%  \parbox{7cm}{\includegraphics[scale = 0.90]{RetTx2.eps}}
%% \caption{Variation des taux de réduction en fonction de
%%\textit{minbond} pour des bases éparses.}\label{tauxminbondBDE}
%%\end{center}
%%\end{figure}

Nous avons ainsi analysé les cardinalités et les taux de réduction des représentations proposées pour différentes bases ``benchmark''. Dans ce qui suit, nous étudions les performances de l'algorithme \textsc{CRPR\_Miner}. Nous soulignons que nous présentons uniquement les temps d'extraction des différentes représentations concises proposées afin de comparer ces différents couts d'extraction.

\section{\'Etude des temps d'extraction des représentations concises proposées}

Nous notons que nous nous limitons à  l'étude des temps d'extraction uniquement pour les deux bases denses \textsc{Mushroom} et \textsc{Pumsb} et pour la base éparse \textsc{T40I10D100K}. Ceci est justifié par le fait que les coûts d'extraction pour les différentes bases évaluées présentent des variations similaires. En effet, nous constatons d'après les résultats obtenus, que pour une base donnée, les temps d'extraction des différentes représentations concises proposées sont proches. Nous désignons par Tps-$\mathcal{RMCR}$,  Tps-$\mathcal{RMM}$$ax$$\mathcal{F}$,
Tps-$\mathcal{RM}$$in$$\mathcal{MF}$ et par
Tps-$\mathcal{RM}$$in$$\mathcal{MM}$$ax$$\mathcal{F}$ les temps d'extraction des représentations
$\mathcal{RMCR}$, $\mathcal{RMM}$$ax$$\mathcal{F}$, $\mathcal{RM}$$in$$\mathcal{MF}$ et
$\mathcal{RM}$$in$$\mathcal{MM}$$ax$$\mathcal{F}$ respectivement.

%\begin{figure}[!t]
%\begin{center}
%  \parbox{7cm}{\includegraphics[scale = 0.90]{MushTp2.eps}}
%  %%\parbox{7cm}{\includegraphics[scale = 0.90]{tcp4Tps1.eps}}
%%\parbox{7cm}{\includegraphics[scale = 0.90]{PbTp2.eps}}
%  \parbox{7cm}{\includegraphics[scale = 0.90]{T40Tp2.eps}}
% \caption{Variation des temps d'extraction des représentations concises proposées.}\label{Analysetps1}
%\end{center}
%\end{figure}

Nous remarquons que les temps d'extraction diminuent avec l'augmentation de \textit{minbond} et augmentent avec l'augmentation de \textit{minsupp} étant donné que les tailles des différentes représentations varient dans ce sens aussi.
Considérons par exemple la base \textsc{Pumsb} pour un seuil fixe \textit{minsupp} = \textit{10}$\%$ et pour \textit{minbond} = \textit{0,10}, nous avons Tps-$\mathcal{RMCR}$ =  Tps-$\mathcal{RMM}$$ax$$\mathcal{F}$ =
Tps-$\mathcal{RM}$$in$$\mathcal{MF}$ = Tps-$\mathcal{RM}$$in$$\mathcal{MM}$$ax$$\mathcal{F}$ =  \textit{1 627} secondes.
Cette durée d'extraction diminue en augmentant le seuil \textit{minbond} de \textit{0,10}, nous avons ainsi Tps-$\mathcal{RMCR}$ =  Tps-$\mathcal{RMM}$$ax$$\mathcal{F}$ =
Tps-$\mathcal{RM}$$in$$\mathcal{MF}$ = Tps-$\mathcal{RM}$$in$$\mathcal{MM}$$ax$$\mathcal{F}$ =
\textit{385} secondes. Nous constatons aussi que les coéts d'extraction des différentes représentations pour les bases \textsc{Mushroom} et \textsc{Pumsb} sont plus élevés que ceux de la base \textsc{T40I10D100K}. Ceci s'explique par le fait que toutes les représentations proposées présentent des cardinalités plus élevées pour les bases denses que pour les bases éparses et le coét d'extraction de la représentation est proportionnelle à  sa taille en nombre de motifs.
Autrement dit, plus la taille de la représentation est élevé plus son extraction est coéteuse.
Nous affirmons ainsi le compromis existant entre le taux de réduction offert par une représentation concise donnée et son coét d'extraction. Dans ce qui suit, nous proposons une application de la  représentation concise exacte $\mathcal{RMCR}$ dans le cadre de la détection d'intrusions.

\section{Application de la représentation concise exacte $\mathcal{RMCR}$ dans le cadre de la détection d'intrusions}
Dans cette section nous présentons le processus d'application de la représentation concise exacte $\mathcal{RMCR}$ sur des données réelles de détection d'intrusions. Décrivons d'abord les bases de données utilisées.
\subsection{Description des bases de données}
Les expérimentations ont été menées sur des bases de données de
détection d'intrusions \textsc{Darpa 1998} 
$^\textsc{(}$\footnote{Ces bases sont disponibles dans l'adresse
	suivante : \textsl{http://www.ll.mit.edu/mission
		/communications/ist/corpora/ideval/data/1998data.html}.}$^\textsc{)}$.
Nous nous sommes principalement basés sur la description de ces bases présentée dans \cite{BImen}. Toutefois, ces bases de données sont issues du trafic réseau \textsc{Darpa 1998}
\textsc{(}Defense Advanced Research Projects Agency\textsc{)} et sont orientées détection d'intrusion.
%%Toutefois, ces données sont les résultats d'une écoute de sept semaine
%%sur un LAN \textsc{(}Local Area Network\textsc{)} simulant le réseau  local de l'armée de l'air américaine et incluant %%une variété d'attaques.
Dans chaque base de données, chaque ligne \textsc{(}ou connexion\textsc{)} code un flot de données \textsc{(}entre deux instants   définis\textsc{)} entre une source et une destination \textsc{(}identifiée chacune par son adresse IP\textsc{)} sous un protocole donné  \textsc{(}TCP, UDP\textsc{)}. Chaque connexion est caractérisée par dix attributs que nous listons dans la table \ref{ATT-darpa}.

%%%%%%%%%%%%%%%%%%%%%%%%%%%%
Les bases de données \textsc{Darpa 1998} recensent 35 types d'attaques. Ces derniéres sont groupées en quatre catégories. Ces catégories sont décrites dans \cite{nahla2006} comme suit.\\
$\bullet$ \textbf{Le déni de service \textsc{(}DOS\textsc{)}} :  Les attaques de type déni de service \textsc{(}Denial of service\textsc{)}. Il s'agit d'empécher les utilisateurs légitimes d'un service de l'utiliser.
Comme exemples d'attaque DOS, nous citons  ``Neptune'', ``Smurf'', ``Apache2'' et ``Pod''.\\
$\bullet$ \textbf{Les attaques de reconnaissance \textsc{(}Probing\textsc{)}} : Ces actions ne sont pas destructrices. En effet, elles permettent d'acquérir des informations importantes afin de mener plutard une vraie attaque.
Un exemple d'outils de reconnaissance est ``Satan'' \textsc{(}Security Administrator Tool for Analyzing Networks\textsc{)} qui est un analyseur de ports TCP/IP et effectue la recherche des failles de sécurité et les défauts de configuration courants.\\
$\bullet$ \textbf{Les attaques de type Remote to Local Access \textsc{(}R2L\textsc{)}} : Ce type d'attaque consiste à  exploiter la vulnérabilité du systéme afin de contréler la machine distante. Comme exemple de ce type d'attaques, il y a les attaques qui visent les failles des protocoles IMAP \textsc{(}Internet Message Acess Protocol\textsc{)}, nous citons aussi ``Httptunnel'' et ``Perl''.\\
$\bullet$ \textbf{Les attaques de type User to Root \textsc{(}U2R\textsc{)}} : Ce type d'attaque se réalise lorsque l'attaquant
essaie d'avoir les droits d'accés à  partir d'un poste afin d'accéder au systéme. Comme exemple de ce type d'attaques, nous citons Rootkit, Ftp$\_$write, Guess$\_$passwd et Guest. Les différentes attaques appartenant à  chacune de ces catégories sont présentées dans \cite{nahla2006}.
%%%%%%%%%%%%%%%%%%%%%%%%%%%%
\subsection{Description du processus de traitement des bases de détection d'intrusion}
\begin {figure}
\hspace{-0.9cm}
\includegraphics[scale = 0.4]{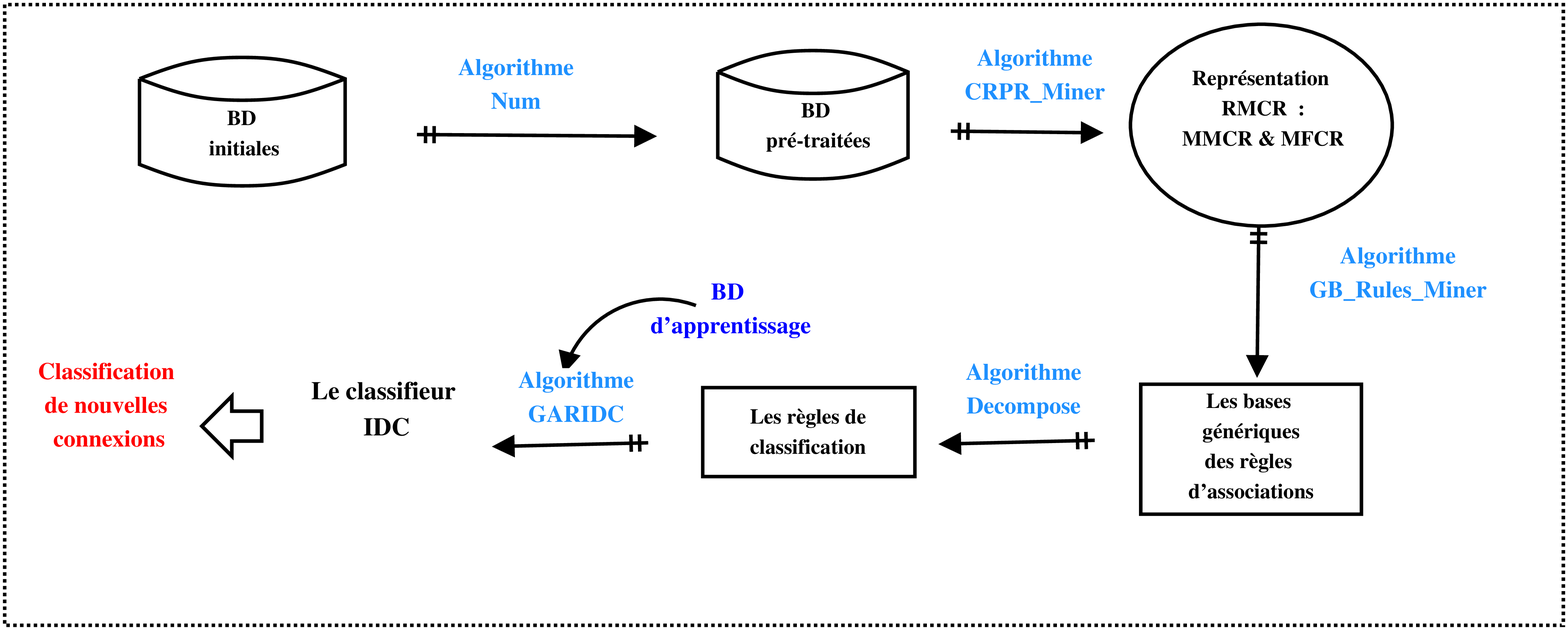}
\caption{Processus de traitement des données depuis l'extraction des motifs jusqu'au la tâche de la classification.}
\label{prcs}
\end{figure}

Nous proposons une application de la représentation concise exacte $\mathcal{RMCR}$  proposée dans le cadre de la détection d'intrusions. Nous décrivons à  cet égard la démarche suivie depuis l'extraction de la représentation concise $\mathcal{RMCR}$ jusqu'à l'étape de la classification basée sur les règles d'association corrélées rares. Ce processus est donné par la figure \ref{prcs} et se réalise en cinq principales phases : \\
\begin{table}
	\footnotesize{
		\begin{center}
			\begin{tabular}{|lcl|}
				\hline
				$1.$& :&Identifiant de la connexion.\\
				$2.$& :&Date de la connexion.\\
				$3.$& :&Instant de début de la connexion.\\
				$4.$& :&Instant de fin de la connexion.\\
				$5.$& :&Service.\\
				$6.$& :&Adresse source de port.\\
				$7.$& :&Adresse destination de port.\\
				$8.$& :&Adresse $IP$ source.\\
				$9.$& :&Adresse $IP$ destination.\\
				$10.$&:&Label de la connexion.\\
				\hline
			\end{tabular}
	\end{center}}
	\caption{Liste des attributs dans les bases \textsc{Darpa 1998}.} \label{ATT-darpa}
\end{table}

\textbf{\'Etape 1 : Pré-traitement des bases de données}\\ 

Nous recourons à  l'algorithme \textsc{Num} \cite{BImen} afin
de pré-traiter les données en les transformant en des données énumératives ordonnées. Par exemple, la discrétisation de l'attribut ``Service'' $\{$http, ftp, tcpmux, private, http$\}$ correspond à  : $\{$1, 2, 3, 4, 1$\}$. Les caractéristiques des bases pré-traitées sont décrites dans la table \ref{Caract_bases_IDS}.\\
\font\xmplbx = cmbx8.5 scaled \magstephalf 
\begin{table}[htbp]
\begin{center}
	\footnotesize{\begin{tabular}{|l||r|r|r|r|} \hline
			\textbf{Base}& \textbf{Nombre} & \textbf{Nombre de}& \textbf{Taille des}\\
			&\textbf{d'items} &\textbf{transactions}& \textbf{transactions}\\
			\hline
			\hline \textsc{Dos}        &{\xmplbx30 128} & {\xmplbx212 774} & {\xmplbx6}\\
			\hline \textsc{Probe}      &{\xmplbx24 353} & {\xmplbx18 018} & {\xmplbx6}\\
			\hline \textsc{Normal}     &{\xmplbx12 743} & {\xmplbx59 481} & {\xmplbx6}\\
			\hline \textsc{U2R}        &{\xmplbx329}    & {\xmplbx466} & {\xmplbx6}\\
			\hline \textsc{R2L}        &{\xmplbx1 294}  & {\xmplbx1 289} & {\xmplbx6}\\
			\hline
	\end{tabular}}
	\caption{Caractéristiques des bases de données \textsc{Darpa 1998} pré-traitées.}
	\label{Caract_bases_IDS}
\end{center}
\end{table}

\textbf{\'Etape 2 : Extraction de la représentation concise exacte $\mathcal{RMCR}$}\\

Ces bases pré-traitées sont communiquées à  l'algorithme \textsc{CRPR\_Miner} afin d'extraire la représentation concise exacte $\mathcal{RMCR}$. Les motifs minimaux et les motifs fermés corrélés rares sont identifiés d'une maniére séparée. Nous désignons par $\mathcal{NR}$-$\mathcal{MFCR}$ l'ensemble des motifs fermés corrélés rares qui ne sont pas des minimaux. Ces itemsets seront utilisés plutard dans la phase de génération des régles génériques. 
Les résultats des expérimentations menées sur les différentes bases utilisées sont donnés par la table \ref{TabXpAr1}. 
\font\xmplbx = cmbx8.5 scaled \magstephalf 
\begin{table}[htbp]
\begin{center}
	\footnotesize{\begin{tabular}{|l||r|r|r|r|r|r|} \hline
			\textbf{Base}& \textbf{Minsupp($\%$)}&\textbf{Minbond} &\textbf{$|\mathcal{MCR}|$}&\textbf{$|\mathcal{MMCR}|$}&
			\textbf{$|\mathcal{MFCR}|$}& \textbf{$|$$\mathcal{NR}$-$\mathcal{MFCR}$$|$}\\
			\hline
			\hline \textsc{Dos}  &{\xmplbx20} &{\xmplbx0,80} & {\xmplbx31 208}&{\xmplbx30 133} & {\xmplbx815}&{\xmplbx801}\\
			\hline \textsc{Probe}&{\xmplbx50} &{\xmplbx0,70} & {\xmplbx39 688} &{\xmplbx25 546}&{\xmplbx4 835}&{\xmplbx3 097}\\
			\hline \textsc{Normal}&{\xmplbx60} &{\xmplbx0,60} & {\xmplbx12 910} & {\xmplbx12 763}&{\xmplbx4 664}&{\xmplbx51}\\
			\hline \textsc{R2L}&{\xmplbx60}    &{\xmplbx0,60}& {\xmplbx1 404} & {\xmplbx1 308}&{\xmplbx200}&{\xmplbx54}\\
			\hline \textsc{U2R}&{\xmplbx60}    &{\xmplbx0,60} & {\xmplbx434} & {\xmplbx348}&{\xmplbx101}&{\xmplbx45}\\
			\hline
	\end{tabular}}
	\caption{\'Evaluation expérimentale du nombre de motifs extraits pour les différentes bases de détection d'intrusions.}
	\label{TabXpAr1}
\end{center}
\end{table}

\textbf{\'Etape 3 :  Extraction des régles d'association génériques}\\

Cette étape consiste à  extraire les régles d'association génériques à  partir des ensembles $\mathcal{MMCR}$ et $\mathcal{NR}$-$\mathcal{MFCR}$ précédement identifiés. Désignons par $Gen$ un motif minimal et par $Ferme$ un motif fermé corrélé rare, toutes les régles d'association de la forme $Gen$ $\Rightarrow$ $Ferme$ $\backslash$ $Gen$ seront extraites gréce à  l'algorithme \textsc{GB$\_$Rules$\_$Miner}. Cet algorithme prend en entrée les ensembles $\mathcal{MMCR}$ et $\mathcal{MFCR}$ et offre en sortie toutes les régles d'association à  prémisse minimale et à  conclusion maximale \textsc{(}en terme de nombre d'items\textsc{)} munie chacune de la valeur de son support et de sa confiance.
Les régles exactes, ayant une valeur de la confiance égale à  $1$ forment  
la base générique des régles exactes, tandis que les régles approximatives, ayant une valeur de la confiance inférieure à  $1$ forment la base informative des régles approximatives. Toutefois, l'extraction des régles génériques a été motivé par les avantages que présentent ces derniéres en terme de compacité et d'informativité \cite{BImen}. Ces régles étant génériques, elles permettent de convoyer le maximum d'informations. Nous proposons un exemple dans ce qui suit.\\
$\bullet$ \textbf{R1} : service = \textit{http} $\Rightarrow$ port-source = \textit{35}.\\
$\bullet$ \textbf{R2} : service = \textit{tcpmux} et port-destination = \textit{80} $\Rightarrow$ attaque  = \textit{Neptune}.\\
$\bullet$ \textbf{R3} : protocole = \textit{tcp} $\Rightarrow$ attaque = \textit{Dos}.\\
$\bullet$ \textbf{R4} : service = \textit{private} et IP-destination = \textit{209.051.071.32}  $\Rightarrow$ IP-source = \textit{172.016.118.48}.\\
$\bullet$ \textbf{R5} : port-destination = \textit{80} $\Rightarrow$ attaque = \textit{Neptune}.\\

\textbf{\'Etape 4 : Génération des régles associatives de classification} \\

Cette étape consiste à  générer les régles de classification à  partir des 
régles associatives génériques moyennant l'algorithme \textsc{Decompose} \cite{cari2008}.
En effet, cet algorithme permet de filtrer les régles associatives générées lors de la phase précédente afin de ne garder que les régles les plus intéressantes \textit{c.-é.-d.} celles ayant le label de la classe dans la partie conclusion. En effet, les régles retenues seront encore filtrées afin de ne garder que les régles de classification ayant la prémisse la plus minimale. Ce genre de régles est plus générique et supporte plus de connexions de la base de données. \`A la fin de cette phase, les régles maintenues correspondent aux régles dont la conclusion englobe la classe voulue et la prémisse est la plus minimale.
Par exemple, parmi les régles \textbf{R1}, \textbf{R2}, \textbf{R3}, \textbf{R4} et \textbf{R5} précédemement proposées, seules les régles \textbf{R2}, \textbf{R3} et \textbf{R5} seront maintenues.\\
$\bullet$ \textbf{R2} : service = \textit{tcpmux} et protocole = \textit{udp} et port-destination = \textit{80} $\Rightarrow$ attaque = \textit{Neptune}.\\
$\bullet$ \textbf{R3} : protocole = \textit{tcp} $\Rightarrow$ attaque = \textit{Dos}.\\
$\bullet$ \textbf{R5} : port-destination = \textit{80} et protocole = \textit{udp} $\Rightarrow$ attaque = \textit{Neptune}.

Nous remarquons que les régles \textbf{R2} et \textbf{R5} ont la même conclusion \textsc{(}attaque = \textit{Neptune}\textsc{)} et que la prémisse de la régle \textbf{R5} est plus minimale que celle de \textbf{R2}. Ainsi, la régle \textbf{R5} est plus générique que \textbf{R2}, elle sera gardée alors que la régle \textbf{R2} sera supprimée.\\

\textbf{\'Etape 5 :  Construction du classifieur à  partir des régles génériques de classification}\\

Cette phase consiste à  constuire le classifieur \textsc{IDC} à  partir des régles génériques de classification moyennant l'algorithme \textsc{GARIDC} \cite{cari2008}. En effet, 
l'ensemble des régles génériques de classification et la base de données d'apprentissage seront communiqués à  l'algorithme \textsc{GARIDC} afin d'alimenter le classifieur.
\`A cet égard, l'ensemble d'apprentissage est parcouru d'une maniére séquentielle et pour chaque connexion nous sélectionons la premiére régle qui couvre cette connexion.
Autrement dit, la régle dont la prémisse est incluse ou égale aux attributs de la connexion courante et dont la conclusion est égale à  l'étiquette de la connexion courante sera ajoutée au classifieur \textsc{IDC}. \`A la fin de cette étape, le classifieur englobe toutes les régles qui couvrent toute la base d'apprentissage.\\

\textbf{\'Etape 6 :  Classification de nouvelles connexions}\\

La phase de la classification de nouvelles attaques consiste à  affecter à  une nouvelle connexion $C$ la classe de l'attaque qui lui est associée. En effet, les régles de classification dont la prémisse est incluse dans la connexion $C$ seront extraites. Ensuite, le score de chacune de ces régles sera calculé \cite{BImen} en fonction des valeurs des mesures du support et du lift. La classe attribuée à  la connexion $C$ correspond à  la conclusion de la régle présentant la valeur maximale du score. Par exemple,
désignons par $C_{1}$ une nouvelle connexion, $C_{1}$ : port-destination = \textit{80} et protocole = \textit{udp} et IP-source = \textit{010.200.030.040}.
En considérant les régles \textbf{R3} et \textbf{R5} extraites lors de la quatriéme phase, la classe \textit{Neptune} sera affectée à  la connexion $C_{1}$. 

\subsection{\'Evaluation expérimentale de l'efficacité de la classification basée sur les régles d'association corrélées rares}

Nous évaluons à  présent expérimentalement l'efficacité de la classification basée sur les 
régles génériques de classification corrélées rares. Les résultats des expérimentations menées sur les différentes bases sont décrites par la table \ref{TabXpAr2} et \ref{TabXpAr3}.

\font\xmplbx = cmbx8.5 scaled \magstephalf 
\begin{table}[htbp]
\hspace{-0.5cm}
\footnotesize{\begin{tabular}{|l||r|r|r|r|r|} \hline
		\textbf{Base}& \textbf{Minsupp($\%$)}&\textbf{Minconf} &\textbf{Nombre de régles}&
		\textbf{Nombre de régles}& \textbf{Nombre de régles}\\
		& & &\textbf{génériques}&\textbf{génériques}& \textbf{génériques}\\
		& & &\textbf{exactes}&\textbf{approximatives}& \textbf{de classification}\\
		\hline
		\hline \textsc{Dos}   & {\xmplbx20}  &{\xmplbx0,80} & {\xmplbx1 601} & {\xmplbx6}    &{\xmplbx5}\\
		\hline \textsc{Probe} & {\xmplbx50}  &{\xmplbx0,60} & {\xmplbx8 514} & {\xmplbx201}  &{\xmplbx7}\\
		\hline \textsc{Normal}& {\xmplbx60}  &{\xmplbx0,90} & {\xmplbx105}   & {\xmplbx17}   &{\xmplbx2}\\
		\hline \textsc{R2L}   & {\xmplbx60}  &{\xmplbx0,70} & {\xmplbx111}   & {\xmplbx15}   &{\xmplbx4}\\
		\hline \textsc{U2R}   & {\xmplbx60}  &{\xmplbx0,80} & {\xmplbx95}    & {\xmplbx11}   &{\xmplbx7}\\
		\hline
\end{tabular}}
\caption{\'Evaluation expérimentale pour les différentes bases de détection d'intrusions.}
\label{TabXpAr2}
\end{table}

Nous constatons d'aprés les résultats donnés par la table \ref{TabXpAr2} que, le nombre de régles génériques de clasification est plus réduit que le nombre des régles génériques approximatives et exactes.
Toutefois, les régles d'association génériques peuvent englober des régles inutiles et non porteuses d'informations. \`A cet égard, seules les régles intéressantes et qui contribuent à  l'amélioration de la téche de classification seront maintenues.

\begin{table}[h]
\begin{center}
	\footnotesize{
		\begin{tabular}{|l||r|r|r|} \hline
			\textbf{Base} & \textbf{Taux de détection TD\textsc{(}$\%$\textsc{)}} &\textbf{Taux de fausses alarmes TFA\textsc{(}$\%$\textsc{)}}\\
			\hline
			\hline \textsc{Dos}   & {\xmplbx98,99}   &{\xmplbx1,00}\\
			\hline \textsc{Probe} & {\xmplbx88,01}   &{\xmplbx11,98}\\
			\hline \textsc{Normal}& {\xmplbx100,00}  &{\xmplbx0,00}\\
			\hline \textsc{R2L}   & {\xmplbx88,05}   &{\xmplbx11,94}\\
			\hline \textsc{U2R}   & {\xmplbx90,12}   &{\xmplbx9,87}\\
			\hline
	\end{tabular}}
\end{center}
\caption{\'Evaluation expérimentale de l'efficacité des régles de classification corrélées rares.}
\label{TabXpAr3}
\end{table}

La table \ref{TabXpAr3} présente les taux de détection ainsi que les taux de fausses alarmes.
Le taux de détection est équivalent au pourcentage de classifications correctes \textsc{(}PCC\textsc{)} et 
est défini ainsi, PCC  = $\displaystyle\frac{\displaystyle NbrCC}{\displaystyle NbrTotal}$ avec 
$NbrCC$ correspond au  nombre d'instances correctement classées et $NbrTotal$ correspond au nombre total d'instances classées. Tandis que le taux de fausses alarmes \textsc{(}TFA\textsc{)} est défini ainsi,  TFA  = $\displaystyle\frac{\displaystyle NbrMalC}{\displaystyle NbrTotal}$ avec $NbrMalC$ correspond au nombre d'instances mal classées. 

Nous constatons d'aprés les résultats donnés par la table \ref{TabXpAr3} que les 
taux de détection de vraies attaques sont intéressants et les taux de fausses alarmes sont faibles pour les cinq catégories de connexions et pour les différentes seuils de \textit{minsupp} et de \textit{minconf}.
Nous présentons dans ce qui suit une comparaison entre les taux de détection fournis par les régles génériques corrélées rares de classification et les différentes approches de la littérature.

Il est clair d'aprés les résultats donnés par la table \ref{TabXpAr4} que les taux offerts par les régles génériques corrélées rares sont plus intéressants que ceux des régles génériques fréquentes pour les bases
\textsc{Dos}, \textsc{Probe} et \textsc{R2L}. Toutefois, pour les bases \textsc{Normal} et \textsc{U2R} les régles génériques fréquentes présentent les mêmes taux de précision que les régles corrélées rares.

Nous remarquons aussi que la classification basée sur la base \textsc{IGB} des régles fréquentes \cite{cari2008} présente des taux de précision trés proches des taux fournis par les régles corrélées rares pour les classes \textsc{Normal}, \textsc{Dos} et \textsc{U2R}. Ces régles fréquentes ont été générées à  partir des bases de données \textsc{Darpa 1998} et différent de notre approche dans le choix des régles à  extraire.

Au cas oé nous faisons abstraction des conditions d'expérimentations et nous nous limitons aux taux de détection pour chacune des classes d'attaques, les résultats obtenus dans nos expérimentations sont globalement meilleurs que ceux des approches proposées dans \cite{farid2010} et dans \cite{nahla2006}. Ces approches traitent la base de données \textsc{Kdd 1999}.
En particulier, notre approche présente de meilleurs résultats que l'approche proposée dans \cite{farid2010}
pour les classes d'attaques \textsc{Normal} et \textsc{Dos}.
Pour les approches discutées dans \cite{nahla2006}, nos résultats sont plus intéressants pour les classes 
d'attaques \textsc{U2R}, \textsc{R2L}, \textsc{Normal} et \textsc{DOS}.

\begin{table}[htbp]
\begin{center}
	\footnotesize{
		\begin{tabular}{|l||r|r|r|r|} \hline
			\textbf{Base}&\textbf{Les régles}&\textbf{Les régles} &\textbf{La base IGB}\\
			&\textbf{génériques} & \textbf{génériques}&\textbf{des régles fréquentes}\\
			& \textbf{corrélées rares}& \textbf{fréquentes}&\cite{cari2008}\\
			\hline
			\hline \textsc{Dos} & {\xmplbx98,99} & {\xmplbx98,60}&{\xmplbx99,28}\\
			\hline \textsc{Probe} & {\xmplbx88,01}&{\xmplbx68,20}&{\xmplbx99,66}\\
			\hline \textsc{Normal}& {\xmplbx100,00}&{\xmplbx100,00} &{\xmplbx100,00}\\
			\hline \textsc{R2L}& {\xmplbx88,05} & {\xmplbx88,00} &{\xmplbx98,99}\\
			\hline \textsc{U2R}& {\xmplbx90,12} & {\xmplbx90,12} &{\xmplbx91,41}\\
			\hline
	\end{tabular}}
\end{center}
\caption{Taux de détection en pourcentage obtenus pour différentes approches de détection d'intrusions.}
\label{TabXpAr4}
\end{table}
Nous présentons aussi le taux de détection moyen offert par l'approche \textsc{Wifi Miner} \cite{wifiminer}. Cette derniére est basée sur les motifs rares pour la détection d'intrusions dans les réseaux informatiques sans fil. Nous constatons d'aprés les résultats donnés par la table \ref{TabXpAr5}, que le taux moyen de détection assuré par les régles corrélées rares est plus intéressant que celui de l'approche \textsc{Wifi Miner}.
Nous concluons ainsi que les régles génériques corrélées rares constituent un outil de classification efficace dans le cadre de la détection d'intrusions dans les réseaux informatiques.

\begin{table}[h]
\begin{center}
	\footnotesize{
		\begin{tabular}{|l||c|c|c|} \hline
			& \textbf{Les régles génériques} &\textbf{L'approche \textsc{Wifi Miner}} \\
			& \textbf{corrélées rares}       & \cite{wifiminer} \\
			\hline              
			\hline \textbf{Taux moyen}                             & \textbf{93,03}  &86,00\\
			\textbf{de détection TD\textsc{(}$\%$\textsc{)}} &                 &\\                         
			\hline \textbf{Taux moyen}                              &\textbf{06,95}   &14,00\\
			\textbf{de fausses alarmes TFA\textsc{(}$\%$\textsc{)}}&      &\\
			\hline
	\end{tabular}}
\end{center}
\caption{Comparaison entre les régles génériques corrélées rares de classification et l'approche \textsc{Wifi Miner} basée sur les motifs rares.}
\label{TabXpAr5}
\end{table}
%%%%%%%%%%%%%%%%%%
\section{Conclusion}
Dans ce chapitre, nous avons mené une étude expérimentale des
deux algorithmes \textsc{CRP\_Miner} et \textsc{CRPR\_Miner} sur des bases ``benchmark''
communément utilisées. Nous avons prouvé expérimentalement que les représentations proposées permettent de réduire considérablement le nombre de motifs corrélés rares générés.
Nous avons par la suite effectué la génération des régles d'association de classification à  partir de ces représentations. La classification basée sur ces régles corrélées rares a présenté de bons résultats et a prouvé l'utilité des représentations concises extraites dans le cadre de la détection d'intrusions.
%%Toutefois, les bases testées ont montré des comportements différents lors de la variation des paramétres %%d'expérimentations. \`A cet égard, l'idée de considérer uniquement un échantillon de la base et d'étudier %%l'adéquation des résultats expérimentaux obtenus sur l'échantillon et ceux obtenus sur toute la base est une %%idée intéressante.

%%%%%%%%%%%%%%%%%%%%%%%%%%%%%%%%%%%%

\addcontentsline{toc}{chapter}{Conclusion générale}
\chapter*{Conclusion générale}\label{chapitre_conclusion}
\markboth{Conclusion générale}{Conclusion générale}

Dans ce mémoire, nous nous sommes intéressés à  la fouille des motifs corrélés rares associés à  la mesure \textit{bond}. Ces motifs résultent de la conjonction de deux contraintes de types opposés, à  savoir la contrainte anti-monotone de la corrélation et la contrainte monotone de la rareté. Cette nature opposée des contraintes traitées dans ce travail rend complexe la localisation de l'ensemble des motifs corrélés rares. Cet ensemble se différe ainsi de l'ensemble des motifs induit par une ou plusieurs contraintes de même type. Cette caractéristique constitue la principale originalité de notre contribution.
\`A cet égard, nous avons proposé, dans ce mémoire, une caractérisation de cet ensemble moyennant la notion de classe d'équivalence et nous avons proposé de nouvelles représentations concises de cet ensemble.
En effet, nous avons entamé ce mémoire avec la présentation des notions préliminaires relatives aux motifs rares et aux motifs corrélés selon la mesure \textit{bond}. Nous avons tout de même décrit les spécificités des classes d'équivalence induites par l'opérateur de fermeture $f_{bond}$ associé à  la mesure \textit{bond}. Ensuite, nous avons étudié les différentes approches de la littérature traitant de l'extraction des motifs rares ainsi que les approches traitant la problématique de la fouille des motifs corrélés sous contraintes. Nous avons, par la suite, enchaéné avec l'analyse et la critique des approches génériques d'extraction des motifs sous la conjonction de contraintes de types opposés. Une revue profonde des approches réductrices des motifs extraits a été également présentée.

Dans ce sens, nous avons défini rigoureusement l'ensemble des motifs rares corrélés selon la mesure \textit{bond}. Nous avons étudié profondément ses propriétés spécifiques. Par la suite, nous nous sommes basés sur
les éléments minimaux et maximaux des classes d'équivalence corrélées rares, afin d'introduire de nouvelles représentations concises des motifs corrélés rares. En effet, nous avons proposé les représentations concises exactes
$\mathcal{RMCR}$, $\mathcal{RMM}$$ax$$\mathcal{F}$ et $\mathcal{RM}$$in$$\mathcal{MF}$ ainsi que la représentation concise approximative $\mathcal{RM}$$in$$\mathcal{MM}$$ax$$\mathcal{F}$. Une étude minutieuse des propriétés théoriques de ces approches a été également menée.
Toutefois, ces représentations permettent d'une part de réduire significativement le nombre de motifs corrélés rares extraits. Elles améliorent aussi leur qualité et ce en évitant la redondance entre les motifs puisqu'elles ne maintiennent qu'un sous-ensemble sans perte d'information de l'ensemble total des motifs corrélés rares. D'autre part, elles assurent la régénération aisée et efficace de l'ensemble des motifs corrélés rares.
Nous avons présenté également l'approche \textsc{CRP\_Miner} d'extraction de l'ensemble total des motifs corrélés rares et l'approche \textsc{CRPR\_Miner} d'extraction des différentes représentations concises proposées. Nous avons aussi démontré les propriétés théoriques de validité, de complétude et de terminaison de l'algorithme \textsc{CRPR\_Miner} et calculé sa complexité théorique.
Par ailleurs, nous avons décrit les deux stratégies de régénération à  partir de la représentation concise exacte $\mathcal{RMCR}$. \`A cet égard, nous avons suggéré l'algorithme \textsc{Regenerate}, permettant l'interrogation de cette représentation ainsi que l'algorithme \textsc{CRPRegeneration} permettant la dérivation de tous les motifs corrélés rares à  partir de la représentation $\mathcal{RMCR}$. Nous avons prouvé, gréce aux expérimentations réalisées sur différentes bases de test, l'apport bénéfique en terme de compacité des différentes représentations concises proposées. Nous avons effectué la génération des régles d'association de classification à  partir de ces représentations. La classification basée sur ces régles corrélées rares a présenté des résultats intéressants et a prouvé l'utilité des représentations concises extraites dans le cadre de la détection d'intrusions.

Les perspectives de travaux futurs concernent :

\begin{itemize}
	\item L'extraction et l'application dans des cas réels des formes généralisées de régles d'association présentant des conjonctions, des disjonctions, et des négations d'items corrélés rares en prémisse ou en conclusion. Dans cette situation, les représentations proposées offrent l'information concernant les différents supports d'un motif en plus de sa mesure \textit{bond}. 
	
	\item L'extension de l'approche proposée dans ce travail pour les motifs corrélés rares selon toute autre mesure de corrélation vérifiant les mêmes propriétés que la mesure \textit{bond}, telle que la mesure \textit{all-confidence} \cite{Omie03} par exemple. L'évaluation sur la base d'une application réelle de la qualité des motifs corrélés rares associés à  chacune de ces mesures est alors une perspective intéressante.
	
	\item L'optimisation des algorithmes proposés et la comparaison de leurs performances avec celles des algorithmes de la littérature.
\end{itemize}

%%%%%%%%%%%%%%%%%%%%%%%%%%%%%%%%%%%%%
\addcontentsline{toc}{chapter}{Bibliographie}
\bibliographystyle{apalike-fr}
%%\bibliography{références-valides}
\markboth{Bibliographie}{Bibliographie}
%%%%%%%%%%%%%%%%%%%%%%%%%%%%%%%%%%%%%%%%%%%%%%%%%

\appendix

\part*{Annexe}

\newpage
\thispagestyle{empty}

\chapter{Tableaux des résultats expérimentaux}\label{appendix}
\setcounter{footnote}{0}
\markboth{Tableaux des résultats expérimentaux}{Tableaux des résultats expérimentaux}

Dans cet annexe, nous présentons les résultats obtenus suite aux différentes expérimentations réalisées dans l'objectif de comparer les cardinalités des différents ensembles et représentations proposés dans ce travail.

%%%%%%%%%%%%%%%%%%%%%%%%%%%%%%%%%%%%%%%%%%%%%%%%%%%%%%%%%%%%%
\font\xmplbx = cmbx8.5 scaled \magstephalf
\begin{table}[!t]
	\parbox{16cm}{
		\hspace{-1.3cm}
		\footnotesize{
\begin{tabular}{|c||r|r|r|r|r|r|r|r|r|r|r|}
				\hline  Base & \textit{minsupp} & \textit{minbond} &
				|$\mathcal{MCR}$|& |$\mathcal{MCF}$|  &|$\mathcal{MMCR}$| & |$\mathcal{MFCR}$|
				&|$\mathcal{MMCRM}in$|
				&|$\mathcal{MFCRM}ax$| \\
				\hline\hline
				\textsc{Mushroom} & {\xmplbx5$\%$}&  {\xmplbx0,15}   &   {\xmplbx361}     &{\xmplbx100 906} &{\xmplbx205} &{\xmplbx167}&{\xmplbx49} &{\xmplbx29} \\
				\hline
				& {\xmplbx10$\%$}&   {\xmplbx0,15}   &    {\xmplbx913}     &{\xmplbx100 354}    & {\xmplbx451}&  {\xmplbx386}&{\xmplbx73} &{\xmplbx33} \\
				\hline
				& {\xmplbx15$\%$}&   {\xmplbx0,15}   &    {\xmplbx3038}     &{\xmplbx98 229}    & {\xmplbx709}&  {\xmplbx558}&{\xmplbx85} &{\xmplbx15} \\
				\hline
				& {\xmplbx20$\%$}&   {\xmplbx0,15}   &    {\xmplbx48 056}     &{\xmplbx53 211}    & {\xmplbx974}&  {\xmplbx578}&{\xmplbx92} &{\xmplbx8} \\
				\hline
				& {\xmplbx25$\%$}&   {\xmplbx0,15}   &    {\xmplbx95 876}     &{\xmplbx5 391}    & {\xmplbx1 223}&  {\xmplbx624}&{\xmplbx85} &{\xmplbx9} \\
				\hline		
				& {\xmplbx30$\%$}&   {\xmplbx0,15}   &    {\xmplbx98 566}     &{\xmplbx2 701}    & {\xmplbx1 318}&  {\xmplbx633}&{\xmplbx91} &{\xmplbx9} \\
				\hline		
				& {\xmplbx35$\%$}&   {\xmplbx0,15}   &    {\xmplbx100 156}     &{\xmplbx1 111}   & {\xmplbx1 412}&  {\xmplbx652}&{\xmplbx95} &{\xmplbx9} \\
\hline
	\end{tabular}}}
	\begin{center}
		\caption{Cardinalités des ensembles extraits en fonction de \textit{minsupp} pour les bases denses \textsc{Chess} et \textsc{Mushroom}.} \label{Tab1}
	\end{center}
\end{table}
%%%%%%%%%%%%%%%%%%%%%%%%%%%%%%%%%%%
\font\xmplbx = cmbx8.5 scaled \magstephalf
\begin{table}[htbp]
	\parbox{16cm}{
		\hspace{-0.9cm}
		\footnotesize{
			\begin{tabular}{|c||r|r|r|r|r|r|r|r|r|}
				\hline  Base & \textit{minsupp} & \textit{minbond} &
				|$\mathcal{MCR}$|& |$\mathcal{RMCR}$|  &|$\mathcal{RMM}$$ax$$\mathcal{F}$|& $|\mathcal{RM}$$in$$\mathcal{MF}|$ &|$\mathcal{RM}$$in$$\mathcal{MM}$$ax$$\mathcal{F}$|\\
				\hline\hline
				\textsc{Mushroom} & {\xmplbx5$\%$}&   {\xmplbx0,15}   &    {\xmplbx361}     &{\xmplbx255}    & {\xmplbx220}&  {\xmplbx184}&{\xmplbx78} \\
				\hline
				& {\xmplbx10$\%$}&   {\xmplbx0,15}   &    {\xmplbx913}     &{\xmplbx650}    &{\xmplbx478}&  {\xmplbx406}&{\xmplbx106}  \\
				\hline
				& {\xmplbx15$\%$}&   {\xmplbx0,15}   &    {\xmplbx3 038}     &{\xmplbx1 061}    & {\xmplbx724}&  {\xmplbx587}&{\xmplbx100} \\
				\hline
				& {\xmplbx20$\%$}&   {\xmplbx0,15}   &    {\xmplbx48 056}     &{\xmplbx1 300}    & {\xmplbx982}&  {\xmplbx608}&{\xmplbx100}  \\
				\hline
				& {\xmplbx25$\%$}&   {\xmplbx0,15}   &    {\xmplbx95 876}     &{\xmplbx1 621}    & {\xmplbx1 232}&  {\xmplbx640}&{\xmplbx94} \\
				\hline		
				& {\xmplbx30$\%$}&   {\xmplbx0,15}   &    {\xmplbx98 566}     &{\xmplbx1 704}    & {\xmplbx1 327}&  {\xmplbx648}&{\xmplbx100} \\
				\hline		
				& {\xmplbx35$\%$}&   {\xmplbx0,15}   &    {\xmplbx100 156}     &{\xmplbx1 810}   & {\xmplbx1 421}& {\xmplbx667}&{\xmplbx104} \\
\hline
	\end{tabular}}}
	\begin{center}
		\caption{Cardinalités des représentations concises extraites en fonction de \textit{minsupp} pour les bases denses \textsc{Mushroom} et \textsc{Chess}.}\label{Tab2}
	\end{center}
\end{table}	
%%%%%%%%%%%%%%%%%%%%%%%
\font\xmplbx = cmbx8.5 scaled \magstephalf
\begin{table}[htbp]
	
	\parbox{16cm}{
		\vspace{-0.5cm}
		\hspace{-1.3cm}
		\footnotesize{
			\begin{tabular}{|c||r|r|r|r|r|r|r|r|r|r|r|}
				\hline Base & \textit{minsupp} & \textit{minbond} &|$\mathcal{MCR}$|&|$\mathcal{MCF}$| &|$\mathcal{MMCR}$|&|$\mathcal{MFCR}$| &|$\mathcal{MMCRM}in$| &|$\mathcal{MFCRM}ax$|\\
				\hline\hline
				\textsc{Pumsb}     & {\xmplbx10$\%$}&   {\xmplbx0,80} & {\xmplbx2 052}   &{\xmplbx143 091}    & {\xmplbx1 993}& {\xmplbx787}&{\xmplbx1 978} &{\xmplbx302}\\
				\hline		
				& {\xmplbx20$\%$}&   {\xmplbx0,80}   &  {\xmplbx2 077}  &{\xmplbx143 876}   &{\xmplbx2 018}&  {\xmplbx812}&{\xmplbx2 002} &{\xmplbx304}\\
				\hline		
				& {\xmplbx30$\%$}&   {\xmplbx0,80}   & {\xmplbx3 260}  &{\xmplbx142 693}   & {\xmplbx2 090}&  {\xmplbx451}&{\xmplbx2 028} &{\xmplbx112}\\
				\hline		
				& {\xmplbx40$\%$}&   {\xmplbx0,80}   &   {\xmplbx3 395}  &{\xmplbx142 558} & {\xmplbx2 126}&  {\xmplbx403}&{\xmplbx2 042} &{\xmplbx20}\\
				\hline		
				&{\xmplbx50$\%$}&   {\xmplbx0,80}   &    {\xmplbx3 521}  &{\xmplbx142 432}   & {\xmplbx2 181}&  {\xmplbx412}&{\xmplbx2 061} &{\xmplbx19}\\
				\hline		
				& {\xmplbx60$\%$}&   {\xmplbx0,80}   &    {\xmplbx3 655}  &{\xmplbx142 298}  &{\xmplbx2 208} & {\xmplbx361} &{\xmplbx2 074} &{\xmplbx20}\\
				\hline		
				&{\xmplbx70$\%$}&  {\xmplbx0,80}     &  {\xmplbx3 736}  &{\xmplbx142 217}  & {\xmplbx2 240}&  {\xmplbx386}&{\xmplbx2 079} &{\xmplbx21}\\
\hline
	\end{tabular}}}
	\begin{center}
		\caption{Cardinalités des ensembles extraits en fonction de \textit{minsupp} pour des bases denses.}\label{Tab3}
	\end{center}
\end{table}

\font\xmplbx = cmbx8.5 scaled \magstephalf
\begin{table}[htbp]
	\parbox{16cm}{
		\vspace{-0.5cm}
		\hspace{-0.9cm}
		\footnotesize{
			\begin{tabular}{|c||r|r|r|r|r|r|r|r|r|}
				\hline  Base & \textit{minsupp} & \textit{minbond} &
				|$\mathcal{MCR}$|& |$\mathcal{RMCR}$|  &|$\mathcal{RMM}$$ax$$\mathcal{F}$|& $|\mathcal{RM}$$in$$\mathcal{MF}|$ &|$\mathcal{RM}$$in$$\mathcal{MM}$$ax$$\mathcal{F}$|\\
				\hline\hline	
				\textsc{Pumsb}    & {\xmplbx10$\%$}&   {\xmplbx0,80}   &    {\xmplbx2 052}     &{\xmplbx2 045}   &{\xmplbx2 041}& {\xmplbx2 045}& {\xmplbx2 036} \\
				\hline		
				&{\xmplbx20$\%$}&   {\xmplbx0,80}   &    {\xmplbx2 077}  & {\xmplbx2 070}   &{\xmplbx2 066}&  {\xmplbx2 070}&{\xmplbx2 061} \\
				\hline		
				&{\xmplbx30$\%$}&   {\xmplbx0,80}   & {\xmplbx3 260}  &{\xmplbx2 126}  & {\xmplbx2 110}& {\xmplbx2 083}&{\xmplbx2 055}\\
				\hline		
				&{\xmplbx40$\%$}&   {\xmplbx0,80}   &   {\xmplbx3 395}  &{\xmplbx2 168}  & {\xmplbx2 136}&  {\xmplbx2 108}&{\xmplbx2 054} \\
				\hline		
				&{\xmplbx50$\%$}&  {\xmplbx0,80}   &    {\xmplbx3 521}  &{\xmplbx2 240}   &{\xmplbx2 192}& {\xmplbx2 149}&{\xmplbx2 074} \\
				\hline		
				&{\xmplbx60$\%$}&  {\xmplbx0,80}   &    {\xmplbx3 655} &{\xmplbx2 267}&{\xmplbx2 220}& {\xmplbx2 162}&{\xmplbx2 088} \\		   		          		          
				\hline		
				&{\xmplbx70$\%$}&   {\xmplbx0,80}   & {\xmplbx3 736}  &{\xmplbx2 315}  & {\xmplbx2 253}&  {\xmplbx2 187}&{\xmplbx2 094} \\
\hline		
	\end{tabular}}}
	\caption{Cardinalités des représentations concises extraites en fonction de \textit{minsupp} pour des bases denses.}\label{Tab4}
\end{table}

\font\xmplbx = cmbx8.5 scaled \magstephalf
\begin{table}[htbp]
	\parbox{16cm}{
		\hspace{-1.3cm}
		\footnotesize{
			\begin{tabular}{|c||r|r|r|r|r|r|r|r|r|r|r|}
				\hline  Base & \textit{minsupp} & \textit{minbond} &
				|$\mathcal{MCR}$|& |$\mathcal{MCF}$|  &|$\mathcal{MMCR}$| & |$\mathcal{MFCR}$|
				&|$\mathcal{MMCRM}in$| &|$\mathcal{MFCRM}ax$|\\
				\hline\hline
				\textsc{Mushroom}& {\xmplbx30$\%$}&  {\xmplbx0,10}  &  {\xmplbx578 359}  &{\xmplbx2 701} & {\xmplbx1 798}&{\xmplbx700} &{\xmplbx91} &{\xmplbx6}\\
				\hline
				& {\xmplbx30$\%$}&  {\xmplbx0,15}      &  {\xmplbx98 566}  &{\xmplbx2 701} & {\xmplbx1 318}&{\xmplbx633} &{\xmplbx91} &{\xmplbx9}\\
				\hline
				& {\xmplbx30$\%$}&  {\xmplbx0,20}      &  {\xmplbx52 237}  &{\xmplbx2 701} & {\xmplbx917}&{\xmplbx522} &{\xmplbx91} &{\xmplbx14}\\
				\hline
				& {\xmplbx30$\%$}&  {\xmplbx0,25}      &  {\xmplbx4 120}  &{\xmplbx2 701} & {\xmplbx611}&{\xmplbx448} &{\xmplbx91} &{\xmplbx30} \\
				\hline
				& {\xmplbx30$\%$}&  {\xmplbx0,30}      &  {\xmplbx674}  &{\xmplbx2 701} & {\xmplbx382}&{\xmplbx322} &{\xmplbx91} &{\xmplbx28}\\
				\hline		
				& {\xmplbx30$\%$}&  {\xmplbx0,35}      &  {\xmplbx319}  &{\xmplbx1 258} & {\xmplbx240}&{\xmplbx230} &{\xmplbx91} &{\xmplbx31}\\
				\hline		
				& {\xmplbx30$\%$}&  {\xmplbx0,40}      &  {\xmplbx223}  &{\xmplbx635} & {\xmplbx176}&{\xmplbx162} &{\xmplbx91} &{\xmplbx20}\\
\hline
	\end{tabular}}}
	\begin{center}
		\caption{Cardinalités des ensembles extraits en fonction de \textit{minbond} pour des bases denses.}\label{Tab5}
	\end{center}
\end{table}
%%%%%%%%%%%%%%%%%%%%%%%
\font\xmplbx = cmbx8.5 scaled \magstephalf
\begin{table}[htbp]
	\parbox{16cm}{
		\hspace{-0.9cm}
		\footnotesize{
			\begin{tabular}{|c||r|r|r|r|r|r|r|r|r|}
				\hline  Base & \textit{minsupp} & \textit{minbond} &
				|$\mathcal{MCR}$|& |$\mathcal{RMCR}$|  &|$\mathcal{RMM}$$ax$$\mathcal{F}$|& $|\mathcal{RM}$$in$$\mathcal{MF}|$ &|$\mathcal{RM}$$in$$\mathcal{MM}$$ax$$\mathcal{F}$|\\
				\hline\hline
				\textsc{Mushroom}& {\xmplbx30$\%$}&  {\xmplbx0,10}  &  {\xmplbx578 359}  &{\xmplbx2 251} & {\xmplbx1 804}&{\xmplbx715} &{\xmplbx97} \\
				\hline
				& {\xmplbx30$\%$}&  {\xmplbx0,15}  &  {\xmplbx98 566}  &{\xmplbx1 704} & {\xmplbx1 327}&{\xmplbx648} &{\xmplbx100} \\
				\hline
				& {\xmplbx30$\%$}&  {\xmplbx0,20}  &  {\xmplbx52 237}  &{\xmplbx1 223} & {\xmplbx931}&{\xmplbx567} &{\xmplbx105} \\		
				\hline
				& {\xmplbx30$\%$}&  {\xmplbx0,25}  &  {\xmplbx4 120}  &{\xmplbx816} & {\xmplbx637}&{\xmplbx463} &{\xmplbx121} \\
				\hline
				& {\xmplbx30$\%$}&  {\xmplbx0,30}  &  {\xmplbx674}  &{\xmplbx480} & {\xmplbx395}&{\xmplbx337} &{\xmplbx119} \\
				\hline		
				& {\xmplbx30$\%$}&  {\xmplbx0,35}  &  {\xmplbx319}  &{\xmplbx285} & {\xmplbx251}&{\xmplbx245} &{\xmplbx121} \\		
				\hline		
				& {\xmplbx30$\%$}&  {\xmplbx0,40}  &  {\xmplbx223}  &{\xmplbx194} & {\xmplbx184}&{\xmplbx177} &{\xmplbx109} \\
\hline				
	\end{tabular}}}
	\caption{Cardinalités des représentations concises extraites en fonction de \textit{minbond} pour des bases denses.}\label{Tab6}
\end{table}	

%%%%%%%%%%%%%%%%%%%%%%%%%%%%%%%%%%%%%%%%%%%%%%%%%%%%%%%%%%%

\font\xmplbx = cmbx8.5 scaled \magstephalf
\begin{table}[htbp]
	\parbox{16cm}{
		\hspace{-1.3cm}
		\footnotesize{
			\begin{tabular}{|c||r|r|r|r|r|r|r|r|r|r|r|}
				\hline  Base & \textit{minsupp} & \textit{minbond} &
				|$\mathcal{MCR}$|& |$\mathcal{MCF}$|  &|$\mathcal{MMCR}$| & |$\mathcal{MFCR}$|
				&|$\mathcal{MMCRM}in$|
				&|$\mathcal{MFCRM}ax$|\\
				\hline\hline	
				\textsc{T10I4D100K}& {\xmplbx0,1$\%$}&  {\xmplbx0,10}  &  {\xmplbx140}  &{\xmplbx2 563} & {\xmplbx137}&{\xmplbx137} &{\xmplbx73}&{\xmplbx36}  \\	
				\hline		
				& {\xmplbx0,5$\%$}&  {\xmplbx0,10}  &  {\xmplbx1 838}  &{\xmplbx865} & {\xmplbx1 297}&{\xmplbx1 233} &{\xmplbx301}&{\xmplbx57}  \\			
				\hline		
				& {\xmplbx1$\%$}&  {\xmplbx0,10}  &  {\xmplbx2 319}  &{\xmplbx384} & {\xmplbx1 635}&{\xmplbx1 500} &{\xmplbx495}&{\xmplbx56}  \\			
				\hline		
				& {\xmplbx5$\%$}&  {\xmplbx0,10}  &  {\xmplbx2 693}  &{\xmplbx10} & {\xmplbx2 006}&{\xmplbx1 866} &{\xmplbx860}&{\xmplbx56}  \\			
				\hline		
				& {\xmplbx10$\%$}&  {\xmplbx0,10}  &  {\xmplbx2 703}  &{\xmplbx0} & {\xmplbx2 016}&{\xmplbx1 876} &{\xmplbx870}&{\xmplbx56}  \\			
				\hline		
				& {\xmplbx15$\%$}&  {\xmplbx0,10}  &  {\xmplbx2 703}  &{\xmplbx0} & {\xmplbx2 016}&{\xmplbx1 876} &{\xmplbx870}&{\xmplbx56}  \\			
				\hline		
				& {\xmplbx20$\%$}&  {\xmplbx0,10}  &  {\xmplbx2 703}  &{\xmplbx0} & {\xmplbx2 016}&{\xmplbx1 876} &{\xmplbx870}&{\xmplbx56}  \\
\hline		
	\end{tabular}}}
	\caption{Cardinalités des ensembles extraits en fonction de \textit{minsupp} pour les bases éparses.}
	\label{TabXpT101}
\end{table}
%%%%%%%%%%%%%%%%
%%%%%%%%%%%%%%%%%%%%%
\font\xmplbx = cmbx8.5 scaled \magstephalf
\begin{table}[htbp]
	\parbox{16cm}{
		\hspace{-0.9cm}
		\footnotesize{
			\begin{tabular}{|c||r|r|r|r|r|r|r|r|r|}
				\hline  Base & \textit{minsupp} & \textit{minbond} &
				|$\mathcal{MCR}$|& |$\mathcal{RMCR}$|  &|$\mathcal{RMM}$$ax$$\mathcal{F}$|& $|\mathcal{RM}$$in$$\mathcal{MF}|$ &
				|$\mathcal{RM}$$in$$\mathcal{MM}$$ax$$\mathcal{F}$|\\
				\hline\hline
				\textsc{T10I4D100K}& {\xmplbx0,1$\%$}&  {\xmplbx0,10}  &  {\xmplbx140}  &{\xmplbx140} & {\xmplbx139}&{\xmplbx137} &{\xmplbx109}\\	
				
				\hline
				& {\xmplbx0,5$\%$}&  {\xmplbx0,10}  &  {\xmplbx1 838}  &{\xmplbx1 689} & {\xmplbx1 351}&{\xmplbx1 233} &{\xmplbx358}\\	
				\hline
				& {\xmplbx1$\%$}&  {\xmplbx0,10}  &  {\xmplbx2 319}  &{\xmplbx2 087} & {\xmplbx1 690}&{\xmplbx1 500} &{\xmplbx551}\\			
				\hline
				& {\xmplbx5$\%$}&  {\xmplbx0,10}  &  {\xmplbx2 693}  &{\xmplbx2 457} & {\xmplbx2 061}&{\xmplbx1 866} &{\xmplbx916}\\			
				\hline
				& {\xmplbx10$\%$}&  {\xmplbx0,10}  &  {\xmplbx2 703}  &{\xmplbx2 467} & {\xmplbx2 071}&{\xmplbx1 876} &{\xmplbx926}\\
				\hline		
				& {\xmplbx15$\%$}&  {\xmplbx0,10}  &  {\xmplbx2 703}  &{\xmplbx2 467} & {\xmplbx2 071}&{\xmplbx1 876} &{\xmplbx926}\\		
				\hline		
				& {\xmplbx20$\%$}&  {\xmplbx0,10}  &  {\xmplbx2 703}  &{\xmplbx2 467} & {\xmplbx2 071}&{\xmplbx1 876} &{\xmplbx926}\\
\hline
	\end{tabular}}}
	\caption{Cardinalités des représentations concises extraites en fonction de \textit{minsupp} pour des bases éparses.}
	\label{TabXpT102}
\end{table}
%%%%%%%%%%%%%%%
\font\xmplbx = cmbx8.5 scaled \magstephalf
\begin{table}[htbp]
	\parbox{16cm}{
		\hspace{-1.3cm}
		\footnotesize{
			\begin{tabular}{|c||r|r|r|r|r|r|r|r|r|r|r|}
				\hline  Base & \textit{minsupp} & \textit{minbond} &
				|$\mathcal{MCR}$|& |$\mathcal{MCF}$|  &|$\mathcal{MMCR}$| & |$\mathcal{MFCR}$|
				&|$\mathcal{MMCRM}in$|
				&|$\mathcal{MFCRM}ax$| \\
				\hline\hline	
				\textsc{T10I4D100K} &{\xmplbx25$\%$}&   {\xmplbx0,05}   &   {\xmplbx7 876}   &{\xmplbx0}    &{\xmplbx3 275}& {\xmplbx2 552}&{\xmplbx870} &{\xmplbx29} \\
				\hline		
				&{\xmplbx25$\%$}&   {\xmplbx0,10}   &   {\xmplbx2 703}   &{\xmplbx0}    &{\xmplbx2 016}& {\xmplbx1 876}&{\xmplbx870} &{\xmplbx56} \\
				\hline		
				&{\xmplbx25$\%$}&   {\xmplbx0,15}   &   {\xmplbx1 548}   &{\xmplbx0}    &{\xmplbx1 435}& {\xmplbx1 402}&{\xmplbx870} &{\xmplbx64}\\
				\hline		
				&{\xmplbx25$\%$}&   {\xmplbx0,20}   &   {\xmplbx1 189}   &{\xmplbx0}    &{\xmplbx1 164}& {\xmplbx1 162}&{\xmplbx870} &{\xmplbx66}\\
				\hline		
				&{\xmplbx25$\%$}&   {\xmplbx0,30}   &   {\xmplbx994}   &{\xmplbx0}    &{\xmplbx989}& {\xmplbx989}&{\xmplbx870} &{\xmplbx62} \\
				\hline		
				&{\xmplbx25$\%$}&   {\xmplbx0,40}   &  {\xmplbx925}   &{\xmplbx0}    &{\xmplbx925}& {\xmplbx925}&{\xmplbx870} &{\xmplbx54} \\		   		          		          
				\hline		
				&{\xmplbx25$\%$}&   {\xmplbx0,50}   &   {\xmplbx901}   &{\xmplbx0}    &{\xmplbx901}& {\xmplbx901}&{\xmplbx870} &{\xmplbx81} \\
\hline		
	\end{tabular}}}
	\caption{Cardinalités des ensembles extraits en fonction de \textit{minbond} pour des bases éparses.}
	\label{TabXpT104}
\end{table}
%%%%%%%%%%%%%%%%%%%%%%%%%%%%
\font\xmplbx = cmbx8.5 scaled \magstephalf
\begin{table}[htbp]
	\parbox{16cm}{
		\hspace{-0.9cm}
		\footnotesize{
			\begin{tabular}{|c||r|r|r|r|r|r|r|r|r|}
				\hline  Base & \textit{minsupp} & \textit{minbond} &
				|$\mathcal{MCR}$|& |$\mathcal{RMCR}$|  &|$\mathcal{RMM}$$ax$$\mathcal{F}$|& $|\mathcal{RM}$$in$$\mathcal{MF}|$ &|$\mathcal{RM}$$in$$\mathcal{MM}$$ax$$\mathcal{F}$|\\
				\hline\hline	
				\textsc{T10I4D100K} &{\xmplbx$25\%$}&   {\xmplbx0,05}   &  {\xmplbx7 876} &{\xmplbx4 387}   &{\xmplbx3 304}    &{\xmplbx2 552}& {\xmplbx2 552}\\
				\hline		
				&{\xmplbx25$\%$}&   {\xmplbx0,10}   &   {\xmplbx2 703}   &{\xmplbx2 467}    &{\xmplbx2 071}& {\xmplbx1 876}&{\xmplbx926}\\
				\hline		
				&{\xmplbx25$\%$}&   {\xmplbx0,15}   &   {\xmplbx1 548}   &{\xmplbx1 539}  &{\xmplbx1 464}& {\xmplbx1 402}&{\xmplbx934}\\
				\hline		
				&{\xmplbx25$\%$}&   {\xmplbx0,20}   &   {\xmplbx1 189}   &{\xmplbx1 189}    &{\xmplbx1 178}& {\xmplbx1 162}&{\xmplbx936}\\
				\hline		
				&{\xmplbx25$\%$}&   {\xmplbx0,30}   &   {\xmplbx994}   &{\xmplbx994}    &{\xmplbx991}& {\xmplbx989}&{\xmplbx932}\\
				\hline		
				&{\xmplbx25$\%$}&   {\xmplbx0,40}   &  {\xmplbx925}   &{\xmplbx925}    &{\xmplbx925}& {\xmplbx925}&{\xmplbx899}\\		   		          		
				\hline		
				&{\xmplbx25$\%$}&   {\xmplbx0,50}   &   {\xmplbx901}   &{\xmplbx901}   &{\xmplbx901}& {\xmplbx901}&{\xmplbx897}\\
\hline		
	\end{tabular}}}
	\caption{Cardinalités des représentations concises extraites en fonction de \textit{minbond} pour des bases éparses.}
	\label{TabXpT105}
\end{table}

\end{document}